\documentclass[longauth]{aa} % for the long lists of affiliations 
\newcommand{\kms}{$\rm km\,s^{-1}$}

\newcommand{\cdo}{$\rm C^{18}O$}

\newcommand{\carbontt}{$\rm ^{12}C/^{13}C$}
\newcommand{\oxigense}{$\rm ^{16}O/^{18}O$}
\newcommand{\oxigenes}{$\rm ^{18}O/^{17}O$}
\newcommand{\sulfurtttf}{$\rm ^{32}S/^{34}S$}
\newcommand{\nitrogenff}{$\rm ^{14}N/^{15}N$}
\newcommand{\silicontetn}{$\rm ^{28}Si/^{29}Si$}
\newcommand{\silicontte}{$\rm ^{28}Si/^{30}Si$}

\usepackage{graphicx}
\usepackage{natbib}
\usepackage{longtable}
\usepackage{threeparttablex}
\usepackage{rotating}
\usepackage[export]{adjustbox}% http://ctan.org/pkg/adjustbox
\usepackage{xcolor}
\usepackage{hyperref}
\usepackage{gensymb}

\usepackage{txfonts}
\begin{document}

   \title{ALCHEMI: an ALMA Comprehensive High-resolution Extragalactic Molecular Inventory.\\ Survey presentation and first results from the ACA array}
   %Molecular complexity towards the starburst NGC~253 with the 7~m array}
   \titlerunning{ALCHEMI: Survey and ACA results}
   %NGC~253 chemistry with 7~m}

   \author{S. Mart\'in \inst{\ref{inst.ESOChile},\ref{inst.JAO}}
          \and J. G. Mangum \inst{\ref{inst.NRAOCV}}
          \and N. Harada \inst{\ref{inst.NAOJ},\ref{inst.ASIAA},\ref{inst.SOKENDAI}}
          \and F. Costagliola \inst{\ref{inst.ONSALA}}
          \and K. Sakamoto \inst{\ref{inst.ASIAA}}
          \and S. Muller \inst{\ref{inst.ONSALA}}
          \and R. Aladro \inst{\ref{inst.MPIfR}}
%----------DATA REDUCTION
          \and K. Tanaka \inst{\ref{inst.KeioUniversity}}
          \and Y. Yoshimura  \inst{\ref{inst.UTokio}}
          \and K. Nakanishi \inst{\ref{inst.NAOJ},\ref{inst.SOKENDAI}}
          \and R. Herrero-Illana \inst{\ref{inst.ESOChile},\ref{inst.ICECSIC}}
          \and S. M\"uhle \inst{\ref{inst.UBonn}}
%----------SCIENCE
          \and S. Aalto \inst{\ref{inst.ONSALA}}
          \and E. Behrens \inst{\ref{inst.UVirginia}}
          \and L. Colzi \inst{\ref{inst.CAB-INTA},\ref{inst.Arcetri}}
          \and K. L. Emig \inst{\ref{inst.NRAOCV}\thanks{Jansky Fellow of the National Radio Astronomy Observatory}}
          \and G. A. Fuller \inst{\ref{inst.JodrellBank},\ref{inst.IAA}}
          \and S. Garc\'ia-Burillo\inst{\ref{inst.OAN}}
          \and T. R. Greve\inst{\ref{inst.CDC},\ref{inst.UCL}}
          \and C. Henkel \inst{\ref{inst.MPIfR},\ref{inst.Abdulaziz},\ref{inst.Urumqi}}
          \and J. Holdship \inst{\ref{inst.Leiden},\ref{inst.UCL}}
          \and P. Humire \inst{\ref{inst.MPIfR}}
          \and L. Hunt \inst{\ref{inst.Arcetri}}
          \and T. Izumi \inst{\ref{inst.NAOJ},\ref{inst.SOKENDAI}}
          \and K. Kohno \inst{\ref{inst.UTokio},\ref{inst.UTokioRCEU}}
          \and S. K\"onig \inst{\ref{inst.ONSALA}}
          \and D. S. Meier \inst{\ref{inst.NMIMT},\ref{inst.NRAOSocorro}}
          \and T. Nakajima \inst{\ref{inst.UNagoya}}
          \and Y. Nishimura \inst{\ref{inst.UTokio},\ref{inst.NAOJ}}
          \and M. Padovani \inst{\ref{inst.Arcetri}}
          \and V. M. Rivilla \inst{\ref{inst.CAB-INTA},\ref{inst.Arcetri}}
          \and S. Takano \inst{\ref{inst.UNihon}}
          \and P. P. van der Werf\inst{\ref{inst.Leiden}}
          \and S. Viti\inst{\ref{inst.Leiden},\ref{inst.UCL}}
          \and Y.T. Yan\inst{\ref{inst.MPIfR}}
%----------SCIENCE 2
          }

   \institute{
\label{inst.ESOChile}European Southern Observatory, Alonso de C\'ordova, 3107, Vitacura, Santiago 763-0355, Chile  \email{smartin@eso.org}
\and\label{inst.JAO}Joint ALMA Observatory, Alonso de C\'ordova, 3107, Vitacura, Santiago 763-0355, Chile
\and\label{inst.NRAOCV}National Radio Astronomy Observatory, 520 Edgemont Road, Charlottesville, VA 22903-2475, USA
\and\label{inst.NAOJ}National Astronomical Observatory of Japan, 2-21-1 Osawa, Mitaka, Tokyo 181-8588, Japan
\and\label{inst.ASIAA}Institute of Astronomy and Astrophysics, Academia Sinica, 11F of AS/NTU Astronomy-Mathematics Building, No.1, Sec. 4, Roosevelt Rd, Taipei 10617, Taiwan   
\and\label{inst.SOKENDAI}Department of Astronomy, School of Science, The Graduate University for Advanced Studies (SOKENDAI), 2-21-1 Osawa, Mitaka, Tokyo, 181-1855 Japan
\and\label{inst.ONSALA}Department of Space, Earth and Environment, Chalmers University of Technology, Onsala Space Observatory, SE-43992 Onsala, Sweden
\and\label{inst.MPIfR}Max-Planck-Institut f\"ur Radioastronomie, Auf dem H\"ugel 69, 53121 Bonn, Germany 
\and\label{inst.KeioUniversity}Department of Physics, Faculty of Science and Technology, Keio University, 3-14-1 Hiyoshi, Yokohama, Kanagawa 223--8522 Japan
\and\label{inst.UTokio}Institute of Astronomy, Graduate School of Science, The University of Tokyo, 2-21-1 Osawa, Mitaka, Tokyo 181-0015, Japan
\and\label{inst.ICECSIC}Institute of Space Sciences (ICE, CSIC), Campus UAB, Carrer de Magrans, E-08193 Barcelona, Spain
\and\label{inst.UBonn}Argelander-Institut f\"ur Astronomie, Universit\"at Bonn, Auf dem H\"ugel 71, D-53121 Bonn, Germany
\and\label{inst.UVirginia}Astronomy Department, University of Virginia, 530 McCormick Road, Charlottesville, VA 22904--4325, USA
\and\label{inst.CAB-INTA}Centro de Astrobiología (CSIC-INTA), Ctra. de Torrej\'on a Ajalvir km 4, 28850, Torrej\'on de Ardoz, Madrid, Spain       
\and\label{inst.Arcetri}INAF Osservatorio Astrofisico di Arcetri, Largo Enrico Fermi 5, I-50125 Firenze, Italy
\and\label{inst.JodrellBank}Jodrell Bank Centre for Astrophysics, Department of Physics \& Astronomy, School of Natural Sciences, The University of Manchester, M13 9PL, UK     
\and\label{inst.IAA}Intituto de Astrof\'isica de Andalucia (CSIC), Glorieta de al Astronomia s/n E-18008, Granada, Spain  
\and\label{inst.OAN}Observatorio Astron\'omico  Nacional (OAN-IGN), Observatorio de Madrid, Alfonso XII, 3, 28014-Madrid, Spain
\and\label{inst.CDC}Cosmic Dawn Center, DTU Space, Technical University of Denmark, Elektrovej 327, Kgs.~Lyngby, DK-2800, Denmark
\and\label{inst.UCL}Department of Physics and Astronomy, University College London, Gower Street, London WC1E6BT, UK
\and\label{inst.Abdulaziz}Astron. Dept., Faculty of Science, King Abdulaziz University, P.O. Box 80203, Jeddah 21589, Saudi Arabia
\and\label{inst.Urumqi}Xinjiang Astronomical Observatory, Chinese Academy of Sciences, Urumqi 830011, P.R. China
\and\label{inst.Leiden}Leiden Observatory, Leiden University, PO Box 9513, NL - 2300 RA Leiden, The Netherlands
\and\label{inst.UTokioRCEU}Research Center for the Early Universe, Graduate School of Science, The University of Tokyo, 7-3-1 Hongo, Bunkyo-ku, Tokyo 113-0033, Japan
\and\label{inst.NMIMT}New Mexico Institute of Mining and Technology, 801 Leroy Place, Socorro, NM 87801, USA
\and\label{inst.NRAOSocorro}National Radio Astronomy Observatory, PO Box O, 1003 Lopezville Road, Socorro, NM 87801, USA
\and\label{inst.UNagoya}Institute for Space-Earth Environmental Research, Nagoya University, Furo-cho, Chikusa-ku, Nagoya, Aichi 464-8601, Japan
\and\label{inst.UNihon}Department of Physics, General Studies, College of Engineering, Nihon University, Tamura-machi, Koriyama, Fukushima 963-8642, Japan
%\and\label{inst.Guangzhou}Center for Astrophysics, Guangzhou University, 510006 Guangzhou, People's Republic of China
         }

   %\date{Received September 15, 1996; accepted March 16, 1997}
   \date{}
 
  \abstract
{
The interstellar medium is the locus of 
physical processes affecting the evolution of galaxies which 
drive or are the result of star formation activity,
supermassive black hole growth and feedback. 
The resulting physical conditions determine the observable chemical abundances that can be explored through molecular emission observations at millimeter/submillimeter wavelengths. }
{
Unveiling the molecular richness of the central region of the prototypical nearby starburst galaxy NGC~253
at an unprecedented combination of sensitivity, spatial resolution, and frequency coverage.}
{We used the Atacama Large Millimeter/submillimeter Array (ALMA), covering a nearly contiguous 289~GHz frequency range between 84.2 and 373.2~GHz, to image the continuum and spectral line emission at 1.6\arcsec ($\sim 28$~pc) resolution down to a sensitivity of $30-50$~mK.
This article describes the ALMA Comprehensive High-resolution Extragalactic Molecular Inventory (ALCHEMI) Large Program.  We focus on the analysis of the spectra extracted from the $15''$
($\sim255$~pc) resolution ALMA Compact Array data.}
{We model the molecular emission assuming local thermodynamic equilibrium with
78 species detected.%contribute to the observed spectrum. 
Additionally, multiple hydrogen and helium recombination lines are identified. 
Spectral lines contribute 5 to 36\% of the total emission in frequency bins of 50~GHz.
We report the first extragalactic detections of C$_2$H$_5$OH, HOCN, HC$_3$HO, and several rare isotopologues. Isotopic ratios of carbon, oxygen, sulfur, nitrogen and silicon were measure with multiple species.
}
{
{Infrared pumped vibrationaly excited HCN, HNC, and HC$_3$N emission, originating in massive star formation locations, is clearly detected at low resolution, while we do not detect it for HCO$^+$. We suggest high temperature conditions in these regions driving a seemingly "carbon-rich" chemistry which may also explain the observed high abundance of organic species close to those in Galactic hot cores. The $L_{vib}/L_{IR}$ ratio is used as a proxy to estimate a $3\%$ contribution from proto super star cluster to the global infrared emission. Measured isotopic ratios with high dipole moment species agree with those within the central kiloparsec of the Galaxy, while those derived from $\rm^{13}C^{18}O$ are a factor of 5 larger,
%and CCH, 
confirming the existence of multiple ISM components within NGC~253 with different degrees of nucleosynthesis enrichment. 
The ALCHEMI data set provides a unique template for studies of star-forming galaxies in the early Universe.}
}
   \keywords{Line: identification - Galaxies: ISM - Galaxies: individual: NGC~253 - Galaxies: starburst - ISM: molecules - Submillimeter: ISM}

\maketitle

\section{Introduction}
\label{Sec.Introduction}

The interstellar medium (ISM) is the location and source of fuel for key phenomena that influence the evolution of galaxies. While star formation is one of the most important of such phenomena, %the nature of 
the ISM 
%makes it 
is sensitive to a large number of processes, such as radiative transfer effects, heating and cooling, and/or active chemistry \citep[see][for a review]{Omont2007}.
%Such a diverse collection of phenomena implies that many parameters influence the ISM properties. 
%Molecular emission stems from the relatively dense molecular gas shielded from UV photodissociation. 
Moreover, the physical properties of the ISM and the effects of such processes imprint their signatures in the many atomic and molecular spectral lines they emit. This fact makes the observation of molecular emission an essential tool in the study of the ISM, where different tracers probe different physical processes within the gaseous component in galaxies \citep[i.e.][]{Meier2005,Meier2012,Takano2014,Meier2015,Martin2015,Harada2019}. Thus 
%the importance emerges of observing 
it is essential to observe as many molecular tracers as observationally feasible to understand the ongoing processes in these regions.

Additionally, it is crucial to evaluate as many different types of environments as possible to capture and understand how the mechanisms at work affect the ISM. Although our own Galaxy is the ideal nearby laboratory, our position within the disk sometimes hinders our ability to have a clear view of the overall ISM properties.  Furthermore, our Galaxy lacks the extreme environments created by star bursting regions, growth of supermassive blackholes, intense feedback by massive outflows, or high/low metallicity gas environments.
%, etc. found in many galaxies. 
The observation of nearby galaxies allows us to probe different environments and study their physical and chemical properties. In this work we probe the ISM under a star bursting environment.

\subsection{The revolution of extragalactic mm observations}

Almost five decades have passed since the first extragalactic detection of carbon monoxide (CO) towards the nearby starburst galaxies NGC~253 and M~82 \citep{Rickard1975}, the two brightest extragalactic IRAS sources beyond the Magellanic clouds \citep{Soifer1989}, and shortly after the first detection of CO in the Galaxy \citep{Wilson1970}.
%and considered as the prototypical nearby starburst galaxies \citep{Rieke1980}.
These CO emission detections had only been preceded by the extragalactic detections of OH \citep{Weliachew1971} and H$_2$CO \citep{Gardner1974} in absorption, and were quickly followed by detections of higher dipole moment species such as HCN \citep{Rickard1977}. Such milestones in extragalactic molecular observations were only possible thanks to improvements in receiver technology. These observations have gone on to shape our current knowledge of galaxy evolution and complexity of processes in the interstellar medium within those galaxies in a fundamental way.

The advent of instruments operating at millimeter wavelengths with larger collecting area, lower noise receivers, broadband spectrometers and being placed at drier locations %\citep[achieving system temperatures more than one order of magnitude lower than that of the 930~K reported by][]{Rickard1975} 
resulted in observing speed improvements of more than three orders of magnitude. Such a technological leap allowed, for example, the detection of the fainter CO isotopologues \citep{Harrison1999} and tentative detections of its double isotopologue $\rm^{13}C^{18}O$ \citep{Mart'in2010a} in extragalactic environments, now routinely achievable with the Atacama Large Millimeter/Submillimeter array \citep[ALMA;][]{Martin2019}. %Subsequent s
Studies using both spectral detection and imaging of dense molecular gas tracers \citep[see][and the subsequent series of papers]{Mauersberger1989a} could be considered the genesis of today's field of extragalactic molecular astrophysics and astrochemistry.

%{\bf Low resolution scans on Galactic sources}
However, it was not until the pioneering systematic large multi-transition and multi-molecule work from \citet{Wang2004}, followed shortly after by the first unbiased extragalactic spectral line surveys \citep{Mart'in2006,Mart'in2011,Muller2011,Aladro2011a}, that the field of extragalactic astrochemistry developed its full power, with dozens of species detected (see Sect.~\ref{sec.newdetections}) and made full use of the bandwidth increase in receiver and spectrometer technology. %allowed for multiple follow-ups towards a variety of extragalactic objects. 
To the best of our knowledge, Table~\ref{tab.Surveys} summarizes every published wide-band ($>10$~GHz) extragalactic spectral line survey conducted with millimeter/submillimeter ground-based observatories.

\begin{table}[!ht]
\centering
%\begin{center}
\caption{Extragalactic spectral scans at mm and submm wavelengths.}
\label{tab.Surveys}
\begin{tabular}{l l l l}
\hline 
\hline 
Source (\#)                &   Frequency                 &  Telescope    & Ref.         \\
                           &   Coverage                  &               &              \\
                           &   (GHz)                     &               &              \\
\hline 
NGC~4945                   &  82-354 \tablefootmark{a}   & SEST~15m      & 1  \\
NGC~253                    &  86-116                     & IRAM~30m      & 2 \\
                           &  86-116 \tablefootmark{b}   & ALMA          & 3 \\
                           &  85-116                     & Nobeyama~45m  & 4, 5  \\
                           &  129-175                    & IRAM~30m      & 6 \\
N113 (LMC)                 &  85-357 \tablefootmark{a}   & SEST~15m      & 7  \\
M~82                       &  86-116                     & IRAM~30m      & 2  \\
                           &  130-175                    & IRAM~30m      & 8 \\
                           &  241-260                    & IRAM~30m      & 8  \\
                           &  190-307 \tablefootmark{c}  & CSO 10.4m     & 9 \\
NGC~1068                   &  86-116                     & IRAM~30m      & 10  \\ 
                           &  85-116                     & Nobeyama~45m  & 4, 5 \\
                           &  161-169 \tablefootmark{b}  & IRAM~30m      & 11  \\
                           &  176-184 \tablefootmark{b}  & IRAM~30m      & 11  \\
                           &  190-307 \tablefootmark{c}  & CSO 10.4m     & 12 \\
PKS~1830-211\tablefootmark{d}&  57-94                      & ATCA          & 13 \\
                           &  59-94                      & Yebes~40m     & 14   \\
Arp\,220                    &  86-116                     & IRAM~30m      & 2  \\
                           &  202-242                    & SMA           & 15  \\
M~51                       &  86-116                     & IRAM~30m      & 2  \\
                           &  83-116                     & IRAM~30m      & 16 \\
                           &  130-148                    & IRAM~30m      & 16 \\
M~83                       &  86-116                     & IRAM~30m      & 2  \\
NGC~4418                   &  84-113                     & ALMA          & 17 \\
                           &  214-294  \tablefootmark{b} & ALMA          & 17 \\ 
IC~342                     &  84-116                     & Nobeyama~45m  & 4, 5 \\
IC~10                      &  84-116                     & Nobeyama~45m  & 18 \\
NGC~3256                   &  85-113                     & ALMA          & 19 \\
                           & 214-273                     & ALMA          & 19 \\
NGC 3627                   &  85-116                     & IRAM~30m      & 20 \\
                           &  140-148                    & IRAM~30m      & 20 \\
                           &  85-115.5                   & Nobeyama~45m  & 20 \\
Survey (10)                &  74-111 \tablefootmark{c}   & FCRAO 14m     & 21 \\
Survey (23)                &  84-92                      & IRAM~30m      & 22 \\
                           &  108-116                    & IRAM~30m      & 22 \\
Survey (4)                 &  85-268 \tablefootmark{b}   & IRAM~30m      & 23 \\
Survey (7, LMC)            &  85-116                     & MOPRA~22m     & 24 \\
\hline
\end{tabular}
%\end{center}
\tablefoot{
The source column indicates the name of the survey target or the number of surveyed sources in multi-source works (labeled Survey). Sources are chronologically ordered according to the year of its first survey publication. Large Magellanic Cloud (LMC) sources are indicated.
\tablefoottext{a}{
Multiple targeted tunings within the indicated frequency coverage.
}
\tablefoottext{b}{
Not fully sampled. Multiple tunings within the indicated frequency coverage with broad band receivers. 
}
\tablefoottext{c}{
Using low spectral resolution high redshift ultra wide band receivers.
}
\tablefoottext{d}{
Molecular absorber system. The frequency refers to rest frequency, which at the redshift of the source corresponds to observed frequecy range of $30-50$ and $31.5-50$~GHz for the two references, respectively.
}
}
\tablebib{
(1) \citet{Wang2004};
(2) \citet{Aladro2015};
(3) \citet{Meier2015};
(4) \citet{Nakajima2018};
(5) \citet{Takano2019};
(6) \citet{Mart'in2006};
(7) \citet{Wang2009};
(8) \citet{Aladro2011a};
(9) \citet{Naylor2010};
(10) \citet{Aladro2013};
(11) \citet{Qiu2020};
(12) \citet{Kamenetzky2011};
(13) \citet{Muller2011};
(14) \citet{Tercero2020};
(15) \citet{Mart'in2011};
(16) \citet{Watanabe2014};
(17) \citet{Costagliola2015}
(18) \citet{Nishimura2016a};
(19) \citet{Harada2018}
(20) \citet{Watanabe2019};
(21) \citet{Snell2011};
(22) \citet{Costagliola2011};
(23) \citet{Li2019};
(24) \citet{Nishimura2016}
}
\end{table}

\subsection{The nearby nuclear starburst galaxy NGC~253}
\label{sec.NGC253}

\begin{figure*}[!h]
\begin{center}
\includegraphics[page=1,trim=4 4 4 4,height=0.41\textwidth]{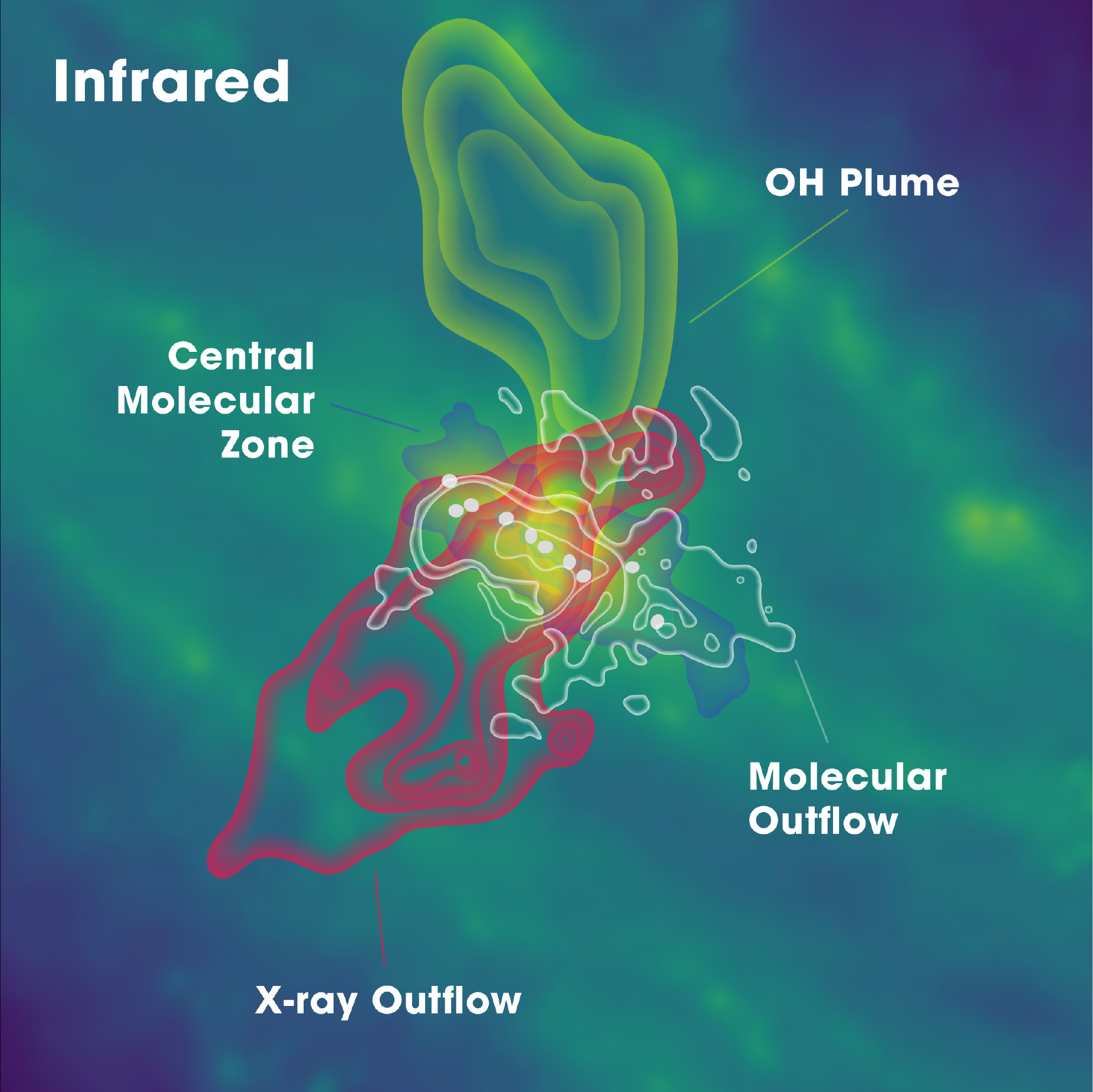}
\includegraphics[page=2,trim=4 4 4 4,height=0.41\textwidth]{ALCHEMISketch.pdf}
\caption{Schematic summary of the activity within central molecular zone of NGC~253.
%where the region targetted by the ALCHEMI project is put into the context of the different physical processes ongoing within this region. 
See Sect.~\ref{sec.NGC253} for a comprehensive summary of the activity in its central region as probed by multi-wavelength observations. 
In both figures, the CO traced CMZ, and the dense gas traced GMCs \citep{Leroy2015} are included as a spatial scale reference.
(Left) 
IRAC $8\mu$m from Spitzer Local Volume Legacy survey \citep{Dale2009} in the background;
Chandra X-ray traced outflow \citep{Strickland2000};
18~cm OH plume \citep{Turner1985};
molecular outflow observed in CO emission \citep{Bolatto2013}.
(Right)
2~cm TH sources \citep{Turner1985a} and HII regions and SNRs \citep{Ulvestad1997};
proto-super stellar clusters traced by vibrationally excited HC$_3$N emission \citep{Rico-Villas2020}; 
star cluster identified from near-IR HST imaging \citep{Watson1996}.
\label{fig.sketch}}
\end{center}
\end{figure*}

The Sculptor galaxy, NGC~253, is a nearby \citep[D$\sim3.5\pm0.2$~Mpc,][]{Rekola2005,Mouhcine2005} almost edge-on barred spiral galaxy \citep{Puche1991,Pence1981,deVacouleurs1991}.
Its central molecular zone (CMZ), about $300\times100$~pc in size \citep{Sakamoto2011}, contains $\sim 10^8$~M$_\odot$ of molecular gas \citep{Canzian1988,Mauersberger1996,Harrison1999,Sakamoto2011}. 
Such a large amount of gas in the NGC~253 CMZ is built up as a result of gas inflow toward the inner Lindblad resonance (ILR) region at $r\sim500$~pc. This inflow appears to be driven by a stellar bar, which has a deprojected length of 2.5~kpc and clearly stands out in near-infrared observations \citep{Scoville1985,Forbes1992,Paglione2004,Iodice2014}, rather than by interaction with the nearby galaxy NGC~247, based on the regularity of the H\,I velocity field and density distribution outside its central region \citep{Combes1977,Puche1991}.
This molecular material is responsible for feeding the burst of star formation of 2~M$_\odot$\,yr$^{-1}$ in the central kiloparsec \citep{Leroy2015,Bendo2015} which accounts for approximately half of the global star formation of $3.6-4.2$~M$_\odot$\,yr$^{-1}$, based on the infrared luminosity of $L_{\rm IR}=2.1\times10^{10}\,L_\odot$ \citep{Sanders2003,Strickland2004}.

%Radio observations reveal at least 64 individual compact sources within the CMZ, half of which might be associated to HII regions \citep{Ulvestad1997}. 
Radio observations reveal at least 64 individual compact continuum sources within the NGC~253 CMZ, 23 of which have spectral indices measured by \cite{Ulvestad1997}. Of these 23 spectral index measurements, 17 have errors less than 0.4. About half the sources in this subset have spectral indices below -0.4, indicating synchrotron emission likely associated with supernovae remnants. The remaining sources have spectral indices spanning ~0.0 – 0.2, which is indicative of free-free emission stemming from HII regions. \cite{ Ulvestad1997} note that the majority of flat-spectrum sources lie along the galaxy disk midline, whereas the steeper-spectrum sources lie farther away from the central axis.
The brightest of these radio sources \citep[TH2,][]{Turner1985a} is associated with the nucleus of the galaxy, within $1''$ of the galaxy's kinematic center \citep{Mueller-Sanchez2010}.
The five giant molecular cloud complexes identified from $1''$ resolution dust and CO observations \citep{Sakamoto2011} are resolved into 14 dust clouds at $0.11''$ resolution \citep[1.9~pc,][]{Leroy2018}.
Only one of these dust clumps is associated with a near-infrared identified super star clusters \citep[SSCs;][]{Watson1996}. These molecular clouds are responsible for the star formation activity within the central region of NGC~253 and appear to be at different stages of evolution, with proto-SSCs identified through vibrational molecular emission \citep{Rico-Villas2020} and molecular outflows \citep{Levy2021}.
Adaptive optics observations resolve 37 IR knots on top of the diffuse emission, eight of which have radio counterparts \citep{Fern'andez-Ontiveros2009}.
Among the many X-ray observations towards NGC~253 \citep[e.g.][and references therein]{Bauer2008}, \citet{Lehmer2013} 
reported the detection of 3 ultra-luminous X-ray sources, one of which is located $~1''$ from the dynamical center, but with no signs of active galactic nucleus (AGN) activity.
A starburst-driven outflow is traced by X-ray and H$\alpha$ emission all the way to 9~kpc from the disk \citep{Dahlem1998,Strickland2000}. The outflow entrains molecular gas away from its base, limiting the star formation activity in NGC~253 by negative feedback \citep{Bolatto2013,Walter2017,Krieger2019}.
The sketch presented in Figure~\ref{fig.sketch} aims at visually summarizing this complex region.

As one of the brightest extragalactic infrared sources \citep{Soifer1989} and the most prominent molecular emitter beyond the Magellanic Clouds, the prototypical local starburst galaxy NGC~253 has been the target of multiple molecular spectral line studies (see Table~\ref{tab.Surveys}). Due to its proximity, high resolution studies can resolve the giant molecular cloud (GMC) %size 
scales of a few tens of parsecs \citep{Sakamoto2011,Ando2017,Leroy2018}. In particular, NGC~253 has been the target of a number of ALMA observations which analyzed the properties of individual molecular clouds and complexes within its CMZ \citep{Sakamoto2011,Meier2015,Leroy2015,Ando2017,Leroy2018,Mangum2019,Martin2019,Rico-Villas2020,Krieger2020}.%, but they were limited to detected species within relatively narrow spectral ranges around their targeted bright molecular lines.

\subsection{The ALCHEMI survey}

The ALMA Comprehensive High-resolution Extragalactic Molecular Inventory (ALCHEMI) is an ALMA large program whose aim is to obtain the most complete spatially resolved unbiased molecular inventory towards a starburst environment. For this purpose we carried out a broad-band spectral scan towards the NGC~253 CMZ with a homogeneous spatial resolution.
%This project made use of the unprecedented combination of sensitivity and resolution now routinely achieved by ALMA. 
Unbiased wide-band spectral line surveys provide immediate advantages over narrow-band spectroscopy since the detection of multiple transitions per molecular species allows us to observationally constrain the excitation conditions \citep[e.g.][]{Aladro2011,Perez-Beaupuits2018,Scourfield2020}.  %Unbiased wide-band spectral line surveys
They also allow for the evaluation of line blending between molecular lines by simultaneously fitting many transitions of a given species rather than fitting individual spectral features.

The main objectives of ALCHEMI are to:
1) Define a uniform molecular template for an  extragalactic starburst environment, where systematic uncertainties are minimized; 
2) accurately constrain the physical conditions of individual star-bursting molecular cloud complexes;
3) study the ISM enrichment by stellar nucleosynthesis through measurement of isotopic ratios;
4) enable a direct comparison of the physical and chemical ISM properties between the Milky Way and an active star-forming environment;
5) explore the chemistry of complex organic molecules (COMs) in the CMZ of NGC~253;
6) evaluate gas processing in galactic outflows.
This study provides a template for the molecular emission in a starburst galaxy that can be compared to future spatially-resolved millimeter/submillimeter studies of more distant galaxies on GMC size scales.% afforded only by future enhancements to current instrumentation.

This paper, the first in a series of articles which will describe the scientific results from ALCHEMI, provides a global presentation of the ALCHEMI survey that includes data obtained with both the main (12~m) and Morita (ACA 7~m) arrays. However, here we focus, as a first step, on the analysis and discussion of the low resolution ($15''\sim255$~pc) 7m~array data. This study aims at describing the global molecular properties of the entire unresolved CMZ of NGC~253. As such, and similar to existing single-dish line surveys but with improved frequency coverage and sensitivity \citep[$5.6\times$ broader and $>2-10$ deeper than][]{Mart'in2006}, the low resolution data analysed here provide a template of a starburst environment for spatially-unresolved targets at larger distances. Already at this low resolution we explore the physical conditions of the gas and the enrichment of the ISM, and peer into the genesis of complex organic molecules in NGC~253.

\section{Observations}
\label{Sec.Observations}

The CMZ in NGC~253 was imaged 
%spectroscopically surveyed
with ALMA %, with the original intent to cover a region of $50\arcsec \times 20\arcsec$ ($850\times340$~pc) at 1\arcsec~(17~pc) resolution,
in frequency Bands~3, 4, 6, and 7 as part of the Cycle~5 Large Program 2017.1.00161.L. The survey was subsequently extended to Band~5 during ALMA Cycle~6 (project code 2018.1.00162.S). A total of 101.5~hours of integration time on source were acquired, 38.8~hours of which were obtained with the 12~m array.

The nominal phase center of the observations is 
$\alpha=00^h47^m33.26^s$, $\delta=-25^\circ17'17.7''$ (ICRS).
Observations were configured to cover a common rectangular area of $50''\times20''$ ($850\times340$~pc) with a position angle of $65^\circ$ (East of North), and a target resolution of $1''$ (17~pc, see Sect.~\ref{sec.imaging}). This targeted region required a single pointing in Band~3, where the 12~m antenna primary beams range between $57''$ and $68''$, and Nyquist-sampled mosaic patterns of 5 to 19 pointings (from the lower frequency end of Band~4 to the upper frequency end of Band~7, respectively) with the 12m array. The average integration time per mosaic pointing to achieve the target sensitivity (Sect.~\ref{sec.sensitivity}) varied from $\sim2.6$~hours in Band~3, $\sim12$~min in Band~4, $\sim9$~min in Band~5, $\sim4$~min in Band~6, and $\sim2.5$~min in Band~7. Additional single pointing observations in a 12~m more compact configuration or 7~m array were performed to achieve a common maximum recoverable scale of 15$''$ across the whole survey (see Sect.~\ref{sec.MRS}).
%B3 TM1 3.8h -1.8 per pointing
%B4 TM1 17 min - 7.5 m
%B5 TM1 13 min - 5 min
%B6 TM1 4.5 min - 2min
%B7 TM1 2.5

\begin{table}
\centering
\caption{Imaged spectral tuning frequency coverage and amplitude scaling.}
\label{tab.observingsetup}
\begin{tabular}{lccc}
\hline
\hline
ID \tablefootmark{a}    & \multicolumn{2}{c}{Rest. Freq. Coverage (GHz)} & a$_i$ \tablefootmark{b} \\
      &      LSB           &     USB         &   Com./Ext.     \\
\hline
B3a & $ 84.172- 87.614$ & $ 96.190- 99.335$  &  0.976/1.108 \\
B3b & $ 87.423- 90.980$ & $ 99.174-102.731$  &  1.003/0.789 \\
B3c & $ 90.777- 94.334$ & $102.531-106.088$  &  0.990/1.107 \\
B3d & $ 94.131- 97.688$ & $105.885-109.441$  &  0.997/1.128 \\
B3e & $ 97.488-101.045$ & $109.242-112.798$  &  1.009/0.948 \\
B3f & $100.848-103.993$ & $112.617-116.005$  &  1.001/0.922 \\
B4a & $125.193-128.804$ & $137.281-140.977$  &  0.996/0.998 \\
B4b & $128.771-132.457$ & $140.934-144.625$  &  1.003/1.026 \\
B4c & $132.424-136.110$ & $144.587-148.273$  &  1.015/1.001 \\
B4d & $136.077-139.763$ & $148.240-151.931$  &  0.980/0.989 \\
B4e & $139.731-143.421$ & $151.894-155.584$  &  0.992/0.968 \\
B4f & $143.384-147.069$ & $155.547-159.237$  &  1.021/1.014 \\
B4g & $147.037-150.723$ & $159.200-162.891$  &  0.992/1.002 \\
B5a & $163.215-166.761$ & $175.306-178.834$  &  1.016/1.147 \\
B5b & $166.566-170.112$ & $178.657-182.203$  &  0.998/1.050 \\
B5c & $169.917-173.469$ & $183.684-185.560$\tablefootmark{c}  &  1.013/0.974 \\
B5d & $173.274-176.820$ & $185.365-188.911$  &  0.993/0.928 \\
B5e & $185.380-188.890$ & $197.441-200.988$  &  0.992/0.856 \\
B5f & $188.713-192.241$ & $200.798-204.338$  &  0.955/0.428 \\
B5g & $192.058-195.598$ & $204.149-207.695$  &  1.033/0.927 \\
B5h & $195.409-198.955$ & $207.506-211.046$  &  1.001/0.974 \\
B6a & $211.286-214.733$ & $226.279-229.766$  &  0.948/0.995 \\
B6b & $214.545-218.096$ & $229.578-233.121$  &  1.000/1.005 \\
B6c & $217.908-221.459$ & $232.941-236.492$  &  1.004/1.034 \\
B6d & $221.271-224.822$ & $236.304-239.847$  &  1.001/1.010 \\
B6e & $224.634-228.185$ & $239.667-243.218$  &  0.995/0.990 \\
B6f & $243.202-246.753$ & $258.235-261.786$  &  0.867/0.991 \\
B6g & $246.565-250.116$ & $261.598-265.157$  &  1.015/1.003 \\
B6h & $249.936-253.479$ & $264.969-268.512$  &  0.924/0.992 \\
B6i & $253.299-256.842$ & $268.332-271.875$  &  0.959/1.006 \\
B6j & $256.662-260.141$ & $271.703-275.126$  &  1.027/1.008 \\
B7a & $275.348-278.726$ & $287.333-290.761$  &  1.020/0.975 \\
B7b & $278.490-281.978$ & $290.536-294.014$  &  1.013/1.008 \\
B7c & $281.753-285.231$ & $293.789-297.267$  &  0.988/1.003 \\
B7d & $284.996-288.484$ & $297.041-300.509$  &  1.257/1.040 \\
B7g & $295.015-298.663$ & $307.050-310.528$  &  0.990/0.984 \\
B7h & $298.257-301.876$ & $310.293-313.781$  &  0.833/0.998 \\
B7e & $300.519-303.997$ & $312.555-316.023$  &  0.993/1.014 \\
B7f & $303.772-307.240$ & $315.808-319.266$  &  0.996/0.994 \\
B7i & $328.313-331.811$ & $340.579-344.097$  &  1.023/0.956 \\
B7j & $331.636-335.184$ & $343.922-347.460$  &  0.989/0.984 \\
B7k & $334.999-338.527$ & $347.275-350.803$  &  0.972/1.017 \\
B7l & $338.372-341.870$ & $350.658-354.136$  &  1.028/0.967 \\
B7o & $348.851-352.469$ & $360.887-364.425$  &  0.922/1.007 \\
B7p & $352.204-355.822$ & $364.239-367.767$  &  1.025/1.034 \\
B7m & $354.156-357.694$ & $366.441-368.298$\tablefootmark{c}  &  1.034/1.044 \\
B7n & $357.509-361.047$ & $369.814-373.202$  &  1.008/0.991 \\
\hline
\end{tabular}
\tablefoot{
Rest frequency coverage of each individual spectral setup in the lower (LSB) and upper (USB) receiver sidebands, in order of increasing frequency.
Each sideband is covered by two slightly overlapping 1.875\,GHz-wide spectral windows (Sect.~\ref{sec.frequencysetup}). Observed frequencies were Doppler corrected to rest frequency assuming a receding velocity of 258.80\,\kms~ (LSR, radio convention) as originally set up in the observations.
\tablefoottext{a}{
The ID of each setup corresponds to the ALMA band and an identifying letter matching that in the ALMA archive \footnote{\url{http://almascience.nrao.edu/aq/}} for archival search purposes.
}
\tablefoottext{b}{
Amplitude scale factors. See Sect.~\ref{sec.relativeflux} for details.
}
\tablefoottext{c}{
This sideband was imaged in only one spectral window since the other suffered from poor atmospheric transmission.
}
}
\end{table}

%\begin{table*}
%\centering
%\caption{Central rest frequencies of spectral windows and frequency coverage of individual setups}
%\label{tab.observingsetup}
%\input{TableObservingParameters.tex}
%{\it Notes:} {\textcolor{red}{\bf DEPRECATED TABLE. Originally we thought of having central frequencies per SPW and %then frequency coverage per band. But do we need such amount of information? Is Table~\ref{tab.observingsetup} enough? I WOULD SAY SO.}}
%{\bf Seb: maybe a figure would do as well, showing the tuning coverage per band, to give a visual feeling complementary for Tab2 ?.}\\
%\end{table*}

\subsection{Frequency setup}
\label{sec.frequencysetup}

The full data set results in a rest-frequency coverage between 84.2 and 373.2~GHz, with 47 individual tunings, each composed of four 1.875~GHz spectral windows from two receiver sidebands.
%Observations were setup in 47 receiver tuning with four 1.875~GHz spectral window.
Table~\ref{tab.observingsetup} compiles the frequency coverage of the final data products per tuning and separated by receiver sideband after homogeneous processing (Sect.~\ref{sec.imaging}).
%A summary of the imaged frequency coverage is presented in Table~\ref{tab.observingsetup}.

The broad frequency range of 289~GHz was continuously covered except for a few narrow regions:
First, the $\sim$9~GHz gap between bands 3 and 4 (from 116 to 125\,GHz avoiding the deep 118.75\,GHz telluric oxygen line) is not observable with the ALMA receivers. Between the other receiver bands, the frequency gaps are significantly narrower: only 325\,MHz around 163\,GHz (bands 4--5), 250\,MHz around 211.1\,GHz (bands 5--6), and 225\,MHz around 275.25\,GHz (bands 6--7).
The spectral window centered close to the 183\,GHz telluric water line was observed, but the data quality was not good enough for calibration. Finally, the $\sim9$\,GHz frequency range from 319.3 to 328.3\,GHz, surrounding the 325\,GHz telluric water line, was intentionally not covered, given the expected poor atmospheric transmission, to reduce the number of tunings necessary to
%as part of the tuning optimization 
cover the large frequency width of ALMA Band 7.

A variable frequency overlap of 50 to 500\,MHz was used between two adjacent spectral windows within a given sideband, and a frequency overlap of 100-200\,MHz between contiguous tunings was adopted, allowing us to check the relative amplitude calibration across the survey (Sect.~\ref{sec.relativeflux}).
%Table~\ref{tab.observingsetup} summarizes the central frequencies of the four spectral windows within each of the individual spectral setups together with the observed frequency coverage within both lower and upper-sideband of each tuning. 

The native spectral resolution was 0.977\,MHz, equivalent to 3.4 to 0.8\,\kms~(with Hanning smoothing) for Bands 3 to 7, respectively. Only a few setups in Band 7 (B7g to B7p)
% p7 to 16 
had some spectral windows set to a resolution of 1.128\,MHz.
Final products were produced at a coarser uniform velocity resolution during imaging (Sect.~\ref{sec.imaging}).

\subsection{Sensitivity}
\label{sec.sensitivity}

The targeted brightness temperature sensitivity of the survey was 50, 50, 40, 30, and 30~mK in 10~\kms~channels across ALMA Bands 3, 4, 5, 6, and 7, respectively. This was a compromise to achieve a deep uniform sensitivity across all bands while keeping an achievable ALMA time request.
These brightness temperature sensitivities can be converted to point source flux density sensitivities using Equation~3.31 in the ALMA Cycle 8 technical Handbook\footnote{
\label{ALMAtechandbook}
\url{https://almascience.nrao.edu/documents-and-tools/cycle8/alma-technical-handbook/view}}:
\begin{equation}
    S_\nu(Jy) = 0.0736\left(\frac{\nu}{300\,GHz}\right)^2 
    \left(\frac{\theta_{max}}{1\arcsec}\right)
    \left(\frac{\theta_{min}}{1\arcsec}\right)
    T_R(K)
    \label{eq:fluxtobrightness}
\end{equation}
%(where we have recast THB Equation 3.31 in terms of flux density and used the equivalence between antenna temperature ($T_A$) and radiation temperature ($T_R \equiv T_A/\eta_{mb}$, where $\eta_{mb}$ is the antenna main beam efficiency). 
For the originally-specified circular beam size of $1''$ the approximate point source flux density sensitivities are $\sim$0.35, 0.9, 1.2, 1.6, and 2.3~mJy %in 10~\kms~channels 
at the ALMA frequency band centers.

\begin{figure}
\begin{center}
\includegraphics[width=0.5\textwidth]{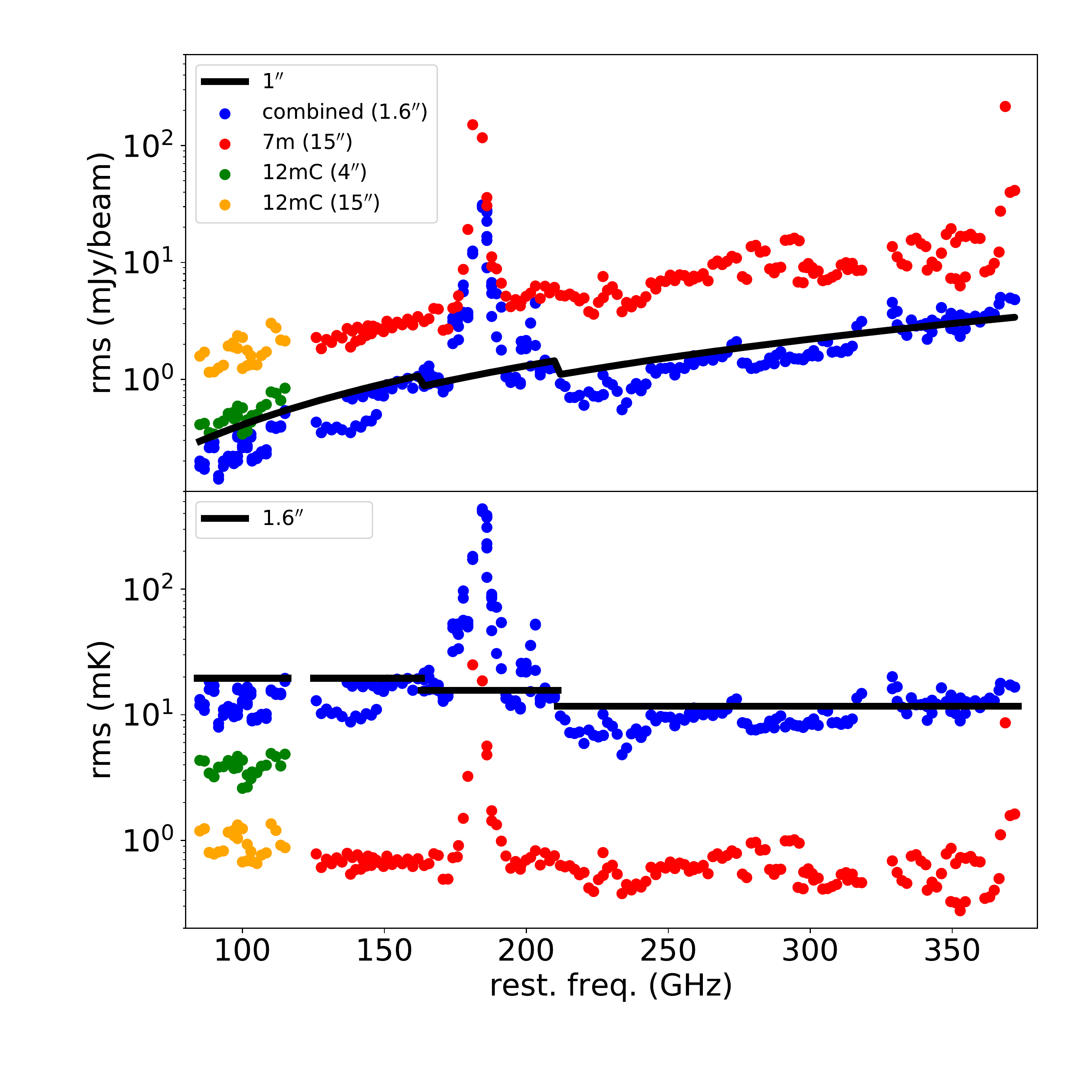}
\caption{Measured RMS flux density (top) and equivalent brightness temperature (bottom) noise level of each individual data cube (spectral windows) imaged for the combined arrays (blue), compact 12~m array band 3 observations (green) and 7~m array band 4 to 7 (red).
Black lines correspond to the target sensitivities requested for $1''$ (top) and $1.''6$ (bottom) resolution imaging. 
Brightness temperature noise levels have been calculated for a beam of $1.6''$, $4''$, and $15''$ for combined, 12~m compact, and 7~m array, respectively. 
For completeness, compact 12~m array band 3 sensitivities are also displayed (orange) for data cubes imaged at 7~m array $15''$ resolution. Target brightness temperature sensitivities are scaled down to that of the common beam of $1.6''$ resolution used during imaging. See Sects.~\ref{sec.sensitivity} and \ref{sec.imaging} for details.
\label{fig.rms}}
\end{center}
\end{figure}

The flux density root mean square (RMS) values of each individual continuum subtracted spectral window were estimated from line free spectral channels. Channels showing bright line emission were masked by hand, while those showing fainter emission were eliminated by using the biweight algorithm in the CASA task {\it imstat} \citep{McMullin2007}. % after masking the regions with obvious molecular emission.
The top panel in Fig.~\ref{fig.rms} displays the measured flux density RMS of each spectral window as a function of frequency. Combined (12m compact plus extended configurations in Band~3 or 12-m plus 7-m array data for higher frequency bands, see Sect.~\ref{sec.MRS}) sensitivities (blue) as well as the 12-m compact configuration (green) and 7-m array (red) sensitivities are shown. Fig.~\ref{fig.rms} shows that for most of the survey, the achieved flux density complies with the project's original sensitivity goals.
Apart from the spectral windows directly affected by the 183\,GHz telluric water line (USB of B5b, B5c, and B5d, and LSB of B5e) and the oxygen and water lines at 368~GHz and 380~GH, respectively (USB of B7m and B7n), the flux density RMS noise ranges from 0.18 to 5.0 mJy~beam$^{-1}$, with average flux density sensitivity of 1.4~mJy~beam$^{-1}$ and median of 1.0~mJy~beam$^{-1}$. This noise is measured in the final $8-9$~\kms~channels (Sect.~\ref{sec.imaging}).

The bottom panel in Fig.~\ref{fig.rms} displays the $1.6''$ beam (Sect.~\ref{sec.imaging}) equivalent brightness temperature noise per spectral window. 
%The target RMS in black is corrected with respect to the values above by a factor of 2.56, corresponding to the equivalent brightness temperature difference between the targeted $1''$ beam and the final $1.6''$ beam. 
The targeted brightness temperature sensitivity with a beam of $1''$ is corrected by a factor of 2.56 to account for the achieved common beam of $1.6''$ (Sect.~\ref{sec.imaging}).
The average sensitivity is 14.8~mK with a median of 10.2~mK.

\subsection{Maximum recoverable scales}
\label{sec.MRS}

Due to the lack of short spacings in interferometric observations, structures larger than the maximum recoverable scale (MRS) are filtered out.
Following Eq.~7.6 in the ALMA Cycle~8 technical handbook $^{\ref{ALMAtechandbook}}$
%$\footnote{
%\label{ALMAtechandbook}
%\url{https://almascience.nrao.edu/documents-and-tools/cycle8/alma-technical-handbook/view}
%}, 
the MRS is defined as $\theta_{MRS}\sim0.6\lambda$/$B_{min}$, where $\lambda$ is the wavelength and $B_{min}$ the shortest projected baseline.

\begin{figure}
\begin{center}
\includegraphics[width=0.5\textwidth]{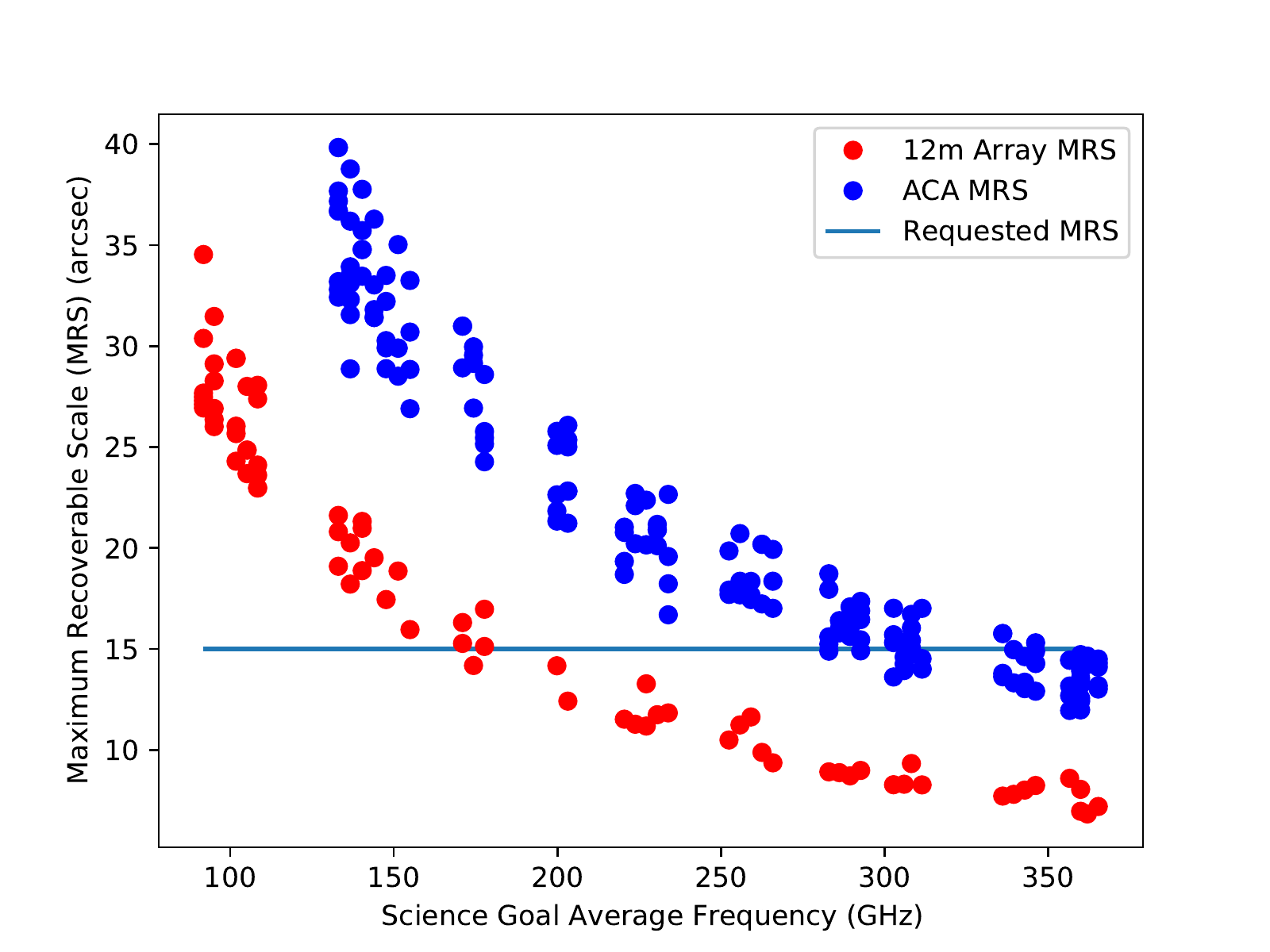}
\caption{
Maximum recoverable scale per channel estimated for each individual 12-m (red) and 7-m (blue) array observations (Sect~\ref{sec.MRS}). 
Each point corresponds to an individual execution centered at the average frequency of all four spectral windows (thus the gaps that appear in some frequencies).
The targeted $15''$ maximum recovered scale is represented by an horizontal black line.
\label{fig.mrs}}
\end{center}
\end{figure}

Based on the extent of the CO $J=1-0$ emission \citep{Meier2015}, ALCHEMI targeted an MRS of $15''$ across its entire frequency coverage. This corresponds to spatial scales of up to $\sim250$~pc, which should recover most of the emission from the GMCs in the NGC~253 CMZ. These spatial scales also correspond to one fourth of the region enclosed within the ILR (Sect.~\ref{sec.NGC253}) and are similar to the length of the $120-320$~pc filaments tracing the molecular outflow \citep{Bolatto2013}. In Band\,3, our required MRS could be achieved with the extra-compact 12-m array configuration, but additional observations with the ACA 7-m array were required for Bands 4 through 7. Figure~\ref{fig.mrs} displays the MRS for each of the individual 12-m and 7-m array observations. 
%This scale is estimated from the smallest baseline length following .
The targeted MRS was achieved across the entire ALCHEMI spectral coverage, ensuring that $\lesssim 15''$ scales are recovered throughout the survey. We note however that at the lower frequencies, scales larger than $15''$ could also contribute to the observed emission. In that sense, the survey is not strictly homogeneous, although this could be corrected, if deemed important for a given science case, by cropping the visibilities within a given $uv$ radius. Based on the results presented in Sect.~\ref{sec.analysis}, we expect
%can confidently assume that very little emission is contributing 
only minor contributions from structures larger than our expected MRS of $15''$, with the exception of CO transitions.

\section{Data calibration, equalization, and imaging}

Calibration and data quality assessment were performed by ALMA staff. For all but a couple of scheduling blocks the ALMA calibration pipeline was used. A summary of the calibrators used within each scheduling block is provided in Table~\ref{tab.ObsSum}. 
%All calibration assessments where within ALMA norms for QA2 qualification.

\begin{table}
\begin{center}
%\centering
\caption{ALCHEMI Calibrator Summary}
\label{tab.ObsSum}
\begin{tabular}{lp{6cm}}
\hline
\hline
Calibrator & Frequency Setup/Array \tablefootmark{a} \\
\hline
\multicolumn{2}{l}{Flux and Bandpass Calibration}\\
%\hline
J0006$-$0623 & B3a, B3bTM2, B3c, B3dTM1, B3eTM2, B3f, B4a,
B4b, B4c, B4d, B4e, B4f, B4g, B5aTM1, B5bTM1, B5d7M, B6aTM1, B6b,
B6c, B6d, B6e, B6fTM1, B6gTM1, B6hTM1, B6iTM1, B6j, B7aTM1, B7bTM1,
B7cTM1, B7d,  B7e, B7f, B7g, B7hTM1, B7i, B7jTM1, B7kTM1, B7l, B7nTM1,
B7p \\
J0237$+$2848 & B5c7M, B3c, B3dTM2 \\
J0334$-$4008 & B4aTM1 \\
J0423$-$0120 & B5b7M, B5f7M, B5g7M, B5h, B7oTM1 \\
J0522$-$3627 & B4b7M, B4e7M, B5b7M, B5e7M, B5h7M, B6a7M,
B6g7M, B6i7M, B7a7M, B7b7M, B7c7M, B7d7M, B7e7M, B7f7M, B7g7M, B7h7M,
B7i7M, B7j7M, B7l7M, B7m7M, B7n7M, B7o7M, B7p7M \\ 
J2253$+$1608 & B5c7M, B5d7M, B5e7M, B6b7M, B6f7M, B6g7M,
B7c7M, B7e7M, B7f7M, B7h7M, B7i7M, B7j7M, B7k7M, B7l7M, B7o7M, B7p7M \\
J2258$-$2758 & B3bTM1, B3eTM1, B3fTM1, B4a7M, B4b7M, B5a7M,
B5b7M, B5c, B5d, B5e, B5f, B5g, B5h, B6a7M, B6d7M, B6e7M, B6g7M,
B6h7M, B6i7M, B7a7M, B7b7M, B7c7M, B7d7M, B7g7M, B7h7M, B7k7M, B7mTM1,
B7n7M, B7p7M \\ 
J2357$-$5311 & B3bTM1, B3fTM1, B5cTM1 \\
\hline
\multicolumn{2}{l}{Phase Calibration}\\
%\hline
J0038$-$2459 & all except B7aTM1, B7m7M \tablefootmark{b} \\
J0106$-$2718 & B4a7M, B4c7M, B4f7M, B4g7M, B6a7M, B6c7M, B7aTM1 \\
J0118$-$2141 & B7m7M, B5h7M, B6e7M, B6j7M, B7b7M, B7f7M,
B7g7M, B7i7M, B7k7M, B7l7M, B7m7M, B7p7M \\ 
J0132$-$1654 & B7b7M, B6i7M, B7a7M, B7n7M, B7p7M \\
J0143$-$3200 & B4e7M, B6a7M, B6g7M, B7m7M, B7n7M \\
\hline
\end{tabular}
\tablefoot{
Calibrator sources used on any of the executions of a given setup.
Information in this table extracted from the ALMA archive using {\texttt astroquery} \citep{Ginsburg2019}.
\tablefoottext{a}{Frequency setup ID (Table~\ref{tab.observingsetup})
%(SB) 
and array combined label. TM refer to 12~m Array, with TM1 and TM2 referring to the extended and compact array, respectively (see Sect.~\ref{Sec.Observations}). Only ID
  listed when both arrays used calibrator.}
\tablefoottext{b}{All spectral setups used J0038$-$2459 as phase calibrator on at least one of their executions but for those indicated.}
}
 
\end{center}
\end{table}

\subsection{Flux calibration accuracy}
\label{sec.relativeflux}

According to the ALMA Cycle~5 Proposer's Guide\footnote{
\label{ALMApropguide}
\url{https://almascience.nrao.edu/documents-and-tools/cycle5/alma-proposers-guide/view}
}, 
delivered absolute flux calibration should be better than $5\%$ for Bands 3, 4 and 5, and $10\%$ for Bands 6 and 7.
The flux calibration method adopted by ALMA \citep[described in][]{Guzman2019} uses regularly-monitored fluxes from a catalog of secondary flux calibrators to set the flux calibration scale for all science measurements. One or more secondary flux calibration sources are measured with each science scheduling block. The absolute flux scale for the secondary calibrators are determined through almost simultaneous measurements of primary flux calibrators (solar system objects, including Uranus, Neptune, Callisto, Ganymede, and Mars) with a monitor cadence of $10-14$~days.
%made during the 10-14 day flux monitoring period for these secondary flux calibration sources.
The accuracy of this flux calibration scheme has not been fully assessed, and recent studies suggest that the ALMA flux calibration uncertainty can be significantly worse than that stated in the ALMA user guidelines \citep{Francis2020,deKleer2021}.

Taking advantage of the multiple contiguous frequency tunings of the ALCHEMI data set across the five covered frequency bands, we are able to further estimate the relative flux calibration accuracy of the individual frequency tunings. Prior to this analysis, data were cleaned and preliminary imaging was performed to a common beam as described in Sect.~\ref{sec.imaging}.
Two independent methods were used to check the relative flux alignment. We derived amplitude scaling factors between tunings (Table~\ref{tab.observingsetup}), based on overlapping channels (Sect.~\ref{sec.OverlapChannel}). The relative continuum level was then used to double check the accuracy of our spectral flux alignment (Sect.~\ref{sec:ContLev}). %As described below, the final data set presented here made use of the first approach.

For a target source with strong continuum and a significant amount of spectral line emission within each independent frequency tuning, absolute flux calibration precision is required.  Accurate absolute flux calibration minimizes amplitude misalignment in the final concatenated spectrum as well as assures a high level of accuracy when comparing spectral line fluxes derived from different frequency tunings.  As shown by \cite{Harada2018}, for spectra with a low density of spectral lines per sampled frequency bandwidth, one can derive and subtract the continuum emission first within individual
tunings.  This continuum information can then be used to perform an amplitude rescaling, followed by concatenation of the continuum-subtracted spectra from each frequency tuning to achieve accurate relative flux scaling and minimal gaps at the frequency tuning boundaries.  Subtraction of a smooth continuum from a spectrum with a high density of spectral lines, on the other hand, cannot extract the necessary continuum emission information in order to use this flux rescaling technique.  We describe a method to improve the flux calibration accuracy of our ALCHEMI spectra when there is a high density of spectral lines in Section~\ref{sec.OverlapChannel}.

In Appendix~\ref{Sec.AppendixFLuxAlignment} we show the unscaled spectra where only the standard ALMA pipeline calibration has been applied to the data. As evidenced by Figures~\ref{fig.unscaledfullspectrum} through \ref{fig.unscaledB7p2} and in the scaling factors in Table~\ref{tab.observingsetup} some of the misalignment between adjacent receiver tunings are beyond nominal calibration uncertainties.
%Based on the analysis processes described in Sections~\ref{sec.OverlapChannel} and \ref{sec:ContLev}, amplitude scaling factors for each ALCHEMI science goal have been derived (Table~\ref{tab.observingsetup}). 
In Appendix~\ref{Sec.AppendixFluxCal} we provide an analysis of the relative and absolute flux calibration uncertainties for all ALCHEMI image cubes. 

The two methods used below assume that flux calibration has no systematic bias, or said otherwise, cannot account for systematic biases in the fluxes accross the spectral scan.
In fact, our analysis suggest that for absolute flux calibration an overall uncertainty of 15\% is justified and appropriate.
However, the derivation and application of the amplitude scaling factors 
%described in Sections~\ref{sec.OverlapChannel} and \ref{sec:ContLev} 
has allowed us to improve the relative flux calibration accuracy beyond that which a single ALMA scheduling block might normally attain. 
The relative flux calibration scaling factors listed in Table~\ref{tab.observingsetup} were applied to the originally calibrated visibilities prior to final imaging.

In the analysis presented in this paper (Sect.~\ref{sec.analysis}), only the statistical uncertainty in the fits due the noise in the spectra are considered and not the absolute flux calibration uncertainty mentioned above,  which is enough for our purposes. 

\subsubsection{Overlapping channels alignment}
\label{sec.OverlapChannel}

As the ALCHEMI spectra toward NGC\,253 in many cases possess a high degree of spectral crowding, we refined our flux calibration by using the target signal itself as reference.
This technique, originally developed for other ALMA spectral scans, is referred to as “flux self-calibration” \citep{Sakamoto2021}. Below we recapitulate the technique details fully described there.

The amplitude re-scaling in this technique is
based on a comparison of the initial spectra at their overlaps, which as described in Section~\ref{sec.frequencysetup} has been built-into our scheduling block tuning setup for this purpose.  For the choice of the reference signal it does not matter whether the emission at that position is dominated by continuum or spectral line emission as long as it is reasonably spatially compact and presents no time variability over the observation period.  This is due to the fact that a pair of frequency tunings is always compared at their overlapped frequencies, using the same emission from each frequency.  Furthermore, the ALCHEMI spectra have been produced with the same spatial resolution across each spectral scan, so that the two observations being compared should possess approximately the same range of baseline {\it uv} lengths at their respective overlapped frequencies. This flux self-calibration through overlapping
tunings can therefore be used on targets with numerous broad lines with a limited number of line-free channels. A single scaling factor is derived for each tuning, shared by all the spectral windows in that tuning. It is important to note that the flux scales for individual spectral windows within a given sideband align to $\lesssim1$\%.

As indicated in Sect.~\ref{sec.frequencysetup} our frequency setup included a $100-200$~MHz overlap among contiguous frequency tunings. In order to derive the flux rescaling factors for each frequency tuning and array configuration, we assigned to each frequency tuning a scaling factor a$_i$, where $i$ is the scheduling block ID (column 1 in Table~\ref{tab.observingsetup}).  For each set of adjacent tunings we then solved (with the least squares method when
necessary) a set of equations given by:
\begin{equation}
    r_{ij} = \frac{a_i}{a_j}
    \label{eq.fluxrescale}
\end{equation}
where r$_{ij}$ is the measured amplitude ratio between the independent frequency tunings $i$ and $j$. The spectra extracted from the TH2 position
($\alpha_{J2000}=00^h47^m33.182^s$, $\delta_{J2000}=-25^\circ17'17.148''$; \citealt{Lenc2006})
within the preliminary imaged data cubes was used as the reference measurement in Equation~\ref{eq.fluxrescale}.  We used the constraint mean(a$_i$) = 1 to set the overall scale of the solutions, resulting in the flux self-calibration rescaling factors listed in Table~\ref{tab.observingsetup}.  These flux rescaling factors were applied to the calibrated visibilities before final imaging using the CASA task {\it gencal} as follows:
\begin{equation}
    A^{final}_{uv} = \frac{A^{0}_{uv}}{a_i}
\label{eq.uvscaling}
\end{equation}
where $A^{0}_{uv}$ and $A^{final}_{uv}$ are initial ALMA delivered and final {\it uv} amplitudes.
As evidenced by the scaled spectra shown in  Figures~\ref{fig.unscaledfullspectrum} through \ref{fig.unscaledB7p2} the rescaled spectral image cubes are in most cases well-aligned in amplitude.

\subsubsection{Continuum level alignment}
\label{sec:ContLev}

%Thanks to the broad contiguous frequency coverage of the ALCHEMI data, we can also use the continuum level to verify the amplitude scaling by assuming a smooth continuum emission across the whole frequency scan.
We can use the continuity of the spectral energy distribution of the continuum emission to verify the relative amplitude scaling of the different tunings. One advantage of this method is that it can make a bridge across bands and gaps in the frequency coverage (Sect.~\ref{sec.frequencysetup}) which is an intrinsic uncertainty when using overlapping channels (Sect.~\ref{sec.OverlapChannel}).
%As described in Sect.~\ref{sec.contsubtraction}, 
The continuum emission is measured on the STATCONT continuum cubes (Sect.~\ref{sec.contsubtraction}), after the first amplitude scaling derived from the overlapping channels alignment process (Sect.~\ref{sec.OverlapChannel}). While the amplitude scaling was applied per tuning, here we measure the continuum emission for each individual spectral window. The position TH2, close to the continuum emission peak, is also used as in Sect.~\ref{sec.OverlapChannel}. At this step, we do not want to introduce a complicated fit of the overall SED, so we simply use a third order polynomial to estimate the standard deviation of the continuum levels with respect to a smooth and continuous function. This strategy allows us to test the robustness of the channel-overlapping scaling by checking "residual" scaling factors (i.e., if the channel-overlapping scaling was perfect across all data, then those new factors would be all equal to one).

We have run this method on both the 7m-array and 12m+7m array cubes separately. After removing a few spectral windows close to the 183-GHz telluric water line (in Band~5), which appear as clear outliers, we find that the standard deviation of the new scaling factors is $2.5\%$ across all bands, for both 12m+7m and 7m array data.
We can thus take this value as the maximum additional error after the channel-overlapping scaling, since the STATCONT cubes may introduce some uncertainty due to line crowding and imperfect continuum estimation. Alternatively, this result suggests that the continuum determination is relatively robust and uniform over spectral windows.
We note that the dispersion increases slightly toward the highest frequency edge for the 7m-array data, as the RMS of the new scaling factors for Band~7 alone goes to $3.5\%$.

\subsection{Imaging}
\label{sec.imaging}

Before imaging, several homogenization corrections were applied to all ALCHEMI data sets in order to produce a uniform science archive.  In addition to the normalization of the amplitude scale in each tuning using the procedure described in Section~\ref{sec.OverlapChannel}, all measurement sets were binned to a common velocity scale.  A common velocity scaling was produced by binning to frequency resolutions of 3, 5, 6, 8, and 10~MHz for Bands 3, 4, 5, 6, and 7, respectively, which is equivalent to an approximately common velocity resolution of $\sim8-9$~\kms\ in the LSRK velocity reference frame.  Once homogeneity in amplitude scaling and velocity resolution was attained, the CASA task {\it tclean} was used to produce image cubes of each tuning spectral windows.  The specific {\it tclean} parameters used for imaging were catered to the needs of individual spectral windows as follows.

The cell and image sizes used for imaging each array and frequency band were as follows: 7m Array observations used $cell=0.4$\,arcsec and  $imsize=[320,320]$ pixels for all frequency bands; 12m Array and the combined 7m+12m data sets used $cell=0.15$\,arcsec for all frequency bands and $imsize=$[800,800], [800,720], [720,648], [640,512], and [640,512] for Bands 3 through 7, respectively.
Automasking was used for clean region selection

Based on {\it tclean} dry-runs of the ALCHEMI measurements containing known strong spectral features (i.e. CO $2-1$), a list of line-free spectral channel RMS values for those spectral windows was developed to use as input for the {\it tclean} parameter {\it threshold}.  This was necessary to allow for the proper cleaning of spectral windows where strong spectral lines amplify imaging artefacts, causing the single channel noise values near strong spectral lines to be anomalously high. 
For spectral windows which contained strong spectral lines, the pre-determined spectral channel RMS values were used to set the {\it tclean} threshold, leaving the {\it nsigma} parameter unset.
For spectral windows which did not contain strong spectral lines, the {\it tclean} parameter {\it nsigma=2} was set, and the {\it threshold} parameter was left unset.
Spectral channel flagging was performed on those spectral windows which contained clear absorption due to telluric oxygen and water (see Section~\ref{sec.frequencysetup}).

The {\it hogbom} deconvolver function was used for all spectral windows.
The {\it mosaic} gridder was used for spectral windows comprised of multiple pointings, while the {\it cube} gridder was used for Band 3 and all ACA spectral windows as these measurements required only a single pointing (Sect.~\ref{Sec.Observations}).
Robust (Briggs) weighting was used with a robust parameter of 0.5 for most spectral windows. 
In order to produce images with resultant spatial resolution of 1.6\arcsec~, a few tunings required alternate robust parameters for each of the four spectral windows, and in three cases {\it uv} range settings: 
B3b used $robust=$[0.4,0.5,0.5,0.5];
B3c used $robust=$[0.25,0.5,0.5,0.5];
B3f used $robuts=$[0.0,0.0,0.5,0.5];
B5d used $robust=[0.0,-2.0,0.5,0.5]$ and $uvrange >20~k\lambda$;
B5e used $robust=[-2.0,-2.0,-2.0,-2.0]$ and $uvrange>25~k\lambda$;
B5f used $robust=[-2.0,-2.0,-2.0,-2.0]$ and $uvrange>50~k\lambda$.

The originally requested angular resolution was $1''$ (17~pc). However, the synthesized beams of each individual spectral window significantly varied between frequency setups and between individual spectral windows in the upper and lower sidebands due to the effective antenna configuration used for each observation. This fact did not allow the imaging of the entire survey at the targeted spatial resolution. In order to produce a homogeneous data set, with data cubes sharing a uniform spatial (and spectral) resolution over all frequencies, a common final spatial resolution of $1.6''$ (28~pc) was selected, corresponding to that of the data cube with the coarsest resolution in the survey. Post-imaging convolution using the CASA task {\it imsmooth} was therefore performed to produce final image cubes with a fixed Gaussian circular beam of $1.6''$.

Additional data cubes were generated for the compact 12-m array observations at Band~3 and the 7-m array observations for Bands 4 through 7. Data were imaged with a common restoring beam of $4''$ and $15''$, respectively, for these two compact array configurations.

\subsection{Continuum subtraction}
\label{sec.contsubtraction}

It was determined that it would be inefficient to subtract the continuum in the $uv$ plane from the individual 188 data cubes constituting the ALCHEMI measurement set. The large velocity gradient across the field of view and the significant cumulative line contribution even in the low resolution data (Sect.~\ref{sec.LineContribution}) makes it difficult to accurately identify line free windows to perform the continuum subtraction in the $uv$ plane for all data cubes. 
%Although continuum subtraction could potentially be performed on the uv-visibilities, the search for continuum free channels, in particular in the high resolution data, is not trivial due to the numerous broad emission lines, mostly in the low resolution data, coupled to the large velocity gradient across the CMZ, which is relevant in both low and high resolution images.
Therefore, for consistency across the entire survey, continuum emission was statistically derived from each individual spectral window using STATCONT \citep{Sanchez-Monge2018} to derive and subtract the continuum on a per pixel basis from each image cube. Sigma-clipping continuum determination was used with the default parameter \citep[$\alpha=1.8$, see][for details]{Sanchez-Monge2018}.

Continuum subtraction was performed on both the initial input data cubes used to feed the flux level alignment and also on the final amplitude-scaled data (Sect.~\ref{sec.relativeflux}).    
The per-pixel continuum estimation was thoroughly tested and provides good overall results.
Since the algorithm was mostly tuned to confusion limited Galactic sources with relatively narrow spectral line emission, we observe that the continuum appears slightly overestimated in the regions of spectra with higher noise levels. For example, the continuum level appears to be overestimated near the 183~GHz telluric water line and at the upper end of Band~7 ($>335$~GHz). More importantly, continuum subtraction is seen not to be optimal for spectral lines close to the noise level. In such cases, continuum subtraction in the $uv$ plane or subtraction of a spectral baseline in the image plane using a narrow window around the lines of interest may be needed for accurate imaging. 

\subsection{Self-calibration}
It is foreseen that the ALCHEMI data set will eventually be improved with self-calibration of both the ACA and 12m Array measurements. However, the data in this article and the data that will initially be publicly released from this ALMA large program will not include self-calibration. The ALCHEMI research collaboration intends on providing a subsequent version of the ALCHEMI image cube archive which includes the application of self-calibration. It is important to note that self-calibration may have an impact on the absolute flux calibration of the ALCHEMI measurements. 
As a result, we emphasize that the amplitude scaling factors derived in Sect.~\ref{sec.relativeflux} will need to be recalculated after self-calibration of the ALCHEMI data.

\section{ACA Data: Analysis and First Results}
\label{sec.analysis}

\begin{figure*}
\includegraphics[width=\textwidth]{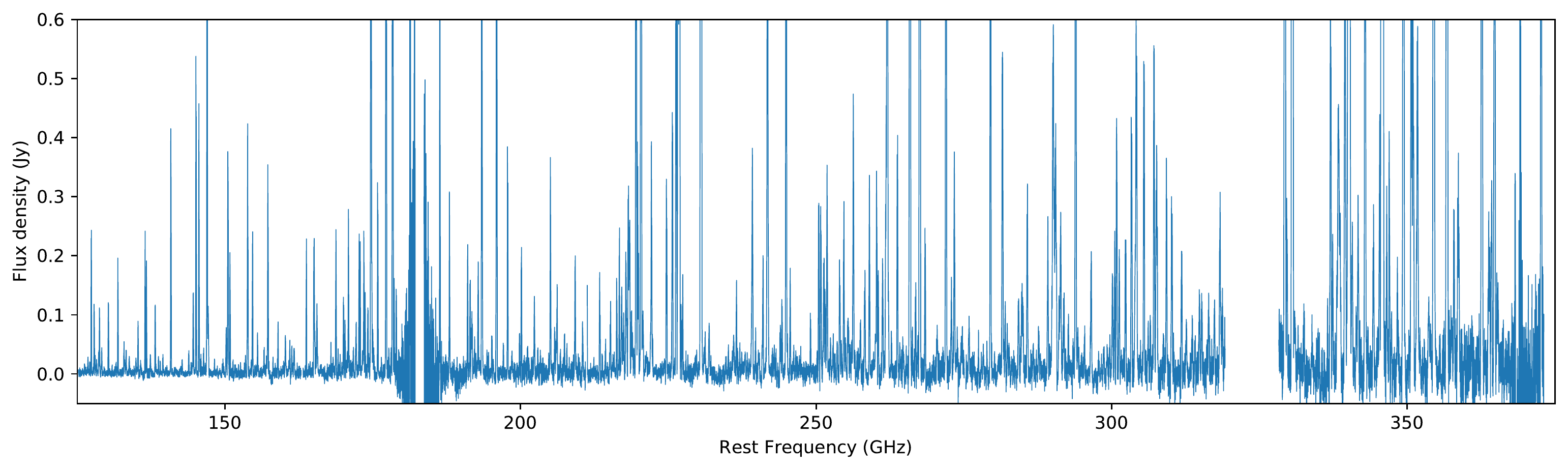}
\caption{Full spectral coverage obtained with the ALMA Compact Array (ACA 7m) alone, extracted from the position of brightest molecular emission (see Section~\ref{Sec.sampleposition}). 
%The observed spectrum is shown in blue and the model (Sect.~\ref{Sec.LTE}) in orange. 
%Individual molecular transitions with intensities higher than 2~mJy are indicated. For the sake of ``clarity'', molecular labels are displayed in rows at different elevations. In different rows starting from the top we label the 744 transitions with flux densities $>500,~>250,>150,~>100,~>70,~>50,~>40,~>30,~>20,~>15$~mJy, respectively, and the 1048 other transitions down to 2~mJy ($1\sigma$ at the lowest frequencies) are indicated with small segments. 
Fig.~\ref{fig.fullspectrum50GHz} in Appendix ~\ref{Sec.AppendixFullSpectrum} presents a zoomed version of this plot in 5 frequency windows 50~GHz wide where the comparison with the modeled emission (Sect.~\ref{Sec.LTE}) and the molecular line identification of each individual feature is included. Figs.~\ref{fig.fullspectrum1}$-$\ref{fig.fullspectrum10} present a further zoomed version in 5~GHz windows. 
\label{fig.fullsurvey}}
\end{figure*}

%We note that in this Section Band 3 observations are also described and displayed in Table~\ref{tab.observingsetup} and Fig.~\ref{fig.mrs}. This is done for the sake of completeness, since in this paper we only report the 7~m observations, which are limited to Band 4 through 7.

While Sect.~\ref{Sec.Observations} describes the observational details of the full ALCHEMI survey, here we focus on the ACA 7-m~(Morita) array alone. The ACA data allow us to probe the global properties of the molecular emission of the CMZ in NGC~253. This provides a template for subsequent analysis of the high resolution data to compare molecular abundances on individual GMC size scales to those on the larger size scales probed by the ACA.
%probe the differences between the local small scale molecular abundances associated with individual GMCs and the large scale abundances probed by the ACA data.

The frequency coverage of the ACA data in this analysis is limited to the 256.7~GHz surveyed with this array; ALMA Bands 4 to 7 between 125.2 and 373.2~GHz. Fig.~\ref{fig.fullsurvey} presents an overview of the survey.
As indicated in Sect.~\ref{sec.imaging}, a homogeneous reconstructing beam of $15''$ was used across the entire frequency range. This resolution is roughly equivalent to 
%the single-dish resolution of the 2~mm atmospheric window (ALMA Band 4) spectral line survey by \citet{Mart'in2006}.
that of the 2~mm (i.e., ALMA Band~4) spectral survey in \cite{Mart'in2006} carried out with the IRAM 30-m single-dish telescope. 
However, in contrast to single-pointing surveys with single-dish telescopes for which the resolution changes as a function of frequency, our ACA interferometric observations allow for uniform spatial resolution across the entire frequency coverage.

The point-source flux density sensitivity of the ACA data alone (Fig.~\ref{fig.rms}, top) is %significantly worse
lower than that of the combined data presented in Sect.~\ref{Sec.Observations}.
If we exclude the noisy data sets at the highest frequencies (i.e., USB of B7m and B7n), sensitivities range between 1.8 and 19.4~mJy~beam$^{-1}$ across the survey, with an average RMS of 7.8~mJy~beam$^{-1}$ and a median of 7.5~mJy~beam$^{-1}$. For the $15''$ synthesized beam, using Equation~\ref{eq:fluxtobrightness}, those values correspond to equivalent point-source brightness temperatures between 0.27 and 1.0~mK, and an average of 0.6~mK.
%rms for ACA is 10mJy with an median of 7.5 mJy.
%0.68 mK, 0.61 mK for 15''

\subsection{Continuum emission}
\label{sec.ContinuumEmission}

Having continuous frequency coverage over $\sim260$~GHz ranging from 2~mm to 850~$\mu$m (resulting in $\Delta\nu/\nu\simeq 1$
%at 248~GHz, 
at the central frequency of the survey) allows us to study the spatially averaged continuum spectral energy distribution (SED). This frequency range is particularly interesting because we can probe the Rayleigh-Jeans (RJ) tail of the dust emission as well as the free-free continuum emission from ionized gas and, to a lesser degree, synchrotron emission from non-thermal sources in the CMZ of NGC~253. 

\begin{figure}
\includegraphics[width=0.5\textwidth]{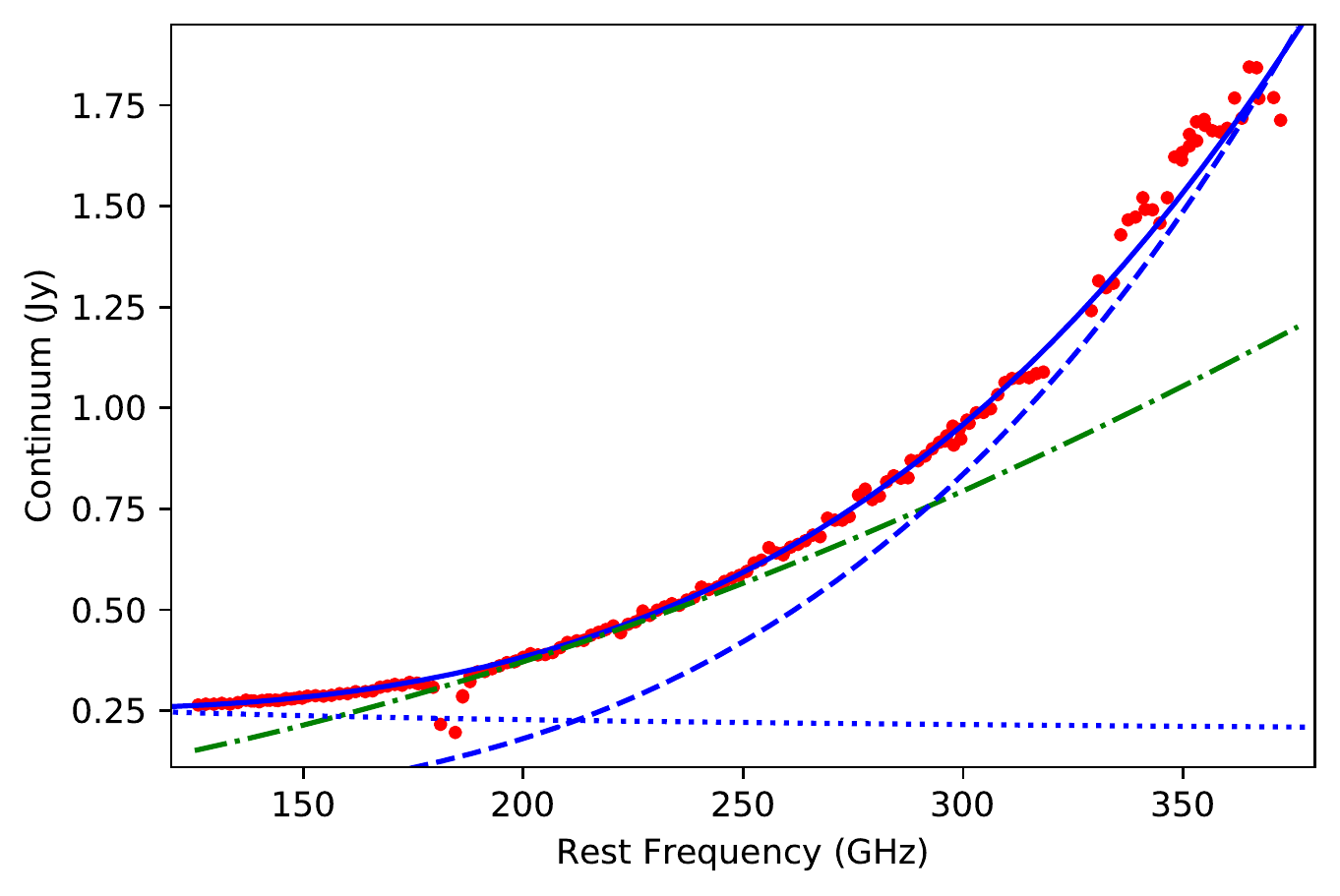}
\caption{
Continuum flux density %derived by subtracting the continuum subtracted spectra from the original spectra. Continuum points have been averaged 
at the peak of emission as derived from each spectral window across the surveyed frequency range (red dots). We note that extended emission in our data might account for up to $\sim20\%$ higher fluxes (Sect.~\ref{sec.ContinuumEmission}). A fit to the data is shown by a continuous blue line, which is the combination of the free-free emission (dotted blue almost horizontal line) and the grey body emission (dashed blue line). See text for further details on parameters used. As a reference to illustrate the deviation from pure black body emission, that for a $T_d=50$~K and $0.5''$ source size is shown as a green dot-dashed line. %(blue solid line), a 200~K  $0.24''$ size source (blue dash dotted line) and a $\nu^2$ power law in the RJ regime (green solid line) are displayed.  
\label{fig.continuumSED}}
\end{figure}

Continuum emission is barely resolved at our $15''$ resolution as derived from the two sample STATCONT continuum product images at 198 and 350~GHz. The 2D Gaussian fit to both continuum images yields a similar $18''\times14''$ ($FWHM; P.A.=55\degree$) emission extent which hints at some elongation along the major axis of the CMZ. In Fig.~\ref{fig.continuumSED}, we show the continuum emission as derived from STATCONT (Sect.~\ref{sec.contsubtraction}) for each spectral window at the pixel position analyzed in this article (Sect.~\ref{Sec.sampleposition}). Although this is not strictly an SED ($\nu~F_\nu$ vs $log(\nu)$) we will refer as such in the following.
Due to the slightly extended emission compared to our spatial resolution, if the continuum is measured integrating over an aperture larger than the beam instead of using the continuum value at the emission peak, a similar SED shape is obtained but with $15-20\%$ larger flux densities.
%would created using integrated intensities instead of values at the peak of emission.
%Rather than display the individual continuum fluxes derived from the use of STATCONT, we show the difference between the original spectra and the continuum subtracted spectra. \textcolor{red}{these should be equivalent, so not sure whether it is even worth mentioning. I did it like that because i had the spectral data in hand}
The SED was calculated from amplitude-aligned data cubes on overlapping channels (Sect.~\ref{sec.OverlapChannel}). Thus, the SED is spectrally smooth except for the regions in which the spectra are noise dominated such that STATCONT did not accurately fit the continuum (Sect.~\ref{sec.contsubtraction}). Such is the case with the apparent drop in continuum intensity due to the higher noise in the measurements around the telluric 183~GHz H$_2$O transition in Figure~\ref{fig.continuumSED}.

%Since we are sampling the RJ tail of the continuum SED, the emission should follow a power law ($S(\nu)=S_{\nu_0}\nu^{\alpha}$) with $\alpha\sim2$. Overlaid on the measured averaged continuum emission of NGC~253, Fig.~\ref{fig.continuumSED} shows the blackbody emission at 50~K of $0.5''$ in size (blue solid line) and at 200~K of $0.24''$ (blue dash dotted line) as modelled with MADCUBA. Also for comparison, a power-law profile is added (green solid line).
%These models are fitted to match the observed emission at 200~GHz.

%Measured continuum is underestimated by the model below $\sim180$~GHz. This is due to free-free emission as well as the high-end tail of synchrotron emission.
%We note that, at frequencies above $\sim250$~GHz, the observed emission departs significantly from that of a blackbody.

%{\bf: Seb: evidence for a two-component dust emission? can we fit with two black-body at two different temperatures? Evidence for different dust grain properties?}

The observed curvature of the continuum SED in Fig.~\ref{fig.continuumSED} fits well to a grey-body with dust temperature $T_d=42\pm1$~K, mass $M_d=8.0\pm0.2\times10^5$~M$_\odot$, emissivity $\beta=1.9$ (i.e., $S\propto\nu^{3.9}$), and a mass opacity coefficient of dust, $\kappa_{\nu}=\kappa_{0}$($\nu$/$\nu_{0}$)$^\beta$, where $\kappa_{0}=$ 0.1\,cm$^{2}$\,g$^{-1}$ and $\nu_{0}=$250\,GHz  \citep[e.g.][]{Cao2019}, plus a free-free component to account for the lower frequency emission with SFR=2.5~M$_\odot$\,yr$^{-1}$ and $T_e=10^4$~K \citep[$S\propto\nu^{-0.1}$, using Eq.3 in][]{DeZotti2019}.

The dust temperature was fit to the data in Fig.~\ref{fig.continuumSED} using Herschel observations (with no aperture correction applied) at high frequencies, finding good agreement with the cold component fit of 37~K by \citet{Perez-Beaupuits2018}. Similarly the derived mass agrees well with the \citet{Perez-Beaupuits2018} value of $1\times10^6$~M$_\odot$ if we correct our estimate to account for the extra $20\%$ flux from extended emission in our data (see above).
The higher temperature components derived by \citet{Perez-Beaupuits2018} are negligible at our observed frequencies.
Dust emissivity ($\beta=1.9$) based on this data agrees well with that derived by \citet{Rodriguez-Rico2006}.

Synchrotron emission was also included in the fit shown in Fig.~\ref{fig.continuumSED} based on Eq.~1 in \citet{DeZotti2019} which considers a steepening of the emission above 20~GHz. Since we assumed a SFR=2.5~M$_\odot$\,yr$^{-1}$ the synchrotron emission at our observed frequencies is negligible compared to that due to free-free emission. However,
%This may explain that 
our assumed SFR is slightly higher than that derived from radio recombination lines \citep[$\sim1.7$~M$_\odot$\,yr$^{-1}$:][]{Kepley2011,Bendo2015} or the SFR of $\sim1.7$~M$_\odot$\,yr$^{-1}$ derived from a fit to the free-free emission by \citet{Rodriguez-Rico2006}.
The power law emission resulting from the combination of synchrotron and free-free emission can only be disentangled with observations at lower frequencies not covered by the ACA data in this article.

\subsubsection{Line contribution to broadband continuum emission}
\label{sec.LineContribution}

\begin{figure}
\includegraphics[width=0.54\textwidth]{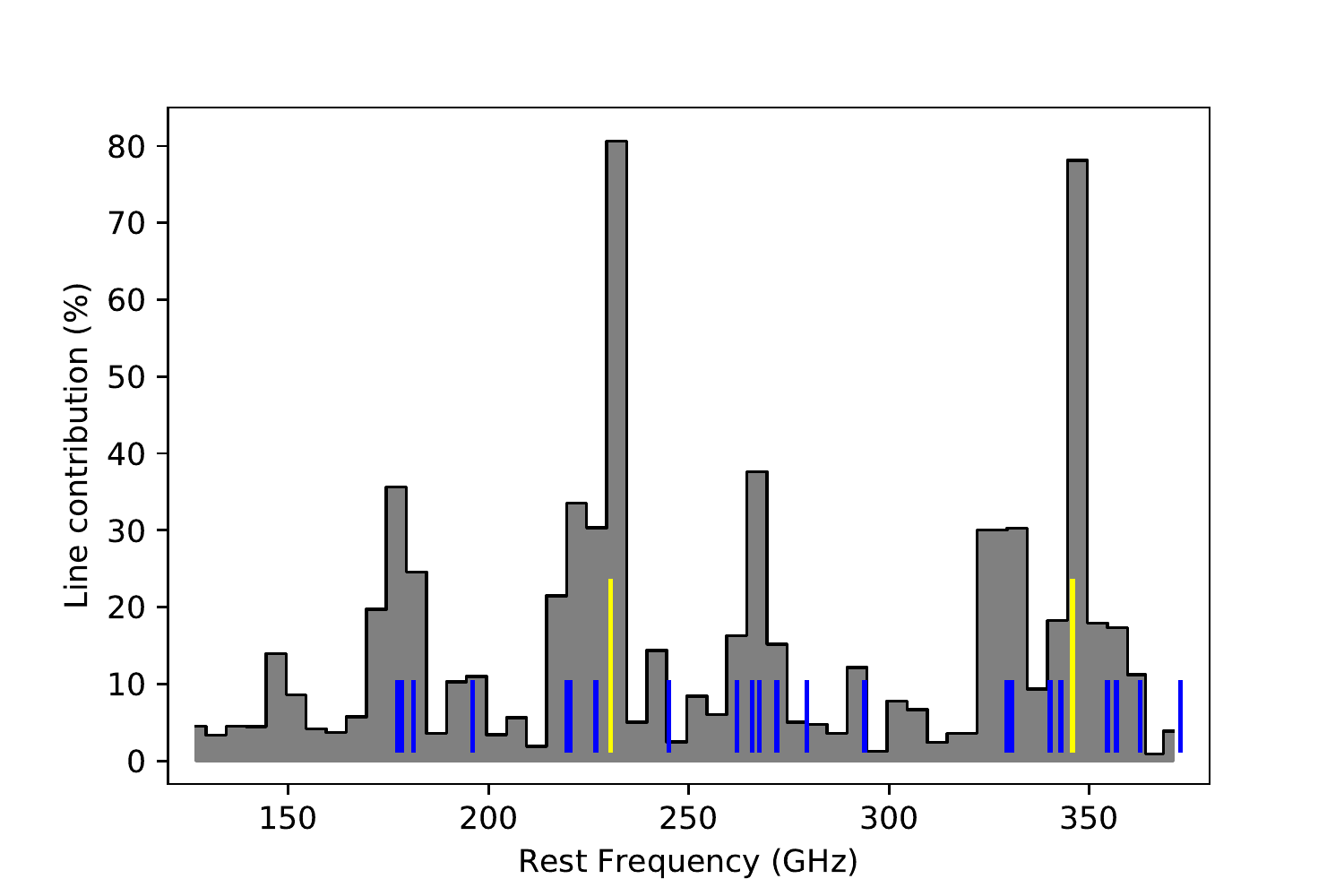}
\caption{
Spectral line contribution to the continuum flux in 5~GHz bins. The position of spectral features brighter than 1~Jy are shown as blue segments at their corresponding frequency, with the two CO $J=2-1$ and $3-2$ transitions displayed in yellow.
\label{fig.linecontribution}}
\end{figure}

Observations with broad, coarse spectral resolution mm/submm continuum detectors such as bolometers may suffer from contamination by spectral line emission.
The 40-GHz-wide spectral scan at 1.3\,mm toward the ULIRG Arp\,220, reported a line contribution to the total observed flux of $\sim28$\% \citep{Mart'in2011}. %Here we analyze the %line contribution in %5~GHz bins, shown in %Fig.~\ref{fig.continuum%SED}, as well the %spectral line %contribution measured %globally per band, in %Table~\ref{tab.linecont%ribution}.
The wide spectral coverage ALCHEMI data set allows us to investigate the typical line contamination for a starburst galaxy like NGC~253.
In order to obtain the spectral line contribution to the total emission per frequency band, we calculated the average flux density over the continuum subtracted spectrum and divided it by the observed average flux density (continuum plus line emission) over the same band. Results are shown in 5~GHz bins in Fig.~\ref{fig.linecontribution} and averaged over 50~GHz bands in Table~\ref{tab.linecontribution}. 
%We note that the negative line contribution in one of the highest frequency bins is the result of a inaccurate continuum subtraction in this noisy part of the data.

Similar to the results for Arp\,220, we observe a
%very significant line contribution to the measured 
significant contribution from spectral lines to the continuum-integrated flux density in 
%the starburst environment of 
NGC~253. 
Most of the %line contribution 
contamination is due to the brightest %spectral features 
lines in the band, as shown by color segments in Fig.~\ref{fig.linecontribution}, but %still 
there is also a significant contribution %across the spectra 
from %other emission lines.
the forest of weaker lines.
The line contribution, both in narrow 5~GHz and broad 50~GHz bins, ranges from a few percent up to a third of the measured flux density, and up to 80\% in narrow ranges containing CO transitions.
We note that the CO contribution over the 50~GHz band considered in Table~\ref{tab.linecontribution} 
would account for ~8\% in both Bands 6 and 7, similar to what was reported towards Arp\,220 \citep{Mart'in2011}, while the remaining contribution of up to $\sim35\%$ corresponds to emission from other species.

Both this work and that on Arp\,220 can now serve as a reference for evaluating the line contribution to broad band continuum observations (i.e., with too coarse spectral resolution to resolve the lines) and considering corrections to broad continuum observations over ALMA Bands 3 to 7 (rest frame), and correspondingly to the spectral index derived from those measurements.
Such corrections may be particularly relevant for high-redshift galaxies showing a nuclear starburst contribution.
%in unresolved objects even in frequency ranges not contaminated by bright CO lines.

\begin{table}
\centering
\caption{Line contribution to the observed flux density.}
\label{tab.linecontribution}
\begin{tabular}{l c c c c }
\hline
\hline
Frequency   &  $S_{TOTAL} \tablefootmark{a}$     &   $S_{MOLEC}$  &  Line               & Line          \\
range       &                  &                &       Contribution  &      density \tablefootmark{b}  \\
(GHz)  &  [$mJy$]         &  [$mJy$]    &   [\%]              &    [GHz$^{-1}$] \\
\hline
$125-175$       &  308 &   23 &  7.6 &  7.7\\
$175-225$       &  437 &   70 & 16.1 &  7.4\\
$225-275$       &  921 &  322 & 35.0 &  8.4\\
$275-325$       &  987 &   53 &  5.3 &  5.8\\
$325-375$       & 2488 &  909 & 36.6 &  6.5\\
\hline
\end{tabular}
\tablefoot{
The numbers in this table are exclusively based on ACA data, thus not including Band 3. See Sect.~\ref{sec.LineContribution} for details.
\tablefoottext{a}{
Total flux density corresponds to the contribution from both continuum and line emission.
}
\tablefoottext{b}{
Only spectral lines $>2$~mJy included in this calculation.
}
}
\end{table}

\subsection{Molecular emission analysis}

\begin{figure*}
\includegraphics[width=\textwidth]{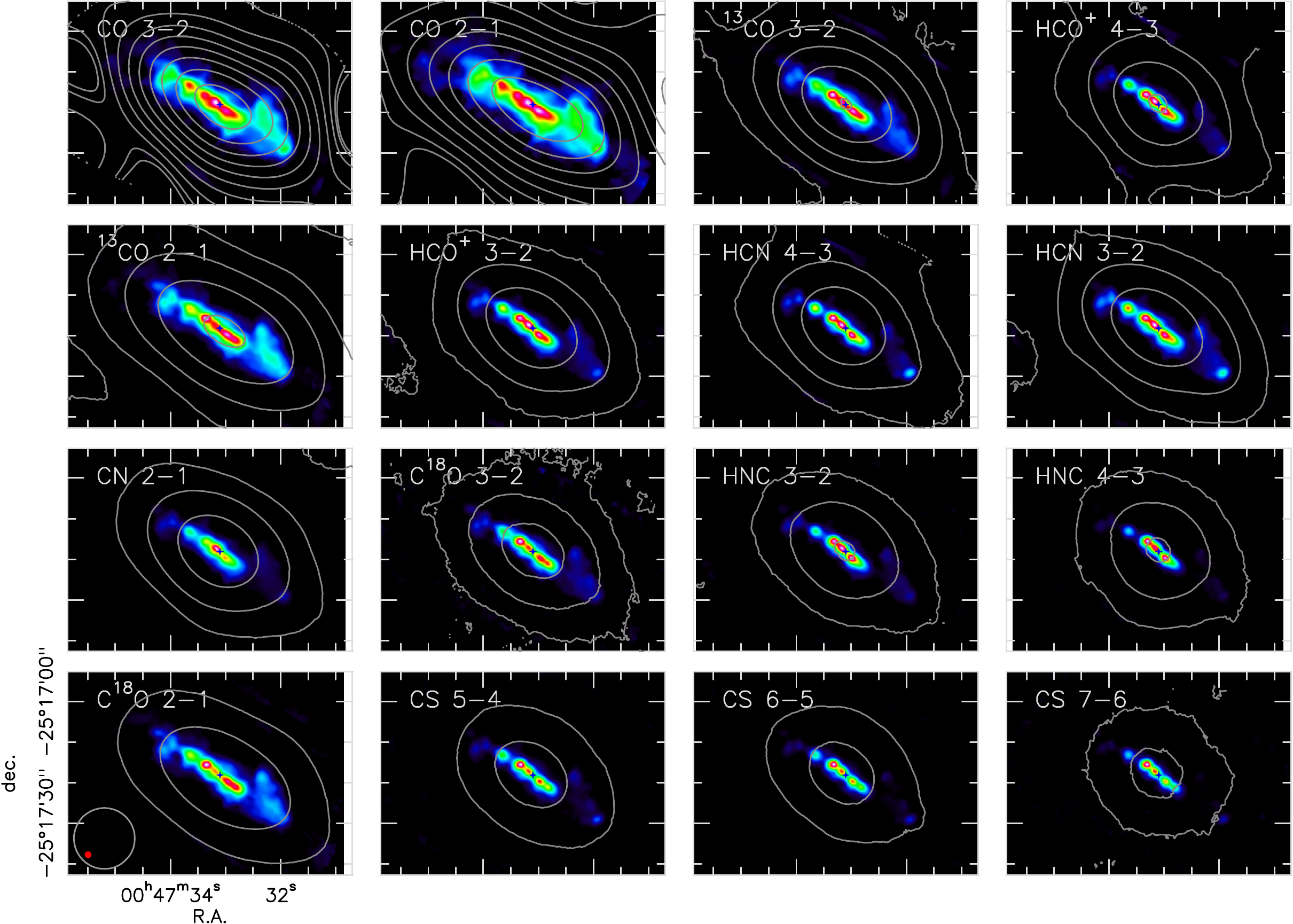}
\caption{Sample of integrated flux density moment 0 maps from 16 of the brightest molecular transitions in the covered frequency band. Each panel is labeled with the corresponding molecular species and transition. We point out that CO $1-0$ is not included since Band 3 is not covered by the ACA data.
In color the combined 12~m+7~m maps are shown where the color coding is adjusted for visibility of each individual species.
Grey contours show the 7~m integrated intensity images where the n-th contour level corresponds to $20~n^3\rm~Jy~km~s^{-1}~beam^{-1}$ for all species.
Species are ordered in decreasing order of integrated flux density from left to right and from top to bottom.
The panel in the lower left shows the reference coordinates and the beam size of the combined 12~m+7~m ($1.6''$, filled red circle) and 7~m data ($15''$, grey line), which are shared by all tunings. 
The blue cross at the center of every map shows the selected position (Sect.~\ref{Sec.sampleposition}) from which we extracted the spectra for modeling (Sect.~\ref{Sec.LTE}) and for presentation in Fig.~\ref{fig.fullsurvey}.
\label{fig.moment0maps}}
\end{figure*}

\subsubsection{Selected sample position}
\label{Sec.sampleposition}
In order to analyze the global molecular emission of the CMZ in NGC~253, we targeted %as representative position that of 
the %brightest 
peak of the molecular emission %from our
in the $15''$ resolution data.
To select this position, the pixel of peak emission was measured for each %To pick up the representative position of the whole CMZ, we searched for the pixel with the maximal emission from 
of the moment 0 maps from 16 of the brightest transitions in the survey shown in Fig.~\ref{fig.moment0maps}. %, as a representative position. 
The spectra analyzed in this article were extracted from the pixel at position 
$\alpha_{J2000}=00^h47^m33.10^s$, $\delta_{J2000}=-25^\circ17'18.1''$, corresponding to the average of all measured peak emission pixels and shown as a blue cross in Fig.~\ref{fig.moment0maps}. This position is just $\sim1.5\arcsec$ away from the giant molecular cloud analog Region 5 from \cite{Leroy2015} and $\sim1.4\arcsec$ from TH2 \citep{Turner1985a}.

We note that despite the different spatial distribution among species observed at the high resolution data, the brightest pixel in these ACA images agrees well among all images. Peak positions deviate from the average of all peak positions within an RMS of $0.8''$ ($\sim$2 pixels), and/or within $0.65''$ ($\sim 1.5$~pixels) if we exclude the two CO transitions, whose emission structure is significantly affected by opacity and overall extended emission.

%Pixel maximo 167-160
%rms 1.5 1.2 pixel
%rms 1.35 0.93 pixel sin CO
%pixel size 0.4"
% 

%Leroy 5
%Leroy2015 published:  00h47m33.2s -25d17m17.42s
%Leroy2015 from Adam:  00h47m33.2112s -25d17m17.916s
%Leroy published - given: ( 0.151901arcsec , -0.496arcsec )
% TH2 00h47m33s.179,−25:17:17.13

\subsubsection{Line identification and LTE modeling}
\label{Sec.LTE}

In this section we describe the overall criteria for molecular line identification and modeling. Further specific details on the fitting of individual species are provided in Appendix~\ref{Sec.AppendixFitDetails}. We emphasize that we did not analyze individual spectral features, but modelled the emission of all spectral lines within the surveyed band at once for each species. Therefore, line identification is done per molecule and not per transition, which is more robust and makes use of the broad frequency coverage in this work. Line flux densities reported in this paper are those from fits to the molecule transitions and accounts for line blending. We do not report the measured flux of each individual spectral feature.

Molecular emission has been identified and modeled under local thermodynamic equilibrium (LTE) conditions using MADCUBA\footnote{MADCUBA VERSION 6.0 (07/05/2018). \url{https://www.cab.inta-csic.es/madcuba/index.html}} \citep{Martin2019a} where physical parameters of column density, excitation temperature, radial velocity, line width, and source size are used to fit a modeled synthetic spectrum to the observations.
Spectroscopic parameters required for LTE modelling in MADCUBA, and therefore all the frequencies reported in this paper, are extracted from the CDMS \citep{Muller2001,Muller2005,Endres2016} and JPL \citep{Pickett1998} catalogs.

One of the visual advantages from fitting through synthetic spectra is that non-LTE emission or just spectral lines not properly fit under the LTE assumption are evidenced by line intensities significantly deviating from the LTE fit, which is often overlooked in the log-log representation in rotational diagrams. As explained in Appendix~\ref{Sec.AppendixFitDetails}, the LTE approximation appears to work well to describe the vast majority of observed spectral features modelled in this analysis, and the most obvious deviations from the fit are also identified.

The spectra extracted at the selected position (Sect.~\ref{Sec.sampleposition}) show a double peak profile which is the result of the convolution of the molecular distribution substructure observed at higher resolution (see Fig.~\ref{fig.moment0maps}). 
In principle, fitting a two component model would allow us to kinematically disentangle the molecular gas from both sides of the nucleus. 
However, the use of a multi-Gaussian fit to the overall spectrum adds little significance to the results (which will be better studied with the high resolution ALCHEMI image cubes), while increasing significantly the complexity of the modeling, even more so given that not all species show double peak profiles. 
For these reasons the modeling described here will consider a single Gaussian emission profile
%.
%Therefore, 
which suits the purpose of this article's focus on the global averaged properties of the molecular emission in the CMZ of NGC~253.

Among the fitted parameters, fitting the source size requires an accurate {\it a priori} excitation temperature for a molecule with enough optically thick transitions covering a wide range of energy levels and not too affected by spectral blending \citep{Martin2019a}. Since the broad line emission in our spectrum does not allow for an accurate constraint of the source size, we assumed a circular Gaussian equivalent source size of $5''$, based on available higher resolution observations \citep{Meier2015,Martin2019}. This parameter is not too relevant for the global relative properties analyzed in this article, as we consider a linear dependence of the derived column density with the source solid angle. 
%\footnote{We note that MADCUBA models the flux density emission as scaling with the solid angle of the source. Thus it does not consider beam dilution due to the coupling of the source extent and the synthesized beam. For our assumed $5''$ source size and the $15''$ beam size, this implies an underestimation of the column density of $4\%$, which can be considered negligible with respect to other uncertainties.}.
It could, however, be significant for lower values of the source size, when opacity starts playing a major role as discussed in Sect.~\ref{sec.isotopologues}. Only in the case of the main $^{12}$C$^{16}$O isotopologue did we find it necessary to assume a larger source size of $10''$ to be able to reproduce the observed flux densities. %\footnote{For a source size of $10''$ the non consideration of beam dilution might result in a $45\%$ underestimation of the CO column density. However, this may be less important than the opacity uncertainty.} 
A difference in the larger molecular emission extent is obvious from the contours in Fig.~\ref{fig.moment0maps} for the CO transitions, and to a lesser extent the $^{13}$CO images.

All other physical parameters mentioned above (column density, temperature, velocity, and width) were kept as free parameters when possible.
For each molecule, all available transitions within the covered frequency band were used for the fit, except those heavily blended or not detected above the noise level ($<3\sigma$).
%dominated by the noise (either because they fall within high noise frequency ranges or because too faint with respect to the noise).
In some cases, for species with many transitions, only a subset of the brightest unblended transitions were used to avoid the fit being dominated by faint transitions too close to the noise level or residual emission from other species.
When line blending or signal-to-noise did not allow fitting the line velocity and width, parameters were fixed to $v_{LSR}=230$~\kms~ and $\Delta~v_{1/2}=150$~\kms, which are the average fitted parameters to the brighter transitions. Similarly, when detected transitions did not allow the excitation temperature to be determined, the excitation temperature was set to $T_{ex}$=15~K, which is the median of the measured temperatures in all species allowing such a fit (ranging between 5 and 60~K). 
%This temperature was chosen because despite the temperatures derived range from 5 to 60~K, the median of these values is $\sim16$~K.
%Despite the considerations above
Additionally, whenever any of these parameters (with the exception of the column density) were derived from a given species, these values were used to obtain more appropriate parameters to be fixed in the fit to
%then fixed to the same values 
their rarer isotopologues or isomers. This allows for a better relative abundance comparison between related species, while no biases resulting from this assumption are obvious in our derived values.

The final model includes 146 species with a total of 42121 transitions. However, only 78 species are considered firmly or tentatively detected, accounting for 1790 transitions with flux densities above 2~mJy in the model. We note that 2~mJy corresponds to $1\sigma$ at the lowest frequencies and $\sim(1/3)\sigma$ for the majority of the survey. However, the sum of faint transitions (even below the noise level) is relevant since in some cases they add up to detectable features or may significantly contaminate other transitions. The detected molecule count includes isotopologues and vibrational states. In addition multiple hydrogen and helium recombination lines from $\rm Hn\alpha$, $\rm Hn\beta$, and $\rm Hen\alpha$ were also detected throughout the survey but are not discussed in this article. 

The criterium for detection of a given species has been based on its LTE model and fit results. A species has been considered detected if all the detectable transitions above $5~\sigma$ (according to the LTE model), which are not blended with brighter transitions, are detected in our data. Additionally we required the convergence of the fitting algorithm within MADCUBA to avoid subjective biases.

Table~\ref{tab.fitmolecparams} shows the result from the fit to all detected or tentatively detected species in this survey. As previously indicated in Sect.~\ref{sec.relativeflux}, reported uncertainties do not include calibration uncertainties but statistical uncertainty on the fit to the spectra.
Values with no errors in Table~\ref{tab.fitmolecparams} represent parameters that were fixed during the fitting process. 

\begin{figure}
\begin{center}
\includegraphics[width=0.5\textwidth]{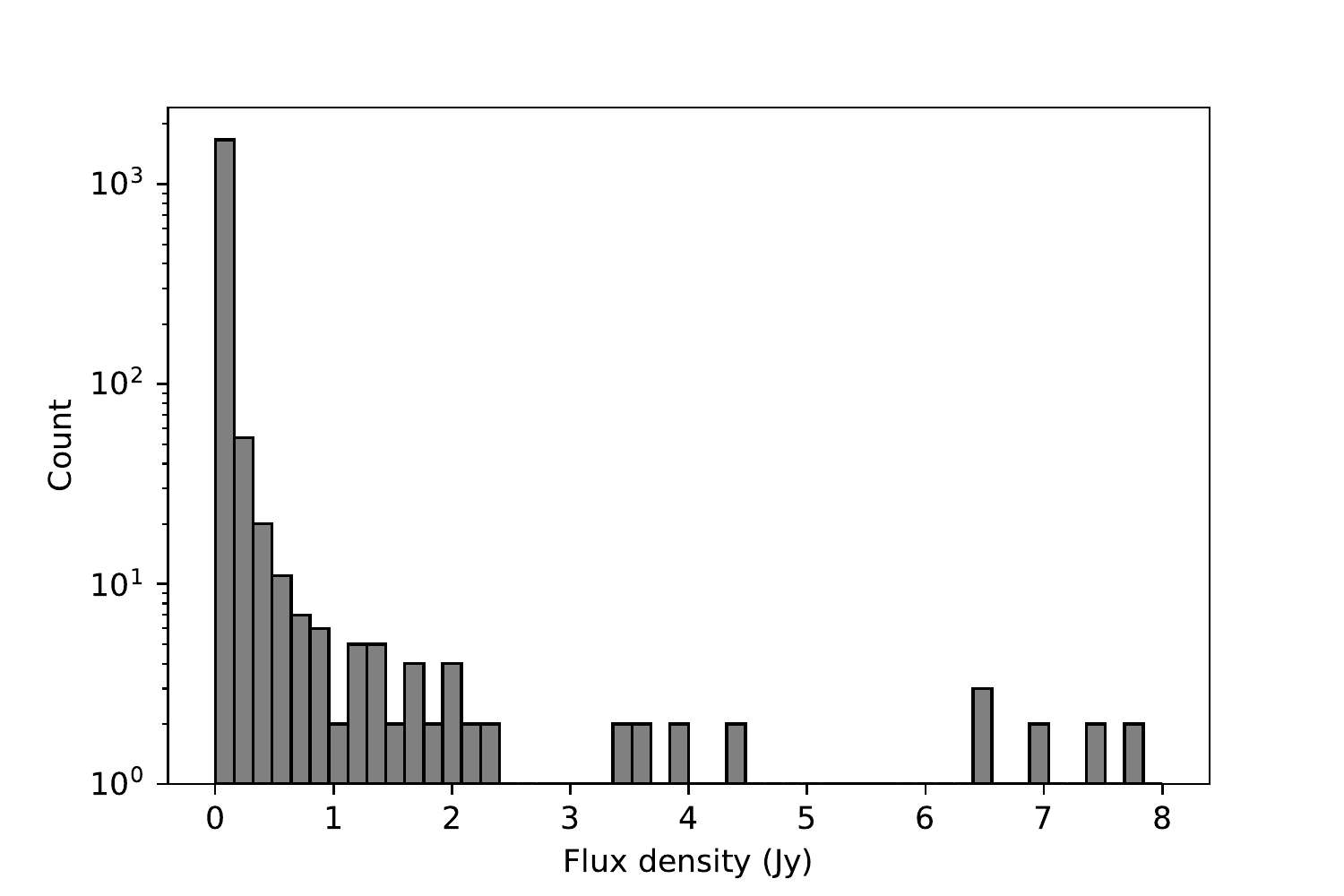}
\includegraphics[width=0.5\textwidth]{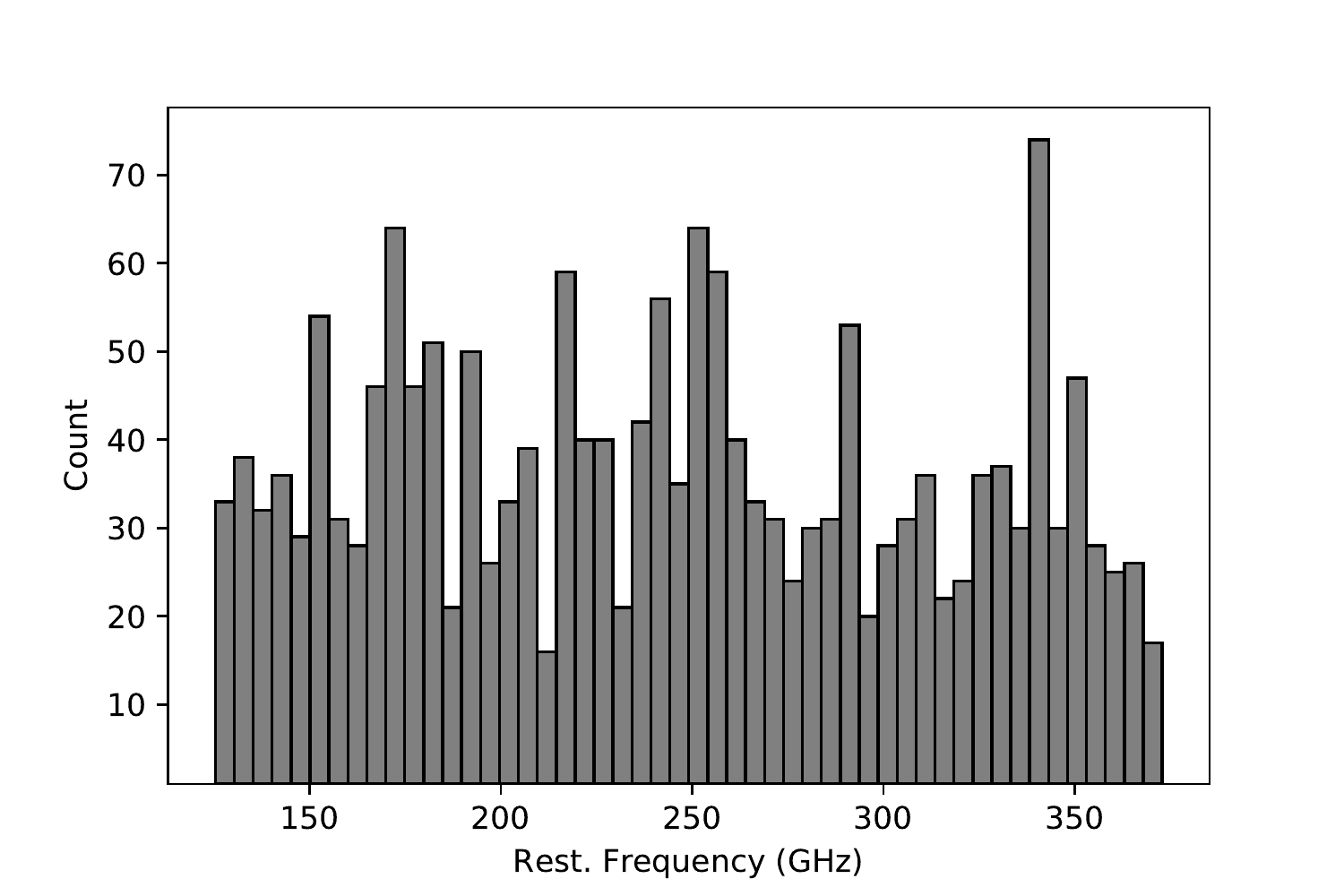}
\caption{
(Top) Histogram showing the number of spectral lines above 2~mJy in the model in bins of flux density. Lines between 2~mJy and 10~Jy are considered. The three spectral features with flux densities above 10~Jy are not included in this diagram.
(Bottom) Histogram showing the number of spectral lines above 2~mJy in frequency bins of $\sim5$~GHz width.
\label{fig.histoflux}}
\end{center}
\end{figure}

Fig.~\ref{fig.histoflux} shows a graphical summary of the number count in flux density bins and in narrow 5~GHz frequency bins of the transitions in the model.
Table~\ref{tab.linecontribution} also includes the density of spectral lines over wider 50~GHz frequency ranges.

Finally, we point out that there are still a number of clearly detected spectral features which are not accounted for by our model as seen in the figures in Appendix~\ref{Sec.AppendixFullSpectrum}. These unidentified features may stem from emission out of LTE or the effect of multiple molecular components \citep{Aladro2011} or from species not included in our model. We evaluated each of these features for emission from different species, but a model to the candidate species could not be found to fit across the whole frequency range covered.

%\begin{table}
%\caption{text}
%\label{table.physicalparameters}
%\centering
%\onecolumn

\onecolumn
%\begin{ThreePartTable}
%\begin{TableNotes}
%\footnotesize
%{\bf Notes.} 
%Reported uncertainties correspond to the statistical uncertainty of the fit.
%Fixed parameters in the fit are presented as values with no errors. 
%Physical parameters for CO are obtained assuming a source size of $10''$, different from the size of $5''$ used %for all other molecules (see Sect.~\ref{Sec.LTE} for details)
%This table was machine generated and values were formatted to a fix number of decimal places and not to %significant digits. 
%\tablefoottext{a}{The reported column density $log_{10} N=A(B)$ correspond to $N=10^A\pm 10^B$.}
%\tablefoottext{b}{Molecules are ordered in descending peak flux density of their brightest transition
%, listed in column ($S_0$)
%.}
%\tablefoottext{c}{Opacity at the line center.}.  
%\end{TableNotes}
\begin{longtable}{llllllr}
\caption{Derived physical parameters from LTE fit to the observed spectra}\label{tab.fitmolecparams}\\
\hline
\hline
Formula & $log_{10}~N$~\tablefootmark{a} &  $T_{ex}$ &  $v_{lsr}$  &  $\Delta v_{1/2}$ &  $S_0$~\tablefootmark{b}           &    $\tau_0$~\tablefootmark{c} \\
        & (cm$^{-2}$)   & (K)       & (\kms)      & (\kms)            & ($\rm Jy~beam^{-1}$) &             \\
\hline
\endfirsthead 
\caption{continued.}\\
\hline
\hline
Formula & $log_{10}~N$~\tablefootmark{a} &  $T_{ex}$ &  $v_{lsr}$  &  $\Delta v_{1/2}$ &  $S_0$\tablefootmark{b}           &    $\tau_0$\tablefootmark{c} \\
        & (cm$^{-2}$)   & (K)       & (\kms)      & (\kms)            & ($\rm Jy~beam^{-1}$) &             \\
        \hline
\endhead
\hline
\multicolumn{7}{l}{{{\bf Notes.} Continued on next page}}\\
\endfoot
\hline
%\insertTableNotes % tell LaTeX where to insert the contents of "TableNotes"
\multicolumn{7}{p{0.9\linewidth}}{{
{\bf Notes.} 
Reported uncertainties correspond to the statistical uncertainty of the fit.
Fixed parameters in the fit are presented as values with no errors. 
Physical parameters for CO are obtained assuming a source size of $10''$, different from the size of $5''$ used for all other molecules (see Sect.~\ref{Sec.LTE} for details)
This table was machine generated and values were formatted to a fix number of decimal places and not to significant digits. 
\tablefoottext{a}{The reported column density $log_{10} N=A(B)$ correspond to $N=10^A\pm 10^B$.}
\tablefoottext{b}{Molecules are ordered in descending peak flux density of their brightest transition
%, listed in column ($S_0$)
.}
\tablefoottext{c}{Opacity at the line center.}  
}}\\
\endlastfoot
%\begin{tabular}{llllllr}
%\toprule
%               Formula &      latexN &       latexT &       latexV &       latexW &          latexS &    tau \\
%\midrule
                    CO &  18.4(17.3) &    19.6(0.5) &   252.7(1.1) &   149.1(3.7) &  112.203(3.943) &  2.608 \\
                $^{13}$CO &  18.1(17.0) &    14.6(0.5) &   237.2(1.4) &  150.0(...) &     13.3(0.789) &  1.347 \\
           HCO$^+$ &  14.8(13.4) &    14.6(0.3) &   233.1(1.0) &   151.0(2.2) &    7.789(0.396) &  0.499 \\
                   HCN &  15.1(13.7) &    12.7(0.2) &   229.2(0.9) &   150.8(2.1) &     6.497(0.35) &  0.547 \\
                 C$^{18}$O &  17.4(16.2) &    15.7(0.9) &   235.1(1.9) &  150.0(...) &    4.462(0.393) &  0.252 \\
                   HNC &  14.7(13.2) &    12.9(0.2) &   227.2(1.0) &   134.8(2.2) &    3.904(0.141) &  0.435 \\
              CN &  15.7(14.4) &    10.2(0.2) &   232.6(1.1) &   148.6(2.0) &    2.129(0.117) &  0.249 \\
                    CS &  15.1(13.5) &    16.0(0.3) &   225.3(1.2) &  150.0(...) &    1.887(0.063) &  0.156 \\
                  N$_2$H$^+$ &  14.0(12.5) &    14.7(0.3) &   226.3(1.1) &   132.3(2.6) &     1.34(0.056) &  0.099 \\
                   C$_2$H &  15.8(14.3) &    12.4(0.2) &   226.9(1.1) &   143.0(2.8) &    1.153(0.042) &  0.124 \\
                 CH$_3$OH &  15.4(13.8) &    24.1(0.6) &   235.3(2.9) &  150.0(...) &    0.728(0.031) &  0.018 \\
                 C$^{17}$O &  16.4(15.2) &    16.4(1.1) &   239.1(2.3) &  150.0(...) &    0.618(0.059) &  0.028 \\
                  H$_2$CO &  14.6(13.3) &    14.4(0.4) &   226.5(2.3) &   163.5(5.4) &    0.567(0.041) &  0.038 \\
                   H$_2$S &  15.1(13.6) &    55.6(1.7) &   225.5(3.2) &  150.0(...) &    0.461(0.025) &  0.004 \\
                    NO &  16.5(15.2) &    54.0(3.1) &   225.2(3.1) &  150.0(...) &     0.374(0.02) &  0.003 \\
              H$^{13}$CO$^+$ &  13.5(12.3) &    11.3(0.4) &   233.7(2.5) &  150.0(...) &    0.339(0.024) &  0.041 \\
               H$^{13}$CN &  13.7(12.1) &   12.4(...) &   224.1(2.2) &  150.0(...) &    0.299(0.007) &  0.021 \\
                    SO &  14.6(13.1) &    25.8(0.6) &   219.0(1.5) &  150.0(...) &      0.27(0.01) &  0.007 \\
                 HC$_3$N &  14.6(14.6) &   33.1(17.7) &  211.3(46.2) &  150.0(...) &    0.249(0.285) &  0.018 \\
                  HOC$^+$ &  13.1(11.8) &    15.2(0.5) &   222.0(2.1) &  150.0(...) &    0.248(0.021) &  0.012 \\
                 C$^{34}$S &  14.1(12.8) &    18.3(0.5) &   220.0(2.2) &  150.0(...) &    0.234(0.012) &  0.015 \\
                c-C$_3$H$_2$ &  14.4(14.2) &    17.8(7.2) &  240.4(42.3) &  150.0(...) &    0.208(0.151) &  0.017 \\
            SiO &  13.7(13.0) &    17.2(1.7) &   207.9(8.3) &  150.0(...) &    0.193(0.045) &  0.012 \\
                 CH$_2$NH &  14.6(13.3) &    16.8(0.9) &   232.9(4.0) &  150.0(...) &     0.191(0.01) &  0.008 \\
                  H$_3$O$^+$ &  15.5(14.5) &   15.0(...) &   207.1(5.7) &  120.3(13.4) &    0.179(0.026) &  0.011 \\
                HN$^{13}$C &  13.4(12.2) &    11.5(0.4) &   224.0(2.1) &  135.0(...) &    0.173(0.012) &  0.020 \\
                HCN$_{v_2=1}$ &  14.8(14.3) &  300.0(...) &  230.0(...) &  150.0(...) &    0.151(0.049) &  $<0.001$ \\
                CH$_3$C$_2$H &  15.6(13.8) &    43.0(0.5) &   243.3(1.0) &   145.0(2.5) &    0.144(0.004) &  0.003 \\
                  HNCO &  15.0(14.0) &    29.4(1.7) &   232.2(5.5) &  150.0(...) &     0.13(0.012) &  0.005 \\
                %Halpha &  16.0(14.5) &   50.0(...) &   233.0(2.5) &   150.9(5.8) &     0.11(0.006) &  0.001 \\
                 CH$_3$CN &  13.9(13.4) &    38.6(8.8) &  246.8(21.1) &  150.0(...) &    0.086(0.026) &  0.002 \\
               HC$^{18}$O$^+$ &  12.8(12.2) &   14.5(...) &  233.0(...) &  150.0(...) &     0.07(0.016) &  0.004 \\
                HC$^{15}$N &  12.9(11.9) &   12.6(...) &  230.0(...) &  150.0(...) &    0.064(0.006) &  0.006 \\
                 %Hbeta &  15.7(14.5) &   50.0(...) &  232.0(...) &  150.0(...) &    0.059(0.004) &  0.001 \\
                  H$_2$CS &  14.1(13.1) &    30.3(3.1) &   224.8(6.6) &  150.0(...) &    0.058(0.009) &  0.002 \\
                   CO$^+$ &  14.1(13.5) &     8.3(0.9) &   231.7(9.9) &  150.0(...) &    0.058(0.016) &  0.014 \\
                $^{13}$CN &  14.4(13.4) &   10.0(...) &  223.8(11.1) &  150.0(...) &    0.057(0.006) &  0.006 \\
                CH$_3$NH$_2$ &  14.6(14.4) &   15.0(...) &  230.0(...) &  150.0(...) &    0.056(0.038) &  0.003 \\
                $^{13}$CS &  13.7(12.7) &    12.2(0.6) &   216.1(4.2) &  150.0(...) &    0.054(0.006) &  0.010 \\
           C$^{33}$S &  13.4(13.1) &   18.0(...) &  225.0(...) &  150.0(...) &    0.047(0.024) &  0.003 \\
             HC$_3$N$_{v_7=1}$ &  13.5(12.5) &  300.0(...) &  212.0(...) &  150.0(...) &    0.047(0.005) &  $<0.001$ \\
              HNC$_{v_2=1}$ &   13.8(...) &  300.0(...) &  230.0(...) &  150.0(...) &      0.046(...) &  $<0.001$ \\
                   OCS &  14.7(13.9) &    60.6(8.3) &  230.0(...) &  130.0(...) &    0.039(0.008) &  0.001 \\
                    NS &  14.2(13.0) &    16.2(0.7) &  225.0(...) &   143.3(6.9) &    0.039(0.003) &  0.004 \\
                 H$_2$C$_2$N &  14.1(13.6) &   15.0(...) &  230.0(...) &  150.0(...) &    0.038(0.012) &  0.007 \\
              H$_2$$^{13}$CO &  13.3(12.7) &    16.5(2.0) &  230.0(...) &  162.0(21.1) &    0.035(0.011) &  0.002 \\
               %Healpha &  15.4(14.6) &   50.0(...) &  232.0(...) &  148.0(...) &    0.035(0.005) &  $<0.001$ \\
                  HCS$^+$ &  13.3(12.5) &   15.0(...) &  230.0(...) &  150.0(...) &    0.029(0.004) &  0.002 \\
                 HOCO$^+$ &  14.0(13.1) &   15.0(...) &  230.0(...) &  150.0(...) &    0.028(0.004) &  0.005 \\
             $^{13}$C$^{18}$O &  15.4(14.3) &   15.6(...) &  230.0(...) &  150.0(...) &    0.026(0.002) &  0.001 \\
                  C$_3$H$^+$ &  13.5(12.9) &   15.0(...) &  230.0(...) &  150.0(...) &    0.025(0.007) &  0.004 \\
                   SO$_2$ &  13.9(13.6) &    11.1(9.0) &  230.0(...) &  150.0(...) &    0.022(0.019) &  0.002 \\
                C$_2$H$_5$OH &  14.4(14.0) &    17.6(6.7) &  230.0(...) &  150.0(...) &    0.021(0.012) &  0.001 \\
                 NH$_2$CN &  13.1(12.6) &   48.5(14.3) &  230.0(...) &  150.0(...) &    0.021(0.006) &  $<0.001$ \\
         $^{29}$SiO &  12.8(11.9) &   17.0(...) &  230.0(...) &  150.0(...) &     0.02(0.003) &  0.001 \\
                   HCO &  14.1(13.7) &     5.1(1.1) &  230.0(...) &  150.0(...) &     0.02(0.009) &  0.020 \\
                $^{34}$SO &  13.5(12.9) &   26.0(...) &  230.0(...) &  150.0(...) &    0.019(0.005) &  0.001 \\
      $^{13}$CH$_3$OH &  13.6(13.4) &   24.0(...) &  235.0(...) &  150.0(...) &    0.018(0.011) &  0.001 \\
           CH$_3$SH &  14.6(14.9) &   15.0(...) &  230.0(...) &  150.0(...) &    0.017(0.031) &  0.001 \\
               H$^{15}$NC &  12.3(11.7) &   13.0(...) &  227.0(...) &  135.0(...) &    0.016(0.004) &  0.001 \\
             %HC$_3$N,v7=2 &   13.5(...) &  300.0(...) &  212.0(...) &  150.0(...) &      0.016(...) &  $<0.001$ \\
                   C$_2$S &  13.7(13.5) &   35.1(14.9) &  230.0(...) &  150.0(...) &    0.016(0.012) &  0.001 \\
                  HOCN &  13.0(12.4) &   29.6(...) &  230.7(...) &  150.0(...) &    0.015(0.003) &  0.001 \\
                 H$_2$C$_2$O &  14.4(14.2) &   15.0(...) &  230.0(...) &  150.0(...) &    0.015(0.009) &  0.002 \\
               $^{13}$CCH &  14.2(13.8) &   12.4(...) &  226.9(...) &  143.0(...) &    0.013(0.005) &  0.001 \\
             %HC$_3$N,v6=1 &   13.5(...) &  300.0(...) &  212.0(...) &  150.0(...) &      0.012(...) &  $<0.001$ \\
                NH$_2$CHO &  13.8(13.6) &   15.0(...) &  230.0(...) &  150.0(...) &    0.012(0.007) &  0.001 \\
               HC$_5$N &  14.6(14.0) &   32.0(...) &  230.0(...) &  150.0(...) &     0.01(0.002) &  0.001 \\
          Si$^{17}$O &  12.5(11.8) &   17.0(...) &  230.0(...) &  150.0(...) &     0.01(0.002) &  0.001 \\
                   SO$^+$ &  13.8(13.1) &   15.0(...) &  230.0(...) &  150.0(...) &     0.01(0.002) &  0.001 \\
                l-C$_3$H$_2$ &  13.2(12.8) &   15.0(...) &  230.0(...) &  150.0(...) &    0.008(0.003) &  0.001 \\
               C$^{13}$CH &  13.9(13.4) &   12.4(...) &  226.9(...) &  143.0(...) &    0.008(0.002) &  0.001 \\
             HC$^{13}$CCN &  13.2(13.4) &   33.0(...) &  212.2(...) &  150.0(...) &    0.007(0.011) &  $<0.001$ \\
                CH$_3$CHO &  14.2(13.2) &   15.0(...) &  230.0(...) &  150.0(...) &    0.007(0.001) &  0.001 \\
            CH$_3$$^{13}$CCH &  14.3(13.8) &   43.0(...) &  242.3(...) &  143.7(...) &    0.007(0.003) &  $<0.001$ \\
             H$^{13}$CCCN &  13.2(13.0) &   33.0(...) &  212.2(...) &  150.0(...) &    0.007(0.004) &  $<0.001$ \\
            CH$_3$C$^{13}$CH &  14.2(13.4) &   43.0(...) &  242.3(...) &  143.7(...) &    0.006(0.001) &  $<0.001$ \\
        %HC$_3$N,v5=1/v7=3 &   13.5(...) &  300.0(...) &  212.0(...) &  150.0(...) &      0.006(...) &  $<0.001$ \\
                   C$_4$H &  15.2(14.5) &   15.0(...) &  230.0(...) &  150.0(...) &    0.005(0.001) &  0.001 \\
             HCC$^{13}$CN &  13.3(13.4) &   33.0(...) &  212.2(...) &  150.0(...) &    0.005(0.006) &  $<0.001$ \\
         $^{30}$SiO &  12.1(11.8) &   17.0(...) &  230.0(...) &  150.0(...) &    0.005(0.002) &  $<0.001$ \\
          %HC$_3$N,v6=v7=1 &   13.5(...) &  300.0(...) &  212.0(...) &  150.0(...) &      0.004(...) &  $<0.001$ \\
                 HCOOH &  13.8(13.4) &   15.0(...) &  230.0(...) &  150.0(...) &    0.003(0.001) &  0.001 \\
            $^{13}$CH$_3$C$_2$H &  13.9(13.7) &   43.0(...) &  242.3(...) &  143.7(...) &    0.003(0.002) &  $<0.001$ \\
             %HC$_3$N,v4=1 &   13.5(...) &  300.0(...) &  212.0(...) &  150.0(...) &      0.002(...) &  $<0.001$ \\
                HC$_3$HO &  13.9(13.6) &   15.0(...) &  230.0(...) &  150.0(...) &    0.002(0.001) &  $<0.001$ \\
     %HC$_3$N,v7=4/v5=v7=1 &   13.5(...) &  300.0(...) &  212.0(...) &  150.0(...) &      0.002(...) &  $<0.001$ \\
             %HC$_3$N,v6=2 &   13.5(...) &  300.0(...) &  212.0(...) &  150.0(...) &      0.001(...) &  $<0.001$ \\
          %HC$_3$N,v4=v7=1 &   13.5(...) &  300.0(...) &  212.0(...) &  150.0(...) &      0.001(...) &  $<0.001$ \\
 %HC$_3$N,v4=1,v7=2/v5=2\textasciicircum 0 &   13.5(...) &  300.0(...) &  212.0(...) &  150.0(...) &        0.0(...) &  $<0.001$ \\
             %HC$_3$N,v3=1 &   13.5(...) &  300.0(...) &  212.0(...) &  150.0(...) &        0.0(...) &  $<0.001$ \\
             %HC$_3$N$_{v_2=1}$ &   13.5(...) &  300.0(...) &  212.0(...) &  150.0(...) &        0.0(...) &  $<0.001$ \\
%\bottomrule
%\end{tabular}

\end{longtable}
%\end{ThreePartTable}
\twocolumn

% NEW DETECTIONS
%\citep{McGuire2018}

\subsection{New extragalactic molecular detections}
\label{sec.newdetections}
Despite the moderate sensitivity of the ACA observations, the broad frequency coverage and the bandpass stability allowed us to probe a number of newly detected species in the extragalactic ISM.

In this paper we report the first extragalactic detections of H$_2^{13}$CO, ethanol (C$_2$H$_5$OH), $^{13}$CCH, C$^{13}$CH, HOCN, the three $^{13}$C isotopologues of CH$_3$CCH, propynal (HC$_3$HO), and tentatively Si$^{17}$O (see discussion in Sect.~\ref{sec.isotopologuesOxygen}). Specific details on the fit to C$_2$H$_5$OH, HOCN, and HC$_3$HO are provided in Appendix~\ref{Sec.AppendixFitDetailsNew}.
Additionally, we confirm previous tentative detections of H$^{15}$NC \citep{Muller2006}, $^{13}$CH$_3$OH (tentatively detected towards NGC~253 by \citealt{Mart'in2009b}, and recently reported towards PKS1830-211 by \citealt{Muller2021}), and HC$_5$N \citep{Aladro2015,Costagliola2015}.
These detections consist of isotopologues and isomers of previously detected species, as well as new complex organic molecules \citep[COMs, 6+ atoms, ][]{Herbst2009}.
Species like formic acid (HCOOH), very recently reported toward an absorption system \citep{Tercero2020}, is detected for the first time in emission towards NGC~253.
We also confirm the detection of the elusive methylamine \citep[$\rm CH_3NH_2$,][]{Bogelund2019}, first detected in the extragalactic ISM in absorption by \citep{Muller2011} and so far only tentatively identified towards NGC~253 in emission by \citet{Meier2015}.

\begin{figure}
\begin{center}
\includegraphics[width=0.5\textwidth]{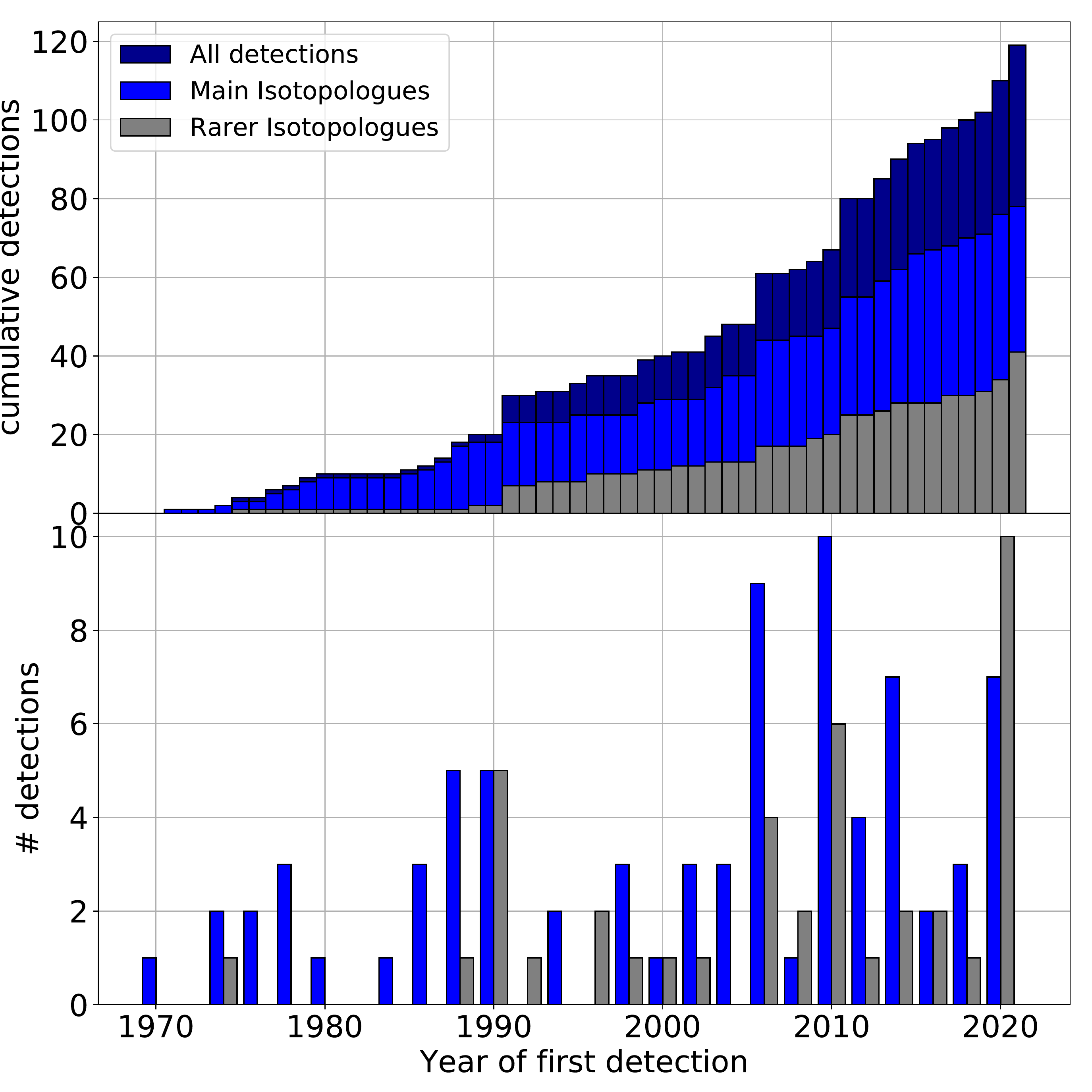}
\caption{Chronology of extragalactic molecular detections including those reported in this work. Detections of main and rarer isotopologue substitutions in blue and grey respectively, with the total number of detections, not considering tentative reports, being displayed in dark blue. One and two year bins are used for top and bottom panel histograms, respectively.
\label{fig.detections}}
\end{center}
\end{figure}

The chronological evolution of the cumulative number of species detected as well as the yearly detections are summarized in Fig.~\ref{fig.detections}, where the detections reported in our work are included. Details on the updated chronology of first extragalactic molecular detections are provided in Appendix~\ref{Sec.AppendixCensus}.

\section{Discussion: NGC~253 as starburst molecular template}

% Give a menu
Even %at low 
with its moderate angular resolution, 
%the information available on 
the ACA data set %are already rich in terms of abundances and excitation. 
provides important information on the abundance and excitation of the gas in the CMZ of NGC~253, which could not be attained by previous surveys with less complete frequency coverage \citep{Mart'in2006,Aladro2015}.
%Thus, in this section we focused on a few highlights that can be discussed based on the abundances analysis directly derived from the LTE fit in Sect.~\ref{Sec.LTE}.
In this section, we highlight some scientific results that make use of the wide frequency coverage of the ALCHEMI data. This unique frequency coverage allows for multi-transition analysis of a variety of molecular species.
%and multi-species results presented here. 
%use of the advantages of the unique ALCHEMI data set in terms of wide frequency coverage  some results based on the abundance analysis derived from the LTE fit described in Sect.~\ref{Sec.LTE}.
%
The upcoming suite of papers based on ALCHEMI data will also make use of this unique wide band data set and will provide a deeper analysis of these and other scientific questions.

%{\bf Skip hereafter}
%1) The detection of vibrational emission of different species is discussed in Sect.~\ref{sec.vibemission} in the context of infrared pumping in a dense molecular environment; 2) Measurements of isotopic atomic ratios from multiple molecular proxies are discussed and put in context with prior literature reports in Sect.~\ref{sec.isotopologues}; 3) The detection of complex organic molecules and a comparison to their Galactic abundances are described in Sect.~\ref{sec.complexorganics}; 4) finally Sect.~\ref{sec.opacity} discuss the determination of opacity based on CN transitions.

%{\bf Seb: I think we can save space and skip the previous list, text after "Skip hereafter". Just say "Here are some highlights".}

\subsection{Extragalactic starburst low resolution molecular template}
\label{sec.template}

One immediate use of the wideband observations in this article is to serve as a molecular template for extragalactic starbursting environments.
The large number of molecules detected in this study is a consequence of both the depth of the ALCHEMI data set and the intrinsic brightness of NGC~253. In fact, we obtained a spectral dynamic range between $\sim60000$ and $\sim6000$, as derived by comparing the flux density of the brightest transition in the survey, CO $3-2$ (see Table~\ref{tab.fitmolecparams}), to the ACA-achieved noise level at Bands 4 and 7 (Sect.~\ref{sec.analysis}), respectively.

% CO 112 Jy
% Sensitivity 1.8 to 19.4 mJy
% Dynamic range 
% 62222 - 5773

\begin{figure*}
\begin{center}
\includegraphics[width=0.49\textwidth]{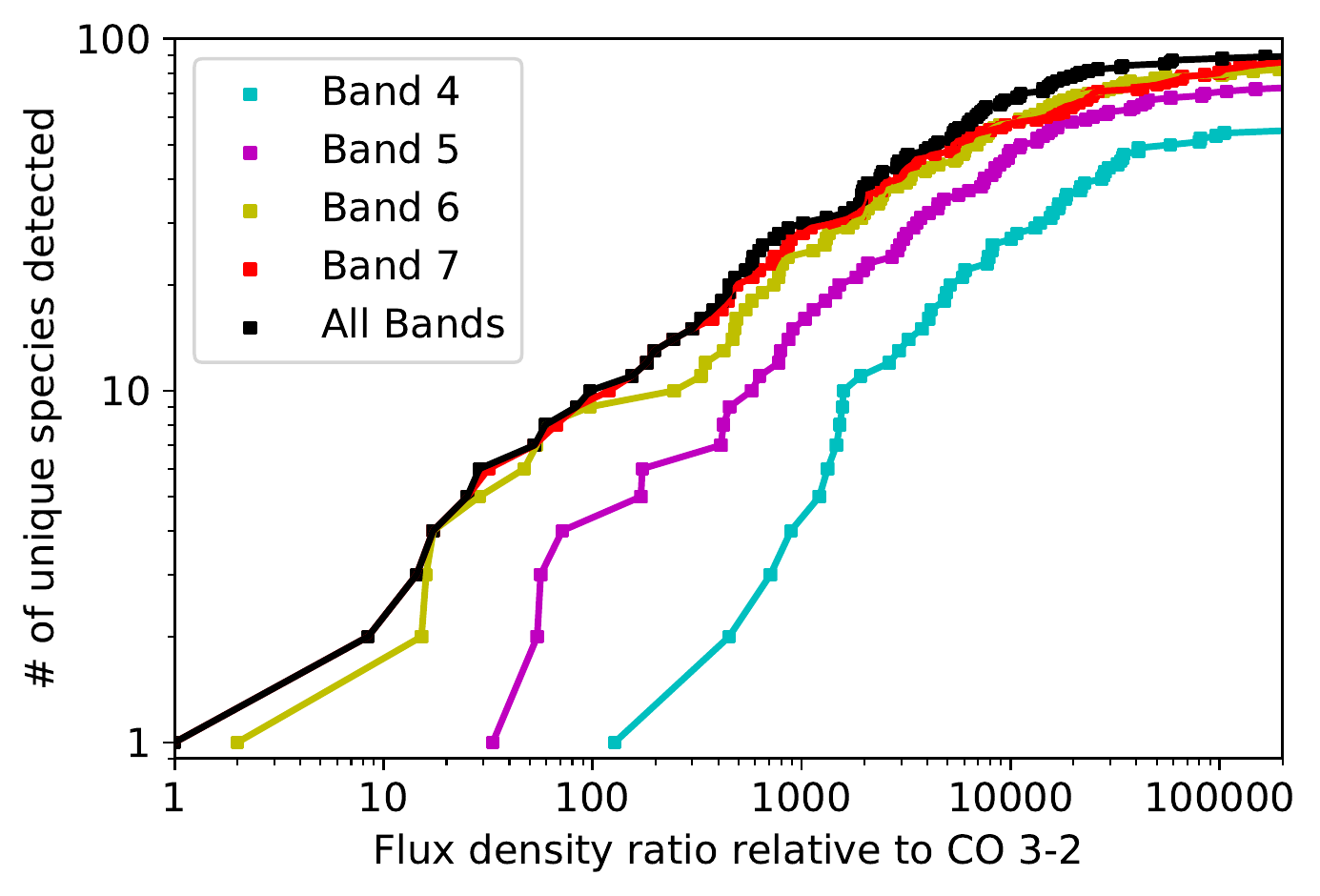}
\includegraphics[width=0.49\textwidth]{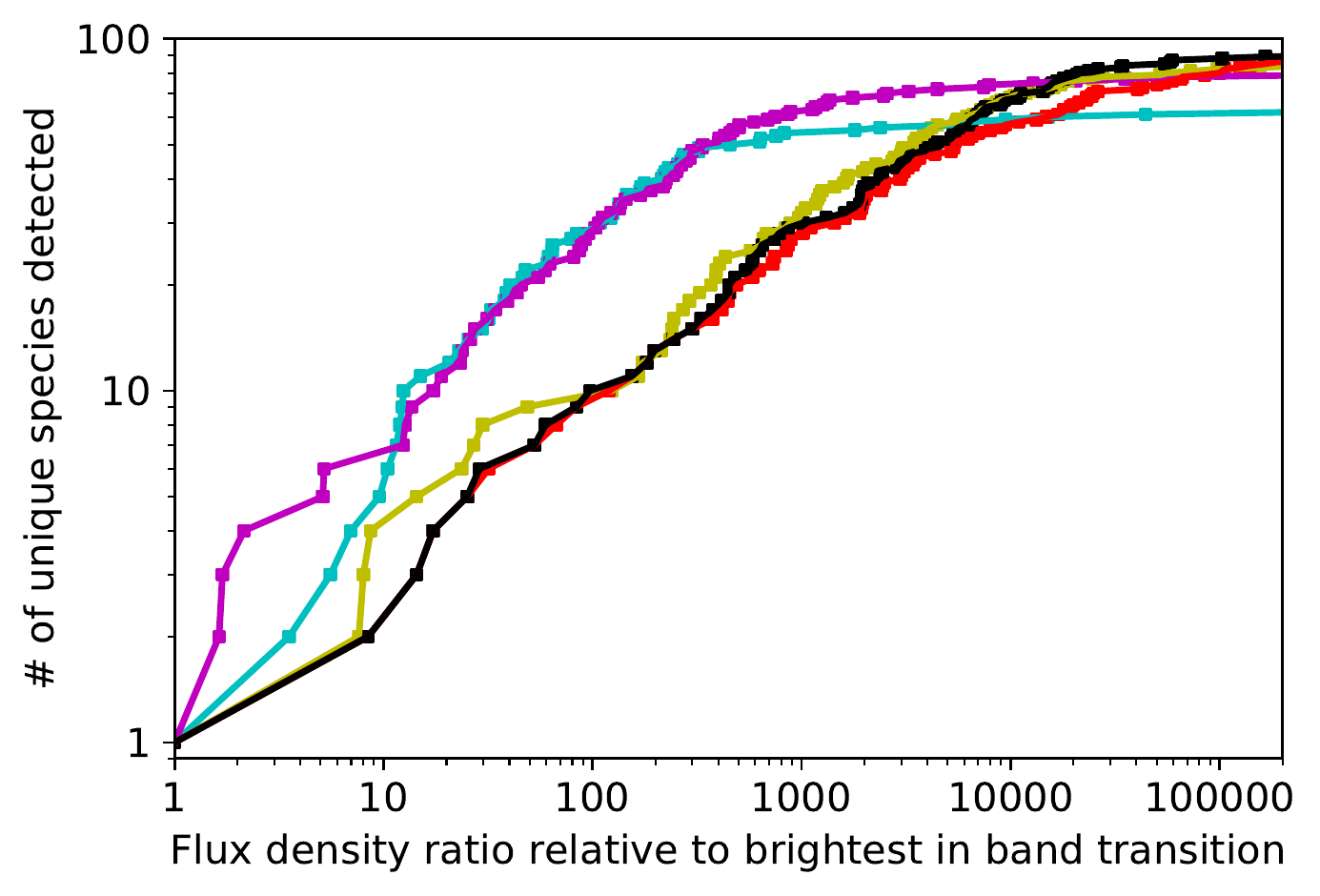}
\caption{Number of individual detectable species as a function of the flux density level relative to the brighter transitions detected in the survey per band.
(Left) Flux density levels are referred to the CO $3-2$ (112 Jy~beam$^{-1}$) as the brightest transition detected in the whole spectral range covered. (Right) Flux densities are referred to the brightest transition detected on each band. This is CS $3-2$ (0.9 Jy~beam$^{-1}$ in Band 4), HCO$^+~2-1$ (3 Jy~beam$^{-1}$ in Band 5), CO $2-1$ (56 Jy~beam$^{-1}$ in Band 6), and CO $3-2$ (in Band 7).
\label{fig.expecteddetections}}
\end{center}
\end{figure*}

Fig.~\ref{fig.expecteddetections} presents the number of detected unique species as a function of the flux level relative to the CO $3-2$ transition (left panel) or to the brightest transition within a given Band (right panel).
The data presented in Fig.~\ref{fig.expecteddetections} are based on modelled intensities from individual transitions and not spectral features (Sect.~\ref{Sec.LTE}). Therefore the number count of species is conservative, since spectral features composed of multiple transitions (e.g., species with unresolved hyperfine structure) will rise above the noise before what is estimated based on the flux density of the brightest transition of any given moleucule. The number of detected species in Fig.~\ref{fig.expecteddetections} goes beyond the number of confirmed detections reported in Sect.~\ref{Sec.LTE}. This is because the number of species in Figure~\ref{fig.expecteddetections} includes recombination lines and species in vibrational states, as they are both considered to be relevant unique detections.

The data presented in Figure~\ref{fig.expecteddetections} can be used to roughly estimate the expected level of molecular complexity achievable in a high redshift "starbursting" object as a function of the sensitivity of the observations. Of course, the main assumption relies on similar abundance and excitation conditions to those in NGC~253, which may not hold for all starburst environments \citep{Aladro2015}. Additionally, larger line widths would hamper the detectability of species due to blending.

% From Spilker2014 Figure 2
% 1 sigma rms 0.15 mJy
% CO 3-2 7.3 mJy
% peak to rms = 50 ~ 3 sigma 16
% from luminosity reported 25

As a test bed to show the template potential of our data set we used the stacked spectrum derived from 22 %SPT 
high-z sources by \citet{Spilker2014}, over the redshift range $z =2.0-5.7$, and covering the frequency range from $\sim250$ to 800~GHz. Based on their Figure 2, the CO $3-2$ line was detected at a signal-to-noise of $\sim50$. That is, observations should be able to detect emission lines $\sim 17x$ fainter if we impose a 3$\sigma$ detection level. Based on our Fig.~\ref{fig.expecteddetections} (left), if we consider lines $\sim17$ times fainter than CO $3-2$ (the level could be even lower considering the integrated line intensity) would result in a detection of 3-4 species. \citet{Spilker2014} reported 6 species above 3~$\sigma$. However, based on the mid panel in their Figure 2, only 3 spectral features actually reach the 3$\sigma$ level at the spectral resolution in their diagram.
The increased noise in the small fraction of rest-frame Band 6 they covered resulted in no detections, also in agreement with what would be expected based on the NGC~253 template. Considering a factor of 2 lower sensitivity in this region, only CO $2-1$ would be expected, and its frequency was actually not covered in their observations.
This comparison shows the predicting potential of NGC~253 for molecular detections towards high-z starbursting galaxies.

\subsection{Vibrational emission}
\label{sec.vibemission}

Rotational transitions in vibrational states (hereafter vibrational emission or transitions) of HCN, HNC, and HC$_3$N, with lower energy levels of 1000, 700, and 500~K above the ground state, respectively, are clearly detected in the ACA data analyzed in this article.
Vibrational emission towards NGC~253 has been recently reported at sub-arcsecond resolution towards individual GMCs with observations of the $J=4-3, v_{2}=1f$ transitions of HNC and HCN \citep{Ando2017,Mangum2018,Krieger2020}, and two rotational transitions in multiple vibrational states of HC$_3$N \citep{Rico-Villas2020}.
However, it was never detected in low resolution spectral scans of NGC~253 \citep{Mart'in2006,Aladro2015}, with spatial resolution similar to that in this work.
While HC$_3$N emission in the $v_7=1$, $v_7=2$, and $v_6=1$ states is clearly detected at $0.2''$ resolution \citep{Rico-Villas2020}, we only detect significant emission from $v_7=1$ states. 
%This is in marked contrast to the compact heavily obscured nuclei within the LIRG NGC~4418 and the ULIRG Arp\,220, where the emission from the $v_7=1$, $v_7=2$, and $v_6=1$ transitions are detected throughout the whole mm spectrum as a comb of emission lines on the side of the pure HC$_3$N rotational transition every ~9~GHz, even in $200-700$~pc spatial resolution studies \citep{Mart'in2011,Costagliola2015}. 
We attribute this difference to
%extent of the vibrational emission where higher 
beam dilution %would be in play towards NGC~253. In fact, in NGC~253 
of the vibrational emission which originates from the compact GMC cores \citep{Rico-Villas2020,Rico-Villas2020a,Krieger2020}.
%, while in the compact (U)LIRGs it will be likely spread over larger regions either surrounding a putative active nuclei or be conformed by many of such star forming associated cores.
This is similar to what is observed within our Galaxy where vibrational emission is solely
%not surprising in a starburst galaxy, since this emission is likely 
arising from hot dense material within star forming cores \citep{deVicente2000,Mart'in-Pintado2005}. 

The wide-band imaging of ALCHEMI data allows us to probe multiple vibrationally excited transitions of these species and to evaluate the contamination by other species.
Fig.~\ref{fig.vibemission} shows the rotational transitions of HCN, HNC, and HCO$^+$ in the $v_2=1f$ vibrational state. 
%Only the $v_2=1f$ transitions are displayed since the frequencies of 
Transitions in the $v_2=1e$ state are too close in frequency to the rotational transitions in the ground vibrational state  \citep[see Fig.3 in][]{Martin2016}. 
The derived LTE fit to the emission of all observed transitions assumes an excitation temperature $T_{ex}=300$~K required to make these high energy transitions detectable (red line in Fig.~\ref{fig.vibemission}). This fit clearly shows that the LTE approximation does not properly reproduce the excitation of these radiatively pumped transitions \citep{Aalto2015a}. %This decoupling between the vibrational temperature measured from vibrational transitions within same $J$ level and the rotational temperature measured for transitions within the same vibrational level has been previously reported \citep{Costagliola2015,Rico-Villas2020}. %Thus, while we assumed a hgh temperature for the LTE fit, so such high energy transitions are populated,
Table~\ref{tab.vibLuminosities} presents the line ratio between the observed rotational transitions in the ground vibrational ($v_2=0$) and $v_2=1f$ vibrational states.
As previously reported, the relative intensities between rotational transitions within a vibrational state follow that measured within the $v=0$ rotational transitions \citep{Costagliola2015,Rico-Villas2020}, with $v_2=0/v_2=1f$ ratios relatively constant. 
We note that fitted column density is not physically meaningful since it requires full radiative transfer modeling to take radiative pumping into account.

Based on the analysis of the full spectrum we estimate that the transitions presented in Fig.~\ref{fig.vibemission} are only marginally blended with fainter transitions from other species, with the exception of HCN $J=2-1~v_2=1f$ and more importantly HNC $J=4-3~v_2=1f$.
%which are more affected by line contamination. 
Although not blended with other species, HCO$^+~J=4-3~v=2f$ falls between two bright features and we therefore consider this line to be tentatively detected.
The other two HCO$^+~v=2f$ transitions, $J=3-2$ and $2-1$, are not detected as shown in Fig.~\ref{fig.vibemission}.
% and the profiles are relevant to probe their emission.

%, with this one being the first report of HCN vibrational emission towards NGC~253.
\subsubsection{High temperature driven "carbon-rich" chemistry}

\begin{figure*}
\begin{center}
\includegraphics[width=0.3\textwidth,valign=t]{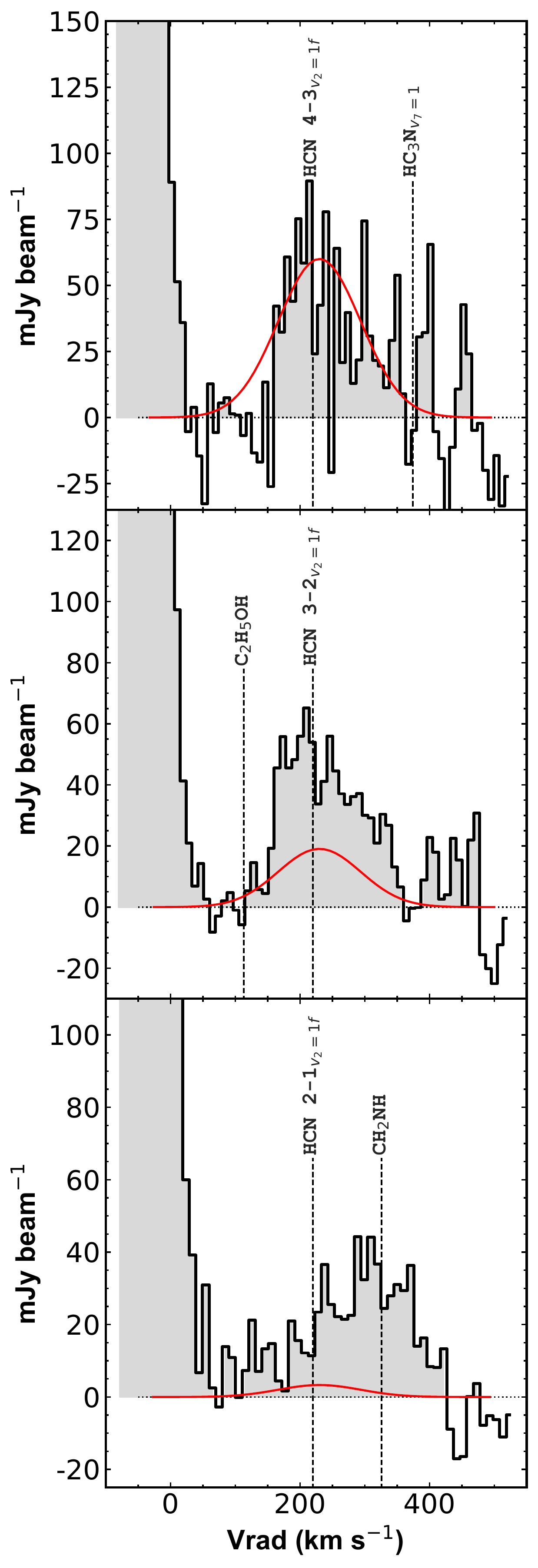}
\includegraphics[width=0.3\textwidth,valign=t]{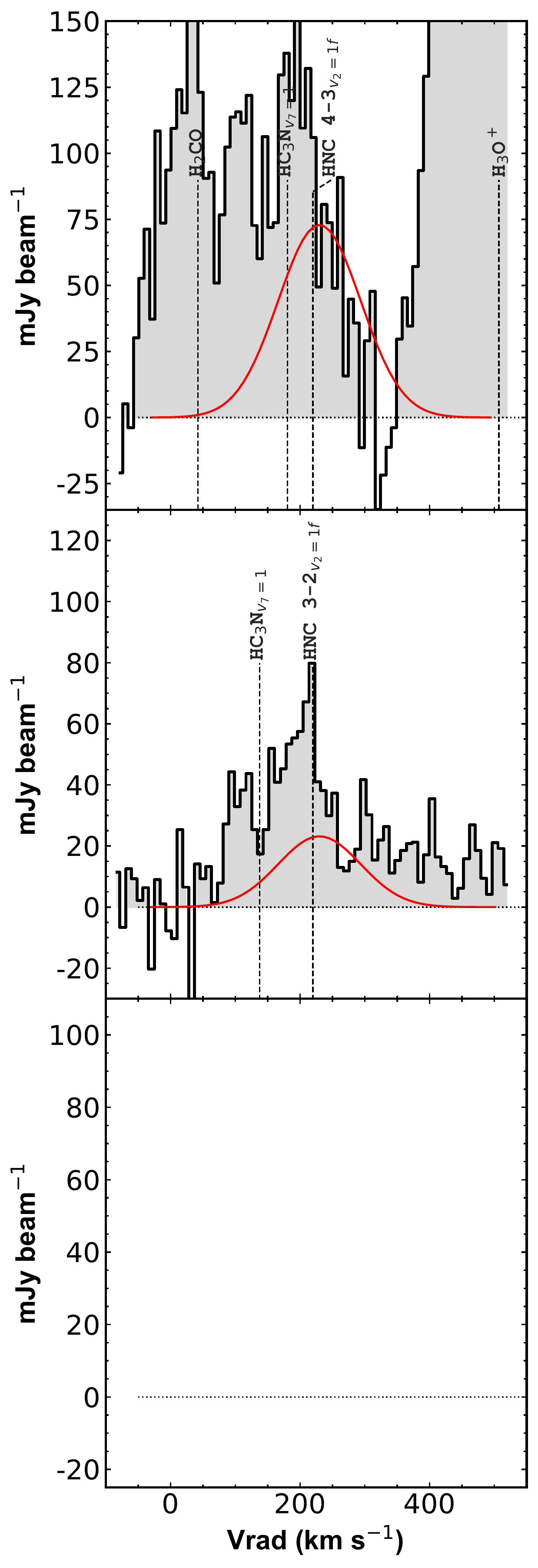}
\includegraphics[width=0.3\textwidth,valign=t]{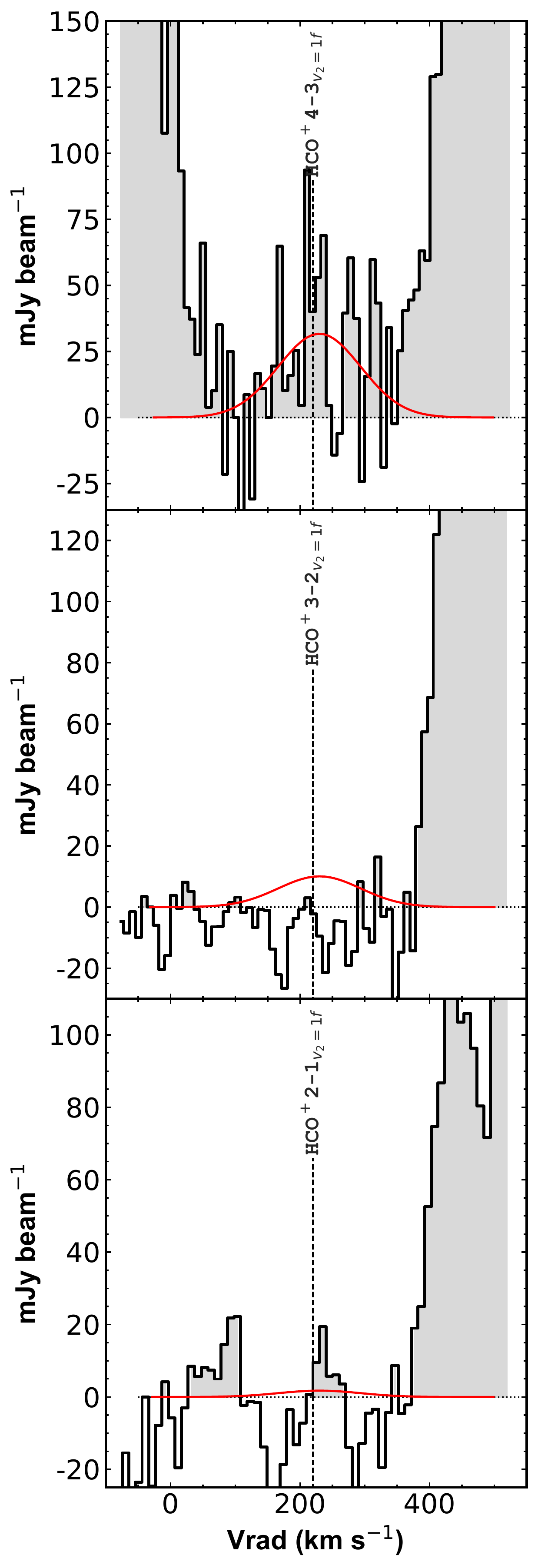}
\caption{Rotational transitions in the vibrational state $v_2=1f$ of HCN, HNC and HCO$^+$ covered within the surveyed frequency range. The box corresponding to HNC $2-1~v_2=1f$ was left intentionally blank since its emission at 182.6~GHz falls within the telluric water transition observation gap (Sect.~\ref{Sec.Observations}). Red lines show an attempt to fit the observed emission under LTE assuming $T_{ex}=300$~K (see text in Sect.~\ref{sec.vibemission} for details).
Nearby transitions from detected species with modelled flux densities $>10\%$ of that of the vibrational transitions are labeled.
\label{fig.vibemission}
}
\end{center}
\end{figure*}

Our data show that the vibrational emission of HCO$^+$ is one order of magnitude fainter both relative to the observed emission of vibrational HCN and HNC as well as relative to the HCO$^+$ ground vibrational state (see Table~\ref{tab.vibLuminosities}).
%Moreover, we also note that the HCO$^+$ vibrational emission is proportionally faint, where we find that the $4-3$ transition is only tentatively detected, relative to the observed flux densities of HCN and HNC. 
The detection of the $J=4-3,v_2=1f$ transition alone might be considered tentative but still relatively faint compared to the corresponding transitions of HCN and HNC.
However, if the same ratio among $J$ transitions within the vibrational level of HCN and HNC would apply to HCO$^+$, we would then expect the $3-2$ and $2-1$ transitions to be significantly above the LTE fit in Fig.~\ref{fig.vibemission}.
Since this is not observed, we argue against the detection of vibrationally excited HCO$^+$.
% This is similar to what was reported by \citet{Imanishi2017} towards the ULIRG IRAS 20551-4250, where vibrational HCO$^+$ was not detected while HCN and HNC were.

\begin{table*}
\centering
\caption{Spectral line emission properties of the
vibrational transitions of HCN, HNC, and HCO$^+$.}
\label{tab.vibLuminosities}
\begin{tabular}{l c c c c c c c}
\hline
\hline
Transition      & Freq.     & $S_{peak}$   & $\int{S\,\delta v}$& $S_{v_2=0}$   & $S_{v_2=0}/S_{v_2=1f}$& $L_{vib}$\tablefootmark{a} & $\frac{L_{vib}}{L_{IR}}$ \tablefootmark{b}\\
    $v_2=1f$    &  (GHz)    &  (mJy)       &  (Jy~\kms)         &    (Jy)      &             & ($L_\odot$)         &     ($\times10^{-9}$)       \\
\hline
HCN~$4-3$    &   356.255 & $50\pm10$    & 8.0         &   $6.70\pm0.08$  &    $130\pm30$    & 36.3   &  1.7  \\
HCN~$3-2$    &   267.199 & $53\pm11$    & 8.5        &   $6.08\pm0.07$  &    $120\pm20$    & 28.9   &  1.4  \\
HCN~$2-1$    &   178.136 & $24\pm10$    & 3.8         &   $3.64\pm0.07$  &    $150\pm60$    & 8.5    &  0.4  \\
HNC~$4-3$    &   365.147 & $68\pm10$    & 10.8        &   $3.66\pm0.04$  &     $53\pm8$    & 50.2   &  2.3  \\
HNC~$3-2$    &   273.869 & $40\pm5$     & 6.4         &   $3.64\pm0.05$  &    $90\pm10$    & 22.3   &  1.0  \\
HCO$^+~4-3$  &   358.242 & $<38$        & $<6.0$      &   $7.9\pm0.1$    & $>200$    & $<27$  &  $<1.3$  \\
% 268.7 GHz
%--RMS of Base Line:8.616814E-3  in 8.1 MHz = 9.0 km/s
%1 SIGMA
%Tmb (mJy):  -2.1090105844398
%3 SIGMA
%Tmb (mJy):  -6.3270317533193
%AREA (Jykm/s):  -0.94905476299789
%v=0 Area = 1285 +- 41
HCO$^+~3-2$  &   268.689 & $<6.3$       & $<0.95$     &   $6.77\pm0.08$  &$>1100$    & $<3.2$ &  $<0.15$  \\
% 179.1 GHz
%--RMS of Base Line:14.81433E-3. in 5.1 MHz = 8.5 km/s
%1 SIGMA
%Tmb (mJy):  -3.5264373888892
%3 SIGMA
%Tmb (mJy):  -10.579312166667
%AREA (Jykm/s):  -1.5868968250001
%v=0 Area = 610 +- 16
HCO$^+~2-1$  &   179.129 & $<3.5$      & $<1.6$       &  $3.58\pm0.09$  &$>1000$     & $<3.6$ &  $<0.17$  \\
\hline
% 150km/s width
\end{tabular}
\tablefoot{
Results from Gaussian fitting individual transitions, not from the LTE model in Fig.~\ref{fig.vibemission}, taking into account contamination from other species. Both the line fit and the upper limit calculations considers a line width of 150~\kms, similar to that fitted to most rotational transitions. 
\tablefoottext{a}{Luminosity estimates of the vibrational emission follow Eq.~3 in \citet{Solomon2005}}
\tablefoottext{b}{Considering an infrared luminosity of NGC~253 is $L_{IR}=2.1\times10^{10}L_\odot$ from \citet{Strickland2004}.}
}
\end{table*}

The wavelengths of photons required to excite the vibrational states range from $\sim22~\mu m$ for HNC to $12-14~\mu m$ for HCN and HCO$^+$ \citep{Aalto2007,Sakamoto2010, Imanishi2017,Gonzalez-Alfonso2019}.
High column densities and dust temperatures are required for the effective photon trapping leading the vibrational excitation of these molecules \citep{Gonzalez-Alfonso2019}.
However, the differences in the conditions for IR pumping of these species may explain the different relative intensities to their respective ground vibrational transitions \citep{Sakamoto2010}. For instance, the one order of magnitude larger HNC Einstein coefficient make it easier to pump than HCN \citep{Aalto2007}. 
However, these excitation differences alone do not explain the non detection of vibrational HCO$^+$ which has relatively similar excitation conditions to HCN.

Detections of vibrational transitions of HCN, HNC, HC$_3$N, and seemingly CH$_3$CN have been reported towards an ever increasing number of (U)LIRGs %\citep{Sakamoto2010,Costagliola2010,Costagliola2013,Costagliola2015,Martin2016,Mart'in2011,Imanishi2013,Imanishi2016,Imanishi2018,Imanishi2019,Aalto2015,Aalto2015a,Aalto2019,
\citep[see the compilation by][]{Falstad2019} at the frequencies covered by ALCHEMI.
However, beside the recent detections of vibrationally excited HCO$^+$ in absorption towards the gas-poor AGN in NGC~1052 \citep{Kameno2020} and the faint emission feature (relative to the global galaxy emission) towards the molecular torus around the luminous AGN in NGC~1068 \citep{Imanishi2020}, no detections of HCO$^+$ in vibrational states have been reported in extragalactic environments.

%Similar to IRAS 20551-4250, towards  we clearly detect  levels with similar flux density, while there appears to be only a weak hint of HCO$^+$ pumping.

%Since the first report of vibrational absorption towards Arp\,220 by \citet{Salter2008} with the Arecibo observatory at cm wavelenghts, a large number of detections in emission have been reported 
%initially 
%towards 
%targeted 
%bright LIRGs and ULIRGs 
%\citep{Sakamoto2010,Costagliola2010,Costagliola2013,Costagliola2015,Martin2016,Mart'in2011,Imanishi2013,Aalto2015}, and now being further extended to a larger number of LIRGS and ULIRGs being sampled at these frequencies \citep{Aalto2015a,Imanishi2016,Imanishi2018,Aalto2019,Falstad2019,Imanishi2019}.

%Vibrational detections in ULIRGs and LIRGs have been reported in HCN, HNC, HC$_3$N, and seemingly in CH$_3$CN (see references above), which has been explained by IR pumping by \citet{Aalto2015a}.

%H3O+ 307.2 = 0.179 +- 0.026
%H3O+ 364.8 = 1.226 +- 0.058
% 364.8/307.2 = 6.8 +-1.0

Similar to what is reported here towards NGC~253, an explicit non-detection of vibrational HCO$^+$ emission was reported towards the compact LIRG NGC~4418 and the ULIRG IRAS 20551-4250 \citep{Sakamoto2010,Imanishi2017}, while clearly detecting HCN and/or HNC vibrational emission. \citet{Imanishi2017} claimed an overabundance of HCN towards IRAS 20551-4250, following the statistically suggested
%, based on a sample of galaxies, 
higher rotational emission HCN/HCO$^+$ ratio in AGN dominated environments \citep{Izumi2013,Privon2015,Imanishi2016}. However, as suggested by \citet{Izumi2013} based on the chemical modeling of \citet{Harada2010} high temperature chemistry could be responsible for such a relative HCN overabundance.

At high temperature \citep{Harada2010}, a "carbon-rich" chemistry can be mimicked when oxygen is locked in the form of H$_2$O. In such conditions, carbon bearing species such as HCN and HNC may be boosted by 1-2 orders of magnitude, while HCO$^+$ would be reduced by a similar amount.
This may actually be the scenario in hot dense gas around the protostars in starburst dominated environments.  In these regions high temperatures and densities are required not only to drive the efficient infrared pumping of vibrational states \citep[$T_d>100$ ~K,][]{Aalto2015a} but also to populate transitions such as those of HC$_3$N to $J=40-39$, with critical density $n_{crit}\sim10^7~\rm cm^{-3}$ and lower level energy $E_l\sim350$~K \citep{Wernli2007}.  Within the ALCHEMI measurements, observed intensities above the LTE-derived fit to HC$_3$N indicate the presence of a high excitation temperature component.
%, with a critical density of $\sim10^7~\rm cm^{-3}$ \citep[at T=40~K][]{Wernli2007} and upper level energy of 340~K.

The proposed scenario, in which high temperatures drive "carbon-rich" chemistry, would be supported by the non-detection of vibrationally excited oxygen-bearing species such as HCO$^+$, while the rotational transitions in the ground vibrational state, not originating in the denser infrared pumped region, may appear to be as bright as those of HCN and HNC.
Similarly it would explain the relatively rich complex carbon chain chemistry detected in NGC~253 (Sect.~\ref{sec.complexorganics}), %and also why vibrationally excited emission has been reported in in carbon rich species like HC$_3$N and CH$_3$CN, the latter only towards the ULIRG Arp\,220 \citep{Mart'in2011}.
%In this scenario, if we adopt objects like NGC~4418 and Arp\,220 are extreme scaled up versions of the starburst environment observed in NGC~253, the dense high temperature driven "carbon-rich" chemistry would explain 
as well as the high abundances of HC$_3$N and HC$_5$N not only in NGC~253 \citep{Aladro2015} but also in the prominent compact obscured nuclei (CON) in the ULIRG Arp~220 and the LIRG NGC~4418 \citep{Mart'in2011,Costagliola2015}, two well known emitters of vibrationally excited rotational lines.
The locking of oxygen into H$_2$O would also explain the bright emission of H$_2$O and its weaker isotopologue H$_2^{18}$O reported towards Arp~220  \citep{Mart'in2011,Koenig2017}.
%assuming this emission to be mostly thermal as supported by both the low variability of the H$_2$O emission and the detection of its weaker isotopologue H$_2^{18}$O \citep{Mart'in2011}. 
Furthermore, the "carbon-rich" scenario might also be supported by bright HCN relative to CO emission reported towards the molecular outflow in Arp~220 \citep{Barcos-Munoz2018}, where the oxygen depleted hot dense gas might have been blown away by the nuclear activity (due to either a starburst or active galactic nuclei).% in the form of molecular outflows, a relatively bright HCN compared to the CO emission as reported by \citet{Barcos-Munoz2018} might also be supported by this "carbon-rich" chemical scenario \citep{Harada2010}.

\subsubsection{Vibrational emission as tracer of global proto-SSC contribution}
\label{sec.LvibLfir}

The ratios between the vibrational emission and the infrared luminosity for HCN, HNC and HCO$^+$ are summarized in Table~\ref{tab.vibLuminosities}. The ratios measured with HCN $J=3-2,v2=1f$ are about an order of magnitude below the $L_{vib}/L_{IR}=10^{-8}$ threshold defining the extreme compact obscured nuclei, such as the above mentioned NGC~4418 and Arp\,220 \citep{Falstad2019}.

We compare NGC~253 CMZ global average $L_{vib}/L_{IR}=1.4\times10^{-8}$ derived from HCN $J=3-2,v2=1f$ with that of $4.2\times10^{-8}$ measured towards the Galactic Center hot core Sgr~B2(N) at a spatial resolution of $0.1-1.2$~pc \citep{Rolffs2011}.
%with  We also compare the average $L_{vib}/L_{IR}$ ratio observed in HCN towards the CMZ of NGC~253 with that measured towards star forming regions within the Galaxy.
%The HCN $J3-2~v_2=1$ observations by \citet{Rolffs2011}, at spatial resolutions ranging $0.1-1.2$~pc, shows variations of this luminosity ratio ranging from $2.9\times10^{-10}$ in W51d up to $4.2\times10^{-8}$ in Sgr~B2(N) with an average/median ratio of $7/3\times10^{-9}$. %/3\times10^{-9}$.
%The average of Sgr B2 (N) and (M) as sample of Galactic CMZ hot cores is $2.3\times10^{-8}$.
%By directly comparing the global luminosity ratio measured towards NGC~253 CMZ with the average Galactic and GC ratio, we calculate a ratio of $20\%$ and $6\%$, respectively. 
Assuming that vibrational emission is mostly contributed by Sgr~B2(N)-like hot cores, this comparison results in a $\sim3\%$ level as a proxy to the contribution from proto-SSCs to the global infrared luminosity in NGC~253.
%This ratio could be as low as $\sim3\%$ if we assume significant contribution from SgrB2(N)-like hot cores to vibrational emission. 
%It could be considered a proxy for the contribution from proto-SSCs to the global infrared luminosity.
%LHCN/LIR = $0.3\times10^{-8}$ from \citep{Aalto2015a} based on \citep{Rolffs2011}.
In fact, this estimate is consistent with the contribution from proto-SSCs of $\sim 3\%$ derived by \citet{Rico-Villas2020} based on high resolution imaging of vibrationally excited HC$_3$N towards NGC~253. We note that \citet{Rico-Villas2020} used a $25\%$ lower IR luminosity from the CMZ of $1.6\times10^{10}L_\odot$, but still both results remain in good agreement.
%{
%\ color{red} (we note, that they used a CMZ NGC~253 to be half %$L_{IR}=3.1\times10^10$).
%ALso it would be 3 \% if we use the SgrB2N Luminosity ratio.
%}

%The observations in this work show that the chemical fingerprints of hot molecular cores are detectable at low (few hundred parsec) resolution.  This observation is in spite of the fact that these chemical signatures are a small scale local effect in the surroundings of massive star forming regions and the location of proto-SSCs \citep{Rico-Villas2020,Rico-Villas2020a}.
Thus, we tentatively show that vibrational emission could be used as a 0th-order proxy to the proto-SSC contribution to the total infrared luminosity in extragalactic environments, under the assumption that there is marginal or no vibrational emission contribution by AGN heating as recently suggested from observations of the Seyfert galaxy NGC~1068 \citep{Rico-Villas2020a,Imanishi2020}.

\subsection{Organic Molecules}
\label{sec.complexorganics}

As described in Sect.~\ref{sec.newdetections} some of the newly detected species in NGC~253 are organic molecules. In the following we will briefly discuss the measured abundances of ethanol ($\rm C_2H_5OH$) and formic acid (HCOOH). These species are selected as ancillary comparison data exists from Galactic Center molecular clouds to provide an initial comparison of organic species abundances towards NGC~253; while we leave further more comprehensive discussions on organic molecules to upcoming publications using the higher resolution data. The detection of these organic species can only be unambiguously claimed thanks to the large spectral coverage which has allowed us to properly account for line blending and to cover multiple transitions from these large molecules.

\begin{figure}
\begin{center}
\includegraphics[width=0.5\textwidth]{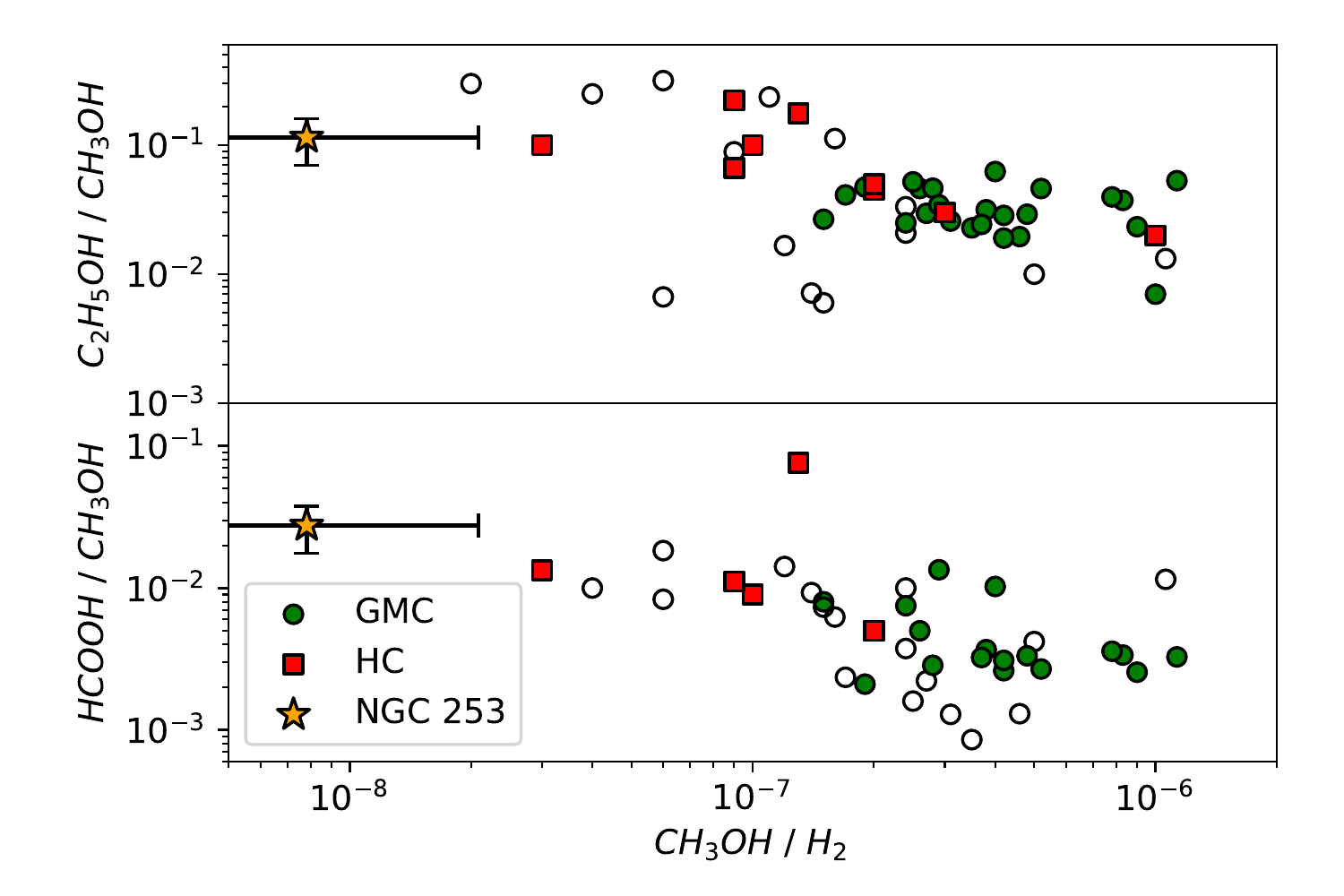}
\caption{Abundances of $\rm C_2H_5OH$ and HCOOH relative to CH$_3$OH towards the NGC~253 CMZ (orange star), compared to those measured towards Galactic hot cores \citep[red squares;][]{Ikeda2001} and Galactic Center giant molecular clouds \citep[green circles;][]{Requena-Torres2006}. Upper limits to $\rm C_2H_5OH$ and HCOOH are represented by open symbols. 
\label{fig.complexmolecules}
}
\end{center}
\end{figure}

Based on a sample of Galactic Center (GC) giant molecular clouds (GMCs) \citet{Requena-Torres2006} claimed an apparently homogeneous COM composition across the Galactic central molecular zone. In their study, the comparison between Galactic hot cores and GMCs showed that organic abundances relative to methanol, and in particular those of $\rm C_2H_5OH$ and HCOOH, agree within a factor of two, although the abundances towards hot cores were systematically found on the high end of the measurements (Fig.~\ref{fig.complexmolecules}).
The detection of these two species over CMZ scales towards NGC~253, allows us to include, for the first time, an extragalactic environment into this comparison. This allows us to explore the claimed grain composition homogeneity outside our Galaxy, and in this case towards a star-bursting environment.

Fig.~\ref{fig.complexmolecules} places the abundances of $\rm C_2H_5OH$ ($1.1\pm0.4\times10^{-1}$) and HCOOH ($2.8\pm1.0\times10^{-2}$) relative to CH$_3$OH towards NGC~253 in the context of the Galactic Center GMCs observed by \citet{Requena-Torres2006} and the Galactic hot cores from \citet{Ikeda2001}. Relative abundances with respect to methanol are commonly used to explore relative COM abundances while avoiding uncertainties associated with an estimate of H$_2$ column densities. In our analysis we have estimated the methanol abundance relative to H$_2$ from \cdo~ and assuming $\rm ^{18}O/^{16}O=100$ (as derived in Sect.~\ref{sec.isotopologuesOxygen}) and CO/H$_2\sim 8\times10^{-5}$ \citep{Frerking1982}, which yields $N_{\rm H_2}=3.1\times10^{23}~\rm cm^{-2}$, in good agreement with previous single dish observations after correcting for the different assumed emission extent \citep{Mart'in2009}. We included a conservative factor of 2 uncertainty in the H$_2$ determination to account for uncertainty in these assumptions.
%, used as reference in Fig.~\ref{fig.complexmolecules} is not relevant to the conclusion which relies on relative abundances among organic molecules. 
As expected, the relative abundance of CH$_3$OH in NGC~253 is significantly lower than in Galactic Center sources, since the clumpier methanol emission is referred to the global extended H$_2$ emission traced by CO over the large scales probed in our low resolution observations. Therefore the CH$_3$OH/H$_2$ in Fig.~\ref{fig.complexmolecules} can be actually considered a lower limit.

The comparison in Fig.~\ref{fig.complexmolecules} shows that the averaged abundances towards NGC~253 are on the high end of GMC measurements, similar to what is observed in the sample of hot cores. In particular the abundance of HCOOH is among the highest in the sample, but still a factor of 3 below the observed one in the Sgr~B2(N) hot core \citep{Requena-Torres2006}.

%The detection of propynal places its global abundance towards the CMZ in NGC~253 at $HC$_3$HO/CH_3OH=3.7\pm1.7\times10^{-2}$ which is an order of magnitude above what is reported by \citep{Requena-Torres2008} towards GC GMCs.

Within the Galactic Center, emission from complex organic molecules is clearly detected towards hot molecular cores associated with massive star formation \citep[e.g.][]{Belloche2013,Belloche2017,Belloche2019}.
However, early surveys on a sample of GC GMCs \citep{Requena-Torres2006,Requena-Torres2008,Mart'in2008a} and the recent reports of an ever increasing COMs richness in GC quiescent clouds \citep{Zeng2018,Rivilla2019,Rivilla2020,Bizzocchi2020,Jimenez-Serra2020} are evidence of the widespread COM emission over GMC scales.

Previous single dish line surveys towards the CMZ of NGC~253 claimed that the observed average chemical abundances resemble that of GC GMCs \citep{Mart'in2006,Aladro2011a}. This chemistry would be mostly driven by the widespread large-scale shocks affecting its CMZ \citep{Garc'ia-Burillo2000,Mart'in2009a,Meier2015}. Indeed, these widespread shocks in NGC~253 would be the responsible for the global ejection of organic molecules from dust grains into the gas phase, similar to what is observed within the Galactic CMZ \citep{Martin-Pintado1997, Mart'in2008}.

Our low resolution observations show how the emission from organic species described above, although consistent with that in GMCs, might imply a significant contribution from hot-core emission based on the observed abundances relative to methanol being closer to those found in GC hot cores. 
Such a hot core contribution should be significant since it is detectable in the averaged low resolution spectrum in this starbursting environment.
The actual distribution and resolved abundances across the CMZ in NGC~253 will be better analyzed with the high resolution ALCHEMI data which will allow the comparison between the individual NGC~253 GMCs and the Galactic Center quiescent GMCs and hot cores.

%Although the abundances measured in our ACA data are uncertain due to limited sensitivity and resolution, %faintness of the observed transitions, 
%the observed abundance ratios are comparable to those measured in Galactic sources. The high resolution ALCHEMI data will further allow 
%for a direct comparison with 
%us to investigate the COM abundances at the scale of individual GMCs within the CMZ of NGC~253.

\subsection{Isotopic ratios}
\label{sec.isotopologues}

% Understanding
Investigating the elemental isotopic ratios and the difficulties to measure them with different molecular proxies in nearby galaxies is key to extend these studies to more distant objects,
%and to accurately 
as well as probing the evolution of isotopic enrichment through cosmic time \citep{Muller2006,Wallstroem2016,Kobayashi2020}.
%In this work, rather than focusing ourselves on a single molecular species, the wide coverage of our observations allows us to estimate the atomic isotopic ratio using all the observable molecular isotopologues detectable within our achieved noise level as probes. 
A major advantage of the wide frequency coverage of the ALCHEMI survey is to allow for the measurement of isotopic ratios based on all the molecular isotopologue pairs detected within the covered bands, coupled with the relatively accurate column density measurement based on multiple transitions observed at the same angular resolution. The use of individual species and/or single transitions are subject to potential uncertainties in opacity (see Sect.~\ref{sec.opacity}), excitation, blending effects, as well as chemical fractionation.

\begin{figure}
\begin{center}
\includegraphics[trim=25 25 50 70,clip,width=0.5\textwidth]{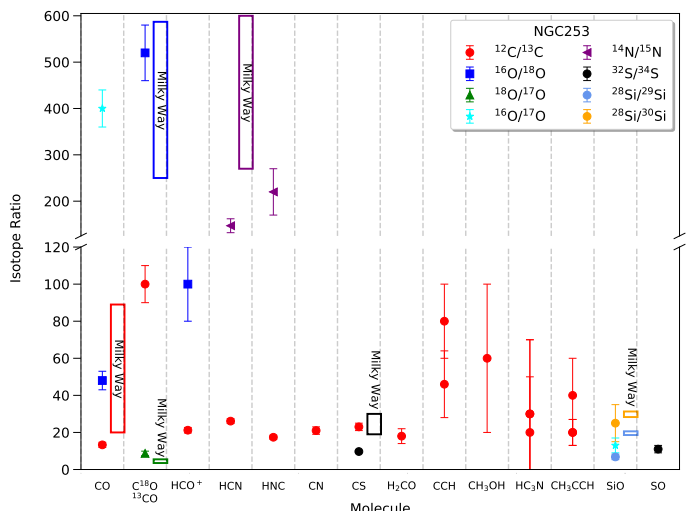}
\caption{Measured NGC\,253 isotope ratios in the ACA data (Table~\ref{tab.isotopicratios}) compared to their equivalent values in the Milky Way \citep{Wilson1999}.  The horizontal axis indicates the molecular species, or in the case of CO the isotopomer, used to measure a given isotope ratio.  Colored rectangles are used to indicate Milky Way isotope ratio value ranges.  As no uncertainties nor range of values was provided for the Si isotope ratios, a range of 10\% of the estimated values was assumed. As the Milky Way $^{16}$O/$^{17}$O value would be $\gtrsim 875$, we have opted to not show this Milky Way isotope ratio value.
\label{fig:IsotopeRatioPlot}
}
\end{center}
\end{figure}

Table~\ref{tab.isotopicratios} shows the isotopic ratios derived from the column densities fitted to all detected isotopologue pairs (Sect.~\ref{Sec.LTE}).  This table is graphically represented in Fig.~\ref{fig:IsotopeRatioPlot} where the equivalent range of values measured in the Milky Way are also displayed for comparison.
%displays the isotope ratios listed in Table~\ref{tab.isotopicratios} and compares them to their equivalent values in the Milky Way.
As explained in Sect.~\ref{Sec.LTE}, a source size of $5''$ was used for all species except for CO where a size of $10''$ was required to match the observed absolute flux density of the CO transitions. In order to provide a meaningful ratio of CO isotopologues when referred to the main isotopologue, the emission of $^{13}$CO and C$^{18}$O fits were recalculated for an equivalent source size of $10''$. In this way the ratio is meaningful, despite opacity considerations, since all the CO isotopologues present resolved structure in the low resolution maps (Fig.~\ref{fig.moment0maps}). We observe how the carbon and oxygen ratios derived with the main CO isotopologue are consistently a factor of $2-10$ below those measured with any other isotopologue pairs in Table~\ref{tab.isotopicratios} (Sect.~\ref{sec.isotopologuesCarbon} and ~\ref{sec.isotopologuesOxygen}). This is evidence of the opacity affecting CO even in the averaged CMZ emission.

The values in Table~\ref{tab.isotopicratios} are optical depth corrected, under the assumptions of LTE and similar source extent. When it was possible to fit the excitation temperatures in the different isotopologues of a given species, we found a good agreement (Sect.~\ref{Sec.LTE}). For the assumed source size of $5''$ most of the main isotopologues show moderate optical depths ($\lesssim1$), so column density ratios are close to the optically thin regime and therefore close to what would be derived from line intensity ratios. The advantage of using column density ratios is that we are considering the contribution from all transitions of a given molecule, rather than relying on a single transition.

In order to evaluate the effect of the assumed source size on the opacity of the main isotopologue and, consequently on the derived isotopic ratio, Table~\ref{tab.opacityHCOp} shows the LTE fit results for various assumed source sizes, where HCO$^+$ is selected as case example. We note that for sizes $<3''$, 
increasing the excitation temperature $T_{ex}$ is requierd to reproduce the observed absolute flux densities, which results in lower opacities, but it then becomes difficult to fit the emission with a single temperature component. Therefore, the ranges explored in Table~\ref{tab.opacityHCOp} are a good representation of the effect of opacity on the derived isotopic ratio, which can become a factor of 2 larger than the optically thin approximation.

To put our results into context, Table~\ref{tab.isotopicratiosliterature} compiles all isotopic ratios reported in the literature towards NGC~253.
%, including some values based on line intensities reported in works that did not actually targeted the study of isotopic ratios. 
The ``nominal'' isotopic ratios commonly used in the literature for local galaxies, as compiled in \citet{Wilson1999}, are \carbontt$ =40$, \oxigense$ =200$, and \oxigenes$ =8$. However, these derived values are subject to the limitations of the species used as proxy as well as the uncertainty of the opacity in the brighter isotopologue (see Sect.\ref{sec.opacity} for a discussion on opacity).
Placing our NGC~253 observations into context of previous detections is key since these standard isotopic ratios are mostly based on measurements toward M~82, the other prototypical nearby starburst galaxy, and our target NGC~253 \citep[see overview and discussions in][]{Henkel2014,Martin2019,Tang2019}

\begin{table}
\centering
\caption{Column density ratios in the CMZ of NGC~253 based on the ACA data}
\label{tab.isotopicratios}
\begin{tabular}{l c c c}
\hline
\hline
Isotopes        &  Ratio\tablefootmark{a}    & Proxy             \\
\hline
\carbontt       &  $13.3\pm1.4$\tablefootmark{b} & CO/$^{13}$CO              \\
% Ratio was 2 with CO 10" and 13CO 5" 
%13CO 18.094744 with 5" and 17.316776 with 10"
% Factor 6 => 12
                &   $21.2\pm1.5$      & HCO$^+$/H$^{13}$CO$^+$    \\
                &   $26.1\pm1.3$      & HCN/H$^{13}$CN            \\
                &   $17.4\pm1.1$      & HNC/HN$^{13}$C            \\
                &   $21\pm2$      & CN/$^{13}$CN              \\
                &   $23\pm2$      & CS/$^{13}$CS              \\
                &   $18\pm4$      & H$_2$CO/H$_2^{13}$CO      \\
                &  $100\pm10$      & C$^{18}$O/$^{13}$C$^{18}$O\\
                &   $46\pm18$      & CCH/$^{13}$CCH            \\
                &   $80\pm20$     & CCH/C$^{13}$CH            \\
                &   $60\pm40$      & CH$_3$OH/$^{13}$CH$_3$OH  \\
                &   $30\pm40$   & HC$_3$N/HC$^{13}$CCN    \\
                &   $30\pm40$   & HC$_3$N/H$^{13}$CCCN    \\
                &   $20\pm30$   & HC$_3$N/HCC$^{13}$CN    \\
%                &$23-27-32$ & HC$_3$N/H$^{13}$C$_3$N    \\
                &   $20\pm7$    &  CH$_3$CCH/CH$_3^{13}$CCH \\             
                &   $22\pm3$    &  CH$_3$CCH/CH$_3$C$^{13}$CH \\             
                &   $40\pm20$    &  CH$_3$CCH/$^{13}$CH$_3$CCH \\             
%                &$20-27-35$ & CH$_3$CCH/$^{13}$CH$_3$CCH \\             
\oxigense       & $48\pm5$\tablefootmark{b}   & CO/C$^{18}$O             \\
% Ratio was 11 with CO 10" and 13CO 5" 
%C18O 17.381025 with 5" and 16.742933 with 10"
% Factor 4.34 => 48
                & $520\pm60$       & $^{13}$CO/ $^{13}$C$^{18}$O   \\
                & $100\pm20$       & HCO$^+$/HC$^{18}$O$^+$   \\
\oxigenes       &  $8.7\pm1.2$      & C$^{18}$O/C$^{17}$O      \\
$\rm^{16}O/^{17}O$&  $400\pm40$\tablefootmark{b}      & CO/C$^{17}$O      \\  %Factor 4
                &  $13\pm4$       & SiO/Si$^{17}$O           \\
\nitrogenff     &  $147\pm15$      & HCN/HC$^{15}$N           \\
                &  $220\pm50$      & HNC/H$^{15}$NC           \\
\sulfurtttf     &  $9.7\pm0.5$       & CS/C$^{34}$S             \\
                &  $14\pm4$       & SO/$^{34}$SO             \\
\silicontetn    &  $9\pm4$        & SiO/$^{29}$SiO           \\ 
\silicontte     &  $40\pm20$       & SiO/$^{30}$SiO           \\ 
\hline
% 150km/s width
\end{tabular}
\tablefoot{
\tablefoottext{a}{Isotopic ratios are based on the column density LTE fits which, though opacity corrected, is close to the optically thin regime. Higher opacity could result in ratios up to a factor of 2 larger (see Sect.~\ref{sec.isotopologues} and Table~\ref{tab.opacityHCOp}).
}
\tablefoottext{b}{This column density ratio has been re-calculated for a common source size of $10''$ in order to match the size required to fit the main CO isotopologue emission (see Sect.~\ref{sec.isotopologues})}
}
\end{table}

\begin{table}
\centering
\caption{Effect of source size on the \carbontt~ ratio derived from HCO$^{+}$}
\label{tab.opacityHCOp}
\begin{tabular}{l c c c c}
\hline
\hline
Source Size   & N(HCO$^+$)     &  $\tau_{3-2}$ & N(H$^{13}$CO$^+ $ )   & \carbontt \\
($''$)  	  & log(cm$^{-2}$) &               & log(cm$^{-2}$) &           \\
\hline
3	          & 15.4(14.1)     &   3.6         &   13.9(12.3)   &  38.2(1.9)  \\
4	          & 15.1(13.2)     &   1.4         &   13.6(12.1)   &  27.2(0.8)  \\
5	          & 14.8(12.9)     &   0.8         &   13.4(12.2)   &  21.1(1.3)  \\
10	          & 14.1(12.2)     &   0.15        &   12.8(11.2)   &  19.2(0.6)  \\
15	          & 13.8(11.9)     &   0.044	   &   12.5(10.9)	&  18.6(0.5)  \\
\hline
\end{tabular}
\tablefoot{Fit with only column density ($N$) as free parameter. All other parameters are fixed to those derived for source size $5''$ to isolate the effect of source size parameter on derived column density and optical depth (see Sect.~\ref{sec.isotopologues}).}
\end{table}

\begin{table}
\centering
\caption{Integrated intensity isotopic ratios towards NGC~253 in the literature}
\label{tab.isotopicratiosliterature}
\begin{tabular}{l c c c c c}
\hline
\hline
Isotopes        &  Ratio    & Proxy                     &  Beam   &  Reference \\
\hline
\carbontt       &  $16\pm 3$       &  HCN/H$^{13}$CN             &  $26''$ & 1  \\   % 16-> $16.2 \pm 3.3$ listed but apparently taken from Nguyen-Q-Rieu et al 1992 and Mauersberger & Henkel 1991
                &  $10.7\pm 1.7$   &  HCO$^+$/H$^{13}$CO$^+$     &  $26''$ & 1  \\ % 11->$10.7 \pm 1.7$ but from Nguyen-Q-Rieu et al 1992 and Mauersberger & Henkel 1991
                &  $28\pm6$        &  HNC/HN$^{13}$C             &  $26''$ & 1  \\ % 29-> $28.2 \pm 6.4$
                &  $14\pm 3$\tablefootmark{a}  &  CS/$^{13}$CS               &  $16''$ & 1  \\ % 15-> $14.7 \pm 3.1$
                &  $>40$    &  CN/$^{13}$CN               &  $22''$ & 1  \\ %ok
                &  $>60$    &  C$^{18}$O/$^{13}$C$^{18}$O &  $23''$ & 2 \\ % ok
                &  $>56$ &  CCH/$^{13}$CCH             &  $14''$ & 2 \\ % separate entries with 13CCH and C13CH
                &  $>46$ &  CCH/C$^{13}$CH             &  $14''$ & 2 \\ %
                &  $36$\tablefootmark{b}  &  CN/$^{13}$CN               &  $22''$ & 3 \\
                &  $13.7\pm0.5$    &  HCN/H$^{13}$CN             &  $24''$ & 8  \\ % 14-> $13.7 \pm 0.5$ beam 28''  from the Ncol values in their Tab.2
                &  $14\pm2$        &  HCO$^+$/H$^{13}$CO$^+$     &  $24''$ & 8  \\ % 15-> $14.5 \pm 2.5$  beam 28''
                &  $17\pm1$        &  HNC/HN$^{13}$C             &  $27''$ & 8  \\ % 17-> $12 \pm 1$ beam 24''->27''  -> Leave it to 17, since we report integrated intensities, not that in Table 2 from Aladro.
                &  $34\pm2$        &  CS/$^{13}$CS               &  $25''$ & 8  \\ % 34->$29 \pm 2$ beam 24''->25''  -> Same as above.
                &  $28\pm2$\tablefootmark{c}   &  CN/$^{13}$CN               &  $24''$ & 8  \\  % 28->$28 \pm 2$
                &  $10-20$  &  HCN/H$^{13}$CN             &  $2''$  & 11 \\  % rather 10-15->$\sim 10-20$
                &  $10-20$  &  HCO$^+$/H$^{13}$CO$^+$     &  $2''$  & 11 \\ % rather 10-15->$\sim 10-20$
                &  21       &  C$^{18}$O/$^{13}$C$^{18}$O &  $3''$  & 4 \\
                &  $19-53$\tablefootmark{d}  &  CN/$^{13}$CN            &  $2.5''$& 5  \\	
\oxigense       & $\sim 150$\tablefootmark{e}  &  $^{13}$CO/C$^{18}$O        &  $12''$ & 6 \\ % no clear value with uncertainty
                & $>300$\tablefootmark{e} &  $^{13}$CO/C$^{18}$O        &  $12$   & 2 \\ % From HarrisonHenkel
                &  130      &  $^{13}$CO/$^{13}$C$^{18}$O &  $3''$  & 4 \\	
                &  $69\pm2$  &  HCO$^+$/HC$^{18}$O$^+$     &  $28''$ & 8  \\ % 69->$45 \pm 2$ beam 24''->28'' ... I take the values in their Tab.2 ...-> Leave it to 69, since we report integrated intensities, not column density ratios from the Table 2.
\oxigenes       &  10       &  C$^{18}$O/C$^{17}$O        &  $16''$ & 7 \\
                &  $\sim 6.5$      &  C$^{18}$O/C$^{17}$O        &  $12''$ & 6       \\ % no clear value with uncertainty
                &  $7.6 \pm 0.5$   &  C$^{18}$O/C$^{17}$O        &  $23''$ & 8  \\ % $7.6 \pm 0.5$   beam 24''->23''
                &  4.5      &  C$^{18}$O/C$^{17}$O        &  $3''$  & 4 \\	
$\rm^{16}O/^{17}O$ &  $\sim300-500$&  CO        &  $2''$  & 11 \\ % rather $350-500$->$\sim 300-500$
%\nitrogenff     &           &            &         &             \\ 
\sulfurtttf     &  16       &  CS/C$^{34}$S               &  $24''$ & 9  \\
                &  $8$      &  CS/C$^{34}$S               &  $16''$ & 9  \\
                &  $8$\tablefootmark{e}   &  $^{13}$CS/C$^{34}$S        &  $16''$ & 10 \\
                &  $>16$\tablefootmark{e} &  $^{13}$CS/C$^{34}$S        &  $16''$ & 2 \\ % From Martin2006
                &  $7.5 \pm 0.3$        &  CS/C$^{34}$S               &  $24''$ & 8  \\  % 8->$7.5 \pm 0.3$   beam 24''->25''
%\silicontetn    &           &            &         &             \\ 
%\silicontte     &           &            &         &             \\ 
\hline
% 150km/s width
\end{tabular}
\tablefoot{
\tablefoottext{a}{With this line ratio, \citet{Henkel1993} (see their Section 5) estimated a column density ratio of $>40$ on opacity grounds, under the assumption that previously measured CS/C$^{34}$S$\sim8$ by \citet{Mauersberger1989a} should actually be equal to the average Galactic and solar system sulfur ratio of 23.
%\sulfurtttf ~of 8 should actually 23 as previously measured in the Galaxy.
}
\tablefoottext{b}{Based on a hyperfine structure derived opacity estimate of 0.5 for the main J=2-1 feature, which results in an opacity corrected ratio of 40.
}
\tablefoottext{c}{Column density ratio estimated using a fit to the CN profiles with MADCUBA \citep{Martin2019a}.
} 
\tablefoottext{d}{When corrected by an average estimated opacity based on the CN hyperfine structure the authors derive a ratio ranging $30-67$, with notable difference between that at the emission peak ($30-40$) and the galaxy outskirts ($47-67$).
}
\tablefoottext{e}{Indirect measurement using the $^{13}$C isotopologue and multiplying by an assumed \carbontt~ ratio which differs among studies.
}
\tablebib{
(1) \citet[][and references therein]{Henkel1993};
(2) \citet{Mart'in2010a};
(3) \citet{Henkel2014};
(4) \citet{Martin2019};
(5) \citet{Tang2019};
(6) \citet{Harrison1999};
(7) \citet{Sage1991};
(8) \citet{Aladro2015};
(9) \citet{Mauersberger1989a};
(10)\citet{Mart'in2006};
(11) \citet{Meier2015};
%(12) This work???
}
}
\end{table}

In the following we discuss individual atomic ratios which are probed by our molecular column density ratios and are globally averaged over the CMZ in NGC~253 at our ACA resolution.

\subsubsection{Carbon}
\label{sec.isotopologuesCarbon}

Observations within the center of our Galaxy show the dependency of the observed isotopic ratios on the molecular species used as proxy. Such variations are observed in \carbontt\ using H$_2$CO, CO, and CN isotopologues \citep{Gardner1982,Langer1990,Milam2005}.
In fact, recent chemical models predict different \carbontt\ ratios for different molecular species depending on the density, on the chemical formation pathways, the temperature and on the cosmic-ray ionisation rate \citep{Colzi2020,Loison2020,Viti2020}.
Therefore, variations among species are expected as they can also trace different gas components.

Under the assumptions described in Sect.~\ref{sec.isotopologues} and excluding the values derived from CO isotopologues (see below), our observations show column density derived \carbontt~ratios ranging from 17 to 60,
%for most $^{13}$C isotopologues detected, which averages to
with an error weighted average $\sim 25\pm 10$. The largest values are derived from the isotopologues with the faintest transitions, close to the noise level and more affected by the uncertainties due to blending to other species. Therefore, the uncertainties in the ratios may be significantly higher than the pure statistical uncertainty from the fit. Still, they are included in Table~\ref{tab.isotopicratios} for the sake of completeness. The weighted average taking into account only the ratios with the lowest uncertainties, that is the brighter species, namely HCO$^+$, HCN, HNC, CN, CS, and H$_2$CO, is $22\pm2$. This value is just marginally lower than the weighted average considering all species and in much narrower agreement among them as can be graphically seen in Fig.~\ref{fig:IsotopeRatioPlot}.

The low average ratio observed in the brighter species of $22\pm2$ is in good agreement to previous low resolution observations (see Table~\ref{tab.isotopicratiosliterature} and notes therein), which were used to derive the nominal extragalactic ratio of 40 based on opacity considerations.
In fact, the only way to reconcile the average \carbontt~ ratio with the literature value of \carbontt$\sim 40$ is by assuming a smaller source size (see discussion in Sect.~\ref{sec.isotopologues}) where the emitting gas within this $850\times340$~pc region is actually confined to a region of $\lesssim 5''$ ($\lesssim 85$~pc). This would imply that all the brightest species showing ratios around $\sim20$ in Table~\ref{tab.isotopicratios} would be affected by approximately the same optical depth. However, based on the opacity of the brightest transition of these species (Table~\ref{tab.isotopicratios}), we do not expect the same opacity effect on all these species, and it would not be supported by the discussion of the opacity of sulfur species (Sect.~\ref{sec.isotopologuesSulfur}). 
%This might indicate that these ratios are actually lower because of other effects, such as isotopic fractionation.

\begin{figure}
\begin{center}
\includegraphics[width=0.45\textwidth]{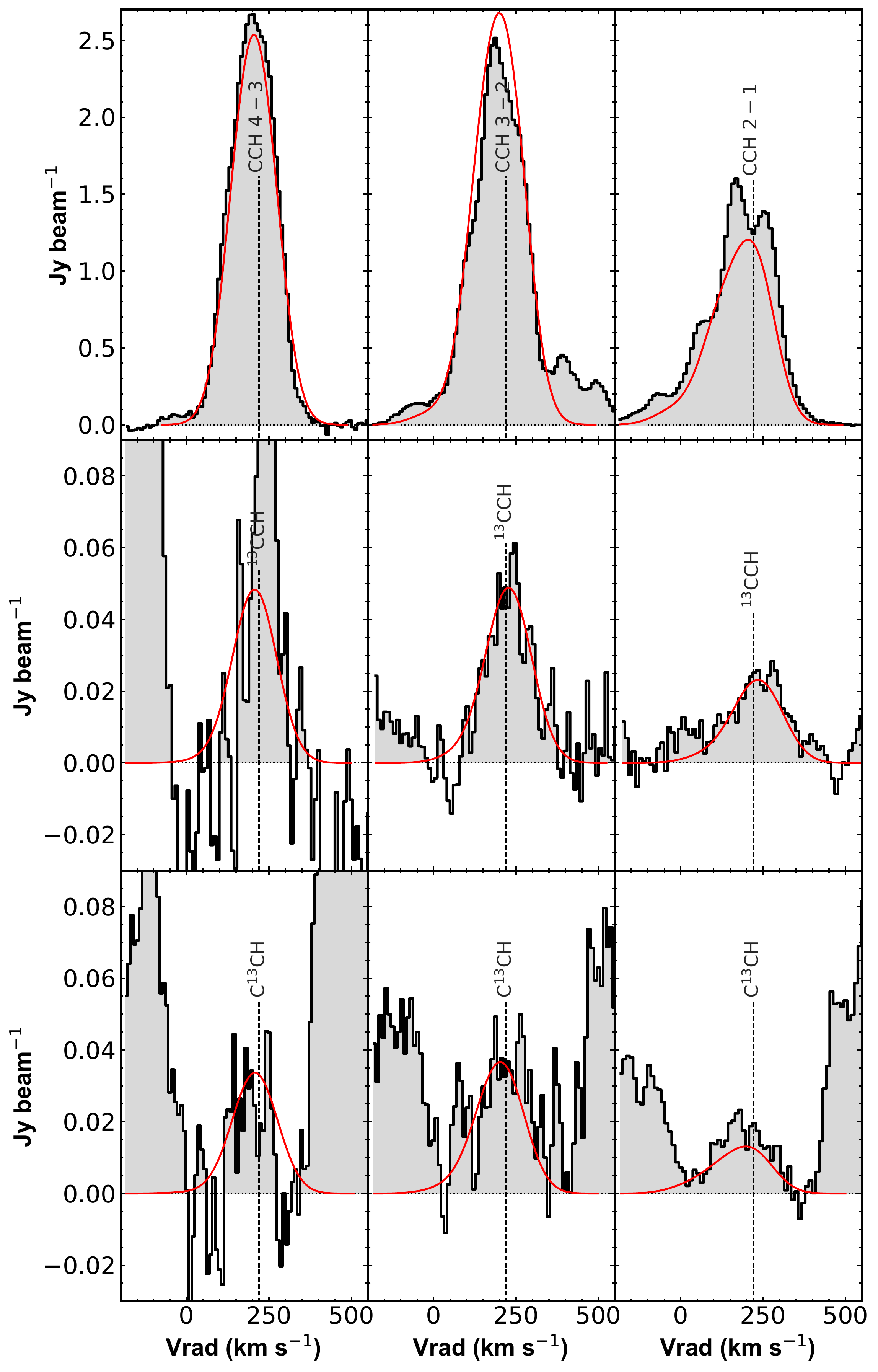}
\caption{Spectral features of the $J=4-3$ (left), $3-2$ (center) and $2-1$ (right) transitions from three of the isotopologues of CCH detected in the survey. Red curves shows the LTE model best fits. Contributions by other species is almost negligible for most transitions according to the global LTE model from the full survey (see Sect.~\ref{Sec.LTE}). However we note that the observed profile of $^{13}\rm CCH~ 4-3$, not following the CCH LTE model and not accounted for by emission from other species, was not used in the fit.
\label{fig.CCHisotopologues}
}
\end{center}
\end{figure}

We note that, $\rm CN/^{13}CN$ observations by \citet{Tang2019} derived opacity corrected ratios ranging $30-67$, and at a similar angular resolution a value of 21 was derived from $\rm C^{18}O/^{13}C^{18}O$ by \citet{Martin2019} .  This discrepancy cannot be explained by opacity effects (see Sect.~\ref{sec.opacity}), but rather by tracing a different molecular component.
Despite the uncertainty on the fainter isotopologues, the ratios derived from CCH isotopologue pairs (Fig.~\ref{fig.CCHisotopologues}) are consistent with the limit of $>50$ derived at single dish resolution based on the non detection of both C$^{13}$CH and the $^{13}$CCH by \citet{Mart'in2010a}. 
%This is due to the fact that although the peak flux ratio we measure of $\sim50$ and $\sim90$ from the $^{13}$CCH and C$^{13}$CH $2-1$ hyperfine group, the resulting column density ratio is the one reported in Table~\ref{tab.isotopicratios}. {\bf Need to double check that spectroscopy of CCH in JPL and CDMS provides exactly the same results.}

The ratios estimated based on CO isotopologues appear to be significantly different than those from higher dipole moment species. As mentioned in Sect.~\ref{sec.isotopologues}, the lower value derived from CO/$^{13}$CO is the result of a large opacity affecting the CO main isotopologue, so this value is not considered meaningful.

\begin{figure}
\begin{center}
\includegraphics[width=0.45\textwidth]{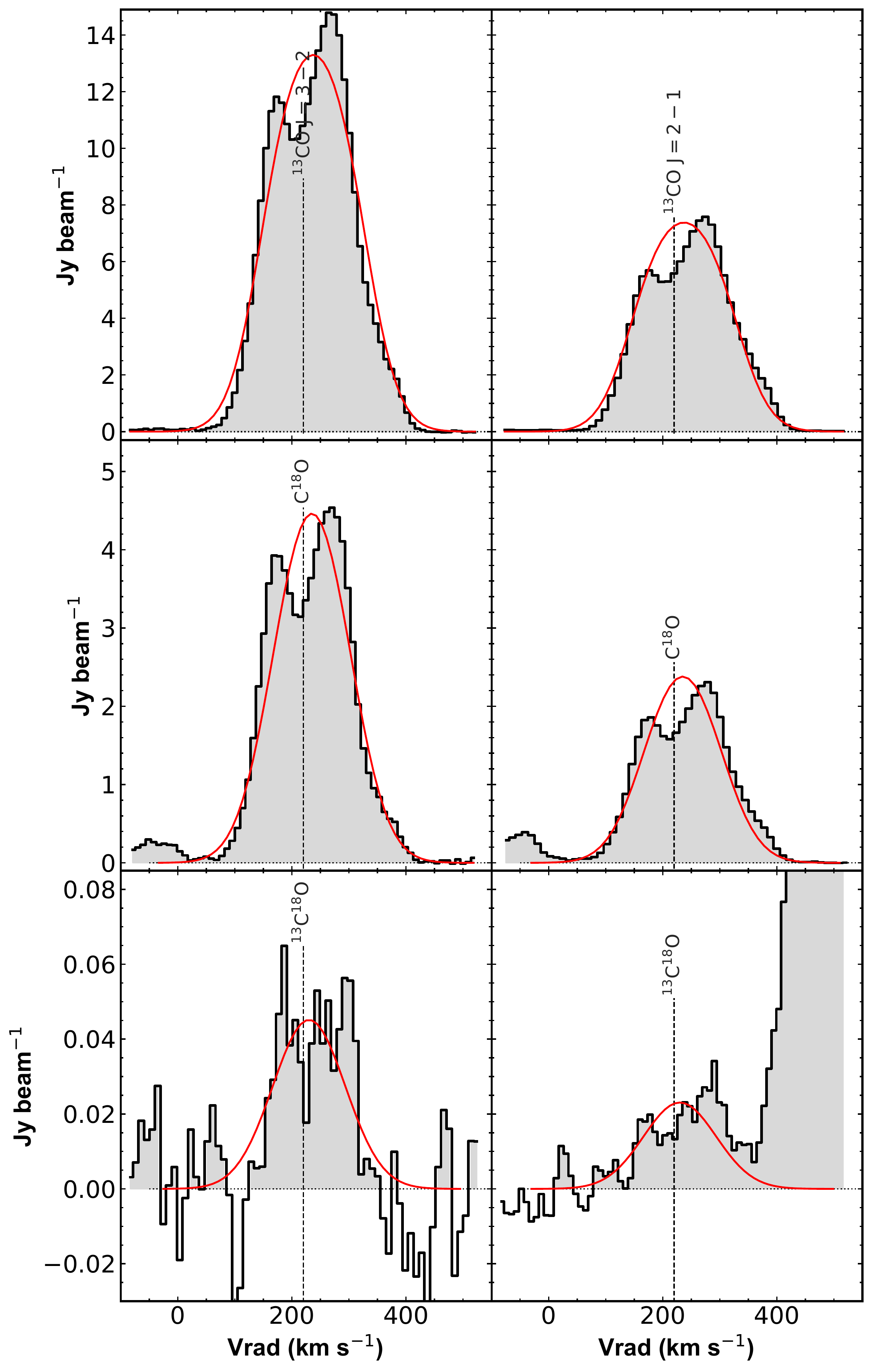}
\caption{Spectral features of the $J=3-2$ (left) and $2-1$ (right) transitions from three of the rarer isotopologues of carbon monoxide detected in the survey. Red curves show the LTE model best fits. Contributions by other species are not shown since all features appear to be free of contamination.
\label{fig.COisotopologues}
}
\end{center}
\end{figure}

Now focusing on the rarer CO isotopologue pairs, our averaged \carbontt~ ratio based on higher dipole moment species agrees well with the value of $\sim 21\pm6$ derived at $3''$ resolution using the C$^{18}$O/$^{13}$C$^{18}$O column density ratio \citep{Martin2019}. In their work, they concluded that there was no obvious signature of high optical depth in the measured 3~mm transitions of the rarer CO isotopologues.
However, our measured C$^{18}$O/$^{13}$C$^{18}$O ratio in the $15''$ resolution data in this work is a factor of 4 above the ratio measured with other species and to that measured with the same isotopologue pair at $3''$ resolution by \citet{Martin2019}. On the other hand this ratio agrees with the limit of $>60$ estimated from single dish observations at $23''$ resolution \citep{Mart'in2010a}. This high ratio is clearly apparent in the spectral features shown in Fig.~\ref{fig.COisotopologues}. We note, for completeness, that both the single dish and high resolution measurements referred here use the $J=1-0$ transitions.  In principle this should not cause differences in the ratios if both transitions originate from the same regions and share similar excitation conditions.
As pointed out by \citet{Martin2019}, the different ratios measured at high and low resolution might be evidence for the existence of two distinct components with different degrees of stellar nucleosynthesis ISM processing, similar to what is observed in the Galactic Center \citep{Riquelme2010}. As indicated by \cite{Viti2020}, the difficulties to reproduce such a high value from chemical modeling is evidence of the large range of carbon isotopic ratios measured being the result of nucleosynthesis and not fractionation.

In this multi-component scenario, the low resolution observations would be tracing extended unprocessed material with a \carbontt~ratio of $\sim100$, similar to what is observed in the outskirts of our Galaxy \citep{Wouterloot1996}. This material must have recently been driven towards the nucleus from the outer regions by the stellar bar in NGC~253 (Sect.~\ref{sec.NGC253}) and not yet enriched by ongoing star formation in the CMZ. On the other hand, the denser and more compact gas dominating the GMC emission would be enriched by the starburst event \citep[e.g.][]{Romano2017} resulting in \carbontt~ratios of $\sim20-30$ as traced by optically thicker high electric dipole moment molecules even at low resolution. The high resolution observations of the optically thinner CO transitions would also trace this processed molecular gas with \carbontt~ ratio of $\sim21$ \citep{Martin2019}.

%Some measured ratios have values significantly outside this range.
%The ratio of 63 measured with CCH/C$^{13}$CH, as shown in Fig.~\ref{fig.CCHisotopologues}, is due to the contamination of the C$^{13}$CH $J=3-2$ transition. If removed from the fit, the fit to the non contaminated transitions $4-3$ and $2-1$ result in a column density $40\%$ higher, and thus a CCH/C$^{13}$CH of 38.
%Similarly to CCH, the measured ratio with methanol is affected by the uncertainty on the fit to the weak $^{13}$CH$_3$OH (see Appendix~\ref{Sec.AppendixFitDetails}).

Tracing the more extended, potentially less processed gas by CO isotopologues agrees with them being the only transitions showing extended emission at the ACA resolution (Fig.~\ref{fig.moment0maps}). Although selective photodissociation might play a more important role in $^{13}$C$^{18}$O, this effect should also affect C$^{18}$O and to a lesser extent $^{13}$CO. We note that despite the uncertainty due to the noise, $^{13}$C$^{18}$O line profiles match very well those of the more abundant CO isotopologues (Fig.~\ref{fig.COisotopologues}). 

\subsubsection{Oxygen}
\label{sec.isotopologuesOxygen}

The situation with oxygen ratios is consistent with that encountered with carbon ratio measurements.
Our \oxigense~ratio of $100\pm20$ measured with HCO$^+$ could be reconciled with the value in the literature of 200 if we account for a possible multiplicative factor of 2 due to opacity as discussed above.
The value derived with the main CO isotopologue is half ($48\pm5$) of that derived with HCO$^+$, once again due to the high opacity of CO (Sect.~\ref{sec.isotopologues}). However, that derived with the $^{13}$C substitutions of CO results in a ratio of $520\pm60$ which, similar to the carbon isotopic ratio discussed above, is a factor of 5 above that measured with HCO$^+$, but still within the range of values observed within the Milky Way as displayed in Fig.~\ref{fig:IsotopeRatioPlot}. 
The \oxigense~ ratio is in reasonable agreement with previous single dish measurements \citep[$69$,][]{Aladro2015} and in good agreement with the value measured at higher resolution with the $^{13}$C substitutions of CO \citep[130,][]{Martin2019} as shown in Table~\ref{tab.isotopicratiosliterature}.

The scenario depicted by the oxygen ratios would be similar to that from carbon isotopic ratios with a low $^{18}$O enriched material traced by the optically thin rarer CO isotopologues.
Regarding $^{17}$O, we find good agreement between our measured \oxigenes$\sim9$~ ratio and the range derived from single dish measurements ($7-10$).
The $\rm^{16}O/^{17}O\sim400$ derived with the main CO isotopologue, similar to the values derived toward individual GMCs by \citet{Meier2015}, would translate into an actual value of $\sim800$ based on the differences observed between CO derived ratios and those with other species above.
Finally the low $\rm^{16}O/^{17}O\sim13$ value deduced from SiO sheds doubts on the actual detection of Si$^{17}$O unless its flux density is boosted due to non-LTE effects, in which case the ratio measured would not be indicative of the actual oxygen ratio. We therefore report its detection as tentative.

\subsubsection{Nitrogen}
\label{sec.isotopologuesNitrogen}

Chemical modeling of nitrogen isotopic ratios has recently been published by \citet{Viti2019}, where models are compared with existing measurements in the extragalactic ISM. Here we present the first $^{15}$N isotopologue detection towards NGC~253 and we place it in the context of previous observations towards other galaxies and high-redshift molecular absorbers.
The $^{14}$N/$^{15}$N nitrogen isotopic ratio measured with both HCN ($147\pm15$) and HNC ($220\pm50$) are consistent with a weighted average of $170\pm20$. This value is also in agreement with the $^{14}$N/$^{15}$N$\sim152\pm27$ measured towards the molecular absorber PKS~1830-211 \citep{Muller2011} and with the $^{14}$N/$^{15}$N$\sim227\pm41$ of towards NGC~4945 \citep{Henkel2018}. In both objects, the HNC ratio is higher, but significantly more uncertain, than that measured with HCN.

The H$^{13}$CN/HC$^{15}$N ratio of $5.6\pm0.6$ is close to that measured towards PKS~1830-211 ($4.8\pm0.2$) and almost twice that of the other molecular absorber B~0218+357 of $3.0\pm0.5$ \citep{Wallstroem2016}. This ratio is $35\%$ lower than that measured towards NGC~4945 at $\sim40$~pc resolution \citep[$2''$,][]{Henkel2018}, where they find a global value of $\sim8.5\pm1.7$, although values of $4-6$ are measured outside the very nuclear region, in contrast with the value of $2.1\pm0.3$ at $60''$ resolution towards the same object \citep{Wang2004}.
On the other hand the ratio HN$^{13}$C/H$^{15}$NC of $13\pm3$ is about twice that measured towards PKS~1830-211 of $\sim7$ \citep{Muller2011}.

Similar to what is discussed with the other isotopic ratios, the effect of opacity could raise the \nitrogenff\;ratio up to 400 if the opacity correction suggested in the carbon isotopic ratio discussion is considered. Based on the updated $^{14}$N/$^{15}$N galactocentric trend by \citet{Colzi2018}, a value of 400 would correspond to a 8~kpc galactocentric distance, while the 4~kpc molecular ring shows a ratio of $\sim 300$, which decreases 200 at $\sim1$~kpc. The latter is consistent with the value derived here, uncorrected for putative high opacity.
%This value would be closer to what is measured in the Galactic 4~kpc molecular ring but still would be lower that what is measured in the Galactic Center \citep{Wilson1999}.

We note that the non detected spectral features of both N$^{15}$NH$^+$ and NN$^{15}$H$^+$ down to our achieved noise level is in agreement with the derived $^{14}$N/$^{15}$N.

\subsubsection{Sulfur}
\label{sec.isotopologuesSulfur}

\sulfurtttf~ ratios measured with both CS and SO ($9.7\pm0.5$ and $14\pm4$, respectively) are in good agreement with previous single-dish measurements based on CS observations of $8\pm2$ \citep{Martin2005} and more than a factor of two above that derived from SO \citep[$5.1\pm1.2$][]{Mart'in2006}. %, both with single-dish observations. 
However, the latter is likely the result of uncertain de-blending from SO$_2$ emission on a very faint feature \citep[see Appendix C.4 in][]{Mart'in2006}

However, if the CS emission is actually optically thick, in order to reconcile the carbon isotopic ratios with the nominal extragalactic value of 40 (see Sect.~\ref{sec.isotopologues}), then the main CS and SO isotopologues should show the same peak optical depth. This way, the derived \sulfurtttf~ would be similar in both species.
Based on the fit values in Table~\ref{tab.fitmolecparams}, the opacity ratio between both species is measured to be 
%would be similarly optically thick despite the abundance ratio of C$^{34}$S/$^{34}$SO$\sim3$ corresponding to an opacity ratio of 
$\tau_{\rm C^{34}S/^{34}SO}\sim15$ which does not support the idea of CS being optically thick, unless both species are tracing very different sulfur isotopic ratio components.
%as can be derived from Table~\ref{tab.fitmolecparams}.

Our sulfur isotopic ratio of \sulfurtttf~$\sim 10$ is about half of 
%Moreover, accounting for a factor of 2 opacity correction to our ratio would set NGC~253 ratio close to 
the Galactic Center value of $\sim 22$ measured with $^{13}$CS/C$^{34}$S (assuming a \carbontt) or $\sim$17, directly obtained from $^{13}$CS/$^{13}$C$^{34}$S \citep{Frerking1980,Wilson1994,Humire2020}.
However, taking into account the \sulfurtttf~ Galactocentric gradient reported by \citet{Yu2020}, there is a good agreement with the trend value within the Galactic central kiloparsec.

Our measured $^{13}$CS/C$^{34}$S=$0.42\pm0.04$ is in between the average value measured in Galactic disk sources and sources in the Galactic Center region \citep[between 0.35 and 0.68][]{Frerking1980}.

Although as discussed above we do not see evidence for opacity effects in the sulfur isotopologue ratios, both opacity uncorrected and corrected values would lie within the range of values observed in other extragalactic sources \citep[see, e.g.,][]{Wallstroem2016}.

\subsubsection{Silicon}
\label{sec.isotopologuesSilicon}

Here we derive the first silicon isotopic ratios measured in emission towards external galaxies through the observed ration between SiO and the rarer $^{29}$SiO, $^{30}$SiO, and Si$^{17}$O substitutions. 

The ratios presented in Table~\ref{tab.isotopicratios} yields $^{29}$Si/$^{30}$Si$\sim4.2$, which is more than a factor of two above the value of 1.5 within the Galaxy \citep{Penzias1981a,Anders1989,Lodders2003}, and the value of 1.9 measured towards PKS~1830-211 \citep{Muller2013}.
The apparent overabundance of $^{29}$Si might imply that this is not a primary nucleus but resulting from some stellar processing. However, observations of silicon isotopic ratios toward a wider variety of environments are required to further analyse the origin of silicon isotopologues, since there appears to be no difference between solar and Galactic Center observations \citep{Wilson1999}.

On the other hand, the ratio \silicontetn=$9\pm2$ is however consistent with the ratio of 11 measured towards the molecular absorber PKS~1830-211 \citep{Muller2011} and about a factor of 2 below the Solar system value of 19.6 \citep{Anders1989,Lodders2003}.

\subsubsection{Isotopologue Ratio Overview}
\label{sec.isotopologuesOverview}

In this section we provide an overview of the results obtained from the different isotopologue ratios measured in this work, which are graphically summarized in Fig.~\ref{fig:IsotopeRatioPlot}.

Carbon, oxygen, nitrogen, and sulfur ratios measured with the brightest species other than CO, namely HCO$^+$, HCN, HNC, CN, CS, and H$_2$CO, are on the low end of the range measured in the Milky Way. In fact, the isotopologue ratios towards the NGC\,253 CMZ are consistent with the values observed within the Galactic CMZ. These high dipole moment species may be tracing molecular gas enriched by the starburst present in the CMZ, which leads to an enrichment in the rarer isotologues, similar to what is derived from the Galactocentric trends observed in the Milky Way Galaxy \citep{Wilson1999,Milam2005,Yu2020}. The emission from these less-abundant species appear unresolved in the $15''$ resolution ACA data (Fig.~\ref{fig.moment0maps}).

Carbon and oxygen isotopic ratios measured with CO appear to have a dependency on the probed scales. The $3''$ resolution ratios measured with the rarer CO isotopologues yielded reported ratios similar to those measured for the high dipole moment species in this work \citep{Martin2019}.
However, the ACA data discussed here, recovering larger scales, yields carbon and oxygen ratios consistent within the limits derived with single-dish observations \citep{Mart'in2010a}, that is a factor of 5 larger than those measured at higher resolution. Carbon monoxide may be tracing a more extended molecular component not yet involved in and enriched by the star formation burst within the CMZ. Indeed, CO and its isotopologues are the only species showing extendend emission at the ACA resolution (Fig.~\ref{fig.moment0maps}). This dependence on the size scales measured highlights the importance of tracing the molecular emission at all scales, and in particular the importance of analyzing the ACA data which probes the largest molecular spatial scales.

Although opacity considerations may result in ratios which differ by up to a factor of $\sim2$ from the higher dipole moment species, in line with the ratios commonly found in the literature, this would imply that all species are affected by the same optical depth correction. Moreover, the high resolution study of CO isotopologues \citep{Martin2019}, resulting in similar ratios to those from high dipole moment species as mentioned above, did not show evidences of optical depth effects except for potentially towards one of the brightest GMCs.
Based on the analysis of CN emission in Sect.~\ref{sec.opacity}, it is likely that the emission from all species contains an optically thick component surrounded by an optically thin envelope within the individual GMCs.
 
 Fractionation is considered to be negligible under the physical conditions in the bulk of the gas in galaxies \citep{Romano2017}. However, some species in our work might be showing some hints of fractionation. This appears to be the case for CCH (Table~\ref{tab.isotopicratios} and Fig.~\ref{fig.CCHisotopologues}), where differences are found between the two $^{13}$C isomers. 
 The relationship of ratios is not consistent with what is observed in dark clouds where $\rm CCH/^{13}CCH>CCH/C^{13}CH$ \citep{Sakai2010}.
 Our results also do not show evidence of the HC$_3$N isotopologue abundance differences that are measured in low-mass star forming regions \citep[e.g.,][]{Araki2016}.
 The uncertainty on the abundance ratios of isotopologue substitution of these more complex species have a relatively large uncertainty, which makes it also difficult to claim fractionation in CH$_3$CCH. Therefore we leave the discussion on the actual effect of fractionation in NGC\,253 to a subsequent publication on the of individual GMCs within the NGC\,253 CMZ measured at higher spatial resolution.
 
\subsection{Opacity analysis with multi-transition CN observations}
\label{sec.opacity}

%{\bf Seb: is this such a strong case that it should be an highlight of ALCHEMI ACA ? Maybe the section could be turned into the glory of multi-transition surveys to constrain opacity: take the case of CN and CCH, and apply to the whole set of lines. And discuss the impact of limited number of transitions on the analysis, with clear examples taken from ALCHEMI. What if we take eg 2 out of N lines from a given molecule. What error in Ncol, Tex, etc?}

One of the main sources of uncertainties in estimating column densities of the emitting molecules, or equivalently relative abundances, is the unknown effect of optical depth \citep{Mangum2015,Martin2019a}. Although the MADCUBA fit accounts for the effect of opacity (Sect.~\ref{Sec.LTE}), the opacity (linearly related to the column density) and source size are partially degenerate (see discussion in Sect.~\ref{sec.isotopologues}).
%As mentioned in Sect.~\ref{Sec.LTE}, MADCUBA fit accounts for the effect of opacity \citep{Martin2019a}. However, the fitted column density, and therefore optical depth, is directly linked to the assumed source size. Since, in the approximation from MADCUBA (see footnote in Sect.~\ref{Sec.LTE}) flux density is proportional to the emitting solid angle, the opacity and column density required to model the observed emission will depend on the source size assumed. Thus requiring a higher column density (optical depth) if a smaller emitting extent is considered to fit the same observed profile.
However, some molecular species present hyperfine structure transitions which are separated enough to constrain the opacity based on either the spectral feature profile or the ratio between the different groups of hyperfine spectral features.
%Many studies in the literature have used molecular hyperfine structure transitions as a direct way to estimate opacity of a transition \citep{Henkel2014,Tang2019}. 
Such is the case of species like CN and CCH.
% whose groups of hyperfine structure transitions are separated enough in frequency so they can be resolved even on large velocity dispersion spectra as that of low resolution extragalactic observations. 
In this section we analyze the hyperfine structure of multiple CN and CCH transitions,
%since the line width at this coarse spatial resolution together with the blending with other species, do not allow to perform this analysis on CCH. Here 
making use of the wide frequency coverage of the ALCHEMI data set.  Through this analysis we highlight the importance of multi-transition observations for an accurate opacity determination.
%Within the covered frequency band in our ACA data, 
%CN presents two very clean and separated groups of transitions. % in the covered frequency range by our survey. 

\begin{figure*}
\begin{center}
\includegraphics[width=0.3\textwidth,valign=t]{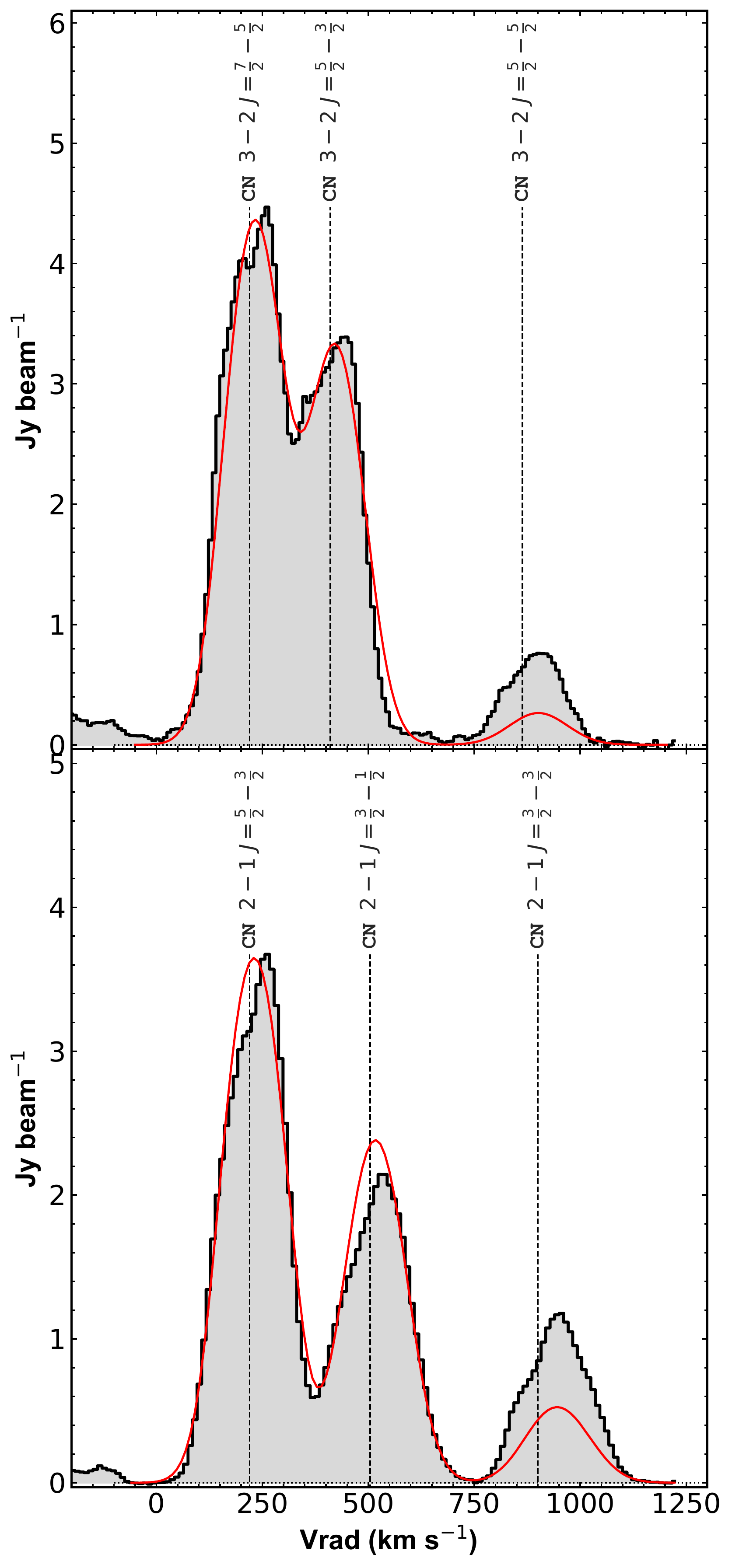}
\includegraphics[width=0.3\textwidth,valign=t]{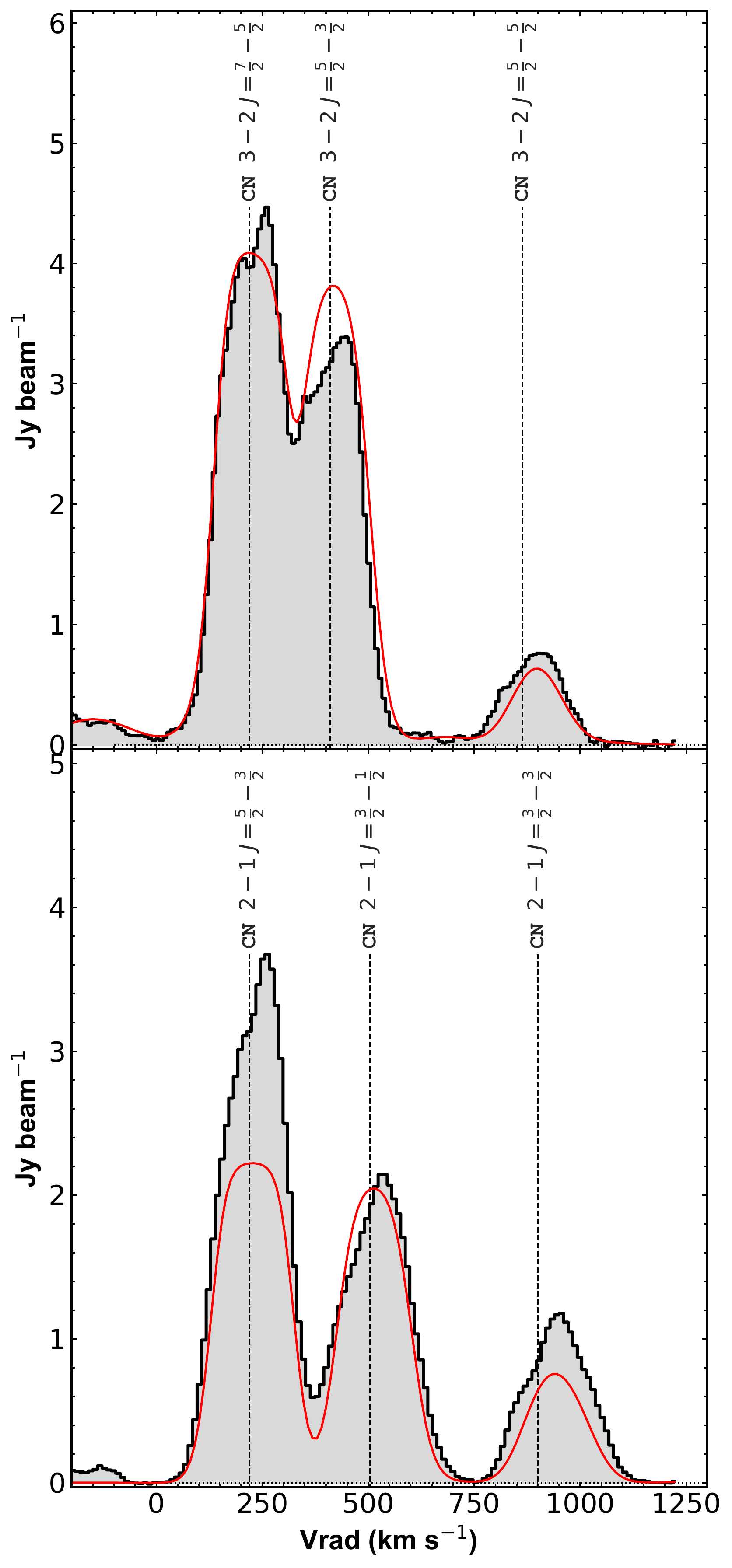}
\includegraphics[width=0.3\textwidth,valign=t]{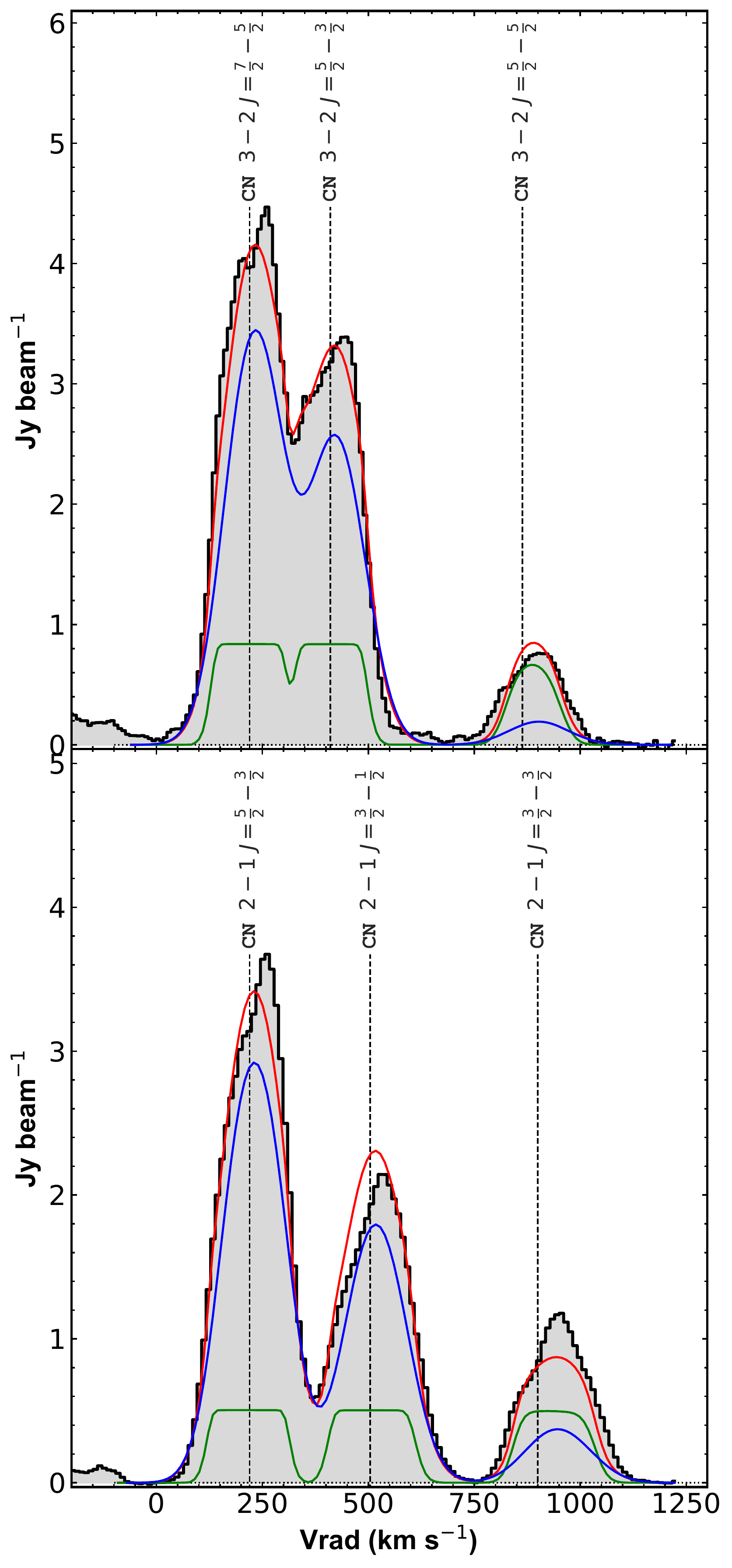}
\caption{CN $N=3-2$ (top) and $N=2-1$ (bottom) emission measured with the ACA data presented in this article. Panels are centered at 340.248 and 226.874~GHz, which is the frequency of the stronger emission hyperfine transition. The data are fitted with three simple models.
($left$) Single component corresponding to the fit presented in Sect.~\ref{Sec.LTE}, where a source size $\theta_s=5''$ is assumed. The central opacity of the brightest transitions are $\tau_{3-2}=0.25$ and $\tau_{2-1}=0.44$, while the cumulative opacity of the stronger hyperfine groups are $\tau^g_{3-2}=0.6$ and $\tau^g_{2-1}=1.0$. See text in Sect.~\ref{sec.opacity} for details.
($center$)
Single component fit assuming a smaller source size of $\theta_s=2.5''$. Physical parameters of the fit are $log(N)=16.43~\rm cm^{-2}$, $T=13.4~\rm K$, $v_{lsr}=229$~\kms, $\delta~v_{1/2}=118$~\kms. Opacity of brighter transitions are $\tau_{3-2}=1.6$ and $\tau_{2-1}=2.1$, and group opacities $\tau^g_{3-2}=3.8$ and $\tau^g_{2-1}=4.7$.
($right$)
Two component model assuming two sources of sizes $\theta_s=5''/1.5''$. Physical parameters of models are $log(N)=15.6/17.3~\rm cm^{-2}$, $T=10.2/10.0$~K, $v_{lsr}=232/220$~\kms, $\delta~v_{1/2}=150/76$~\kms. Opacity of brighter transitions are $\tau_{3-2}=0.18$ and $\tau_{2-1}=0.31$ for the first component, while the thick component is completely saturated. Blue and green lines represent the two components, respectively, with sum of both models is represented by the red line.
\label{fig.CNopacity}
}
\end{center}
\end{figure*}

Previous studies, both at single dish resolution \citep{Henkel2014} and ALMA high-resolution observations \citep{{Tang2019}}, made use of the CN $N=1-0$ hyperfine split spectral line, consisting of two groups, the $J=\frac{3}{2}-\frac{1}{2}$ and the $J=\frac{1}{2}-\frac{1}{2}$ to estimate the optical depth of this transition.
Within the frequency coverage of the ACA data presented in this article, we observe the $N=3-2$ and $N=2-1$, each consisting of three groups of hyperfine transitions as shown in Fig.~\ref{fig.CNopacity}. The brighter transitions of these groups, for reference, emit at 340.248 and 226.874~GHz, respectively. Here we present the analysis of these profiles using three simple LTE models (Figure~\ref{fig.CNopacity}) to explain their observed emission and to estimate the optical depth affecting these lines.

Figure~\ref{fig.CNopacity} (left panels) shows how the single component optically thin fit used in this work (for a source size of $5''$) cannot reproduce the intensity of the fainter hyperfine groups, namely the $J=\frac{5}{2}-\frac{5}{2}$ and $J=\frac{3}{2}-\frac{3}{2}$ for the $N=J=3-2$ and $N=2-1$ groups, respectively. On the other hand, the two brighter hyperfine groups would have appeared to fit well to the optically thin regime if not considering the third group of transitions.

As a second approach (Fig.~\ref{fig.CNopacity} center panels), we assumed a smaller ($2.5''$) and therefore thicker emission which accounts better for the fainter hyperfine groups. However, it does not manage to properly reproduce the emission from the two brighter groups. The three $N=3-2$ line component groups could be considered reasonably fit within the model uncertainties if we would only be observing at this frequency. However, it clearly underestimates the emission of the brighter $N=2-1$ group.

Finally, the right panels in Fig.~\ref{fig.CNopacity} show a third simple model that considers two components. The model consists of one optically thin component with a $5''$ source size, together with a much smaller ($1.5''$) optically thick component. The %figure clearly shows how this 
model fits well all three hyperfine groups in both CN transitions.
%spectral line groups. 
In this model, the second component, $\sim10\times$ larger in mass and accounting for just $\sim25\%$ of the integrated line emission, would consist of a comb of saturated line profiles for the brighter hyperfine groups. Since the emission is saturated, the fit is %too 
degenerate and therefore the mass contained in this component (column density) is subject to a large uncertainty.

This exercise shows that, as we know from high resolution observations \citep{Tang2019} the integrated emission of NGC~253 is a convolution of molecular components with different degrees of obscuration. Indeed, our unresolved observations point to the presence of a heavily obscured compact emitting region, very likely associated with the GMC dense cores in the NGC~253 CMZ \citep{Sakamoto2011,Leroy2015,Rico-Villas2020} which might contain a large fraction of the mass, surrounded by a thinner likely more widespread component.

In the following we analyze the effect of such differences in size/opacities on the 3~mm transition used in the literature.
By extrapolating the first two models above (left and center in Fig.~\ref{fig.CNopacity}) to the band 3 N=1-0 hyperfine groups used in the literature, we obtain integrated flux density ratios between the $J=\frac{3}{2}-\frac{1}{2}$ and $J=\frac{1}{2}-\frac{1}{2}$ hyperfine groups of 1.95 ($\theta=5''$~optically thin model) and 1.8 ($\theta=2.5''$ optically thick model). Thus, the two models differ only by $8\%$ in observed integrated intensity ratio of the N=1-0 hyperfine groups. Therefore, this transition is less sensitive to optical depth effects than the combination of the two transitions consisting of 3 hyperfine groups observed in the ALCHEMI survey.

%{\bf Final conclusion on this. Extrapolate to Band 3 to compare with previous works}
% BAND 3
% MODEL 1
% 77.0  - 150.7    -> 1.95
% MODEL 2
% 86.6 - 156.893   -> 1.8

%{\color{red} revise if something can be done with CCH}
On the other hand CCH does not show evidence of the existence of such an optically thick component. 
However, the CCH hyperfine structure is not as well separated as in CN, and only the $2-1$ and $3-2$ groups at 174.7 and 262.0~GHz show an asymmetric profile at lower velocity due to this hyperfine structure. Unfortunately the large line width at this coarse spatial resolution together with the blending with other species, does not allow us to perform an analysis on CCH as detailed as on CN with the ACA data.

\section{Summary and conclusions}

%As described in Sect.~\ref{Sec.Introduction},
The central molecular zone of the starburst galaxy NGC\,253 is a complicated but interesting multi-component system with various heating mechanisms playing a role at different scales.
Previous low-resolution single-dish observations of %the molecular composition of
NGC\,253 showed that the averaged molecular abundances towards its CMZ resembles that of Galactic GMCs, and its observable chemistry would be mostly driven by low-velocity cloud-cloud collisions \citep{Garc'ia-Burillo2000,Mart'in2006}.

The ALCHEMI spectral survey of NGC\,253 brings new insights into the chemical composition and physical conditions of the CMZ of NGC\,253. 
In this article we present the ALCHEMI survey covering the spectral range from 84.2 to 372~GHz. 
Our large frequency coverage allows us to accurately align the flux scale of the individual tunings, where we observed significant deviations. Here we present the analysis of a subset of the ALCHEMI survey, consisting of the ACA observations covering the 256.7~GHz wide frequency band (i.e., full ALMA Bands 4 to 7) between 125.2 and 373.2~GHz down to a sensitivity of 0.27 to 1.0~mK.
Even at the moderate resolution of the ACA observations ($15''\sim255$~pc) we observe a rich molecular complexity.

Here we summarize the main conclusions from our analysis of the ALCHEMI ACA data:
\begin{itemize}
    \item 
    Continuum emission in the Rayleigh-Jeans tail can be modelled with a 42~K dust emission temperature with emissivity $\beta=1.9$ plus a free-free component with SFR=2.5~M$_\odot$\,yr$^{-1}$. This SFR is an upper limit since we cannot determine the contribution due to synchrotron emission with the frequencies analyzed in this work. 
    \item 
    The line contribution to the observed continuum emission varies across the bands between 5 and 36$\%$ when splitting the survey into five 50~GHz frequency bins. Continuum flux emission in high-z sources with a potentially significant starburst contribution might need to be corrected in certain frequency ranges according to our findings.
    \item 
    Spectral line identification was performed through LTE modelling per molecule, and not per spectral feature. 78 molecular species, including isotopologues and vibrational states are detected. Additionally, multiple emission features from radio recombination lines, namely H$n\alpha$, H$n\beta$ and He$n\alpha$, are identified throughout the survey.
    \item 
    Newly detected species in the extragalactic ISM include complex organic species and isotopologues, namely H$_2^{13}$CO, ethanol (C$_2$H$_5$OH), $^{13}$CCH, C$^{13}$CH, HOCN, the three $^{13}$C isotopologues of CH$_3$CCH, propynal (HC$_3$HO), and tentatively Si$^{17}$O.
    \item 
    The ALCHEMI survey also provides a useful template for observations of high-redshift galaxies that can be used to estimate the number of individual molecular species that are potentially detectable in a starburst environment as a function of the depth of the observations. Our estimate is based on the stacked spectrum from 22 high-z sources at $z=2.0-5.7$ by \citep{Spilker2014} yields $3-4$ expected species detections, in agreement with their reported identification.
    \item
    Emission from infrared-pumped vibrational states of HCN, HNC, and HC$_3$N is detected for the first time in low resolution observations. However we do not detect vibrational emission of HCO$^+$, similar to what is reported towards the ULIRG IRAS~20551-4250 \citep{Imanishi2017}. We postulate the existence of a "carbon-rich" chemistry as result of oxygen depletion into H$_2$O according to high-temperature chemistry models \citep{Harada2010}. This explains both the rich carbon chemistry observed as well as the apparent lack of emission from vibrationally excited HCO$^+$. This would be similar to what is observed in local ULIRGs, where high abundances of carbon chains \citep[HC$_3$N and HC$_5$N]{Mart'in2011,Costagliola2015, Aladro2015} and water \citep{Mart'in2011,Koenig2017} are reported, as well as HCN rich outflows \citep{Barcos-Munoz2018}.
    \item
    The global averaged $L_{vib}/L_{IR}$ ratio that we measure of $1.4\times10^{-9}$ (from HCN $3-2$) in NGC~253 is an order of magnitude below what is observed in compact obscured nuclei \citep{Falstad2019}. We propose that this ratio is a good proxy of the proto-stellar cluster contribution to the infrared luminosity. Based on this ratio we estimate the vibrational emission originating in Sgr~B2(N) like hot cores would contribute $3\%$ to the total infrared luminosity of NGC\,253, in agreement with previous estimates by \citet{Rico-Villas2020}.
    \item
    Organic molecules, in particular $\rm C_2H_5OH$ and HCOOH, show relative abundances consistent with those found in Galactic Center GMCs, but on the high end, similar to those measured in Galactic hot cores. Although complex organic molecules are observed to be widespread within the Galactic Center, the global abundances in the central starburst environment in NGC~253 may be significantly contributed by hot core chemistry.
    \item
    We report the measurement of isotopic ratios of carbon, oxygen, nitrogen, sulfur, silicon with all the isotopologue pairs detected in our survey. $^{14}$N/$^{15}$N$=170\pm20$ is measured for the first time in NGC~253. $^{28}$Si/$^{29}$Si$=9\pm2$ and $^{29}$Si/$^{30}$Si$=4.2$ are measured for the first time in emission in the extragalactic ISM. Based on the analysis of all these ratios, we do not find evidence for opacity effects globally affecting the derived isotopic ratios.
    \item
    High dipole moment species, namely HCO$^+$, HCN, HNC, CN, CS, and H$_2$CO, show consistent isotopic ratios of \carbontt$=24\pm8$ and \oxigense$=100\pm20$ which is half the standard extragalactic values \citep{Wilson1999} and consistent with the ratios observed within the central kiloparsec of our Galaxy. Nitrogen and sulfur isotopic ratios of \nitrogenff$=170\pm20$ and \sulfurtttf$\sim10$ are also consistent with those in the Galactic CMZ.
    \item
    Carbon and oxygen isotopic ratios derived from the rarer CO isotopologues result in values a factor of 5 larger than those measured with high dipole moment species and with the same CO isotopologues observed at higher spatial resolution. This result appears to confirm the multi-component scenario where CO would be tracing the widespread gas recently funneled towards the CMZ from the galactic outskirts and therefore not yet processed by the prominent nuclear star formation.  In this scenario higher dipole moment species would trace more compact dense gas clumps in the GMCs already enriched in secondary isotopologues.
    \item
    Multi-transition analysis of the hyperfine structure of CN reveals the presence of a likely saturated molecular component which could account for a significant fraction of the molecular mass and which is likely associated with the optically thick cores of GMCs.
\end{itemize}

The forthcoming series of papers based on the high resolution ALCHEMI data set will further investigate each of the topics presented in this paper to peer into the physical conditions that drive the observed averaged molecular abundances. The results in this article are therefore a reference for low linear resolution molecular observations of distant extragalactic sources and for follow-up studies of NGC\,253 with higher angular resolution.

\begin{acknowledgements} 
The authors want to specially thank David Fern\'andez from the JAO ALMA Education and Public Outreach department for his contribution in creating the visuals presented in Figure~\ref{fig.sketch}.
This paper makes use of the following ALMA data: ADS/JAO.ALMA\#2017.1.00161.L and ADS/JAO.ALMA\#2018.1.00162.S. ALMA is a partnership of ESO (representing its member states), NSF (USA) and NINS (Japan), together with NRC (Canada), MOST and ASIAA (Taiwan), and KASI (Republic of Korea), in cooperation with the Republic of Chile. The Joint ALMA Observatory is operated by ESO, AUI/NRAO and NAOJ.  The National Radio Astronomy Observatory is a facility of the National Science Foundation operated under cooperative agreement by Associated Universities Inc.
N.H. is supported by JSPS KAKENHI Grant Number JP21K03634.
K.S. is supported by MOST grants 108-2112-M-001-015  and 109-2112-M-001-020.
L.~C. acknowledges financial support from the Spanish State Research Agency (AEI) through the project No. ESP2017-86582-C4-1-R. 
L.C. and V.M.R. acknowledge support from the Comunidad de Madrid through the Atracci\'on de Talento Investigador Senior Grant (COOL: Cosmic Origins Of Life; 2019-T1/TIC-15379). 
G.A.F acknowledges financial support from the State Agency for Research of the Spanish MCIU through the AYA2017-84390-C2-1-R grant (co-funded by FEDER) and through the ``Center of Excellence Severo Ochoa'' award for the Instituto de Astrof\'isica de Andalucia (SEV-2017-0709). 
L.H. and M.P. are grateful for funding from the INAF PRIN-SKA 2017 program 1.05.01.88.04.
K.K. acknowledges the support from the JSPS KAKENHI Grant Number JP17H06130.
K.K. and Y. N. are supported by the NAOJ ALMA Scientific Research Grant Number 2017-06B.
%Finally, we would like to thank Francesco Costagliola who originated this project as its first lead Principal Investigator.
\end{acknowledgements} 

\bibliographystyle{aa}
\bibliography{ACAALCHEMI}

\onecolumn
\appendix

%\section[Section Name]{Section Name\footnote{Whatever}}

\section[Unscaled and Scaled Spectra]{Unscaled and Scaled Spectra\footnote{Analysis and diagrams in these appendices make extensive use of {\texttt astropy} \citep{AstropyCollaboration2013,AstropyCollaboration2018}.}}
\label{Sec.AppendixFLuxAlignment}

\begin{figure}
\includegraphics[width=\textwidth]{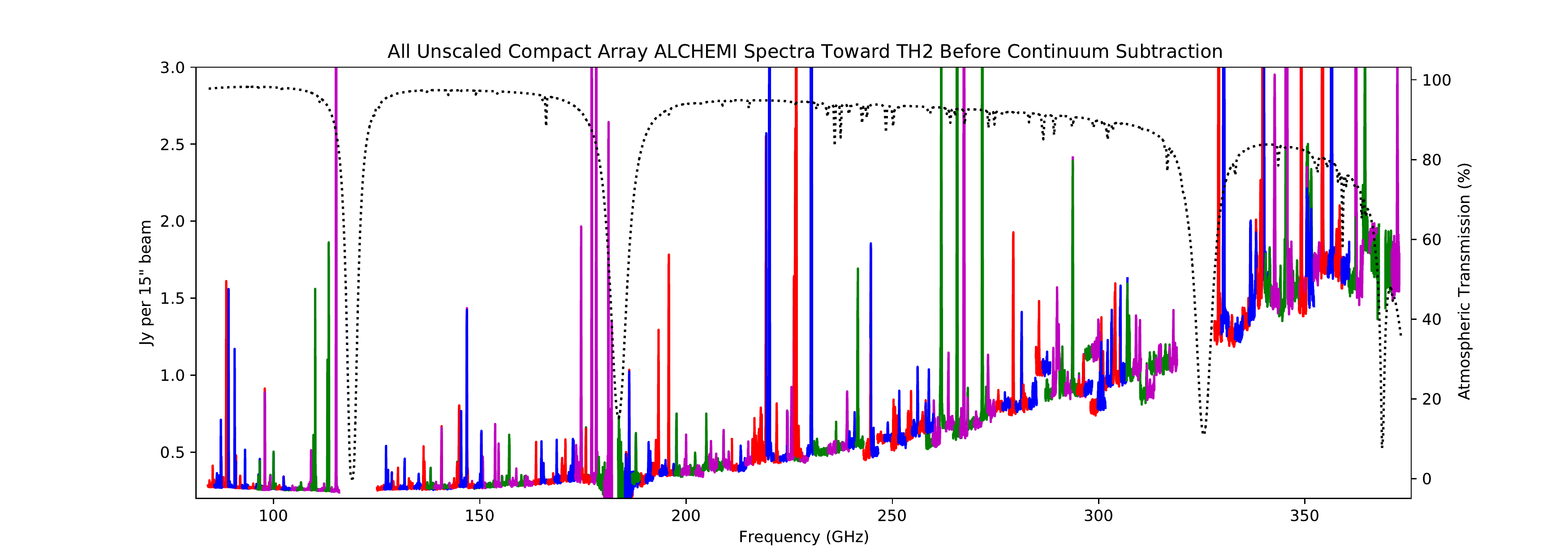}
\includegraphics[width=\textwidth]{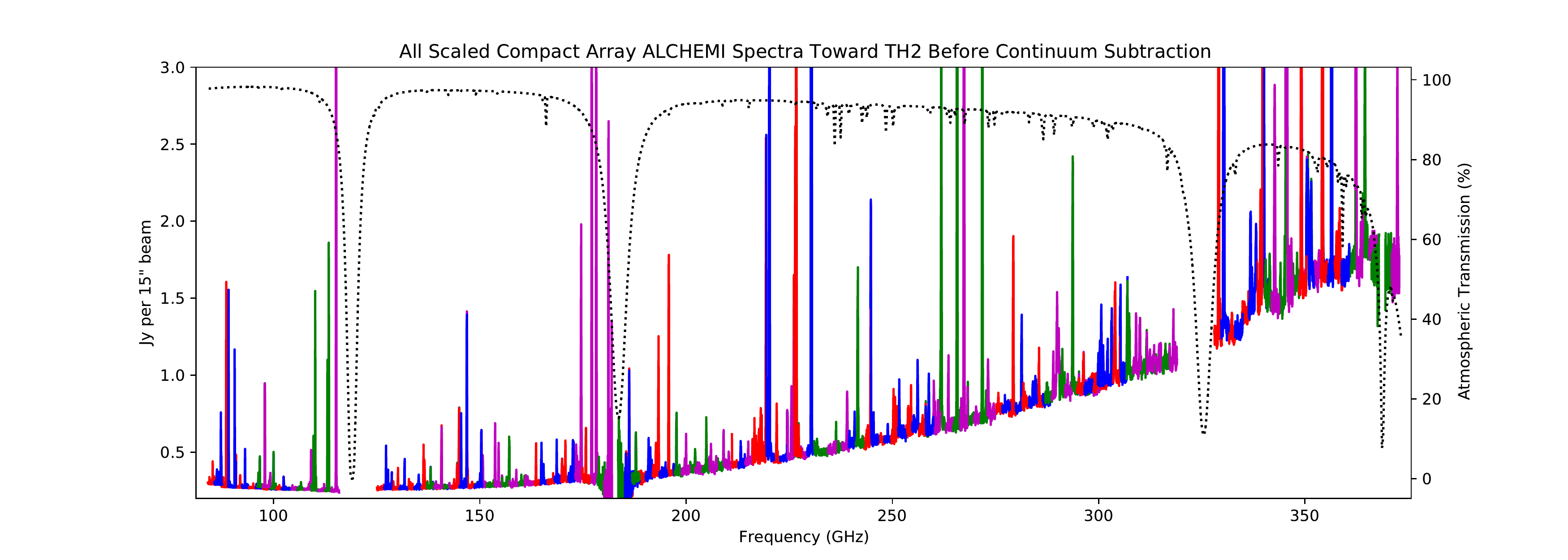}
\caption{Unscaled (top) and scaled (bottom) full spectral scans. The spectra were extracted from the TH2 position in the low resolution data (see Sect.~\ref{Sec.AppendixFLuxAlignment}). The colors in the spectra represent the spectral windows from the lower side band (red and blue) and the upper side band (green and violet). A dotted line indicates the atmospheric transmission for a PWV of 1\,mm. \label{fig.unscaledfullspectrum}}
\end{figure}

\begin{figure}
\includegraphics[width=\textwidth]{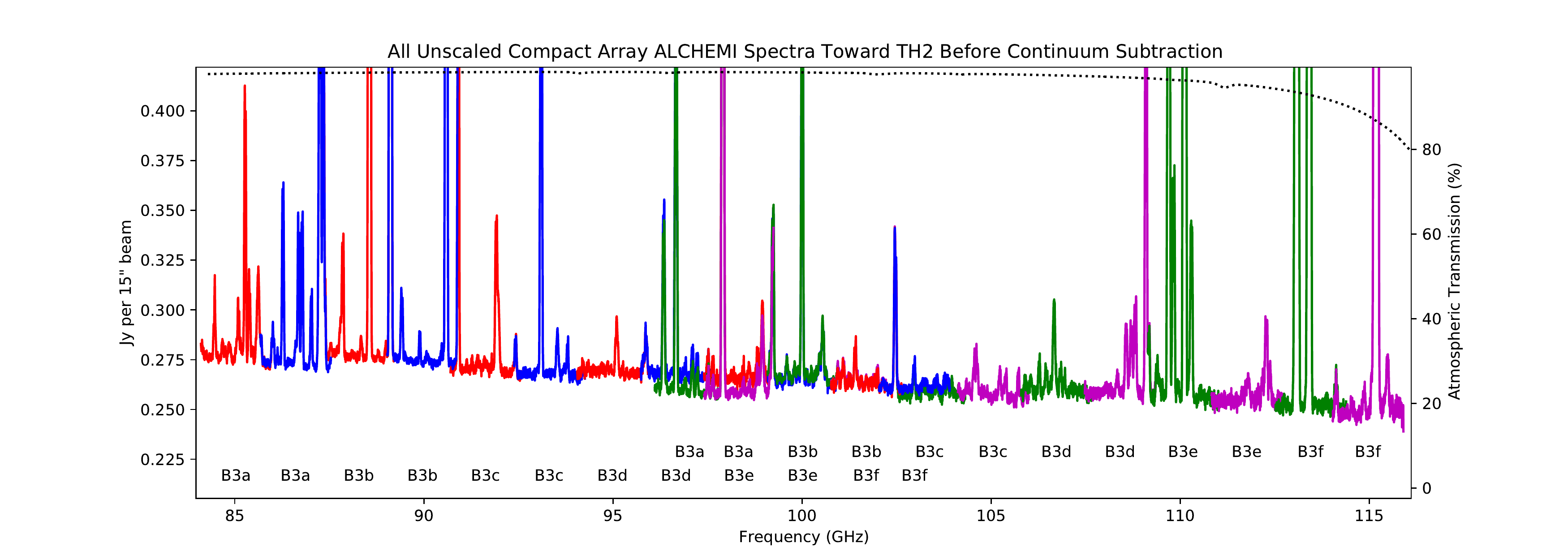}
\includegraphics[width=\textwidth]{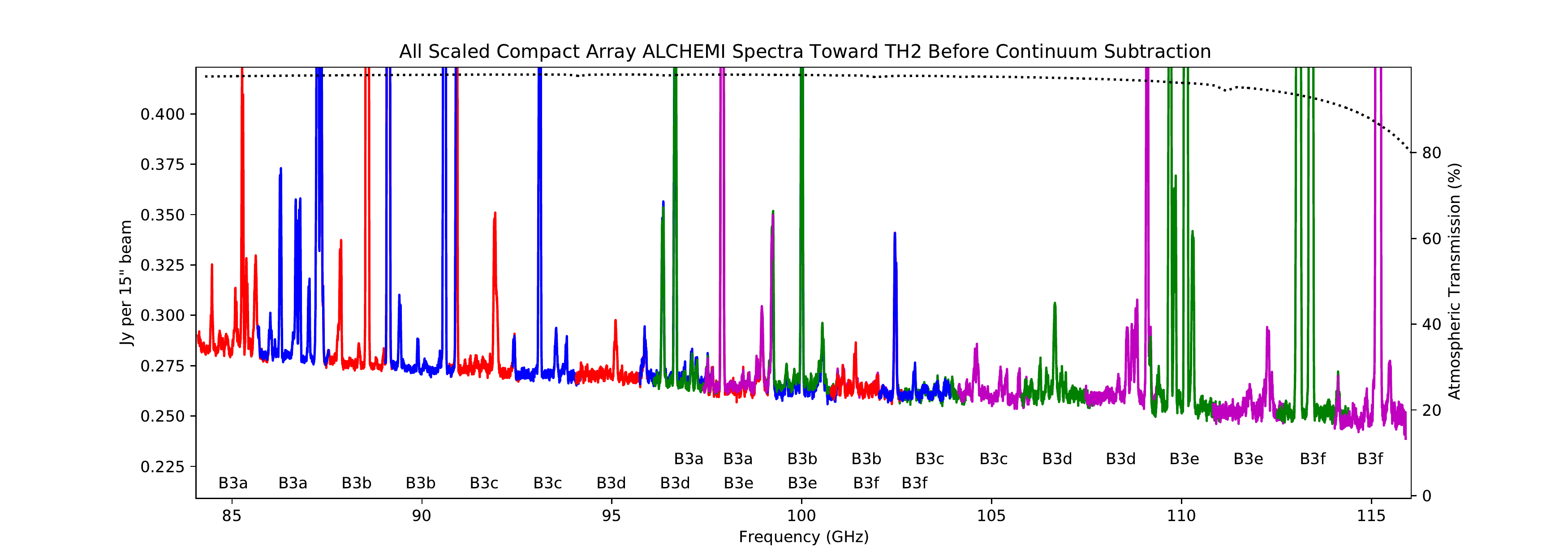}
\caption{Same as Fig.~\ref{fig.unscaledfullspectrum} but for Band 3 unscaled (top) and scaled (bottom). Additionally the label for each of the individual spectral setups (see Sect.~\ref{Sec.Observations}) are displayed. \label{fig.unscaledB3}}
\end{figure}

\begin{figure}
\centering
\includegraphics[width=\textwidth]{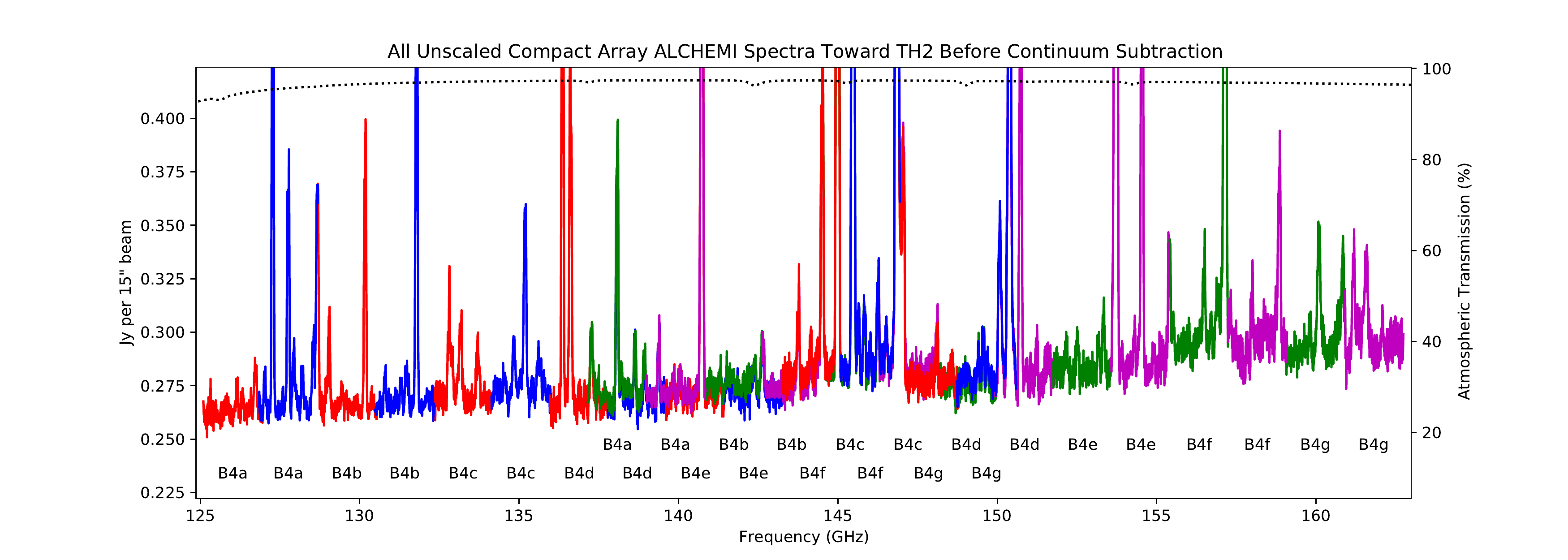}
\includegraphics[width=\textwidth]{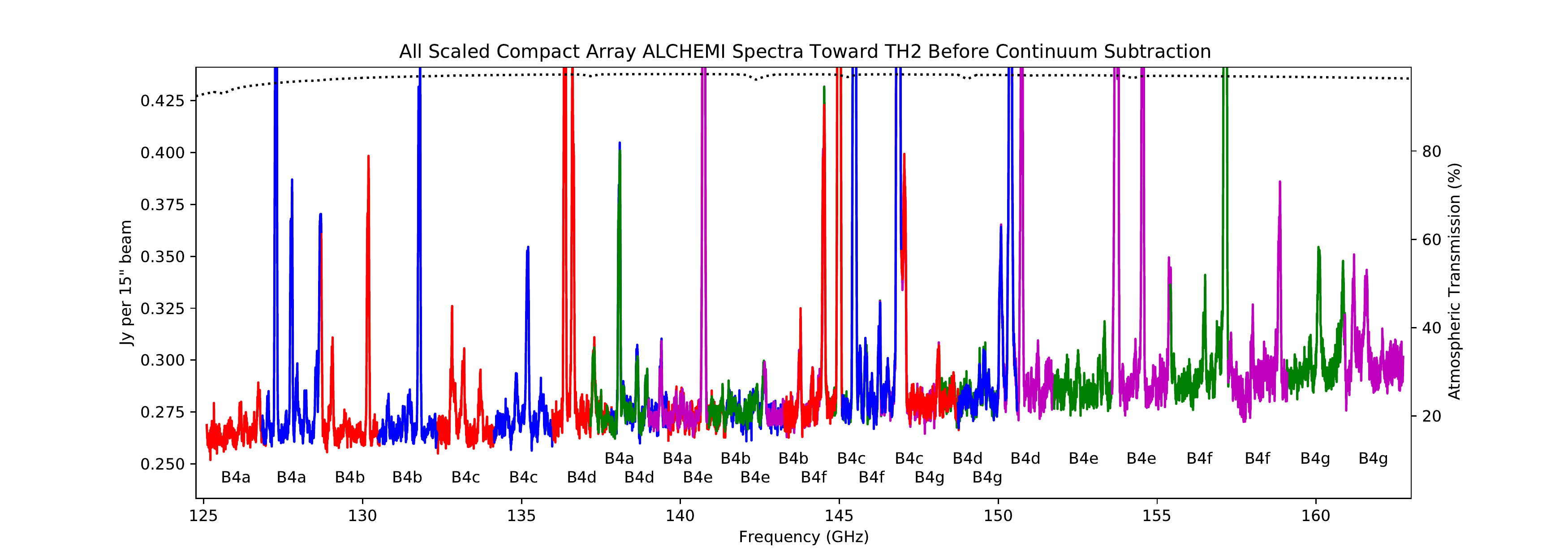}
\caption{Same as Fig.~\ref{fig.unscaledB3} but for Band 4 unscaled (top) and scaled (bottom). \label{fig.unscaledB4}}
\end{figure}

\begin{figure}
\centering
\includegraphics[width=\textwidth]{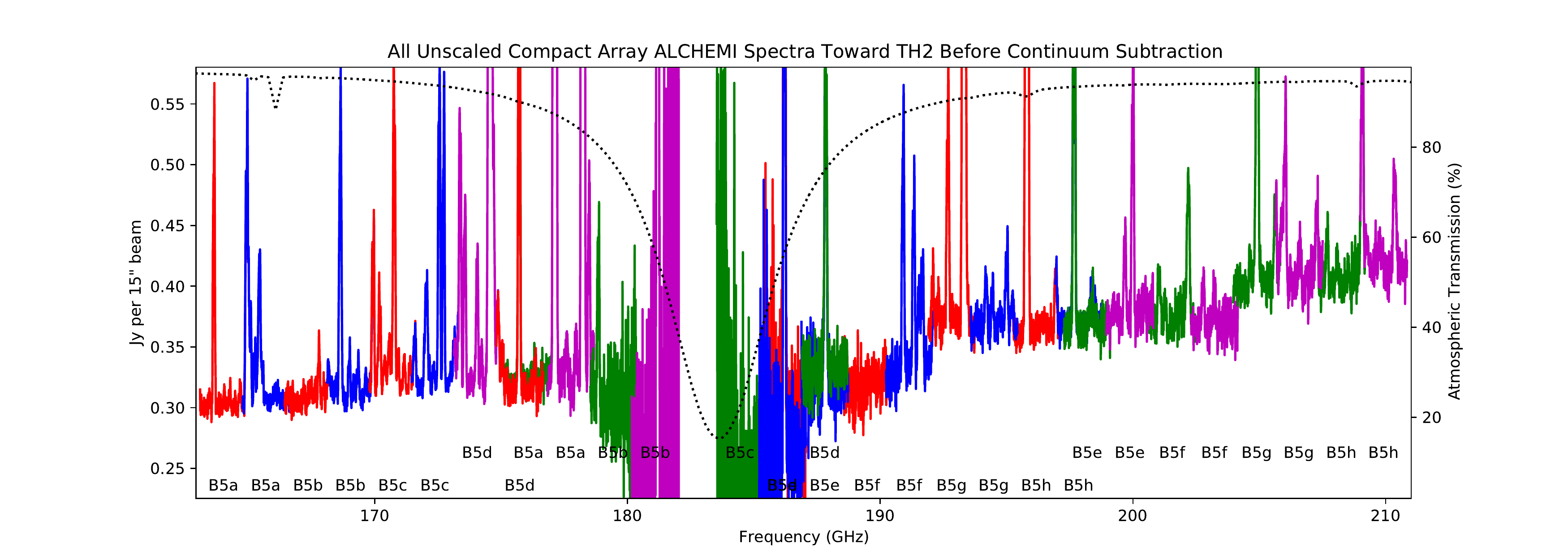}
\includegraphics[width=\textwidth]{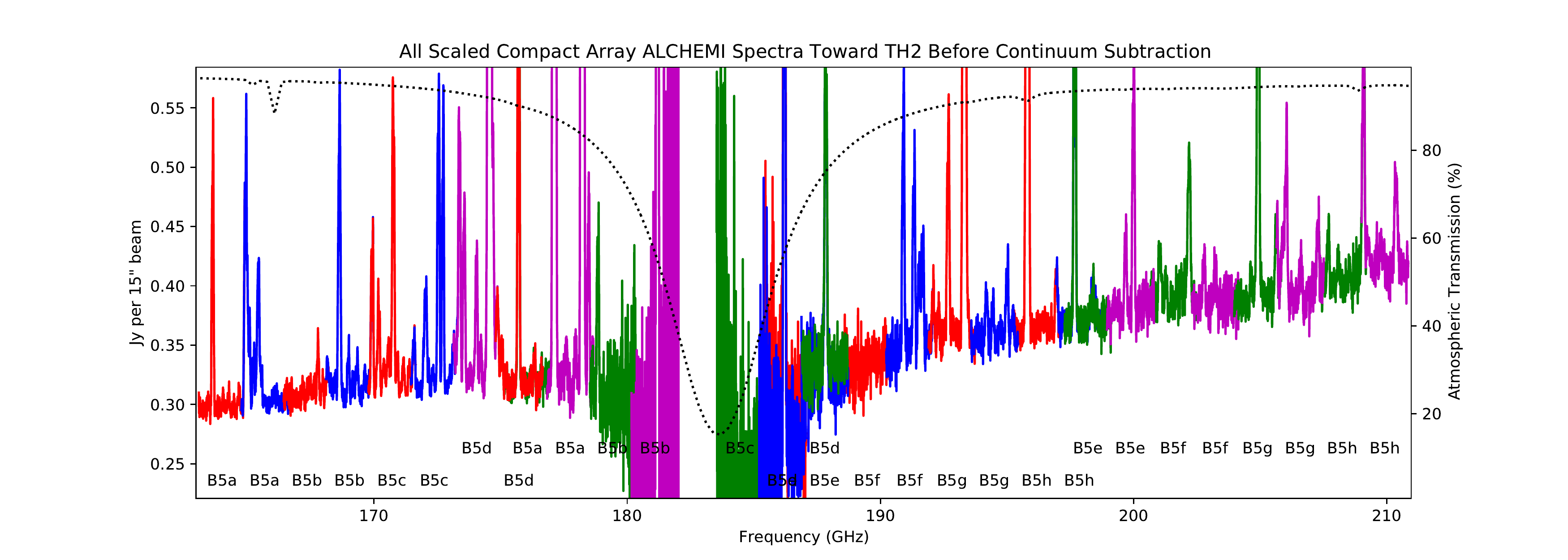}
\caption{Same as Fig.~\ref{fig.unscaledB3} but for Band 5 unscaled (top) and scaled (bottom). \label{fig.unscaledB5}}
\end{figure}

\begin{figure}
\includegraphics[width=\textwidth]{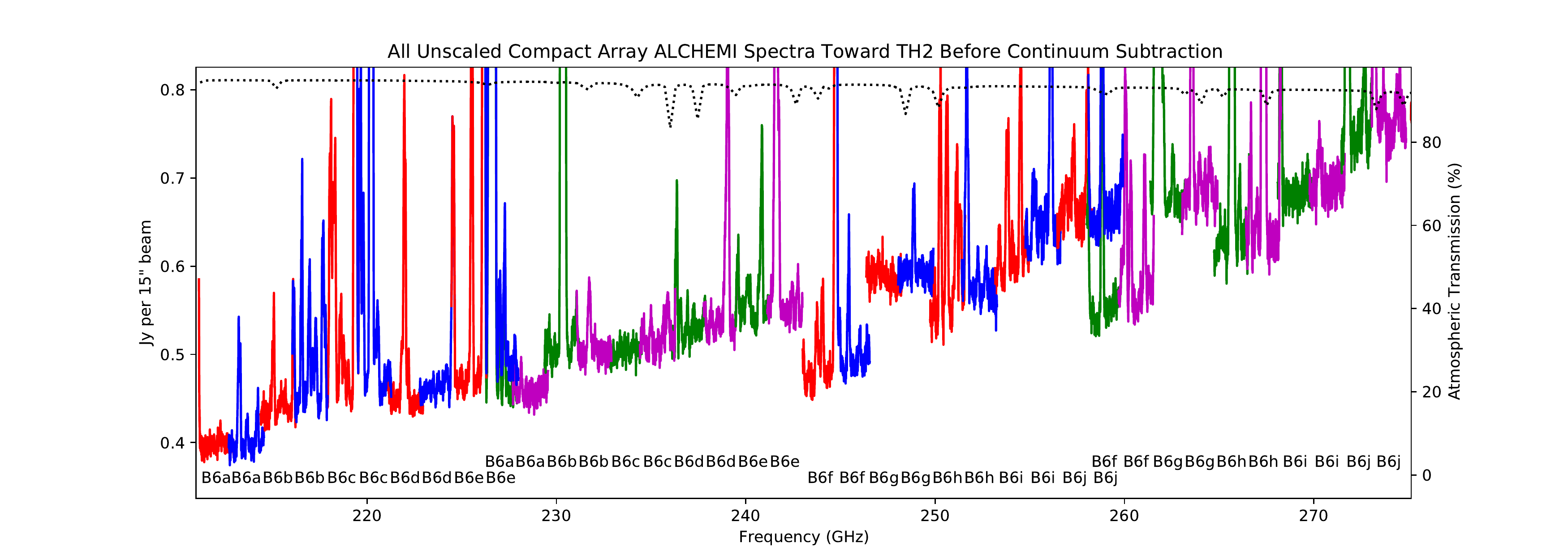}
\includegraphics[width=\textwidth]{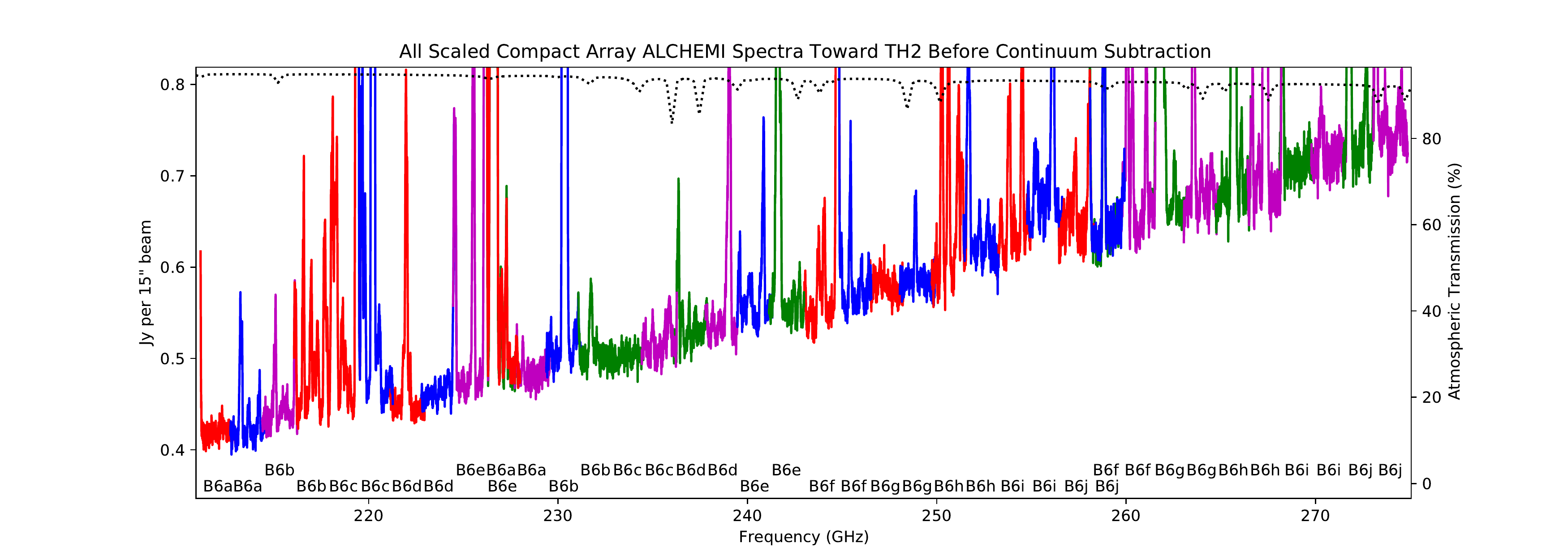}
\caption{Same as Fig.~\ref{fig.unscaledB3} but for Band 6 unscaled (top) and scaled (bottom).\label{fig.unscaledB6}}
\end{figure}

\begin{figure}
\centering
\includegraphics[width=\textwidth]{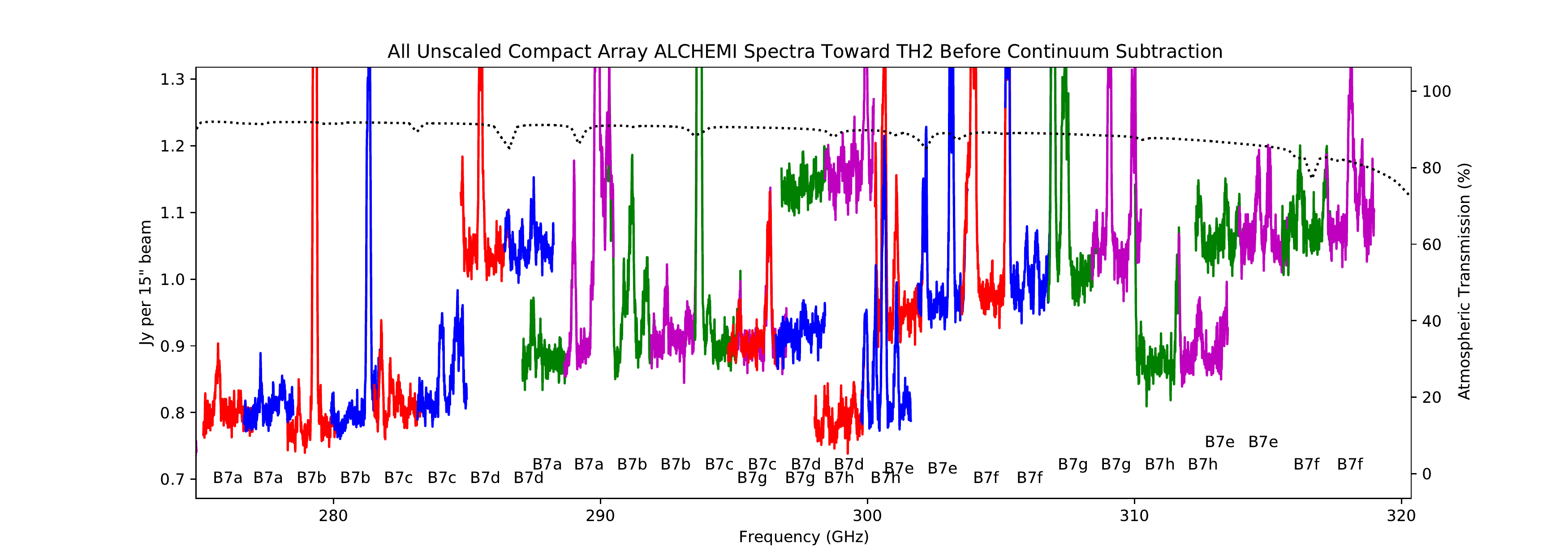}
\includegraphics[width=\textwidth]{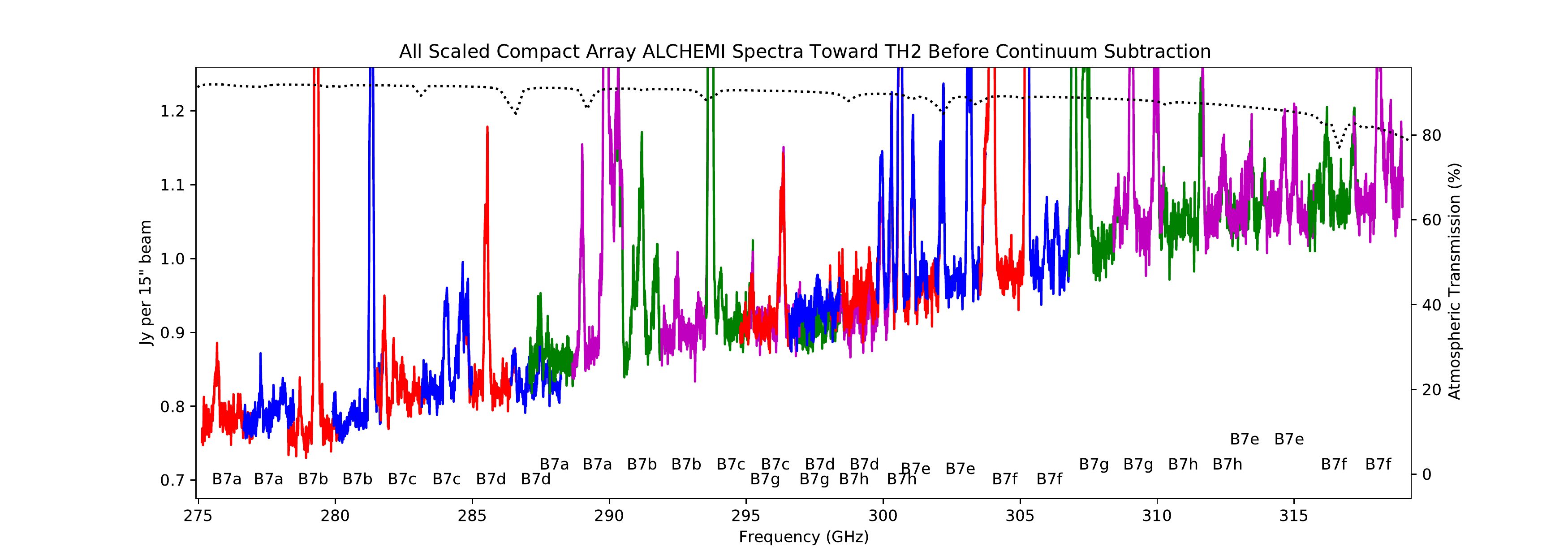}
\caption{Same as Fig.~\ref{fig.unscaledB3} but for the first-half of Band 7 unscaled (top) and scaled (bottom).\label{fig.unscaledB7p1}}
\end{figure}

\begin{figure}
\centering
\includegraphics[width=\textwidth]{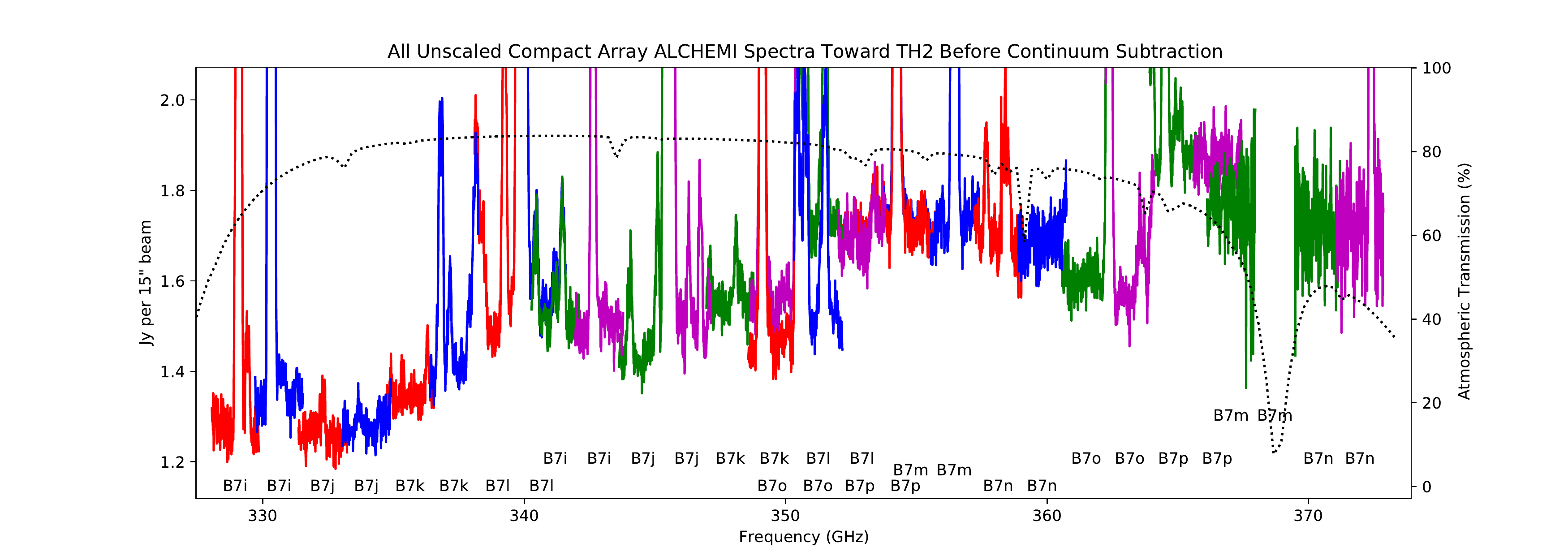}
\includegraphics[width=\textwidth]{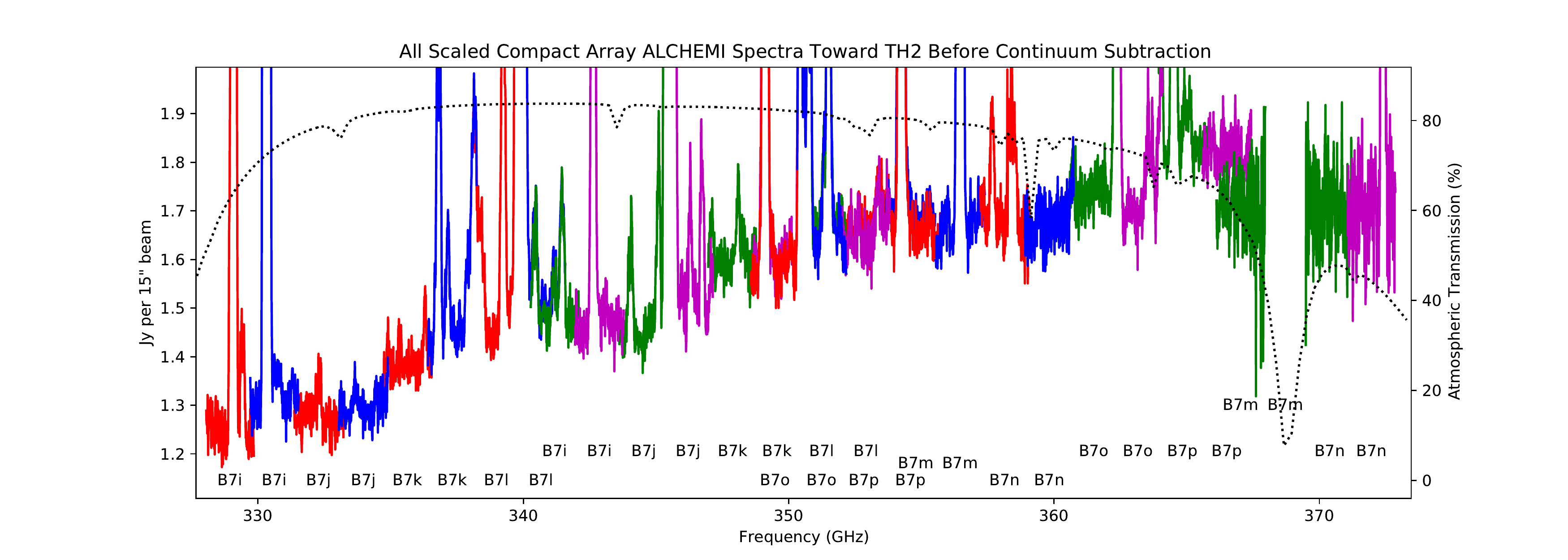}
\caption{Same as Fig.~\ref{fig.unscaledB3} but for the second half of Band 7 unscaled (top) and scaled (bottom).\label{fig.unscaledB7p2}}
\end{figure}

Sect.~\ref{sec.relativeflux} and Appendix~\ref{Sec.AppendixFluxCal} provide details on the absolute and relative flux calibration accuracy for the ALCHEMI measurements. As noted, relative flux calibration was improved by utilizing the information provided by overlapped receiver tunings throughout the ALCHEMI survey.
In Figure~\ref{fig.unscaledfullspectrum} we present, for completeness, the unscaled and scaled spectra prior to continuum subtraction (Section~\ref{sec.contsubtraction}) as extracted from the TH2 position within the compact configuration measurements, 12mC (Band 3) and ACA (Band 4, 5, 6, and 7) observations and imaged to $15''$ resolution.
Additionally, Figs.~\ref{fig.unscaledB3} through ~\ref{fig.unscaledB7p2} present zoomed versions for each receiver Band where Band 7 is split into two figures.

\section{ALCHEMI Flux Calibration}
\label{Sec.AppendixFluxCal}

The ALMA flux calibration process includes a number of contributions to its uncertainty, including systematic errors within a given measurement calibrated with a single primary flux reference such as that due to:
\begin{enumerate}
    \item The primary flux calibrator model used to set the absolute flux calibration scale. 
    \item The primary flux calibrator measurement used to define the reference point for the secondary flux calibrator flux.
\end{enumerate}
as well as random uncertainties due to:
\begin{enumerate}
    \setcounter{enumi}{3}
    \item The bootstrapping from the primary to the secondary flux calibrator.
    \item The bootstrapping from the secondary flux calibrator to the target source.
    \item The lack of a proper elevation-dependent opacity correction during any of the bootstrapping steps.
\end{enumerate}
Furthermore, even though extra mitigation measures can be done during an observation to account for items 4 and 5 above, there is nothing that can allow one to attain an absolute flux calibration error that is better than the error associated with the primary flux calibrator measurement and model.

Early in the process of imaging the ALCHEMI measurements, amplitude offsets between overlapping receiver tunings were noted (see Sect.~\ref{sec.relativeflux} and Appendix~\ref{Sec.AppendixFLuxAlignment}). The analysis of these offsets has given ALCHEMI the ability to correct for \textit{at least one component} of the flux calibration uncertainty. This flux calibration alignment, though, does not allow for the determination of the absolute flux calibration reference. At best we have corrected the flux scales within our individual scheduling blocks to a common value. The ALCHEMI image cubes have been corrected for these flux rescaling factors, which we believe has corrected for noise introduced as part of the flux calibration process after the primary flux calibrator measurement.

In the following we address two levels of flux calibration uncertainty in the ALCHEMI data:
\begin{itemize}
    \item \textbf{Relative Flux Calibration:} This represents the flux calibration uncertainty to be used when comparing fluxes within the ALCHEMI image cubes for a given set of imaging inputs (i.e., array, spatial, and spectral resolution).
    \item \textbf{Absolute Flux Calibration:} This represents the flux calibration uncertainty to be used when comparing fluxes derived from the ALCHEMI image cubes for comparison with other (non-ALCHEMI) measurements.
\end{itemize}

\subsection{Relative Flux Calibration Uncertainty}
\label{Sec.AppendixRelativeFluxCal}

The spectral flux normalization that we have applied to all ALCHEMI measurement sets has effectively normalized all ALCHEMI data to a common flux calibration scale for each array configuration\footnote{By array configuration we mean to differentiate among the Compact Array, either ACA (7m) or 12m Compact (12mC), the Extended Array (12m/12mE) and Combined (12m7m/12mE12mC) that were used for each spectral setup. Each of these observations consisted of a number of individual observations with slightly varying array configurations depending on the antenna availability at the time of the observation.} measured in the ALCHEMI survey. This implies that we have normalized all ALCHEMI data measured with a given array configuration to the same relative flux calibration scale. This furthermore implies that comparison of spectral lines within the ALCHEMI survey can be compared using a relative flux calibration uncertainty.

The statistics of the amplitude scaling factors given in Table~\ref{tab.observingsetup} are provided in Table~\ref{tab:StatScaleFactors}.
It is noteworthy that for our Band~3 measurements the 12mE configuration shows significant deviations from the average values of the scaling factors for that receiver band and configuration.

\begin{table}[h]
\centering
\caption{Statistics by bands and arrays of the relative flux scaling factors ($a_i$) \label{tab:StatScaleFactors}}
%\begin{center} 
\begin{tabular}{lcccccc}
\hline
\hline
 Band & Array & N$_{\rm obs}$\tablefootmark{a} & Average    & RMS            & min(a$_i$) / max(a$_i$) & max(a$_i$)-min(a$_i$) \\
    &       &                & $\overline{a_i}$ & $\sigma_{a_i}$ (\%) &                     & (\%) \\ 
\hline

B3 & 12mC     &  6 & 0.989 &  2.4 & 0.937 / 1.009 &  7 \\
B3 & 12mE     &  6 & 1.000 & 12.4 & 0.789 / 1.123 & 34 \\
B4 & 7m & 7  & 1.000 & 1.3 & 0.980 / 1.021 &  4 \\
B4 & 12m      &  7 & 1.000 &  1.7 & 0.968 / 1.026 &  6 \\
B5 & 7m & 8  & 1.000 & 2.3 & 0.955 / 1.033 &  8 \\
B5\tablefootmark{b} & 12m & 8  & 0.911 & 2.3 & 0.856 / 1.147 &   29  \\
B6 & 7m & 10 & 0.974 & 4.7 & 0.867 / 1.027 & 16 \\
B6 & 12m      & 10 & 1.003 &  1.2 & 0.990 / 1.034 &  4 \\
B7 & 7m & 16 & 1.006 & 8.1 & 0.833 / 1.257 & 42 \\
B7 & 12m      & 16 & 1.001 &  2.5 & 0.956 / 1.044 &  9 \\
\hline
\end{tabular}
% \mbox{\,} \vskip -.25cm
% Notes: $(a)$ Number of 12\,m-antennas in the array;
\tablefoot{
\tablefoottext{a}{Total number of scheduling block measurements included in the subsequent statistics shown.}
\tablefoottext{b}{Excludes 12m Array SB B5f as the scaling factor for this SB appears to include amplitude calibration errors inherent in the delivered calibrated data products.}
}
%\end{center}
\end{table}

An estimate of the relative flux calibration uncertainty associated with a given receiver band and configuration can be derived from the scatter in the scaling factors (Table~\ref{tab:StatScaleFactors}) that we have applied to our measurement sets to normalize them to the same mean flux scale. The RMS values for the flux scale normalization factors for each Band and array configuration(s) (listed as Compact Array / Extended Array / Combined)  are:
\begin{itemize}
    \item Band 3: 2\% / 12\% / 12\%
    \item Band 4: 1\% / 2\% / 2\%
    \item Band 5: 2\% / 2\% / 2\%
    \item Band 6: 5\% / 1\% / 5\%
    \item Band 7: 8\% / 3\% / 9\%
\end{itemize}

The relative flux calibration uncertainty for the ALCHEMI image cubes which are combinations of the Compact Array and Extended Array measurements will in reality be a complex combination of the relative flux calibration uncertainties, which itself depends upon the contribution of each array measurement to a given flux in the ALCHEMI survey. The contribution of each array to the combined measurement is dependent upon numerous factors such as relative visibility contributions and time-dependent variations between the individual array measurements. In the above, we conservatively estimate the relative flux calibration uncertainty for the combined values as the root-sum-square of the Compact Array and Extended Array scaling factor RMS values.

Be aware that the actual "flux uncertainty” to be used in a line-ratio analysis, for example, is not determined solely by the “flux calibration uncertainty” in complex sources such as NGC~253. The line flux one measures depends on how the imaging process (robust parameter, clean mask, clean depth, selfcal, \textit{etc.\/}) has been performed and on the properties of the spectral line itself. The same imaging parameters can have different effects on different spectral lines. For example, when different spectral lines have distinct spatial extents or when one spectral line is bright and the other is faint because the brighter spectral line has a larger fraction of its flux cleaned. For these reasons, the real flux calibration uncertainty is likely to be larger than that assumed from an assessment of the quality of the flux calibration process alone.

\subsection{Absolute Flux Calibration Uncertainty}
\label{Sec.AppendixAbsoluteFluxCal}

The absolute flux calibration uncertainty starts with the relative flux calibration uncertainty and includes contributions due to the measurement, model, and application of the primary flux calibration source (see above). To estimate this additional contribution, we assume that the measurement of the relatively bright sources used as primary flux calibrators provide a negligible contribution to the primary flux calibration uncertainty. This uncertainty is then dominated by that associated with the primary flux calibrator model. Models used by ALMA within CASA are unlikely to be accurate to <5\% \citep[][ priv. comm.]{Butler2012}. We use then the following recommendations for flux calibration accuracy, from the ALMA Proposer's Guide for Cycle 5 $^{\ref{ALMApropguide}}$ (Section A.9.2 on "Flux Accuracy"), as an estimate of the primary flux calibrator model uncertainty:
\begin{itemize}
    \item Bands 3, 4, 5: $<5$\%
    \item Bands 6, 7, 8: $<10$\%
    \item Bands 9 and 10: $<20$\%
\end{itemize}

The absolute flux calibration uncertainty, including the relative flux calibration uncertainty derived for the ALCHEMI image cubes, is given by the root-sum-squared of the relative and primary flux standard model calibration uncertainties (listed as in Sect.~\ref{Sec.AppendixRelativeFluxCal}):
\begin{itemize}
    \item Band 3: 5\% / 13\% / 13\%
    \item Band 4: 5\% / 5\% / 5\%
    \item Band 5: 5\% / 5\% / 5\%
    \item Band 6: 11\% / 10\% / 11\%
    \item Band 7: 13\% / 10\% / 13\%
\end{itemize}

Therefore, for the sake of simplicity, we recommend the usage of a conservative 15\% uncertainty for the absolute flux calibration within the ALCHEMI image cubes, at any frequency and configuration.

\subsection{A Search for the ALCHEMI Flux Normalization Anomalies}
\label{Sec.AppendixFluxCalAnomalies}

We find that 11 of the 47
tunings ($\sim23$\%) which comprise the ALCHEMI data set have
amplitude scale factors which are larger than the ALMA amplitude
calibration specification of 5\%, with the two most discrepant tunings being
26\% (Band 7) and 21\% (Band 3). Table~\ref{tab:BadScaleFactors}
lists the measurement information for the 11 discrepant tunings.

\begin{table*}[h!]
\centering
\caption{ALCHEMI tunings with discrepant amplitude calibration \label{tab:BadScaleFactors}}
\begin{tabular}{llp{6.5cm}lll}
\hline
\hline
SG\tablefootmark{a} & Array & Flux Calibrators & SpixAge\tablefootmark{b} &
$100\times\left|\frac{\sigma_{gain}}{S_{gain}}\right|$\tablefootmark{c} & $a_i$\tablefootmark{d}\\
\hline
B7d & 7m & J2258$-$2758,J0006$-$0623,J0522$-$3627 & (3,-4,20),(-4,20),20 & 9.98 &
1.257\\
B3b & 12mE & J2258$-$2758 & 142,144 & 0.86 & 0.789\\
B7h & 7m & J2258$-$2758,J2253$+$1608,J0522$-$3627 & (3,4),25,4 & 9.87 &
0.833\\
B6f & 7m & J2253$+$1608 & -1,0,6 & 1.98 & 0.867\\
B3d & 12mE & J0006$-$0623 & 7 & 1.57 & 1.128\\
B3a & 12mE & J0006$-$0623 & -5,4 & 0.05,0.68 & 1.108\\
B3c & 12mE & J0006$-$0623 & -3,-6,0,1 & 0.48 & 1.107\\
B7o & 7m & J2253$+$1608,J0522$-$3627 & (7,0,2,3),2 & 2.06 & 0.922\\
B3f & 12mE & J2258$-$2758,J2357$-$5311,J0006$-$0623 & -27,-22,4 & 1.84 &
0.922\\
B6h & 7m & J2258$-$2758 & 5,-3,0,-2 & 3.71,1.03,1.20,4.30 & 0.924\\
B3a & 12mC & J0006$-$0623 & 23,27 & 0.24 & 0.937\\
\hline
\end{tabular}
\tablefoot{
\tablefoottext{a}{Refers to Science Goals within ALMA project nomenclature}
\tablefoottext{b}{Number of days since the most recent ALMA-derived spectral index for a given calibrator.  Multiple executions which used a given calibrator are grouped within parentheses.  Negative values indicate spectral indices derived before a given calibrator measurement. Calibrator spectral indices are derived using measured Band 3 and Band 6 or Band 7 fluxes.}
\tablefoottext{c}{Normalized gain calibrator flux uncertainty averaged
  over all gain calibrator measurements for observation dates within
  five days. Multiple entries indicate multiple observation time
  ranges. Target for all gain calibration measurements was J0038$-$2459.}
\tablefoottext{d}{Amplitude scale factors are applied to the data as indicated by Equation~\ref{eq.uvscaling}.}
}
\end{table*}

In an attempt to determine the source(s) of
the thirteen science goals with poor amplitude calibration, we have
investigated how ALMA calibrates amplitude within the limits of the
information provided to investigators.
ALMA flux calibration is made within a given observation (or ``execution block'') by measurement of a standard quasar selected from a
list of monitored quasars, referred to as the ``grid sources''.
ALMA strives to measure these standard quasars at least once every 14
days at bands 3 and 7, and calibrates their fluxes to
an absolute scale through reference measurements of primary flux
calibrators (\textit{i.~e.\/} Uranus). The measured
absolute fluxes for the grid source calibrators are available from the
\href{https://almascience.nrao.edu/sc/}{ALMA calibrator archive}.
We have extracted the
flux scaling information from the calibration pipeline
weblog file ``flux.csv'' associated with each  observing execution block (EB). This flux
scaling information includes the ``spectral index age'' and time since
the standardized flux for each calibrator was derived.

\subsubsection{Possible Source of Error: Large Spectral Index Age (spixAge)}
\label{sec:SpixAge}

\begin{figure}
\centering
%trim option's parameter order: left bottom right top
\includegraphics[scale=0.9]{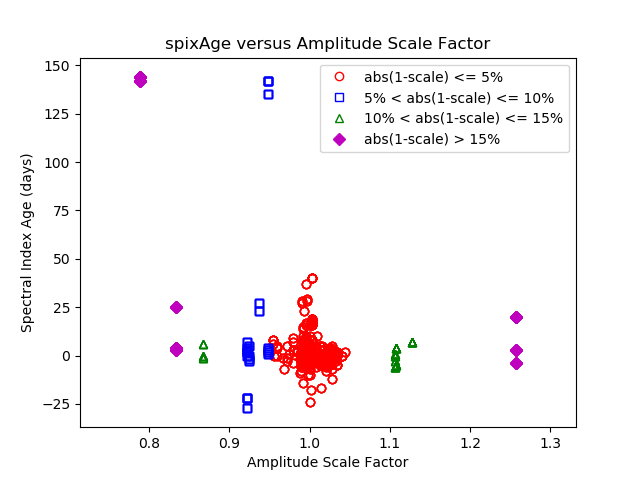}
\caption{Correlation between spectral index age and amplitude scale
  factor for all ALCHEMI measurements. Coloring is used to designate
  ranges of amplitude scaling.}
\label{fig:SpixScaleFactor}
\end{figure}

As the spectral index age (spixAge) is one of the factors used by
\href{https://safe.nrao.edu/wiki/bin/view/ALMA/GetALMAFlux}{getALMAFlux}
to extrapolate measured flux calibrator fluxes to target frequencies,
there was a concern that perhaps large spixAge factors were causing
the large flux calibration errors. Using the spectral index age
information extracted from the pipeline calibration process we show
the correlation between spectral index age and amplitude scale factor
in Figure~\ref{fig:SpixScaleFactor}. There is no correlation between spixAge and the amount of the flux calibration error.

\subsubsection{Possible Source of Error: Calibrator Catalog Band 3 Age}
\label{sec:Band3Age}

By inspecting the age of the absolutely calibrated Band 3 flux used by
\href{https://safe.nrao.edu/wiki/bin/view/ALMA/GetALMAFlux}{getALMAFlux}
(extracted from the flux.csv files associated with the pipeline
calibration process), we find that there is no apparent correlation
between the "Age"/"Band3Age" and large amplitude scaling factors. The "Age" or
"Band3Age" is in almost all cases between 0 and 3 days for our most
discrepant scale factor EBs. In one case it was 6 days, and another
was 5 days, but neither of these were from our "worst
cases". Band 3 age does not appear to be a likely source for the discrepant amplitude scale factors.

\subsubsection{Possible Source of Error: Flux Monitoring Time Gaps}
\label{sec:TimeGaps}

Many, though not all, of the science goals (SGs) with discrepant amplitude scale
factors occur just after a gap in the respective flux calibrator
measurements. Specifically: 
\begin{itemize}
  \item For 8 of the 13 discrepant Band 3 SGs, there was a significant flux calibrator
    measurement time gap just before these SGs were observed.
  \item This correlation between flux calibrator measurement time
    gap and discrepant scale factor does not exist for Band 6
    (green) or Band 7 (black) SGs with discrepant amplitude scale
    factors.
\end{itemize}
Although this may be an explanation for the excessive Band 3
  flux calibration uncertainties, a time gap in the flux monitoring
  for our flux calibrators does not appear to be consistent for all
  SGs with discrepant flux calibration.

\subsubsection{Possible Source of Error: Large Time Span Between System Temperature Measurements}
\label{sec:Tsys}

By perusing the weblogs associated with the ALCHEMI measurements we
know that the scaling from raw amplitude to a temperature scale
(otherwise known as the ``system temperature'' measurement), was
routinely made only every $\sim 11$ minutes at Band 3 and every $\sim
8$ minutes at Band 7. As these system temperature measurements are required
to track the changes in sky emission as a function of time, it could
be that these basic amplitude scaling factors have not been sampled
well enough in time, especially at the higher frequency bands.
However this is an observatory trade-off between enhanced calibration accuracy and observing efficiency.

\subsubsection{Possible Source of Error: Noisy Gain Calibrator Measurement}
\label{sec:GainCal}

For the SG with the most discrepant flux
calibration, B7d 7m (ngc253\_d\_07\_7M) there may be an issue
with gain calibrator phase stability. Three out of five execution blocks (EBs) for the scheduling block (SB) were taken on the same day (2018-01-21). J0038$-$2459 was observed
as the gain calibrator for all the three EBs. Although the gain
calibrator flux is expected to be stable over the three EBs, because
the flux is very likely to be stable over the short time scale between
these three EBs (about six hours), the derived flux densities in the
pipeline calibration changed about 28\% (peak-to-peak). The following
is a summary of the pipeline-derived flux densities of the phase calibrator
(reference: the weblog, stage 15). 
\begin{verbatim}
# EB                      start date/time (UT) flux density (spw 16)
uid___A002_Xc96f17_X8658  2018-01-21 19:06:14  1024.0+-9.979 mJy
uid___A002_Xc96f17_X8ec1  2018-01-21 22:17:40   855.121+-12.804 mJy
uid___A002_Xc96f17_X92b9  2018-01-21 23:53:08   797.500+-5.496 mJy
\end{verbatim}
It is possible that poor atmospheric phase stability is the cause
of the large flux calibration uncertainty, as all the EBs were taken
in the late afternoon to early evening, which is the part of the day when the
atmospheric phase stability tends to be very the poorest. In fact, in the
calibrated visibility amplitude vs time plot of the EB (weblog stage
17) one can see frequent amplitude drops in the Bandpass/Flux
calibrator scan. 

\begin{figure}
\centering
%trim option's parameter order: left bottom right top
\includegraphics[scale=0.3]{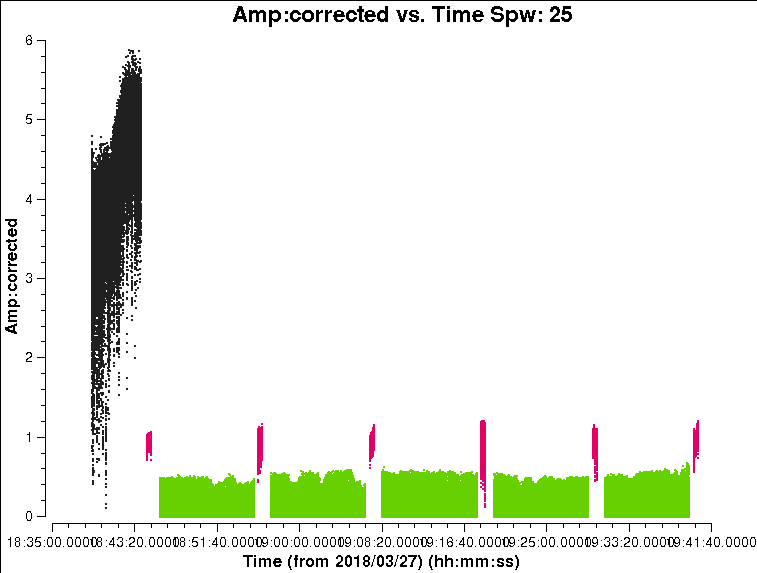}
\includegraphics[scale=0.3]{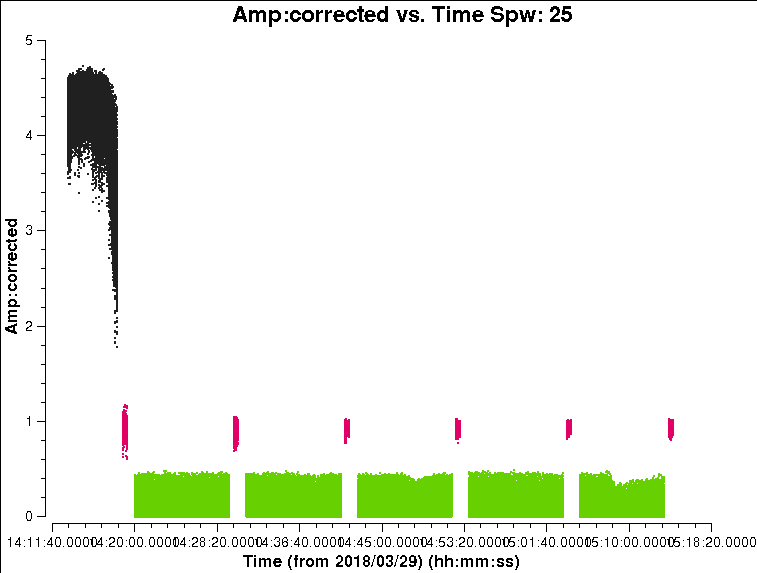}\\
\includegraphics[scale=0.3]{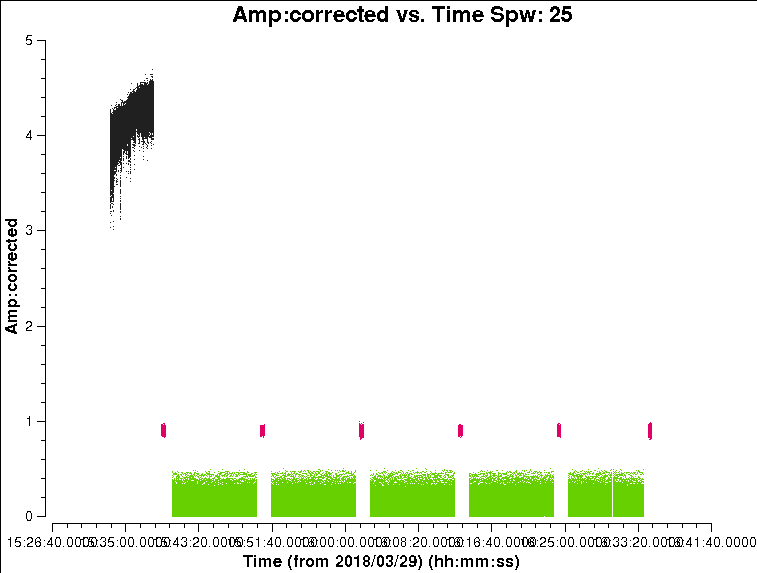}
\includegraphics[scale=0.3]{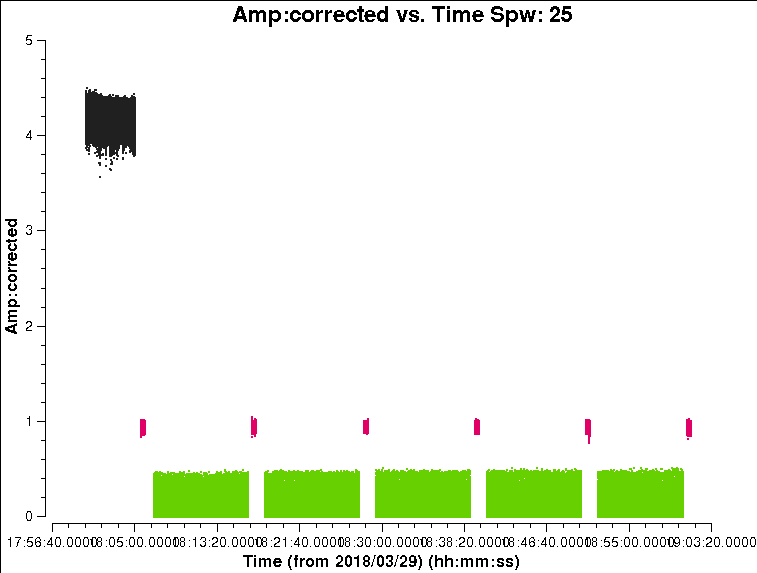}
\caption{B3b 7m amplitude/bandpass calibrator flux versus time weblog
  plots for the four EBs in this SB.}
\label{fig:B3b12mEAmpvsTime}
\end{figure}

Another example of a correlation between poor phase stability and poor
flux calibration is B3b 12mE (ngc253\_b\_03\_TM1). As was the case
for B7d 7m, the EBs for this SB were also affected by large phase
fluctuations, and that they were also executed in daytime (late
afternoon). 
Figure~\ref{fig:B3b12mEAmpvsTime} shows the pipeline plots of
amplitude vs time for all the EBs for B3b 7m. 
Significant amplitude drops in the bandpass/amplitude calibrator
(J2258$-$2758) can be seen especially in EB uid\_\_\_A002\_Xcb1740\_X94c9,
and they could affect the flux scaling of the gain calibrator.

The following is a summary of the pipeline-derived flux density of
gain calibrator J0038$-$2459 (reference: the weblog, stage 15).
\begin{verbatim}
spw=25
uid___A002_Xcb1740_X94c9  2018-03-27 18:36:03  955.851+-14.748 mJy
uid___A002_Xcb339b_X600b  2018-03-29 14:10:14  938.721+-7.809 mJy
uid___A002_Xcb339b_X633f  2018-03-29 15:30:21  903.452+-9.069 mJy
uid___A002_Xcb339b_X68ab  2018-03-29 17:57:08  941.212+-1.904 mJy
\end{verbatim}
The change of delivered flux density looks relatively large
(5.8\% in peak-to-peak) even though the EBs were taken
within three days. This could be an explanation for the discrepant
flux calibration (second worst of all SGs) for this SG.

\begin{figure}
\centering
%trim option's parameter order: left bottom right top
\includegraphics[scale=0.8]{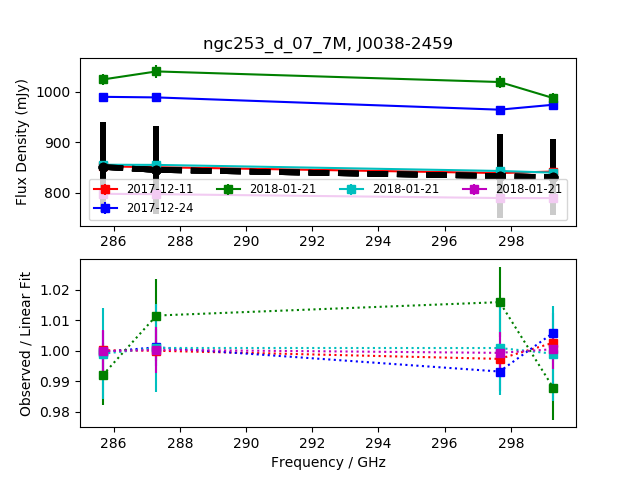}
\caption{Sample gain calibration results for all EBs associated with B7d 7m.  The bottom panel shows the residual from a linear fit for each measurement date to the measured flux densities displayed in the top panel.}
\label{fig:GainCalB7d7m}
\end{figure}

\begin{figure}
\centering
%trim option's parameter order: left bottom right top
\includegraphics[scale=0.8]{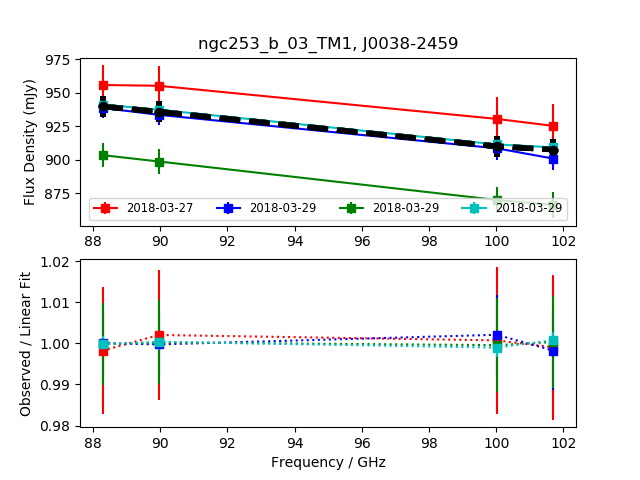}
\caption{Sample gain calibration results for all EBs associated with B3b TM1.  Same diagram style as used in Figure~\ref{fig:GainCalB7d7m}.}
\label{fig:GainCalB3bTM1}
\end{figure}

\begin{figure}
\centering
%trim option's parameter order: left bottom right top
\includegraphics[scale=0.8]{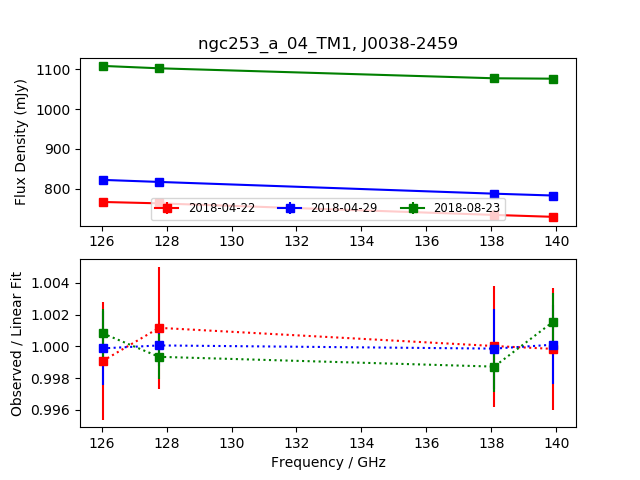}
\caption{Sample gain calibration results for all EBs associated with B4a TM1.  Same diagram style as used in Figure~\ref{fig:GainCalB7d7m}.}
\label{fig:GainCalB4aTM1}
\end{figure}

\begin{figure}
\centering
%trim option's parameter order: left bottom right top
\includegraphics[scale=0.8]{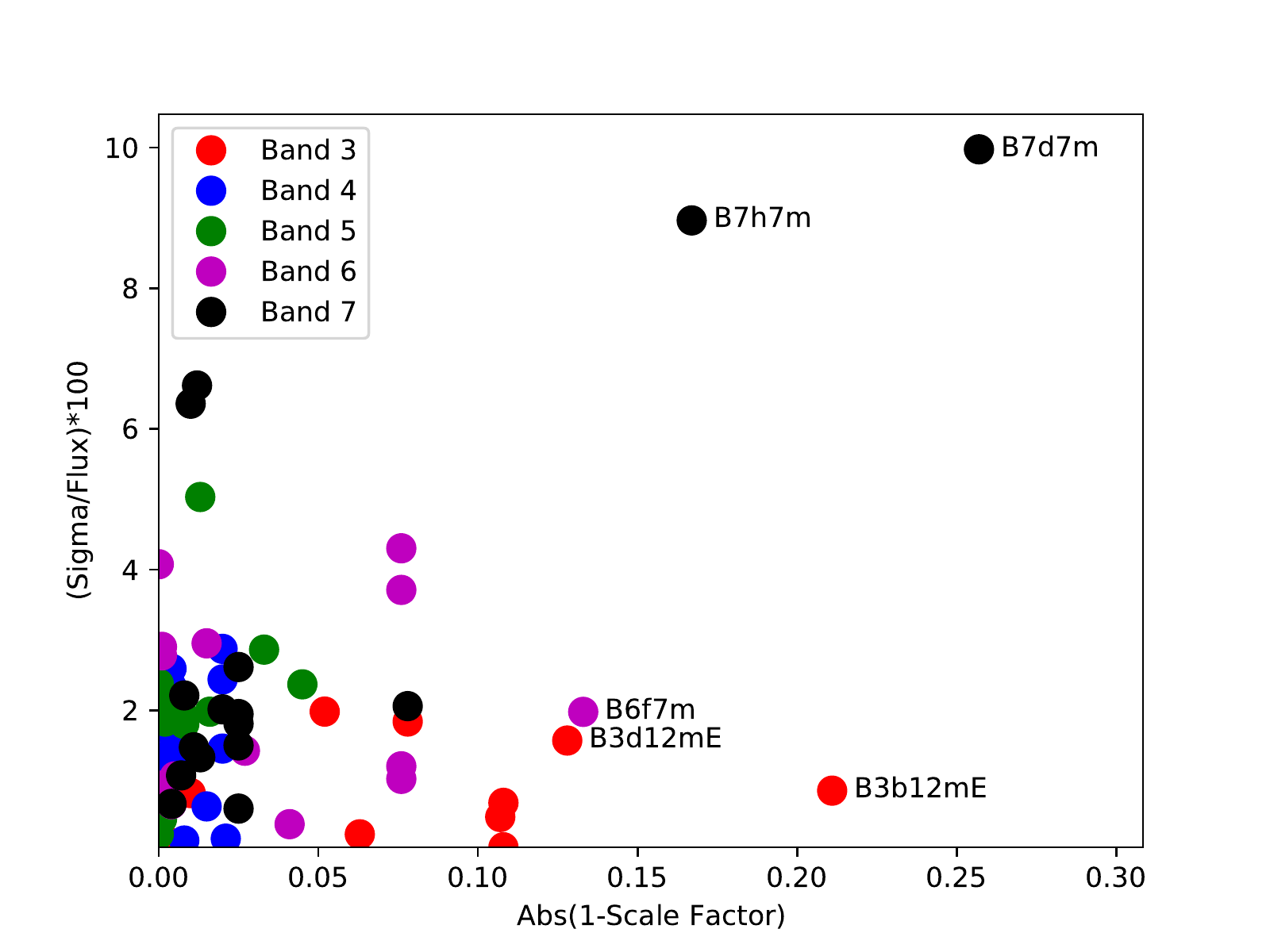}
\caption{Correlation between normalized gain calibrator standard deviation (as a percentage) versus the associated amplitude calibration scale factor (calculated as a difference from a perfect amplitude scale factor of 1.0).  Tunings with larger than 12\% scale factors are annotated.  See above description for calculation details.}
\label{fig:GainCalvSF}
\end{figure}

In order to assess the effect of a large gain calibrator flux
uncertainties on the derivation of our flux calibration uncertainty for
all of the ALCHEMI SGs, we have extracted all derived gain calibrator
fluxes and uncertainties from the ALCHEMI weblogs. We have derived a
normalized gain calibration error for each SG by doing the following:
\begin{itemize}
  \item Calculate the weighted uncertainty ($\sigma_{gain}$) for all
    gain calibrator measurements.
  \item Average normalized gain calibrator uncertainties
    ($\frac{\sigma_{gain}}{S_{gain}}$) over all spectral windows and
    measurements taken within 5 days of each other.  This time window
    is expected to be shorter than any changes in the absolute flux of
    the gain calibrator.
  \item By using a normalized gain calibrator uncertainty, we are
    attempting to smooth-out any changes in gain calibrator flux
    measurement uncertainty due to differences in gain calibrator
    integration time.
\end{itemize}
Figures~\ref{fig:GainCalB7d7m}, \ref{fig:GainCalB3bTM1}, and
\ref{fig:GainCalB4aTM1} show examples of gain calibrator 
measurements, with weighted uncertainties, and their associated
weighted average as a function of frequency.
Figure~\ref{fig:GainCalvSF} shows the correlation between normalized  
gain calibrator standard deviation (as a percentage) versus the
associated amplitude calibration scale factor (calculated as a
difference from a perfect amplitude scale factor of 1.0). Even though
in a few cases large normalized gain calibration errors are associated
with large amplitude scale factors, there is no systematic correlation
between these measures.

\subsection{Conclusion to Search for ALCHEMI Flux Normalization Anomalies}
\label{FluxNormAnomalyConclusion}

In Sections~\ref{sec:SpixAge} through \ref{sec:GainCal} we have investigated
%\begin{itemize}
%\item Large spectral index age
%\item Large calibrator catalog Band 3 age
%\item Flux monitoring time gaps
%\item Large time span between system temperature measurements
%\item Noisy gain calibrator measurements
%\end{itemize}
%as 
the possible sources of the discrepant flux calibration uncertainties
derived for 13 of the ALCHEMI SBs. We find that with the exception of
possible errors in T$_{sys}$ measurement, which we are unable to
properly analyze, none of the above potential sources of error appear
to explain all of our discrepant amplitude calibration.

\section{Fitting details of individual species}
\label{Sec.AppendixFitDetails}

Sect.~\ref{Sec.LTE} provided details on the procedure used to fit the modelled synthetic spectra to the low resolution ACA observations. In summary, the procedure consists on a human-supervised automatic fit, where the only intervention aims at ensuring convergence of the fit.
Additional criteria used for the newly detected species (Sect.~\ref{sec.newdetections}) are provided below in Sect.~\ref{Sec.AppendixFitDetailsNew}. Thus, parameters were fixed or transitions were masked in the fit when convergence could not be achieved. The results are reported in Table.~\ref{tab.fitmolecparams}.
Here we note that in some cases, the large uncertainty in the fit to the temperature resulted in column density errors of the same order as the value fitted. 
Such is the case of H$^{13}$CN, despite being a clear detection with bright features, extra masking and fixing of the temperature was required to provide a reasonably narrow error on the column density (with only marginal change in the fitted column density value). However in most cases we decided to maintain such fits for the sake of consistency with the procedure applied to all species.

Although we know from previous studies that there is evidence for the presence of multiple components with different excitation temperatures \citep{Aladro2011}, there are no really obvious deviations from the LTE fit. As an example the higher energy transitions of HC$_3$N appear to be underestimated, which could support the presence of a higher temperature component.
Similarly H$_2$CO is not that well fitted under LTE, showing flat topped spectral features. For the sake of simplicity in the presentation of these data, we assumed a single LTE component for all species. A detailed analysis of the excitation of some of these species will be presented in future publications which will make use of the higher resolution ALCHEMI 12m~Array data.

For the sake of completeness, a list of all transitions above a peak flux density of 15~mJy 
%(those labeled in Fig.~\ref{fig.fullsurvey}) 
are presented in Table~\ref{tab.fitmolecintensities}. We note that these intensities correspond to those of the model fit to the observed spectra, and therefore parameters of velocity and width are those corresponding to the species as presented in Table~\ref{tab.fitmolecparams}. As indicated in Sect.~\ref{Sec.LTE}, no measurement or analysis of individual spectral features has been performed. For the reasons presented there, analysis has been performed per molecular species.

%One of the visual advantages from fitting through synthetic spectra is that non-LTE emission is evidenced by line intensities significantly deviating from the LTE conditions, which can be overlooked in the log-log representation in rotational diagrams. 
Below we provide some cases of transitions significantly deviating from LTE together with other special details on the fitting of individual species.
%are provided in the following.

$H_3O^+:$ The fit to H$_3$O$^+$ emission has been performed only on the 307.2~GHz transition, which is one of the two transitions covered by our survey. Although the 307.2~GHz is blended with CH$_3$OH, this contribution is accounted for based on the CH$_3$OH fit to the whole survey. The 364.8~GHz transition, on the other hand, is the most obvious case of non-LTE emission, and is observed to be more than an order of magnitude brighter than predicted by LTE (Fig.~\ref{fig.fullspectrum10})
%as mentioned in Sect.~\ref{sec.vibemission}, and is a sign of pumping by infrared radiation or tracing high density ($10^{6-7}\rm cm^{-3}$) gas \citep{Phillips1992}.
The line ratio between these two transitions %given in Sect.~\ref{sec.vibemission} 
has been calculated based on the individual integrated intensity fit to each transition and not on the LTE intensities in Table.~\ref{tab.fitmolecintensities}. To explain the relatively large observed flux density ratio of $S_{364.8}/S_{307.2}=6.8\pm1.0$ between the two transitions, non-LTE models from \citet{Phillips1992} suggest volume densities $<10^7\rm cm^{-3}$ and effective excitation by dust emission. Together with the vibrational emission reported in Sect.~\ref{sec.vibemission}, $H_3O^+$ is also probing the importance of infrared pumping in NGC~253 GMCs. However, the 396~GHz transition of H$_3$O$^+$, not covered in our survey, is key to constrain the non-LTE physical conditions of the emitting gas.
All other species below that show transitions not well fitted by the LTE approximations, are not as extreme as the case of H$_3$O$^+$ where, as mentioned above, the LTE approximation is underestimating the 364.8~GHz by more than an order of magnitude.

%{\bf text to be merged}
% IR pumped and n<10^7 cm-3 so that the 304 is fainter
%In fact, infrared pumping affecting the densest gas in NGC~253 is not only implicated by the presence of vibrational emission of multiple species sensitive to mid-IR photon pumping, but also by non-LTE effects on the observed emission. Such could be the case of the relative non-LTE line strength of the 307.2 and 364.8~GHz transitions of H$_3$O$^+$, where the latter is observed to be more than an order of magnitude brighter than predicted by LTE (Fig.~\ref{fig.fullspectrum10}). To explain the relatively large observed flux density ratio of $S_{364.8}/S_{307.2}=6.8\pm1.0$ between the two transitions, non-LTE models from \citet{Phillips1992} suggest volume densities $<10^7\rm cm^{-3}$ and effective excitation by dust emission. However, the 396~GHz transition of H$_3$O$^+$, not covered in our survey, is key to constrain the non-LTE physical conditions of the emitting gas. 
%{\bf text to be merged}

$H_2S:$ The transition of H$_2$S at 216.71~GHz is approximately two times brighter than predicted by the LTE fit. 

$H_2CO:$ The transitions at 140.83 and 150.48~GHz are twice as bright as predicted by the LTE fit. A number of other transitions also deviate from the fit, but to a lesser extent. We note that we did not fit the ortho- and para-H$_2$CO separately, but we assumed the ortho-to-para ratio of 3.
% % % % %%H2CO blighter flux in 150.48 and 140.83
% though fainter also brighter 291.948, 291.24 and 292.38, 265.37, 363.95, 218.78 and 218.47

$HC_3N:$ All transitions above 270~GHz are brighter than the LTE fit estimate. This could be the signature of a warmer component and partially to the effect of varying opacity across the transitions.

$HNCO:$ The brighter transitions of HNCO show an obvious double peak profile which is likely due to the distribution of this species at high resolution \citep{Meier2015}. This double peak profile, similarly observed in CH$_3$OH, is more apparent than in other species where the velocity components are more blended and the profile is a single peak. On the other hand, many of the fainter transitions of HNCO are overestimated by the LTE fit.

$CH_3^{13}CCH:$ Despite being almost uniformly blended with other species, the fit of $\rm CH_3^{13}CCH$ is consistent within the errors to that of the other two isotopologues, being $\sim0.1$ dex above $\rm CH_3C^{13}CH$ and $\sim0.4$ dex above $\rm ^{13}CH_3CCH$. Although the uncertainties in the column density determination of the latter two isotopologues are of the same order as the value fitted, an independent fit to the three isotopologues are in good agreement, which supports the detection and fit to those species. Moreover, the ratio to the main isotopologue (Table~\ref{tab.isotopicratios}) is consistent with what is observed with other species as discussed in Sect.~\ref{sec.isotopologuesCarbon}). 

$CH_3OH:$ Two of the transitions of methanol, the $1_{1,0}-2_{0,2}$ at 205.79 GHz and $7_{1,7}-6_{1,6}$ at 335.58~GHz, show observed fluxes which are less than half of the LTE fit flux, while the $6_{1,5}-5_{1,4}$ at 292.7\,GHz transition is about half of the LTE fit flux. Similar to what is observed in HNCO, the brighter transitions show a very clear double peak profile.
% bright transitions very peaky double peak profile
% globally fit could be more uncertain than that of the statistical one

$^{13}CH_3OH:$ There is only one transition of $\rm ^{13}CH_3OH$ which is bright enough and is not blended with other species. Therefore this is the only species in which the fit value should be considered with caution and probably an upper limit. This LTE fit has been discussed in Sect.~\ref{sec.isotopologuesCarbon}.

$CH_3CN:$ The $J=7-6$ and $8-7$ groups of transitions at 128.7 and 147.1~GHz, respectively, show significantly brighter emission than that from the LTE fit, being up to a factor of two brighter on the former.

$CH_2NH:$ While the $3_{1,1}-1_{0,1}$ transition at 166.85~GHz is not detected, though predicted by the LTE fit, the $2_{0,2}-1_{0,1}$ transition at 127.85~GHz is brighter than predicted.

%High freq transitions showing higher intensities 
%?????C3H+ 359.71
%?????C4H 247.37 lines (no lines above?, maybe purged)
% good fit but potential Non-LTE (was this CH3CN or CH2NH)

\subsection{Fitting details of newly detected species}
\label{Sec.AppendixFitDetailsNew}

On top of the fit criteria explained in Sect.~\ref{Sec.LTE} on detection of the brightest spectral features and requiring convergence of the fit, we enforced extra criteria to claim newly detected species.
This is, at least one of the brightest (according to the LTE prediction) spectral features needs to be un-blended or marginally contaminated by emission from other species based on the LTE modelling to other transitions from the contaminant species. More importantly, all other blended transitions should be consistent with the residual spectra after subtraction of all other modelled species. In other words, any bright emission line predicted by the LTE model should be consistent with the observed spectra and no big outliers should be present. This may not be the case with previously reported species showing some out of equilibrium transitions reported above. 

\begin{figure}
\begin{center}
\includegraphics[width=\textwidth]{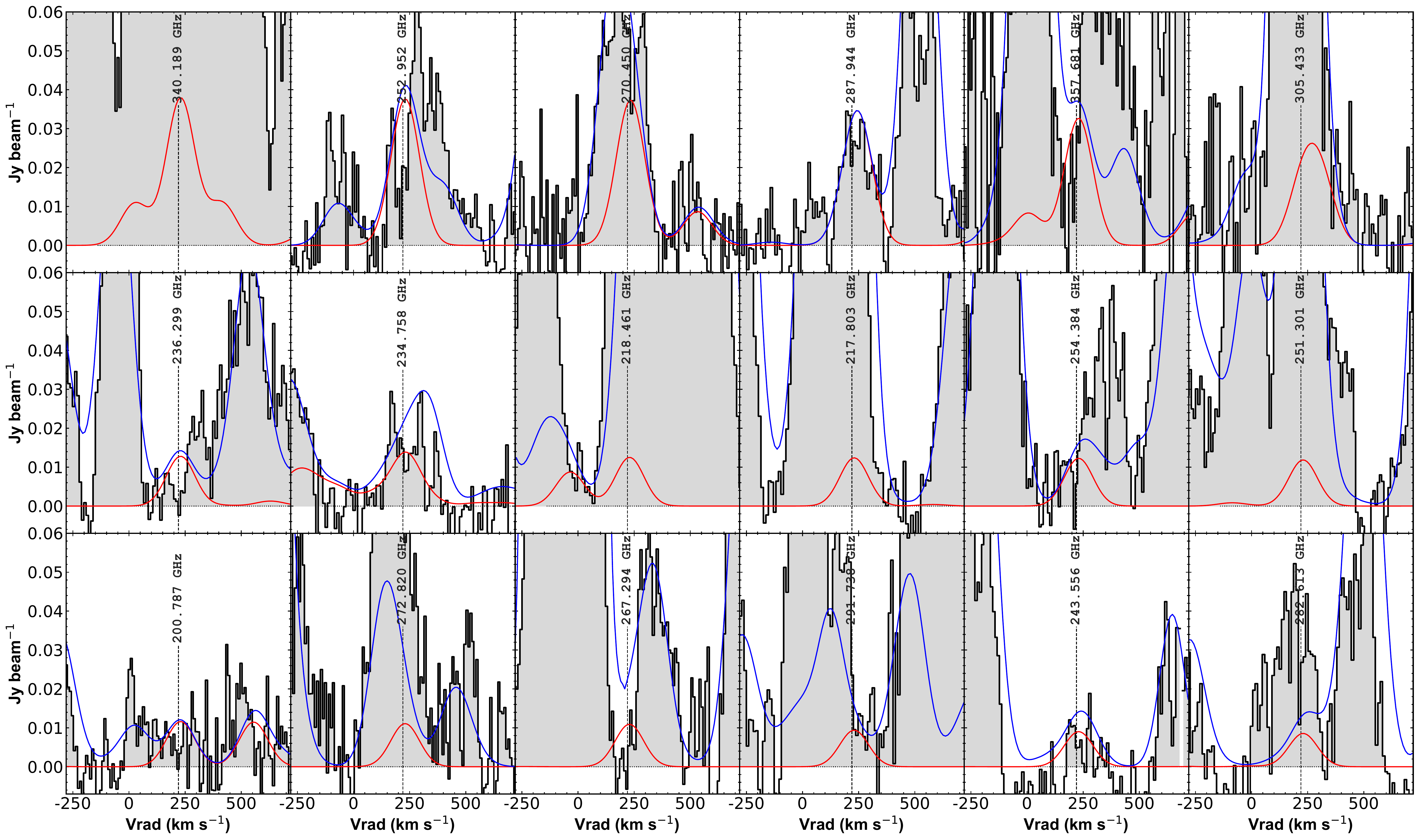}
\caption{Modelled emission of C$_2$H$_5$OH in red, overlaid over the observed spectrum (filled black histogram) and the global model including all molecular transitions in this study.  Only the 18 brightest transitions or group of transitions according to the LTE model of  C$_2$H$_5$OH are displayed, ordered in descending order of brightness from left to right and top to bottom. The model generated with MADCUBA makes use of the spectroscopic parameters in JPL catalog entry 46004. The frequency of the brightest transition in each panel are displayed for reference.
\label{fig.newC2H5OH}}
\end{center}
\end{figure}

\begin{figure}
\begin{center}
\includegraphics[width=\textwidth]{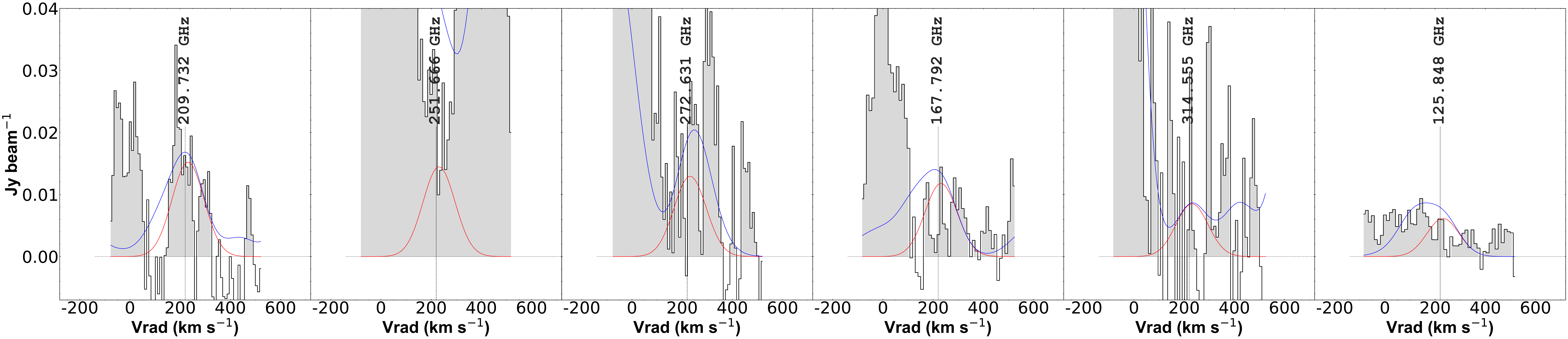}
\caption{Same as Fig.~\ref{fig.newC2H5OH} showing the model of HOCN, using the spectroscopic parameters in CDMS catalog entry 43510.
\label{fig.newHOCN}}
\end{center}
\end{figure}

\begin{figure}
\begin{center}
\includegraphics[width=\textwidth]{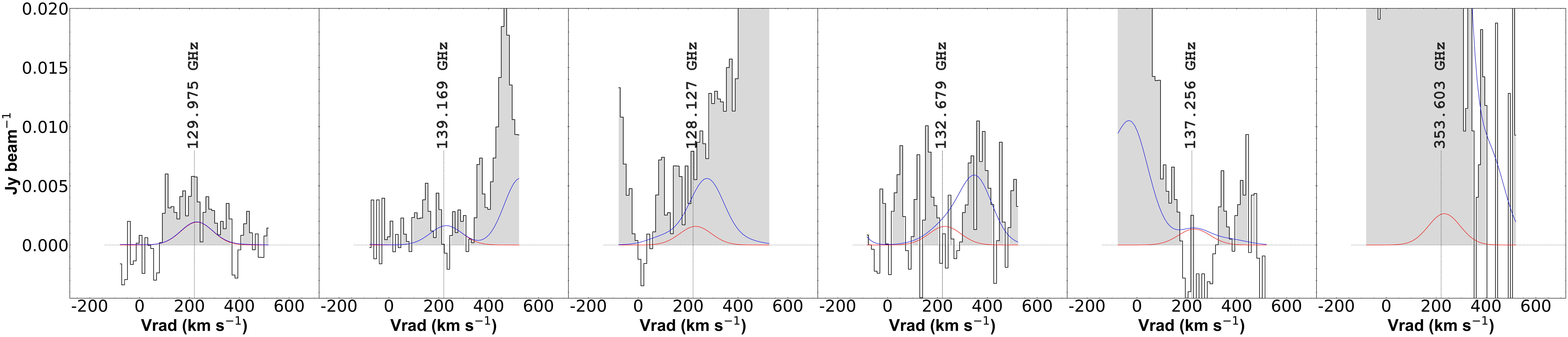}
\caption{Same as Fig.~\ref{fig.newC2H5OH} showing the model of HC$_3$HO, using the spectroscopic parameters in JPL catalog entry 54007.
\label{fig.newHC3HO}}
\end{center}
\end{figure}

In this section we present details of the modelling of C$_2$H$_5$OH, HOCN and HC$_3$HO, since most relevant details regarding the newly detected isotopologues are discussed in Sect.~\ref{sec.isotopologues}. Figs.~\ref{fig.newC2H5OH},~\ref{fig.newHOCN}, and ~\ref{fig.newHC3HO} show the fit results to these species where the spectral features are ordered by the brightness of the LTE modelled emission, thus showing only the brightest spectral features of each molecule.

We do not list the spectroscopic parameters of the detected transitions since these are directly extracted from the catalog entries indicated in the figures and more importantly these species have been previously identified in the Galactic ISM.

$C_2H_5OH$: All transitions above 3~mJy modeled emission were used to fit the emission of 
C$_2$H$_5$OH, where those falling within the spectral features of significantly brighter transitions were masked, adding up to a total of $\sim50$ transitions considered.
Among its brightest transitions only the transitions at 287.944~GHz ($6_{4,2}-5_{3,3}$) and 287.917~GHz ($6_{4,3}-5_{3,2}$) form an clearly unblended spectral feature. Other spectral lines at 252.952, 270.450, and 234.758~GHz, though partially blended, confirm the detection of C$_2$H$_5$OH, together with the fainter feature at 243.556~GHz also marginally blended as displayed in Fig.~\ref{fig.newC2H5OH}. The identified transitions appear to show a double peak, similar to what is observed for CH$_3$OH, further supporting this detection.

$HOCN$: The brightest expected transition of HOCN in the whole frequency coverage of ALCHEMIS ($10_{0,10}-9_{0,9}$ at 209.732~GHz) is unambiguously detected and just marginally blended. The brightness of all other transitions drop quickly below the detection limit or are blended to other brighter species as shown in Fig.~\ref{fig.newHOCN}. All spectral features shown in this figure, but for the one at 251.666~GHz were used in the fit.

$HC_3HO$: The two brighest transitions of HC$_3$HO ($14_{0,14}-13_{0,13}$ @ 129.975~GHz and $15_{0,15}-14_{0,14}$ @ 139.169~GHz) appear unblended in our survey while most other transitions fall close to the noise level of our obsrevations. We conservatively included transitions down to $\sim 1$~mJy in the fit to this species which may have resulted in an understimate of the brightest transitions. A fit performed exclusively with the two brightest transitions would have resulted into a column density 80\% higher yielding $ N_{\rm HC_3HO}=10^{14.2}\pm10^{13.6}~\rm cm^{-2}$.

\begin{center}
\begin{longtable}{llr|llr|llr|llr}
\caption{Intensities from the 744 transitions with $>15$~mJy from the LTE model fit to the data \label{tab.fitmolecintensities}}\\
\hline
\hline
Formula &      $\nu$ &       $S$ &
Formula &      $\nu$ &       $S$ &
Formula &      $\nu$ &       $S$ &
Formula &      $\nu$ &       $S$ \\
 &      (GHz) &       (mJy) &
 &      (GHz) &       (mJy) &
 &      (GHz) &       (mJy) &
 &      (GHz) &       (mJy) \\
\hline
\endfirsthead 

\caption{continued.}\\
\hline
\hline
Formula &      $\nu$ &       $S$ &
Formula &      $\nu$ &       $S$ &
Formula &      $\nu$ &       $S$ &
Formula &      $\nu$ &       $S$ \\
 &      (GHz) &       (mJy) &
 &      (GHz) &       (mJy) &
 &      (GHz) &       (mJy) &
 &      (GHz) &       (mJy) \\
\hline
\endhead
\endfoot
%\multicolumn{7}{l}{{\bf Note.} Molecules are ordered in descending intensity of their brightest transition, which} \\ 
%\multicolumn{7}{l}{is listed in column ($S$), and the corresponding opacity at the line center in column ($\tau_0$).} \\ 
\endlastfoot
\tiny
H$\beta$	&	126.794	&	24	&	HNCO	&	198.529	&	25	&	HCS$^+$	&	256.028	&	28	&	CH$_2$NH	&	317.405	&	64	\\
HC$_3$N	&	127.368	&	213	&	HNCO	&	198.529	&	28	&	CH$_3$C$_2$H	&	256.293	&	60	&	CH$_2$NH	&	317.405	&	78	\\
CH$_3$CN	&	128.779	&	29	&	HNCO	&	198.529	&	22	&	H$\alpha$	&	256.302	&	86	&	c-C$_3$H$_2$	&	318.294	&	34	\\
SO	&	129.139	&	29	&	CH$_2$NH	&	199.823	&	34	&	CH$_3$C$_2$H	&	256.317	&	71	&	CH$_3$OH	&	318.319	&	74	\\
SiO	&	130.269	&	71	&	HC$_3$N	&	200.135	&	189	&	CH$_3$C$_2$H	&	256.332	&	120	&	c-C$_3$H$_2$	&	318.482	&	93	\\
HNCO	&	131.886	&	59	&	H$_2$C$_2$N	&	201.144	&	31	&	CH$_3$C$_2$H	&	256.337	&	142	&	c-C$_3$H$_2$	&	318.79	&	30	\\
HNCO	&	131.886	&	49	&	CH$_3$CN	&	202.34	&	35	&	He$\alpha$	&	256.406	&	27	&	HC$_3$N,v7=1	&	319.576	&	36	\\
HNCO	&	131.886	&	41	&	CH$_3$CN	&	202.356	&	77	&	$^{29}$SiO	&	257.255	&	16	&	NH$_2$CN	&	321.93	&	19	\\
CH$_3$OH	&	132.891	&	15	&	H$_2$CS	&	202.924	&	39	&	CH$_3$CN	&	257.508	&	39	&	CH$_3$OH	&	322.239	&	35	\\
C$_3$H$^+$	&	134.933	&	22	&	c-C$_3$H$_2$	&	204.789	&	19	&	CH$_3$CN	&	257.527	&	83	&	C$_2$H$_5$OH	&	322.691	&	21	\\
H$\beta$	&	135.249	&	26	&	CH$_3$C$_2$H	&	205.045	&	53	&	HC$^{15}$N	&	258.157	&	63	&	C$_2$H$_5$OH	&	322.691	&	21	\\
H$\alpha$	&	135.286	&	50	&	CH$_3$C$_2$H	&	205.065	&	64	&	SO	&	258.256	&	149	&	CH$_3$NH$_2$	&	323.462	&	16	\\
He$\alpha$	&	135.341	&	15	&	CH$_3$C$_2$H	&	205.077	&	108	&	H$^{13}$CN	&	259.012	&	74	&	$^{13}$CS	&	323.685	&	19	\\
HC$_3$N	&	136.464	&	231	&	CH$_3$C$_2$H	&	205.081	&	129	&	H$^{13}$CN	&	259.012	&	110	&	CH$_3$C$_2$H	&	324.607	&	42	\\
CH$_3$C$_2$H	&	136.705	&	21	&	H$\beta$	&	205.76	&	38	&	H$^{13}$CN	&	259.012	&	158	&	CH$_3$C$_2$H	&	324.638	&	49	\\
CH$_3$C$_2$H	&	136.718	&	27	&	CH$_3$OH	&	205.791	&	272	&	H$\beta$	&	260.033	&	46	&	CH$_3$C$_2$H	&	324.657	&	81	\\
CH$_3$C$_2$H	&	136.725	&	47	&	H$_2$CS	&	205.987	&	21	&	H$^{13}$CO$^+$	&	260.255	&	339	&	CH$_3$C$_2$H	&	324.663	&	97	\\
CH$_3$C$_2$H	&	136.728	&	56	&	SO	&	206.176	&	102	&	SiO	&	260.518	&	193	&	CH$_3$NH$_2$	&	325.531	&	50	\\
SO	&	138.179	&	80	&	OCS	&	206.745	&	35	&	CH$_3$NH$_2$	&	261.219	&	34	&	$^{13}$CN	&	325.943	&	30	\\
$^{13}$CS	&	138.739	&	34	&	CH$_2$NH	&	207.38	&	15	&	HN$^{13}$C	&	261.263	&	173	&	$^{13}$CN	&	325.956	&	21	\\
H$_2$CO	&	140.84	&	135	&	NS	&	207.436	&	30	&	NH$_2$CN	&	261.594	&	20	&	$^{13}$CN	&	325.958	&	31	\\
H$_2$C$_2$N	&	140.84	&	36	&	NS	&	207.835	&	39	&	CH$_3$OH	&	261.806	&	128	&	$^{13}$CN	&	326.119	&	42	\\
CH$_3$OH	&	143.866	&	33	&	NS	&	207.838	&	24	&	SO	&	261.844	&	267	&	$^{13}$CN	&	326.125	&	19	\\
H$\beta$	&	144.474	&	28	&	C$_2$S	&	208.216	&	15	&	C$_2$H	&	262.004	&	1153	&	$^{13}$CN	&	326.142	&	57	\\
C$^{34}$S	&	144.617	&	92	&	H$_2$CS	&	209.2	&	41	&	C$_2$H	&	262.006	&	876	&	$^{13}$CN	&	326.143	&	33	\\
c-C$_3$H$_2$	&	145.09	&	48	&	HC$_3$N	&	209.23	&	167	&	C$_2$H	&	262.065	&	837	&	CH$_3$OH	&	326.631	&	15	\\
CH$_3$OH	&	145.094	&	40	&	HOCN	&	209.732	&	15	&	C$_2$H	&	262.067	&	555	&	CH$_3$OH	&	326.961	&	15	\\
CH$_3$OH	&	145.097	&	49	&	H$\alpha$	&	210.502	&	74	&	C$_2$H	&	262.079	&	76	&	OCS	&	328.298	&	25	\\
CH$_3$OH	&	145.103	&	69	&	He$\alpha$	&	210.588	&	23	&	C$_2$H	&	262.208	&	70	&	HC$_3$N,v7=1	&	328.701	&	38	\\
CH$_3$OH	&	145.126	&	15	&	H$_2$CO	&	211.211	&	410	&	HNCO	&	262.77	&	26	&	CH$_3$NH$_2$	&	329.199	&	51	\\
CH$_3$OH	&	145.132	&	26	&	CH$_3$OH	&	213.427	&	56	&	HNCO	&	262.77	&	24	&	C$^{18}$O	&	329.331	&	4461	\\
HC$_3$N	&	145.561	&	243	&	HOCO$^+$	&	213.813	&	18	&	HNCO	&	262.77	&	22	&	HNCO	&	329.665	&	63	\\
H$_2$CO	&	145.603	&	147	&	CH$_3$NH$_2$	&	215.108	&	26	&	HNCO	&	263.749	&	117	&	HNCO	&	329.665	&	58	\\
C$^{33}$S	&	145.756	&	19	&	SO	&	215.221	&	117	&	HNCO	&	263.749	&	108	&	HNCO	&	329.665	&	55	\\
OCS	&	145.947	&	17	&	$^{34}$SO	&	215.839	&	18	&	HNCO	&	263.749	&	99	&	$^{13}$CO	&	330.588	&	13300	\\
CH$_3$OH	&	146.368	&	34	&	c-C$_3$H$_2$	&	216.279	&	155	&	HC$_3$N	&	263.792	&	52	&	CH$_3$CN	&	331.046	&	23	\\
CS	&	146.969	&	880	&	H$_2$S	&	216.71	&	98	&	CH$_3$NH$_2$	&	264.172	&	32	&	CH$_3$CN	&	331.072	&	50	\\
H$\alpha$	&	147.047	&	54	&	CH$_3$OH	&	216.946	&	19	&	HNCO	&	264.694	&	26	&	SO2	&	332.505	&	19	\\
He$\alpha$	&	147.107	&	17	&	SiO	&	217.105	&	191	&	HNCO	&	264.694	&	22	&	CH$_2$NH	&	332.573	&	39	\\
CH$_3$CN	&	147.163	&	19	&	$^{13}$CN	&	217.303	&	44	&	HNCO	&	264.694	&	24	&	CH$_2$NH	&	332.573	&	31	\\
CH$_3$CN	&	147.175	&	42	&	$^{13}$CN	&	217.436	&	19	&	HC$_3$N,v7=1	&	264.817	&	22	&	CH$_3$NH$_2$	&	333.839	&	17	\\
HOCO$^+$	&	149.676	&	27	&	$^{13}$CN	&	217.467	&	41	&	CH$_3$OH	&	265.29	&	27	&	CH$_3$NH$_2$	&	334.712	&	17	\\
NO	&	150.176	&	23	&	c-C$_3$H$_2$	&	217.822	&	207	&	HCN,v2	&	265.853	&	47	&	CH$_3$OH	&	335.134	&	31	\\
H$_2$CO	&	150.498	&	158	&	c-C$_3$H$_2$	&	217.822	&	69	&	HCN	&	265.886	&	6477	&	H$\beta$	&	335.207	&	55	\\
NO	&	150.547	&	23	&	c-C$_3$H$_2$	&	217.94	&	157	&	CH$_2$NH	&	266.27	&	39	&	CH$_3$OH	&	335.582	&	136	\\
c-C$_3$H$_2$	&	150.852	&	126	&	c-C$_3$H$_2$	&	218.16	&	52	&	CH$_2$NH	&	266.27	&	30	&	$^{13}$CCH	&	336.564	&	18	\\
HNCO	&	153.292	&	18	&	H$_2$CO	&	218.222	&	369	&	H$^{15}$NC	&	266.588	&	16	&	$^{13}$CCH	&	336.588	&	15	\\
HNCO	&	153.292	&	16	&	HC$_3$N	&	218.325	&	144	&	CH$_3$OH	&	266.838	&	117	&	C$^{17}$O	&	337.061	&	617	\\
CH$_3$C$_2$H	&	153.791	&	30	&	CH$_3$NH$_2$	&	218.409	&	23	&	HCN,v2	&	267.199	&	48	&	CH$_3$NH$_2$	&	337.119	&	18	\\
CH$_3$C$_2$H	&	153.805	&	37	&	CH$_3$OH	&	218.44	&	93	&	OCS	&	267.53	&	37	&	C$^{34}$S	&	337.397	&	175	\\
CH$_3$C$_2$H	&	153.814	&	63	&	NH$_2$CN	&	218.462	&	16	&	HCO$^+$	&	267.558	&	7021	&	HC$_3$N,v7=1	&	337.825	&	40	\\
CH$_3$C$_2$H	&	153.817	&	76	&	OCS	&	218.903	&	37	&	HOC$^+$	&	268.451	&	240	&	H$_2$CS	&	338.081	&	49	\\
HNCO	&	153.865	&	84	&	HNCO	&	218.981	&	24	&	C$_2$H$_5$OH	&	270.444	&	18	&	CH$_3$OH	&	338.124	&	147	\\
HNCO	&	153.865	&	62	&	HNCO	&	218.981	&	29	&	C$_2$H$_5$OH	&	270.451	&	18	&	c-C$_3$H$_2$	&	338.204	&	100	\\
HNCO	&	153.865	&	72	&	HNCO	&	218.981	&	26	&	H$_2$CS	&	270.521	&	56	&	CH$_3$OH	&	338.345	&	197	\\
HNCO	&	154.415	&	16	&	C$^{18}$O	&	219.56	&	2380	&	HNC	&	271.981	&	3903	&	CH$_3$OH	&	338.409	&	254	\\
HNCO	&	154.415	&	19	&	HNCO	&	219.798	&	130	&	HC$_3$N	&	272.885	&	40	&	CH$_3$OH	&	338.513	&	49	\\
H$\beta$	&	154.557	&	29	&	HNCO	&	219.798	&	117	&	CH$_3$C$_2$H	&	273.373	&	58	&	CH$_3$OH	&	338.541	&	26	\\
HC$_3$N	&	154.657	&	249	&	HNCO	&	219.798	&	106	&	CH$_3$C$_2$H	&	273.399	&	68	&	CH$_3$OH	&	338.543	&	26	\\
c-C$_3$H$_2$	&	155.518	&	26	&	SO	&	219.949	&	226	&	CH$_3$C$_2$H	&	273.415	&	114	&	CH$_3$OH	&	338.56	&	15	\\
CH$_3$OH	&	156.602	&	53	&	$^{13}$CO	&	220.399	&	7383	&	CH$_3$C$_2$H	&	273.42	&	136	&	CH$_3$OH	&	338.583	&	28	\\
CH$_3$OH	&	156.829	&	15	&	HNCO	&	220.585	&	29	&	HC$_3$N,v7=1	&	273.945	&	24	&	CH$_3$OH	&	338.615	&	106	\\
CH$_3$OH	&	157.049	&	28	&	HNCO	&	220.585	&	24	&	H$_2$CS	&	274.521	&	29	&	CH$_3$OH	&	338.64	&	49	\\
CH$_3$OH	&	157.179	&	44	&	HNCO	&	220.585	&	26	&	H$_2$$^{13}$CO	&	274.762	&	29	&	CH$_3$OH	&	338.722	&	92	\\
CH$_3$OH	&	157.246	&	60	&	CH$_3$CN	&	220.73	&	38	&	CH$_3$CN	&	275.894	&	36	&	CH$_3$OH	&	338.723	&	80	\\
CH$_3$OH	&	157.271	&	50	&	CH$_3$CN	&	220.747	&	83	&	CH$_3$CN	&	275.916	&	78	&	SO	&	339.341	&	19	\\
CH$_3$OH	&	157.272	&	70	&	H$_2$C$_2$N	&	221.254	&	20	&	$^{13}$CS	&	277.455	&	37	&	CN	&	339.447	&	52	\\
CH$_3$OH	&	157.276	&	68	&	NH$_2$CN	&	221.361	&	17	&	CH$_3$OH	&	278.305	&	22	&	CN	&	339.476	&	74	\\
OCS	&	158.107	&	21	&	H$\beta$	&	222.012	&	40	&	CH$_3$OH	&	278.342	&	31	&	CN	&	339.517	&	117	\\
SO	&	158.972	&	49	&	CH$_3$C$_2$H	&	222.129	&	58	&	H$_2$CS	&	278.886	&	58	&	CH$_3$NH$_2$	&	339.724	&	56	\\
H$\alpha$	&	160.212	&	58	&	CH$_3$C$_2$H	&	222.15	&	69	&	N$_2$H$^+$	&	279.512	&	1339	&	$^{34}$SO	&	339.858	&	17	\\
He$\alpha$	&	160.277	&	18	&	CH$_3$C$_2$H	&	222.163	&	117	&	OCS	&	279.685	&	36	&	CN	&	340.008	&	214	\\
H$_2$C$_2$N	&	160.949	&	38	&	CH$_3$C$_2$H	&	222.167	&	139	&	H$_2$CO	&	281.527	&	544	&	CN	&	340.02	&	214	\\
NS	&	161.298	&	19	&	C$^{17}$O	&	224.714	&	323	&	NH$_2$CN	&	281.708	&	20	&	CN	&	340.032	&	1639	\\
NS	&	161.697	&	26	&	C$_3$H$^+$	&	224.868	&	15	&	HC$_3$N	&	281.977	&	30	&	CN	&	340.035	&	651	\\
HC$_3$N	&	163.753	&	247	&	CH$_2$NH	&	225.554	&	51	&	H$\beta$	&	282.333	&	49	&	CN	&	340.035	&	1068	\\
CH$_3$OH	&	165.05	&	38	&	CH$_2$NH	&	225.556	&	17	&	c-C$_3$H$_2$	&	282.381	&	93	&	C$^{33}$S	&	340.053	&	33	\\
CH$_3$OH	&	165.061	&	53	&	H$_2$CO	&	225.698	&	456	&	HC$_3$N,v7=1	&	283.072	&	26	&	CH$_3$OH	&	340.141	&	32	\\
CH$_3$OH	&	165.099	&	55	&	CN	&	226.287	&	57	&	H$_2$$^{13}$CO	&	283.442	&	22	&	C$_2$H$_5$OH	&	340.189	&	18	\\
CH$_3$OH	&	165.19	&	48	&	CN	&	226.299	&	46	&	H$\alpha$	&	284.251	&	94	&	C$_2$H$_5$OH	&	340.189	&	18	\\
CH$_3$OH	&	165.369	&	36	&	CN	&	226.303	&	46	&	CH$_2$NH	&	284.253	&	27	&	CN	&	340.248	&	2129	\\
CH$_3$CN	&	165.556	&	25	&	CN	&	226.315	&	110	&	CH$_2$NH	&	284.254	&	152	&	CN	&	340.248	&	1615	\\
CH$_3$CN	&	165.569	&	55	&	CN	&	226.333	&	51	&	CH$_2$NH	&	284.255	&	27	&	CN	&	340.249	&	1201	\\
H$\beta$	&	165.601	&	31	&	CN	&	226.342	&	53	&	He$\alpha$	&	284.366	&	29	&	CN	&	340.262	&	155	\\
CH$_3$OH	&	165.679	&	23	&	CN	&	226.36	&	265	&	HNCO	&	284.662	&	23	&	CN	&	340.265	&	154	\\
H$_2$S	&	168.763	&	265	&	CN	&	226.617	&	60	&	HNCO	&	284.662	&	21	&	CH$_2$NH	&	340.353	&	16	\\
H$_2$CS	&	169.114	&	25	&	CN	&	226.632	&	461	&	HNCO	&	284.662	&	19	&	CH$_2$NH	&	340.354	&	191	\\
CH$_3$OH	&	170.061	&	57	&	CN	&	226.66	&	1399	&	c-C$_3$H$_2$	&	284.805	&	50	&	CH$_2$NH	&	340.355	&	16	\\
OCS	&	170.267	&	25	&	CN	&	226.664	&	458	&	c-C$_3$H$_2$	&	284.998	&	132	&	CH$_3$NH$_2$	&	340.598	&	56	\\
HC$^{18}$O$^+$	&	170.323	&	30	&	CN	&	226.679	&	565	&	CH$_3$OH	&	285.111	&	32	&	HC$^{18}$O$^+$	&	340.633	&	69	\\
HCS$^+$	&	170.692	&	24	&	CN	&	226.874	&	1416	&	HNCO	&	285.722	&	101	&	SO	&	340.714	&	146	\\
CH$_3$C$_2$H	&	170.876	&	39	&	CN	&	226.875	&	2077	&	HNCO	&	285.722	&	93	&	CH$_3$OH	&	341.416	&	137	\\
CH$_3$C$_2$H	&	170.893	&	47	&	CN	&	226.876	&	890	&	HNCO	&	285.722	&	86	&	CH$_3$C$_2$H	&	341.683	&	35	\\
CH$_3$C$_2$H	&	170.902	&	80	&	CN	&	226.887	&	299	&	HNCO	&	286.747	&	23	&	CH$_3$C$_2$H	&	341.715	&	41	\\
CH$_3$C$_2$H	&	170.906	&	96	&	CN	&	226.892	&	297	&	HNCO	&	286.747	&	19	&	CH$_3$C$_2$H	&	341.735	&	68	\\
HOCO$^+$	&	171.056	&	27	&	HC$_3$N	&	227.419	&	122	&	HNCO	&	286.747	&	21	&	CH$_3$C$_2$H	&	341.741	&	81	\\
$^{29}$SiO	&	171.513	&	17	&	CH$_3$OH	&	229.759	&	25	&	CH$_3$OH	&	287.671	&	142	&	NH$_2$CN	&	342.038	&	17	\\
HC$^{15}$N	&	172.108	&	32	&	CH$_3$OH	&	230.027	&	29	&	C$_2$H$_5$OH	&	287.918	&	17	&	CS	&	342.883	&	1138	\\
SO	&	172.181	&	71	&	CO	&	230.538	&	56226	&	C$_2$H$_5$OH	&	287.945	&	17	&	H$_2$CS	&	342.944	&	26	\\
H$^{13}$CN	&	172.677	&	21	&	OCS	&	231.061	&	38	&	C$^{34}$S	&	289.209	&	228	&	CH$_3$SH	&	343.048	&	15	\\
H$^{13}$CN	&	172.677	&	16	&	$^{13}$CS	&	231.221	&	53	&	CH$_3$OH	&	289.939	&	155	&	H$_2$$^{13}$CO	&	343.326	&	27	\\
H$^{13}$CN	&	172.678	&	90	&	H$\alpha$	&	231.901	&	80	&	CH$_3$OH	&	290.07	&	207	&	HC$^{15}$N	&	344.2	&	55	\\
H$^{13}$CN	&	172.678	&	48	&	He$\alpha$	&	231.995	&	25	&	CH$_3$OH	&	290.111	&	268	&	SO	&	344.311	&	133	\\
H$^{13}$CN	&	172.68	&	16	&	CH$_3$OH	&	234.683	&	19	&	CH$_3$OH	&	290.185	&	50	&	CH$_3$SH	&	345.021	&	15	\\
HC$_3$N	&	172.849	&	239	&	CO$^+$	&	235.79	&	32	&	CH$_3$OH	&	290.19	&	25	&	H$^{13}$CN	&	345.34	&	308	\\
HCO	&	173.377	&	19	&	CO$^+$	&	236.063	&	57	&	CH$_3$OH	&	290.191	&	25	&	CO	&	345.796	&	112202	\\
H$^{13}$CO$^+$	&	173.507	&	194	&	HC$_3$N	&	236.513	&	102	&	CH$_3$OH	&	290.21	&	15	&	NS	&	345.823	&	20	\\
SiO	&	173.688	&	141	&	H$_2$CS	&	236.727	&	50	&	CH$_3$OH	&	290.213	&	27	&	NS	&	345.824	&	15	\\
HN$^{13}$C	&	174.179	&	98	&	HC$_3$N,v7=1	&	237.432	&	16	&	CH$_3$OH	&	290.249	&	112	&	NS	&	346.221	&	17	\\
H$_2$CS	&	174.345	&	27	&	NH$_2$CN	&	238.316	&	18	&	CH$_3$OH	&	290.264	&	50	&	SO	&	346.528	&	221	\\
C$_2$H	&	174.663	&	658	&	CH$_3$CN	&	239.12	&	39	&	CH$_3$OH	&	290.307	&	81	&	HC$_3$N,v7=1	&	346.949	&	42	\\
C$_2$H	&	174.668	&	446	&	CH$_3$CN	&	239.138	&	85	&	CH$_3$OH	&	290.308	&	94	&	H$^{13}$CO$^+$	&	346.998	&	252	\\
C$_2$H	&	174.722	&	382	&	CH$_3$C$_2$H	&	239.211	&	60	&	CH$_3$C$_2$H	&	290.452	&	54	&	SiO	&	347.331	&	100	\\
C$_2$H	&	174.728	&	165	&	CH$_3$C$_2$H	&	239.234	&	72	&	CH$_3$C$_2$H	&	290.48	&	63	&	HN$^{13}$C	&	348.34	&	130	\\
C$_2$H	&	174.733	&	103	&	CH$_3$C$_2$H	&	239.248	&	121	&	CH$_3$C$_2$H	&	290.497	&	106	&	H$_2$CS	&	348.532	&	50	\\
C$_2$H	&	174.807	&	90	&	CH$_3$C$_2$H	&	239.252	&	143	&	CH$_3$C$_2$H	&	290.502	&	125	&	C$_2$H	&	349.337	&	934	\\
H$\alpha$	&	174.996	&	63	&	CH$_3$OH	&	239.746	&	119	&	H$_2$CO	&	290.623	&	449	&	C$_2$H	&	349.339	&	751	\\
He$\alpha$	&	175.067	&	20	&	H$\beta$	&	240.021	&	43	&	HC$_3$N	&	291.068	&	22	&	C$_2$H	&	349.399	&	737	\\
HNCO	&	175.189	&	24	&	H$_2$CS	&	240.266	&	27	&	C$^{33}$S	&	291.486	&	45	&	C$_2$H	&	349.4	&	553	\\
HNCO	&	175.189	&	18	&	HNCO	&	240.876	&	29	&	OCS	&	291.84	&	34	&	CH$_3$CN	&	349.427	&	19	\\
HNCO	&	175.189	&	21	&	HNCO	&	240.876	&	26	&	HC$_3$N,v7=1	&	292.199	&	29	&	CH$_3$CN	&	349.454	&	40	\\
HNCO	&	175.844	&	82	&	HNCO	&	240.876	&	24	&	CH$_3$OH	&	292.673	&	144	&	CH$_3$OH	&	350.688	&	255	\\
HNCO	&	175.844	&	106	&	C$^{34}$S	&	241.016	&	233	&	CH$_3$OH	&	293.464	&	35	&	NO	&	350.689	&	373	\\
HNCO	&	175.844	&	94	&	NH$_2$CN	&	241.479	&	19	&	CS	&	293.912	&	1674	&	NO	&	350.691	&	274	\\
HNCO	&	176.472	&	21	&	CH$_3$OH	&	241.7	&	133	&	CH$_3$CN	&	294.28	&	32	&	NO	&	350.695	&	199	\\
HNCO	&	176.472	&	24	&	CH$_3$OH	&	241.767	&	175	&	CH$_3$CN	&	294.302	&	70	&	NO	&	350.73	&	24	\\
HNCO	&	176.472	&	18	&	HNCO	&	241.774	&	128	&	SO	&	296.55	&	161	&	NO	&	350.737	&	24	\\
HCN	&	177.26	&	423	&	HNCO	&	241.774	&	116	&	NH$_2$CN	&	297.869	&	20	&	CH$_3$OH	&	350.905	&	728	\\
HCN	&	177.26	&	557	&	HNCO	&	241.774	&	106	&	$^{34}$SO	&	298.258	&	21	&	NO	&	350.963	&	24	\\
HCN	&	177.261	&	1175	&	CH$_3$OH	&	241.791	&	230	&	NS	&	299.7	&	32	&	NO	&	350.99	&	24	\\
HCN	&	177.261	&	1987	&	CH$_3$OH	&	241.833	&	18	&	NS	&	299.701	&	23	&	NO	&	351.044	&	374	\\
HCN	&	177.262	&	29	&	CH$_3$OH	&	241.833	&	18	&	NS	&	300.099	&	27	&	NO	&	351.051	&	275	\\
HCN	&	177.263	&	423	&	CH$_3$OH	&	241.842	&	40	&	HC$_3$N	&	300.16	&	16	&	NO	&	351.052	&	199	\\
H$\beta$	&	177.723	&	33	&	CH$_3$OH	&	241.844	&	20	&	c-C$_3$H$_2$	&	300.192	&	52	&	SO2	&	351.257	&	15	\\
HCO$^+$	&	178.375	&	3369	&	CH$_3$OH	&	241.879	&	94	&	H$_2$S	&	300.506	&	189	&	c-C$_3$H$_2$	&	351.523	&	64	\\
SO	&	178.605	&	154	&	CH$_3$OH	&	241.888	&	40	&	H$_2$CO	&	300.837	&	567	&	HNCO	&	351.633	&	45	\\
HOC$^+$	&	178.972	&	108	&	CH$_3$OH	&	241.904	&	66	&	SO	&	301.286	&	153	&	HNCO	&	351.633	&	43	\\
C$_3$H$^+$	&	179.903	&	25	&	CH$_3$OH	&	241.905	&	76	&	HC$_3$N,v7=1	&	301.325	&	31	&	HNCO	&	351.633	&	40	\\
CH$_2$NH	&	180.627	&	39	&	HNCO	&	242.64	&	29	&	CH$_3$OH	&	302.37	&	157	&	H$_2$CO	&	351.769	&	426	\\
CH$_3$OH	&	181.296	&	23	&	HNCO	&	242.64	&	24	&	CH$_3$OH	&	303.367	&	496	&	c-C$_3$H$_2$	&	351.782	&	20	\\
HNC	&	181.325	&	2064	&	HNCO	&	242.64	&	26	&	SiO	&	303.927	&	154	&	c-C$_3$H$_2$	&	351.966	&	56	\\
CH$_3$OH	&	181.771	&	15	&	C$^{33}$S	&	242.914	&	47	&	OCS	&	303.993	&	31	&	c-C$_3$H$_2$	&	352.194	&	59	\\
HC$_3$N	&	181.945	&	226	&	OCS	&	243.218	&	39	&	SO	&	304.078	&	263	&	H$\alpha$	&	353.623	&	110	\\
OCS	&	182.427	&	29	&	CH$_3$OH	&	243.916	&	122	&	CH$_3$OH	&	304.208	&	635	&	CO$^+$	&	353.741	&	26	\\
C$_2$S	&	182.553	&	15	&	H$_2$CS	&	244.048	&	52	&	H$_2$CS	&	304.306	&	55	&	He$\alpha$	&	353.767	&	34	\\
CH$_3$CN	&	183.949	&	31	&	c-C$_3$H$_2$	&	244.222	&	39	&	C$_2$H$_5$OH	&	305.354	&	15	&	H$_2$$^{13}$CO	&	353.812	&	19	\\
CH$_2$NH	&	183.957	&	20	&	CS	&	244.936	&	1886	&	C$_2$H$_5$OH	&	305.434	&	15	&	CO$^+$	&	354.014	&	37	\\
CH$_2$NH	&	183.957	&	29	&	CH$_2$NH	&	245.126	&	35	&	CH$_3$OH	&	305.473	&	513	&	c-C$_3$H$_2$	&	354.143	&	22	\\
CH$_3$CN	&	183.963	&	67	&	HC$_3$N	&	245.606	&	83	&	CH$_2$NH	&	306.172	&	31	&	HCN,v2	&	354.46	&	149	\\
c-C$_3$H$_2$	&	184.33	&	61	&	HC$_3$N,v7=1	&	246.561	&	18	&	CH$_2$NH	&	306.172	&	47	&	HCN	&	354.505	&	6497	\\
$^{13}$CS	&	184.982	&	54	&	CH$_3$OH	&	247.229	&	22	&	HNCO	&	306.554	&	18	&	CH$_3$NH$_2$	&	354.844	&	24	\\
N$_2$H$^+$	&	186.345	&	648	&	NO	&	250.437	&	138	&	HNCO	&	306.554	&	17	&	HC$_3$N,v7=1	&	356.072	&	44	\\
CH$_3$C$_2$H	&	187.961	&	47	&	NO	&	250.441	&	87	&	HNCO	&	306.554	&	16	&	HCN,v2	&	356.256	&	151	\\
CH$_3$C$_2$H	&	187.979	&	56	&	NO	&	250.449	&	51	&	CH$_3$OH	&	307.166	&	453	&	HCO$^+$	&	356.734	&	7789	\\
CH$_3$C$_2$H	&	187.99	&	96	&	NO	&	250.475	&	16	&	H$_3$O$^+$	&	307.192	&	178	&	HC$_3$N,v7=2	&	356.937	&	15	\\
CH$_3$C$_2$H	&	187.994	&	114	&	NO	&	250.483	&	16	&	H$\beta$	&	307.258	&	52	&	CH$_3$NH$_2$	&	357.44	&	24	\\
HC$_3$N	&	191.04	&	209	&	NO	&	250.708	&	16	&	CH$_3$C$_2$H	&	307.53	&	48	&	C$_2$H$_5$OH	&	357.681	&	16	\\
H$\beta$	&	191.057	&	35	&	NO	&	250.753	&	16	&	CH$_3$C$_2$H	&	307.56	&	56	&	C$_2$H$_5$OH	&	357.682	&	16	\\
CH$_2$NH	&	191.463	&	39	&	NO	&	250.796	&	138	&	CH$_3$C$_2$H	&	307.577	&	94	&	HOC$^+$	&	357.922	&	248	\\
CH$_2$NH	&	191.463	&	56	&	NO	&	250.816	&	87	&	CH$_3$C$_2$H	&	307.583	&	112	&	CH$_3$OH	&	358.606	&	280	\\
H$\alpha$	&	191.657	&	68	&	NO	&	250.817	&	52	&	HNCO	&	307.694	&	82	&	CH$_3$C$_2$H	&	358.757	&	28	\\
He$\alpha$	&	191.735	&	21	&	c-C$_3$H$_2$	&	251.314	&	193	&	HNCO	&	307.694	&	76	&	CH$_3$C$_2$H	&	358.791	&	33	\\
CH$_3$OH	&	191.811	&	76	&	c-C$_3$H$_2$	&	251.509	&	53	&	HNCO	&	307.694	&	71	&	CH$_3$C$_2$H	&	358.811	&	56	\\
CH$_2$NH	&	192.212	&	27	&	CH$_3$OH	&	251.738	&	23	&	CH$_3$NH$_2$	&	307.792	&	32	&	CH$_3$C$_2$H	&	358.818	&	66	\\
HOCO$^+$	&	192.435	&	24	&	CH$_3$OH	&	251.812	&	33	&	H$_2$CS	&	308.748	&	29	&	CH$_3$OH	&	361.852	&	25	\\
C$^{34}$S	&	192.818	&	178	&	SO	&	251.826	&	145	&	HNCO	&	308.798	&	18	&	NH$_2$CN	&	362.143	&	15	\\
CH$_3$OH	&	193.415	&	87	&	CH$_3$OH	&	251.867	&	39	&	HNCO	&	308.798	&	17	&	HNC,v2=1	&	362.554	&	45	\\
CH$_3$OH	&	193.442	&	112	&	CH$_3$OH	&	251.891	&	33	&	HNCO	&	308.798	&	16	&	HNC	&	362.63	&	3526	\\
CH$_3$OH	&	193.454	&	151	&	CH$_3$OH	&	251.896	&	23	&	CH$_3$OH	&	309.29	&	346	&	H$_2$CO	&	362.736	&	329	\\
CH$_3$OH	&	193.488	&	24	&	CH$_3$OH	&	251.9	&	39	&	SO	&	309.502	&	17	&	CH$_3$OH	&	363.74	&	101	\\
c-C$_3$H$_2$	&	193.489	&	55	&	CH$_3$OH	&	251.906	&	31	&	CH$_3$OH	&	310.193	&	213	&	HNC,v2=1	&	365.147	&	45	\\
CH$_3$OH	&	193.507	&	61	&	CH$_3$OH	&	251.917	&	31	&	HC$_3$N,v7=1	&	310.451	&	33	&	HC$_3$N,v7=1	&	365.195	&	46	\\
CH$_3$OH	&	193.511	&	24	&	$^{13}$CCH	&	252.424	&	17	&	CH$_3$NH$_2$	&	310.528	&	31	&	HC$_3$N,v7=2	&	365.933	&	15	\\
CH$_3$OH	&	193.511	&	38	&	$^{13}$CCH	&	252.449	&	17	&	CH$_3$OH	&	311.853	&	233	&	HC$_3$N,v7=2	&	366.235	&	15	\\
CH$_3$OH	&	193.511	&	44	&	C$_2$H$_5$OH	&	252.951	&	18	&	CH$_3$CN	&	312.664	&	28	&	H$_2$$^{13}$CO	&	366.27	&	26	\\
C$^{33}$S	&	194.337	&	36	&	C$_2$H$_5$OH	&	252.952	&	18	&	CH$_3$CN	&	312.688	&	60	&	H$\beta$	&	366.653	&	59	\\
OCS	&	194.586	&	32	&	NS	&	253.57	&	33	&	SO2	&	313.28	&	21	&	CH$_2$NH	&	367.072	&	28	\\
CH$_3$OH	&	195.147	&	79	&	NS	&	253.971	&	33	&	CH$_3$OH	&	313.597	&	29	&	CH$_2$NH	&	367.072	&	33	\\
C$_2$S	&	195.375	&	15	&	CH$_3$OH	&	254.015	&	60	&	H$_2$CS	&	313.715	&	56	&	CH$_3$NH$_2$	&	367.681	&	17	\\
CS	&	195.954	&	1566	&	CH$_2$NH	&	254.685	&	63	&	$^{13}$C$^{18}$O	&	314.12	&	18	&	CH$_3$CN	&	367.834	&	31	\\
HNCO	&	197.085	&	22	&	CH$_2$NH	&	254.685	&	83	&	$^{13}$C$^{18}$O	&	314.12	&	25	&	H$_2$S	&	369.101	&	460	\\
HNCO	&	197.085	&	24	&	HC$_3$N	&	254.7	&	66	&	CH$_3$OH	&	314.86	&	139	&	H$_2$S	&	369.127	&	32	\\
HNCO	&	197.085	&	27	&	c-C$_3$H$_2$	&	254.988	&	45	&	CH$_3$OH	&	315.267	&	119	&	CH$_3$SH	&	369.394	&	17	\\
HNCO	&	197.821	&	98	&	OCS	&	255.374	&	38	&	OCS	&	316.146	&	28	&	CH$_3$NH$_2$	&	370.166	&	51	\\
HNCO	&	197.821	&	110	&	HC$^{18}$O$^+$	&	255.48	&	67	&	H$\alpha$	&	316.415	&	101	&	H$_2$CS	&	371.844	&	40	\\
HNCO	&	197.821	&	123	&	HC$_3$N,v7=1	&	255.689	&	20	&	He$\alpha$	&	316.544	&	32	&	N$_2$H$^+$	&	372.673	&	1281	\\

\hline
\end{longtable}
\end{center}

\section{Extragalactic molecular census}
\label{Sec.AppendixCensus}

The continuous growth of new species detected during the last two decades has resulted in various publications reporting up to date listings of the extragalactic molecular census including conference proceedings \citep{Mart'in2009a,Mart'in2011a}, refereed publications \citep{Mart'in2006, Mart'in2011,McGuire2018}, as well as online resources such as that hosted at CDMS\footnote{\url{https://cdms.astro.uni-koeln.de/classic/molecules}}.
Each of these reports have had different formats and criteria depending on the scope of the publication but all aiming to maintain updated information on first molecular extragalactic detections. Despite these available resources, given the legacy value of ALCHEMI for the extragalactic molecular content, and the relevance of isotopologue detections in this work, which are not included in most of the references above, we provide here a detailed and updated extragalactic molecular census.

In this appendix we provide a comprehensive listing of all molecular species and isotopologues detected in the extragalactic ISM according to the chronology of detections (Table~\ref{tab.chronologydetections}) and grouped by the number of atoms in the molecule (Table~\ref{tab.census}).
Graphical representations of these lists can be found in Sect.~\ref{sec.newdetections}.

\begin{table}
\centering
\caption{Chronology of extragalactic detections.}
\label{tab.chronologydetections}
\begin{tabular}{l l l | l l l | l l l}
\hline
Year	 &Molecule	 &Reference	 &Year	 &Molecule	 &Reference	 &Year	 &Molecule	 &Reference \\
\hline
\hline
1971	 &OH	 &1	 &2003	 &SO$_2$	 &38	 &2013	 &C$_2$	 &59	 \\
1974	 &H$_2$CO	 &2	 &2003	 &NO	 &38	 &2013	 &C$_3$	 &59	 \\
1975	 &CO	 &3	 &2003	 &NS	 &38	 &2013	 &HCS$^+$	 &60	 \\
1975	 &$^{13}$CO	 &4	 &2003	 &$^{34}$SO	 &38	 &2013	 &NH$_2$CHO	 &60	 \\
1977	 &H$_2$O	 &5	 &2004	 &HOC$^+$	 &39	 &2013	 &$^{30}$SiO	 &60	 \\
1977	 &HCN	 &6	 &2004	 &NH	 &40	 &2014	 &H$_2$Cl$^+$	 &61	 \\
1978	 &H$_2$	 &7	 &2004	 &OH$^+$	 &41,42	 &2014	 &H$_2^{37}$Cl$^+$	 &61	 \\
1979	 &NH$_3$	 &8	 &2006	 &C$_2$S	 &43	 &2014	 &NH$_2$	 &62	 \\
1979	 &HCO$^+$	 &9	 &2006	 &NH$_2$CN	 &44	 &2014	 &H$_2^{17}$O	 &62	 \\
1980	 &CH	 &10	 &2006	 &HOCO$^+$	 &44	 &2014	 &$^{13}$CN	 &63	 \\
1985	 &CS	 &11	 &2006	 &c-C$_3$H	 &44	 &2015	 &ArH$^+$	 &64	 \\
1986	 &c-C$_3$H$_2$	 &12	 &2006	 &DNC $\dagger$	 &44	 &2015	 &$^{38}$ArH$^+$	 &64	 \\
1987	 &CH$^+$	 &13,14	 &2006	 &N$_2$D$^+$ $\dagger$	 &44	 &2015	 &HC$_5$N	 &65,66,75	 \\
1987	 &CH$_3$OH	 &15	 &2006	 &CH$_2$NH	 &44,45	 &2015	 &CH$_3$SH	 &67,68	 \\
1988	 &CN	 &16	 &2006	 &HC$^{18}$O$^+$	 &44,46	 &2016	 &CF$^+$	 &69	 \\
1988	 &C$_2$H	 &16	 &2006	 &HC$^{17}$O$^+$	 &46	 &2017	 &$^{13}$CH$^+$	 &70	 \\
1988	 &HNC	 &16	 &2006	 &H$^{15}$NC	 &46,75	 &2017	 &SH$^+$	 &70	 \\
1988	 &HC$_3$N	 &16,17	 &2006	 &H$_2^{34}$S	 &46	 &2017	 &$^{34}$SH$^+$	 &70	 \\
1989	 &HNCO	 &18,19	 &2006	 &H$_3^+$	 &47	 &2018	 &CH$_3$OCH$_3$	 &71	 \\
1989	 &C$^{34}$S	 &20	 &2006	 &C$_4$H$_2$	 &48	 &2018	 &CH$_3$OCHO	 &71	 \\
1991	 &C$^{18}$O	 &21	 &2006	 &C$_6$H$_2$	 &48	 &2019	 &HCl	 &72	 \\
1991	 &C$^{17}$O	 &21	 &2006	 &C$_6$H$_6$	 &48	 &2019	 &HCl	 &72	 \\
1991	 &SO	 &22,23	 &2008	 &H$_3$O$^+$	 &49	 &2020	 &O$_2$ $\dagger$	 &73	 \\
1991	 &N$_2$H$^+$	 &24	 &2009	 &C$^{33}$S	 &50	 &2020	 &l-C$_3$H$^+$	 &68	 \\
1991	 &SiO	 &24	 &2009	 &$^{13}$CH$_3$OH	 &51,52,75	 &2020	 &C$_3$N	 &68	 \\
1991	 &H$^{13}$CO$^+$	 &24	 &2010	 &HF	 &53	 &2020	 &CH$_2$CHCN	 &68	 \\
1991	 &HN$^{13}$C	 &24	 &2010	 &H$_2$O$^+$	 &53,54	 &2020	 &H$_2$CN	 &68	 \\
1991	 &H$^{13}$CN	 &24	 &2010	 &$^{13}$C$^{18}$O	 &55,56	 &2020	 &HCOOH	 &68,75	 \\
1991	 &CH$_3$CCH	 &25	 &2010	 &C$_{60}$ $\dagger$	 &57	 &2020	 &ND	 &74	 \\
1991	 &CH$_3$CN	 &25	 &2011	 &H$^{13}$CCCN	 &58	 &2020	 &NH$_2$D	 &74	 \\
1993	 &$^{13}$CS	 &26	 &2011	 &HC$^{13}$CCN	 &58	 &2020	 &HDO	 &74	 \\
1995	 &OCS	 &27	 &2011	 &HCC$^{13}$CN	 &58	 &2021	 &H$_2^{13}$CO	 &75	 \\
1995	 &HCO	 &28	 &2011	 &H$_2^{18}$O	 &58	 &2021	 &C$_2$H$_5$OH	 &75	 \\
1996	 &DCO$^+$	 &29	 &2011	 &$^{29}$SiO	 &58	 &2021	 &$^{13}$CCH	 &75	 \\
1996	 &DCN	 &29	 &2011	 &CH$_2$CO	 &58,45	 &2021	 &C$^{13}$CH	 &75	 \\
1998	 &LiH $\dagger$	 &30,31	 &2011	 &SO$^+$	 &45	 &2021	 &HOCN	 &75	 \\
1999	 &HC$^{15}$N	 &32	 &2011	 &l-C$_3$H	 &45	 &2021	 &CH$_3^{13}$CCH	 &75	 \\
1999	 &H$_2$S	 &33	 &2011	 &l-C$_3$H$_2$	 &45	 &2021	 &CH$_3$C$^{13}$CH	 &75	 \\
1999	 &H$_2$CS	 &33,34	 &2011	 &CH$_2$CN	 &45	 &2021	 &$^{13}$CH$_3$CCH	 &75	 \\
1999	 &C$_2$H$_2$	 &35	 &2011	 &C$_4$H	 &45	 &2021	 &HC$_3$HO	 &75	 \\
2000	 &CO$^+$	 &36	 &2011	 &CH$_3$NH$_2$	 &45	 &2021	 &Si$^{17}$O $\dagger$	 &75	 \\
2001	 &HD	 &37	 &2011	 &CH$_3$CHO	 &45	 &	 &	 &	 \\
\hline
\end{tabular}
\tablefoot{
Year of first detection of each indiviudal molecular species, were first detections of isotopologues are also included. When first detection was tentative, confirmation is also included. ($\dagger$) Species where only tentative detection have been reported. \tablebib{(1) \citet{Weliachew1971}; (2) \citet{Gardner1974}; (3) \citet{Rickard1975}; (4) \citet{Solomon1975}; (5) \citet{Churchwell1977}; (6) \citet{Rickard1977}; (7) \citet{Thompson1978}; (8) \citet{Martin1979}; (9) \citet{Stark1979}; (10) \citet{Whiteoak1980}; (11) \citet{Henkel1985}; (12) \citet{Seaquist1986}; (13) \citet{Magain1987}; (14) \citet{Falgarone2017}; (15) \citet{Henkel1987}; (16) \citet{Henkel1988}; (17) \citet{Mauersberger1990}; (18) \citet{Nguyen-Q-Rieu1989}; (19) \citet{ Nguyen-Q-Rieu1991}; (20) \citet{Mauersberger1989a}; (21) \citet{Sage1991}; (22) \citet{Johansson1991}; (23) \citet{ Petuchowski1992}; (24) \citet{Mauersberger1991a}; (25) \citet{Mauersberger1991}; (26) \citet{Henkel1993}; (27) \citet{Mauersberger1995}; (28) \citet{Sage1995}; (29) \citet{Chin1996a}; (30) \citet{Combes1998}; (31) \citet{ Friedel2011}; (32) \citet{Chin1999}; (33) \citet{Heikkila1999}; (34) \citet{Martin2005}; (35) \citet{vanLoon1999}; (36) \citet{Fuente2000}; (37) \citet{Varshalovich2001}; (38) \citet{Mart'in2003}; (39) \citet{Usero2004}; (40) \citet{Gonz'alez-Alfonso2004}; (41) \citet{GonzalezAlfonso2004}; (42) \citet{ vanDerWerf2010}; (43) \citet{Mart'in2006}; (44) \citet{Mart'in2006}; (45) \citet{Muller2011}; (46) \citet{Muller2006}; (47) \citet{Geballe2006}; (48) \citet{Bernard-Salas2006}; (49) \citet{vanDerTak2008}; (50) \citet{Wang2009}; (51) \citet{Mart'in2009b}; (52) \citet{ Muller2021}; (53) \citet{vanDerWerf2010}; (54) \citet{Weiss2010}; (55) \citet{Mart'in2010a}; (56) \citet{Martin2019}; (57) \citet{Garcia-Hernandez2010}; (58) \citet{Mart'in2011}; (59) \citet{Welty2013}; (60) \citet{Muller2013}; (61) \citet{Muller2014a}; (62) \citet{Muller2014}; (63) \citet{Takano2014}; (64) \citet{Mueller2015}; (65) \citet{Aladro2015}; (66) \citet{Costagliola2015}; (67) \citet{Meier2015}; (68) \citet{Tercero2020}; (69) \citet{Muller2016a}; (70) \citet{Muller2017}; (71) \citet{Sewilo2018}; (72) \citet{Wallstroem2019}; (73) \citet{Wang2020}; (74) \citet{Muller2020}; (75) This work.}}

\end{table}

\begin{table}
\begin{center}
\caption{Census of extragalactic molecular species and isotopologues detected.}
\label{tab.census}
%\begin{tabular}{l @{} l @{} l @{\,\,\,} l @{\,\,\,} l @{\,\,\,} l}
\begin{tabular}{ lllllll }
\hline
{\bf 2 atoms} & {\bf 3 atoms} & {\bf 4 atoms} & {\bf 5 atoms} & {\bf 6 atoms} & {\bf 7 atoms} & {\bf >7 atoms}\\
\hline
ArH$^+$, {\tiny $^{38}$ArH$^+$}       & 
                C$_2$H 
                {\tiny \hspace{-5pt} $\Bigg\{ \hspace{-5pt} 
                \begin{array}{l} 
                \rm ^{13}CCH \\
                \rm C^{13}CH 
                 \end{array}$
                 }           &
                                C$_2$H$_2$    & 
                                                C$_4$H        &
                                                                C$_4$H$_2$    & 
                                                                                 CH$_2$CHCN    &
                                                                                                C$_2$H$_5$OH \\

C$_2$         &
                C$_2$S        &
                                C$_3$N        &
                                                c-C$_3$H$_2$   &
                                                                 CH$_3$CN
                                                                              &  CH$_3$CCH 
                                                                                {\tiny \hspace{-5pt} $\Bigg\{ \hspace{-5pt} 
                                                                                \begin{array}{l} 
                                                                                \rm ^{13}CH_3CCH \\
                                                                                \rm CH_3^{13}CCH \\
                                                                                \rm CH_3C^{13}CH \\
                                                                                 \end{array}$
                                                                                 }    &
                                                                                                C$_6$H$_2$ \\ 
CF$^+$        &
                C$_3$         &
                                c-C$_3$H      &
                                                l-C$_3$H$_2$   &
                                                                 CH$_3$OH, {\tiny $^{13}$CH$_3$OH}
                                                                              &
                                                                                CH$_3$CHO  &
                                                                                                C$_6$H$_6$ \\
CH            &
                H$_2$Cl$^+$, {\tiny H$_2^{37}$Cl$^+$}   &
                                l-C$_3$H      & 
                                                CH$_2$CN       &
                                                                 CH$_3$SH     &
                                                                                CH$_3$NH$_2$ &
                                                                                                C$_{60}$ ~\tablefootmark{$\dagger$}\\  
CH$^+$, {\tiny $^{13}$CH$^+$ }        & 
                H$_2$O {\tiny \hspace{-5pt} $\Bigg\{ \hspace{-5pt} 
                \begin{array}{l} 
                \rm H_2^{18}O \\ 
                \rm H_2^{17}O \\ 
                \rm HDO \\ 
                \end{array} $}& 
                                l-C$_3$H$^+$  & 
                                                CH$_2$CO       &
                                                                 HC$_3$HO     & 
                                                                                HC$_5$N &    
                                                                                            CH$_3$OCH$_3$         \\                 
CN, {\tiny $^{13}$CN }            & 
                H$_2$O$^+$    & 
                                H$_2$CN       & 
                                                CH$_2$NH       &
                                                                 NH$_2$CHO    & &
                                                                                                 \\
CO {\tiny \hspace{-5pt} $\Bigg\{ \hspace{-5pt} 
\begin{array}{l} 
\rm ^{13}CO \\ 
\rm C^{18}O \\ 
\rm C^{17}O \\ 
\rm ^{13}C^{18}O \\ 
\end{array} $}            & 
                H$_2$S, {\tiny H$_2^{34}$S}        & 
                                H$_2$CO, {\tiny H$_2^{13}$CO}       &
                                                HC$_3$N {\tiny \hspace{-5pt} $\Bigg\{ \hspace{-5pt} 
                                                \begin{array}{l} 
                                                \rm H^{13}CCCN \\ 
                                                \rm HC^{13}CCN \\ 
                                                \rm HCC^{13}CN \\ 
                                                \end{array} $} & & &
                                                                                                CH$_3$OCHO \\
CO$^+$        & 
                H$_3^+$       &
                                H$_2$CS       &
                                                HCOOH   & & & \\
CS {\tiny \hspace{-5pt} $\Bigg\{ \hspace{-5pt} 
\begin{array}{l} 
\rm ^{13}CS \\ 
\rm C^{34}S \\ 
\rm C^{33}S \\ 
\end{array} $}            & 
                HCN {\tiny \hspace{-5pt} $\Bigg\{ \hspace{-5pt} 
                \begin{array}{l} 
                \rm H^{13}CN \\ 
                \rm HC^{15}N \\ 
                \rm DCN ~\tablefootmark{$\dagger$}\\ 
                \end{array} $}&
                                H$_3$O$^+$    &
                                                NH$_2$CN & & & \\
H$_2$, {\tiny HD}     & 
                HCO           &
                                HNCO          & & & & \\
HF            &
                HCO$^+$ {\tiny \hspace{-5pt} $\Bigg\{ \hspace{-5pt} 
                \begin{array}{l} 
                \rm H^{13}CO^+ \\ 
                \rm HC^{18}O^+ \\ 
                \rm HC^{17}O^+ \\ 
                \rm DCO^+ \\ 
                \end{array} $}&
                                HOCN          & & & & \\
LiH ~\tablefootmark{$\dagger$} &
                HCS$^+$       &
                                HOCO$^+$      & & & & \\
NH, {\tiny ND} &
                HNC {\tiny \hspace{-5pt} $\Bigg\{ \hspace{-5pt} 
                \begin{array}{l} 
                \rm HN^{13}C \\ 
                \rm H^{15}NC \\ 
                \rm DNC \\ 
                \end{array} $}&
                                NH$_3$, {\tiny NH$_2$D} & 
                                              & & & \\
NO            &
                HOC$^+$       & & & & & \\
NS            &
                N$_2$H$^+$, {\tiny N$_2$D$^+$ ~\tablefootmark{$\dagger$}} &
                              & & & & \\
O$_2$  ~\tablefootmark{$\dagger$}  &
                NH$_2$        &
                              & & & & \\
OH            &
                OCS           &
                              & & & & \\
OH$^+$        &
                SH$^+$, {\tiny $^{34}$SH$^+$}        &
                              & & & & \\
SiO{\tiny \hspace{-5pt} $\Bigg\{ \hspace{-5pt} 
\begin{array}{l} 
\rm ^{29}SiO \\ 
\rm ^{30}SiO \\ 
\rm Si^{17}O ~\tablefootmark{$\dagger$} \\ 
\end{array} $} &
                SO$_2$ & & & & & \\
SO, {\tiny $^{34}$SO}            & & & & & & \\ 
SO$^+$        & & & & & & \\ 
                              
\hline
\end{tabular}
\tablefoot{Species are alphabetically ordered in each column.
\tablefoottext{$\dagger$} Sppecies where only tentative detections have been reported.
The table is updated from \citet{Mart'in2011a} according to the list of detections in Table~\ref{tab.chronologydetections}, were references for each detection are provided.
}
\end{center}
\end{table}

\section{Full spectrum and model}
\label{Sec.AppendixFullSpectrum}
\begin{figure}
\includegraphics[width=\textwidth]{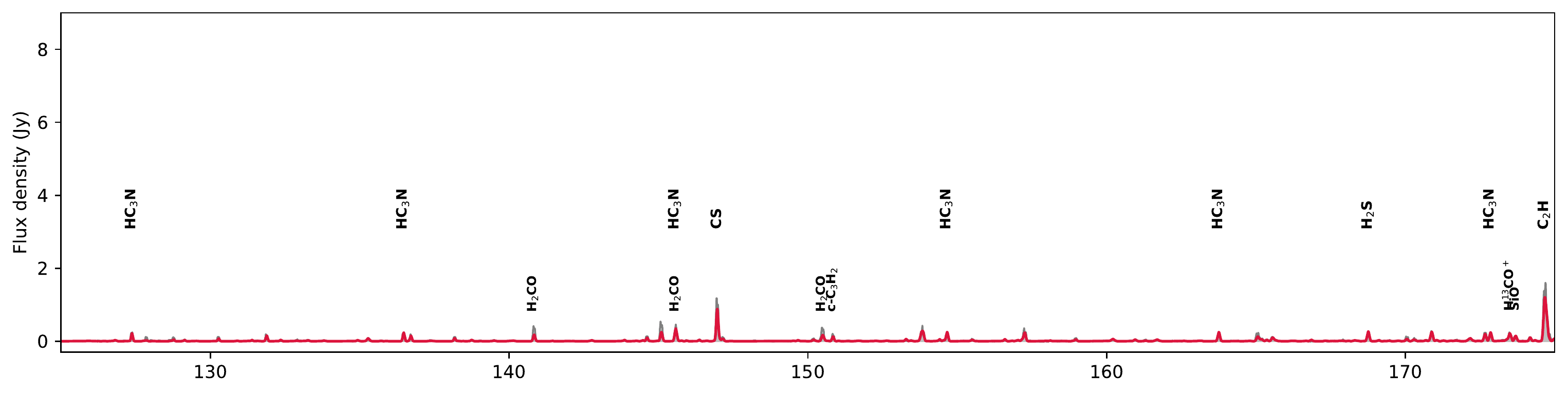}
\includegraphics[width=\textwidth]{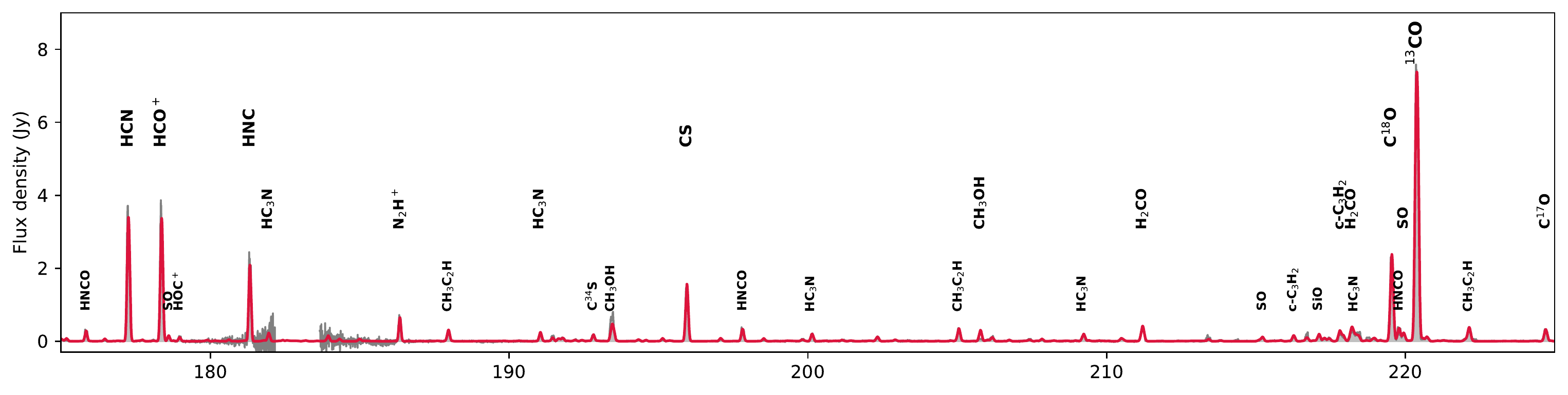}
\includegraphics[width=\textwidth]{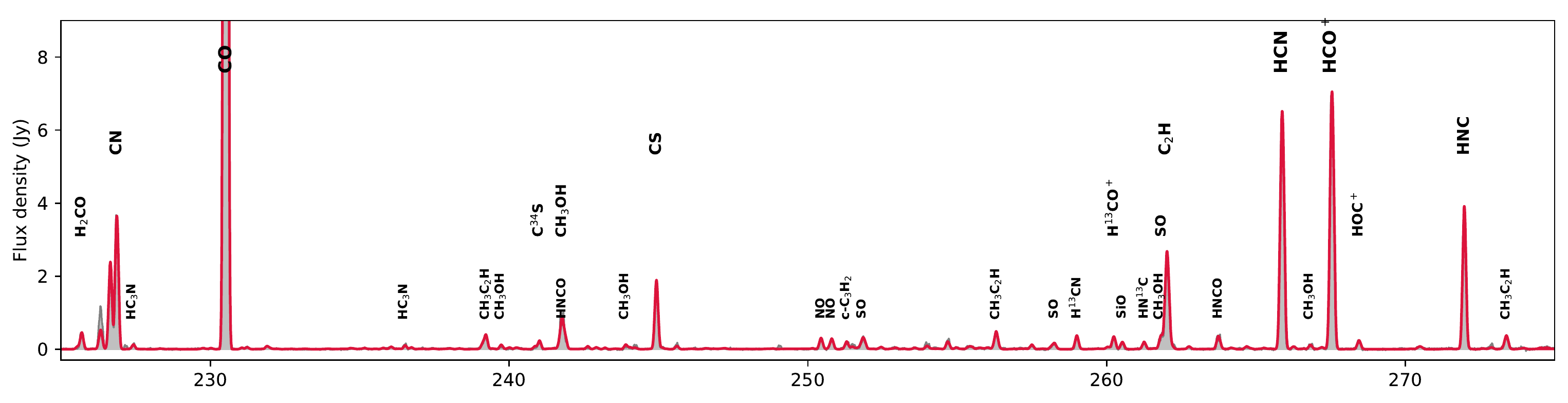}
\includegraphics[width=\textwidth]{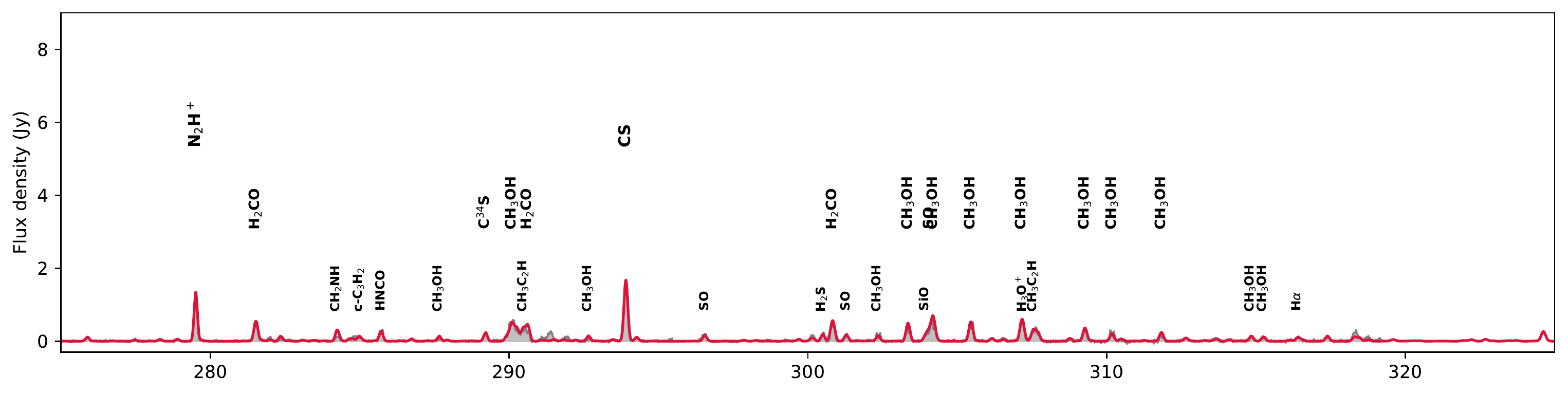}
\includegraphics[width=\textwidth]{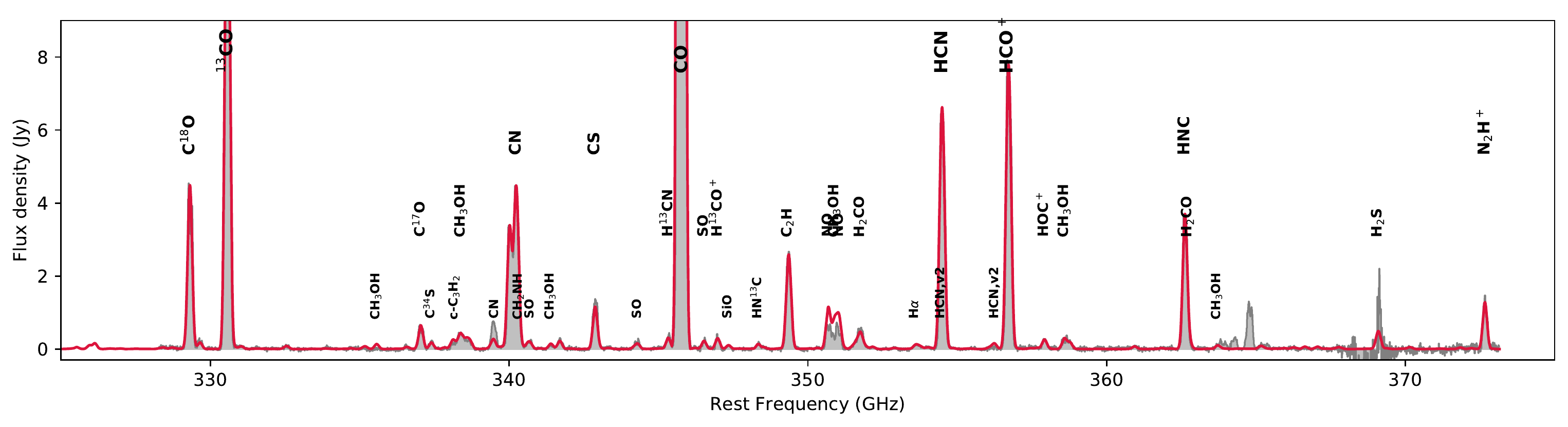}
\caption{Full spectral coverage as in Fig.\ref{fig.fullsurvey} zoomed to 50~GHz frequency windows. The observed spectrum is shown in grey histogram and the model (Sect.~\ref{Sec.LTE}) in red line. Only the brighter individual molecular transitions with intensities higher than 100~mJy are labeled with different y-axis position and character size depending on the modelled intensity for $>5$, $>1$,$>0.2$, and $>0.1$~Jy. \label{fig.fullspectrum50GHz}}
\end{figure}

\begin{figure}
\includegraphics[width=\textwidth]{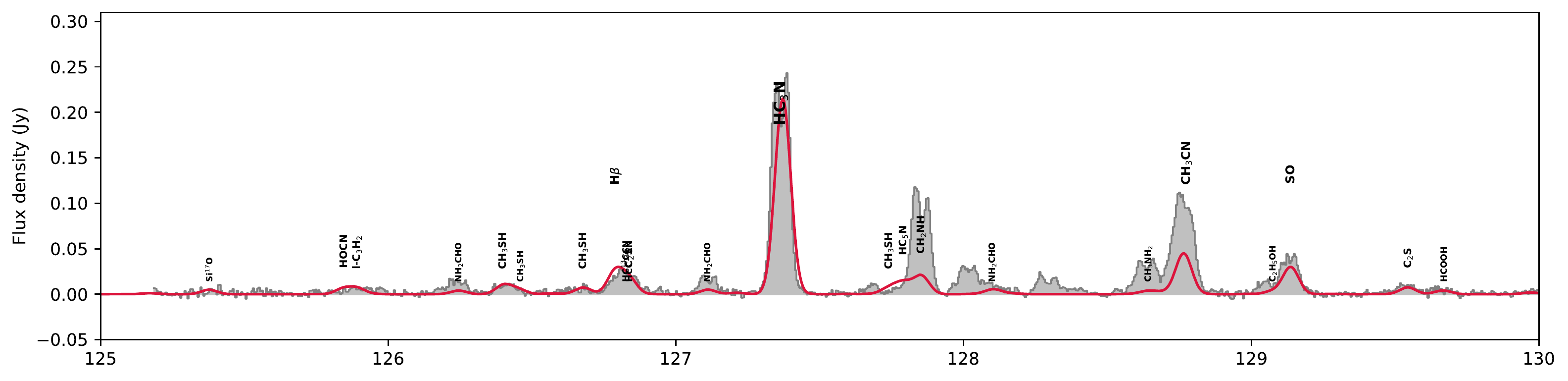}
\includegraphics[width=\textwidth]{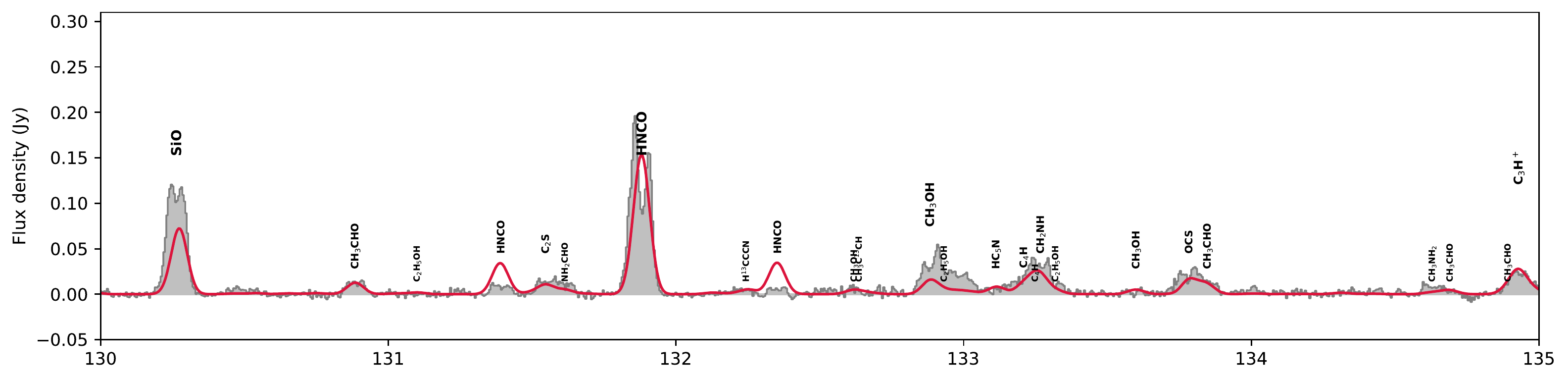}
\includegraphics[width=\textwidth]{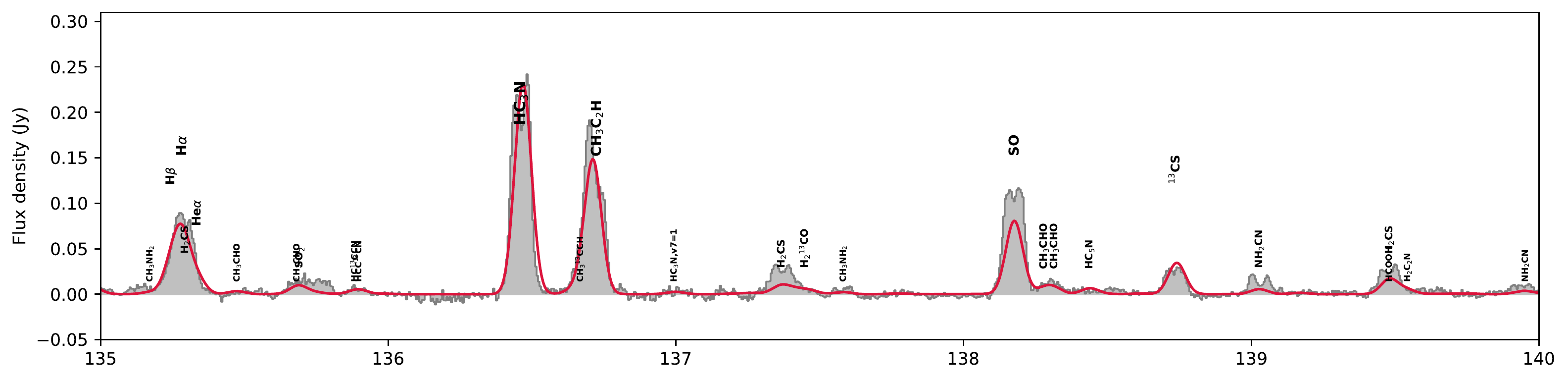}
\includegraphics[width=\textwidth]{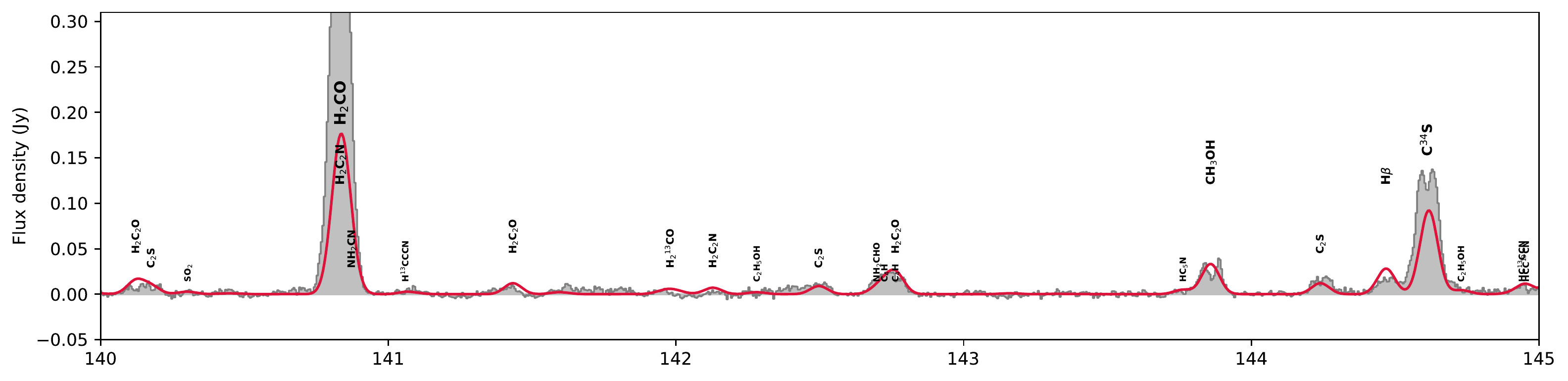}
\includegraphics[width=\textwidth]{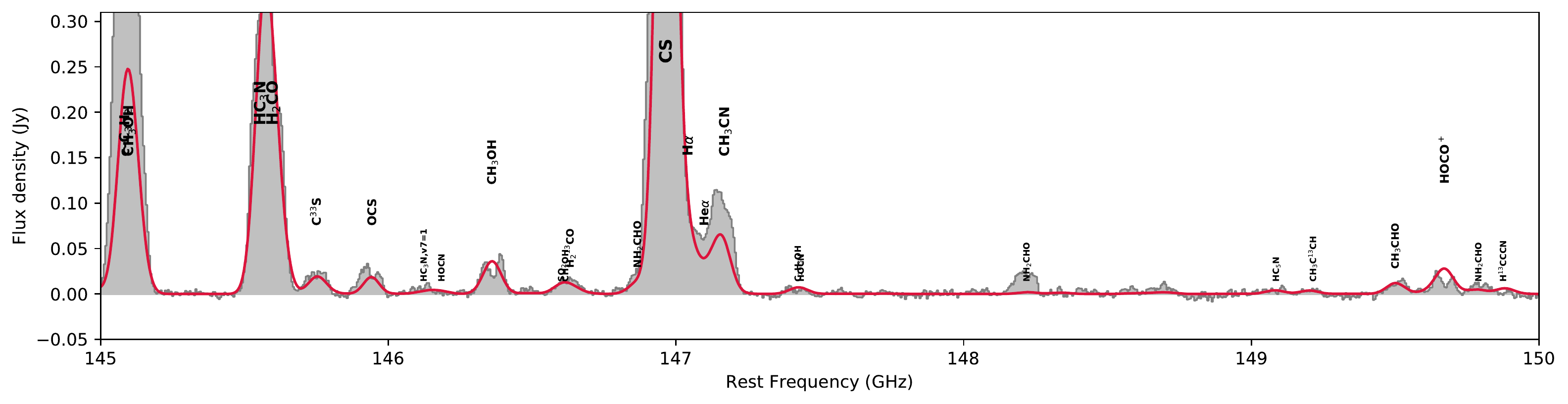}
\caption{Full spectral coverage as in Fig.\ref{fig.fullsurvey} zoomed to 5~GHz frequency windows. The observed spectrum is shown in grey histogram and the model (Sect.~\ref{Sec.LTE}) in red line. Individual molecular transitions with intensities higher than 2~mJy are labeled with different y-axis position and character size depending on the modelled intensity  for $>270$,  $>95$,  $>40$,  $>20$, $>15$, $>10$, $>5$, and $>2$~mJy. \label{fig.fullspectrum1}}
\end{figure}

\begin{figure}
\includegraphics[width=\textwidth]{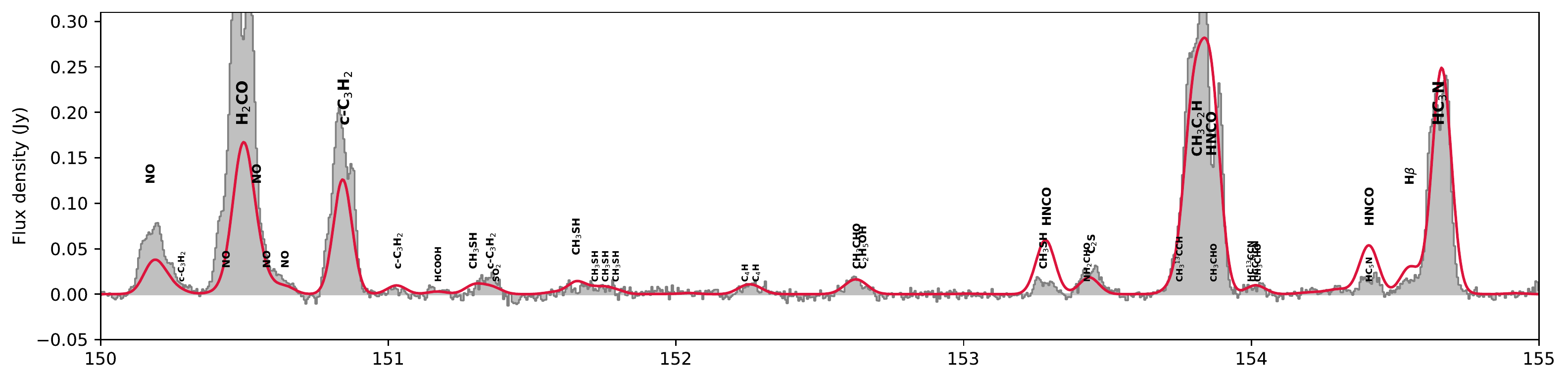}
\includegraphics[width=\textwidth]{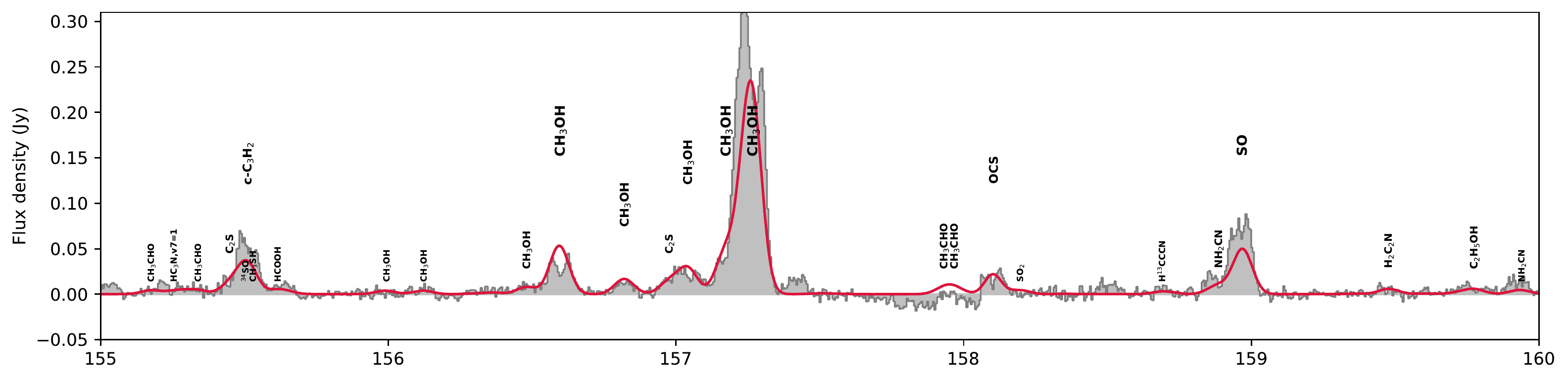}
\includegraphics[width=\textwidth]{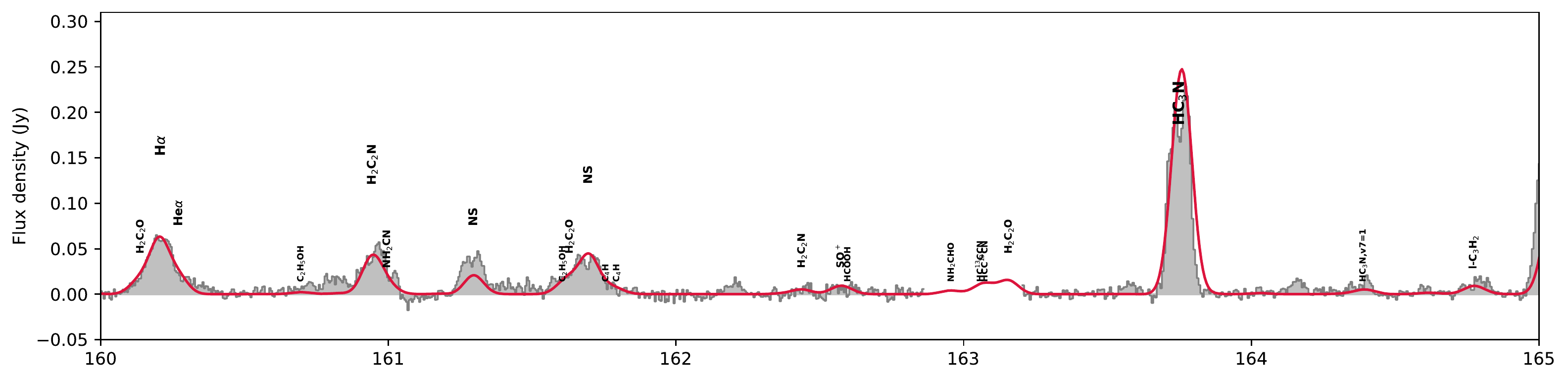}
\includegraphics[width=\textwidth]{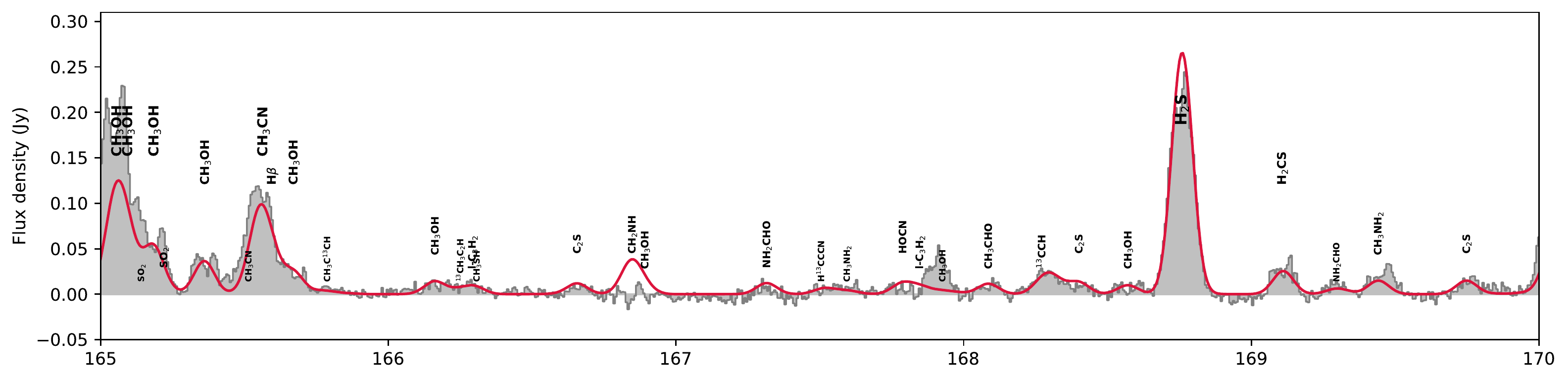}
\includegraphics[width=\textwidth]{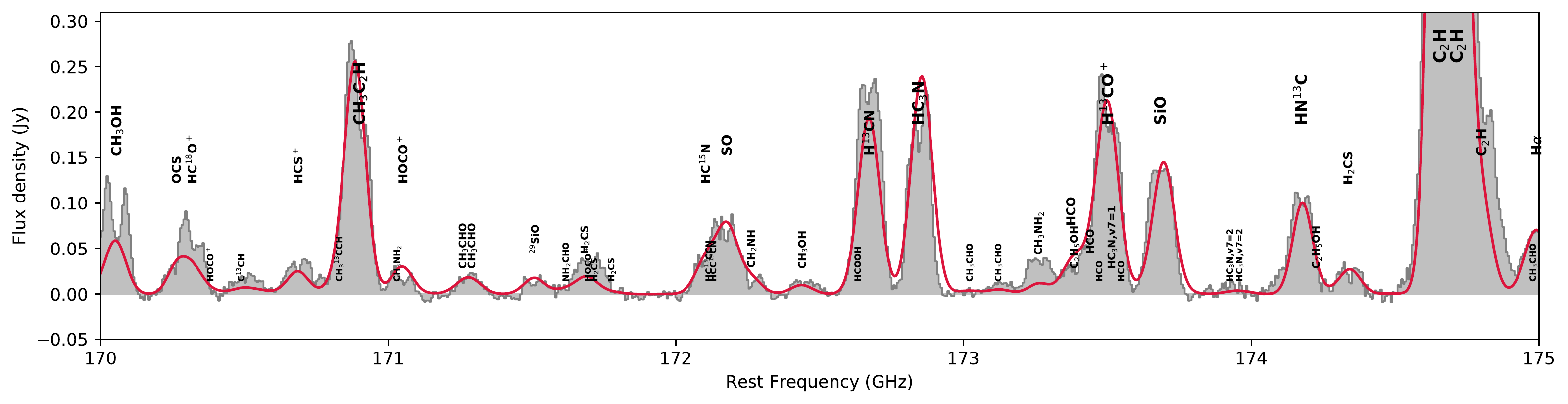}
\caption{Same as Fig.~\ref{fig.fullspectrum1}. \label{fig.fullspectrum2}}
\end{figure}

\begin{figure}
\includegraphics[width=\textwidth]{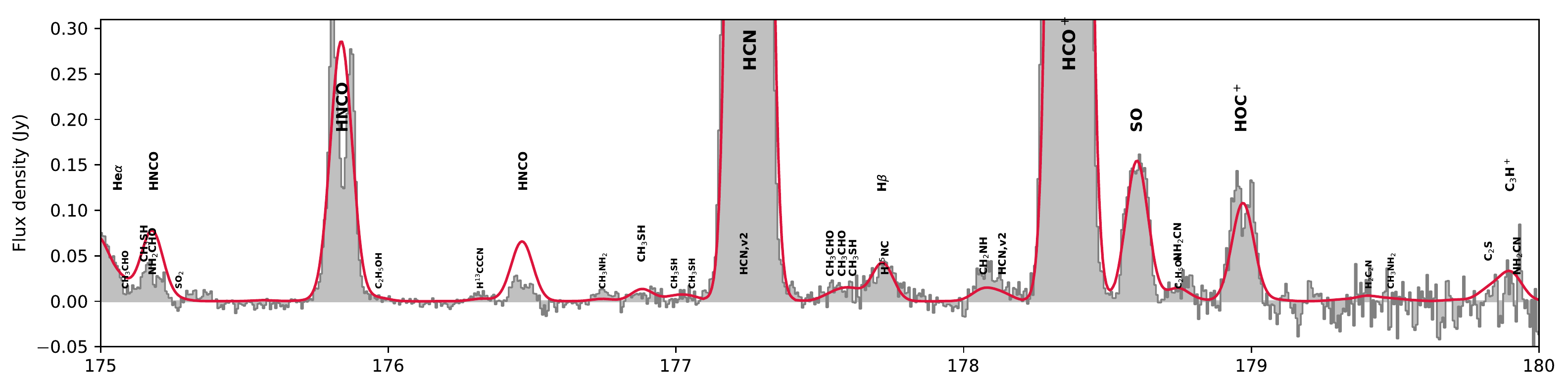}
\includegraphics[width=\textwidth]{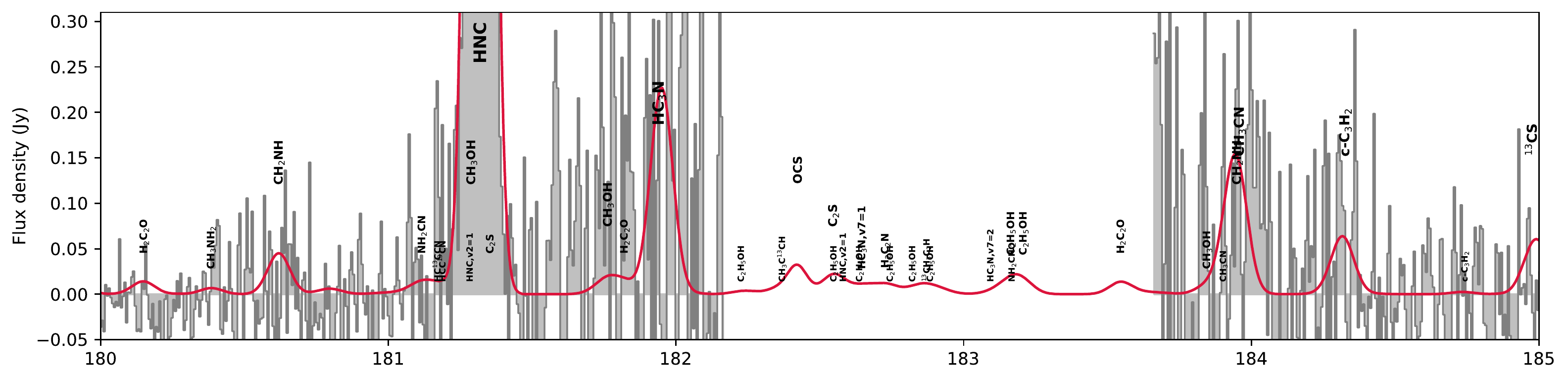}
\includegraphics[width=\textwidth]{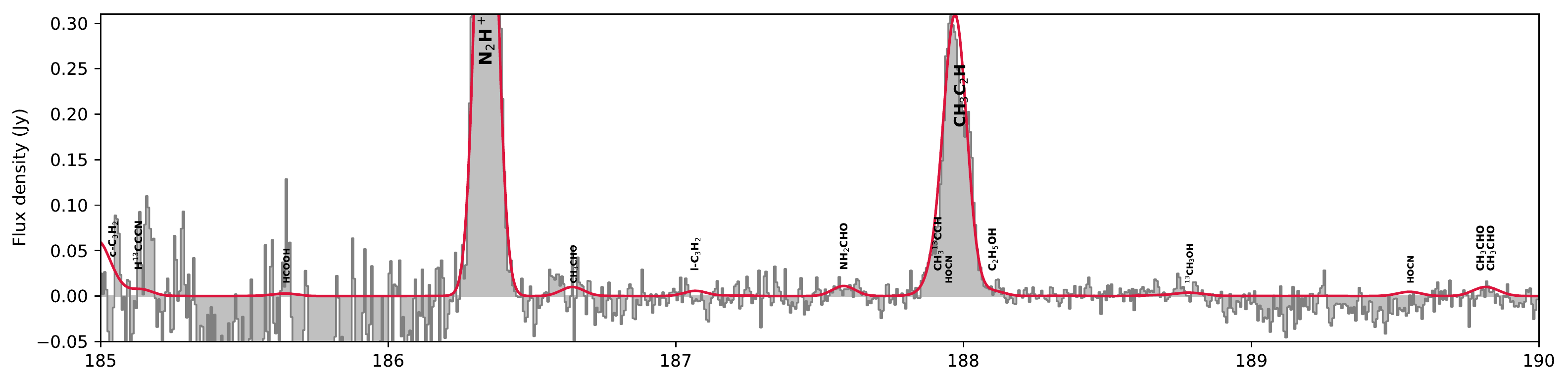}
\includegraphics[width=\textwidth]{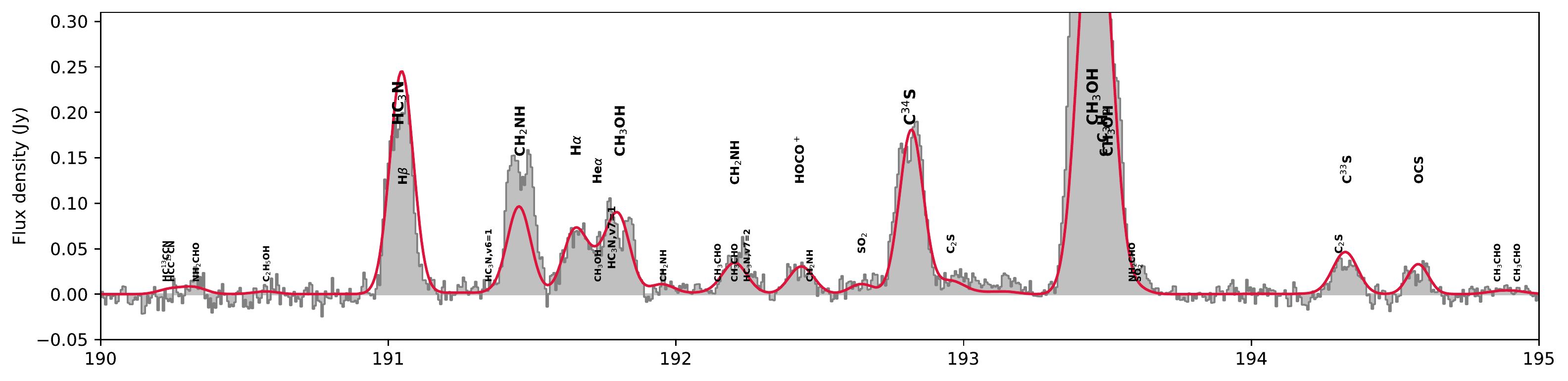}
\includegraphics[width=\textwidth]{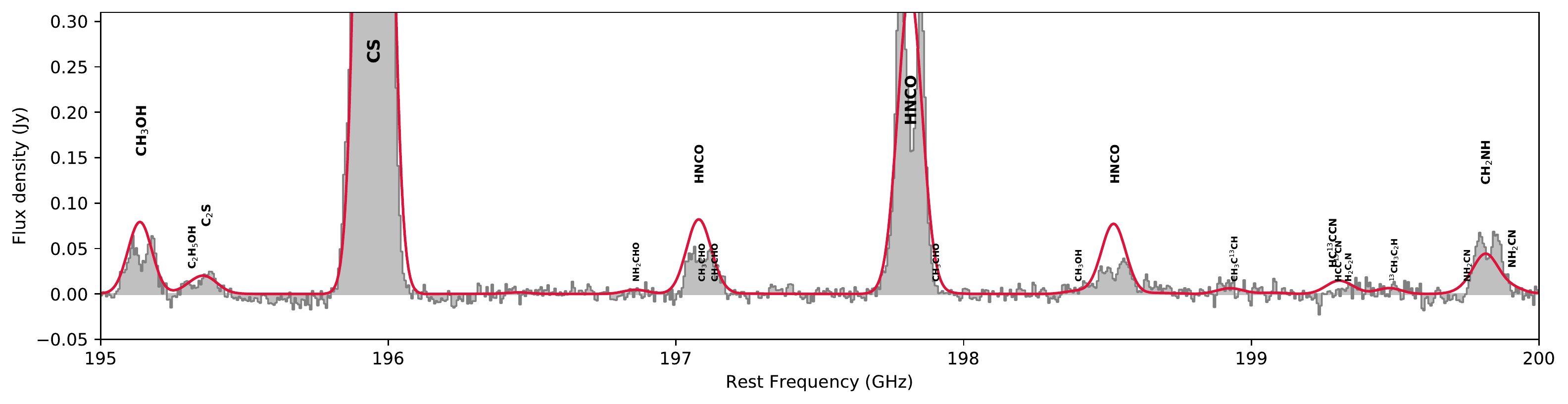}
\caption{Same as Fig.~\ref{fig.fullspectrum1}. \label{fig.fullspectrum3}}
\end{figure}

\begin{figure}
\includegraphics[width=\textwidth]{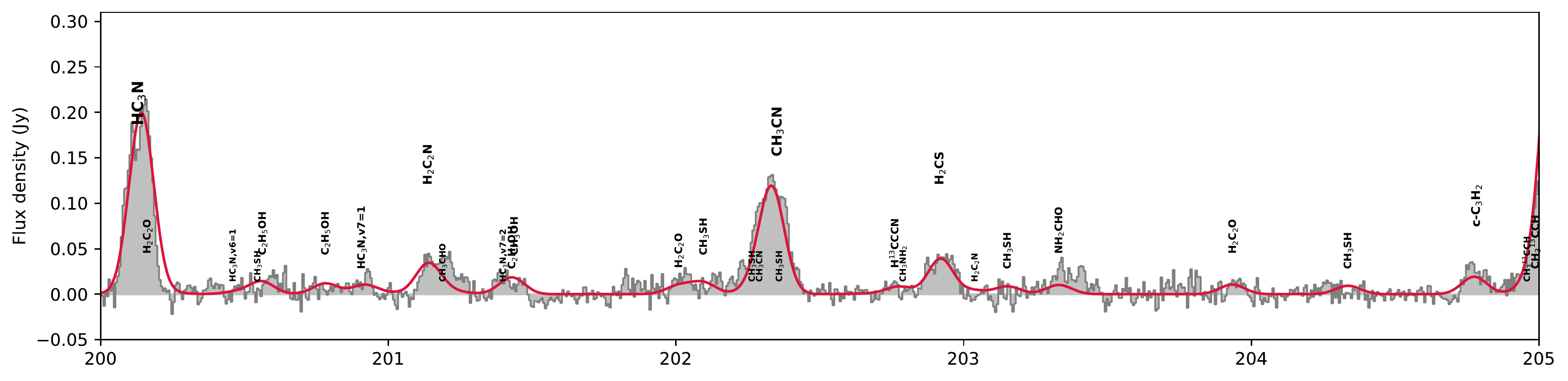}
\includegraphics[width=\textwidth]{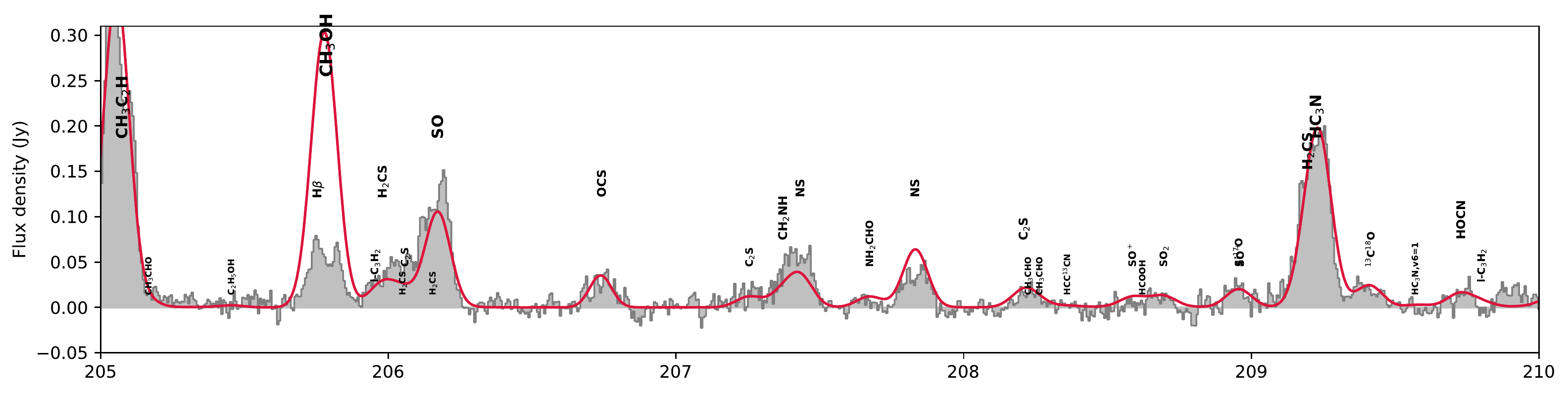}
\includegraphics[width=\textwidth]{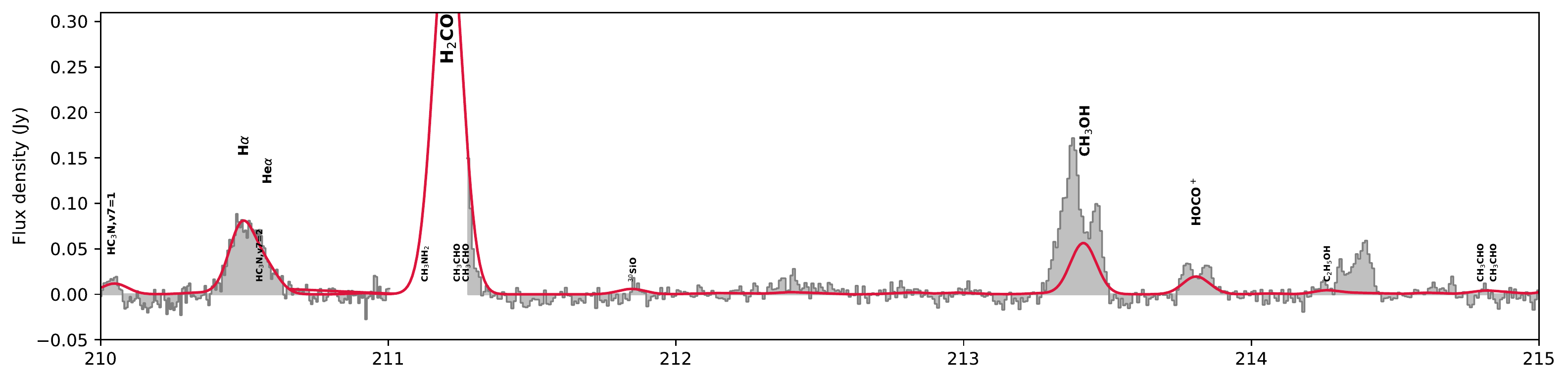}
\includegraphics[width=\textwidth]{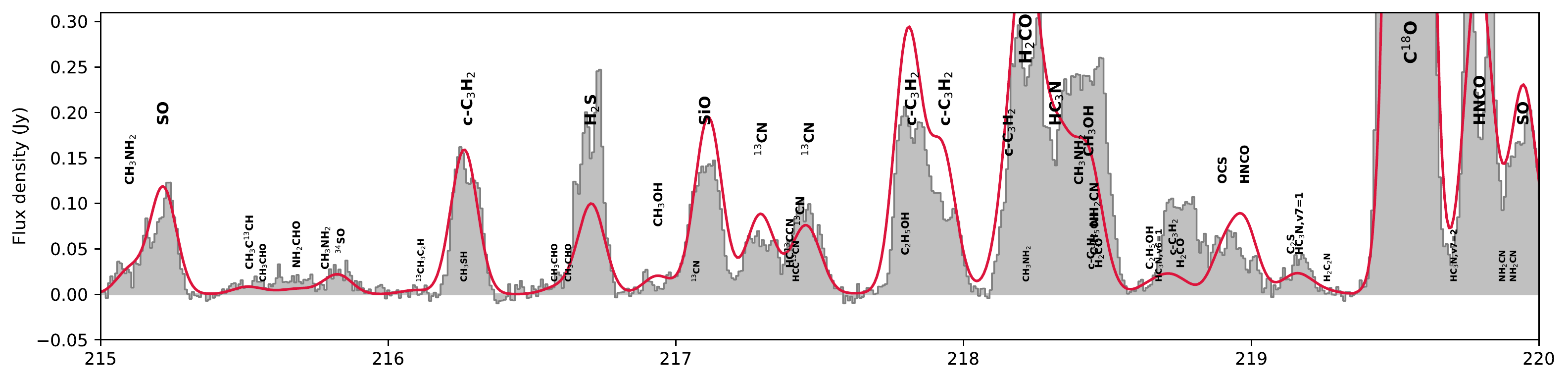}
\includegraphics[width=\textwidth]{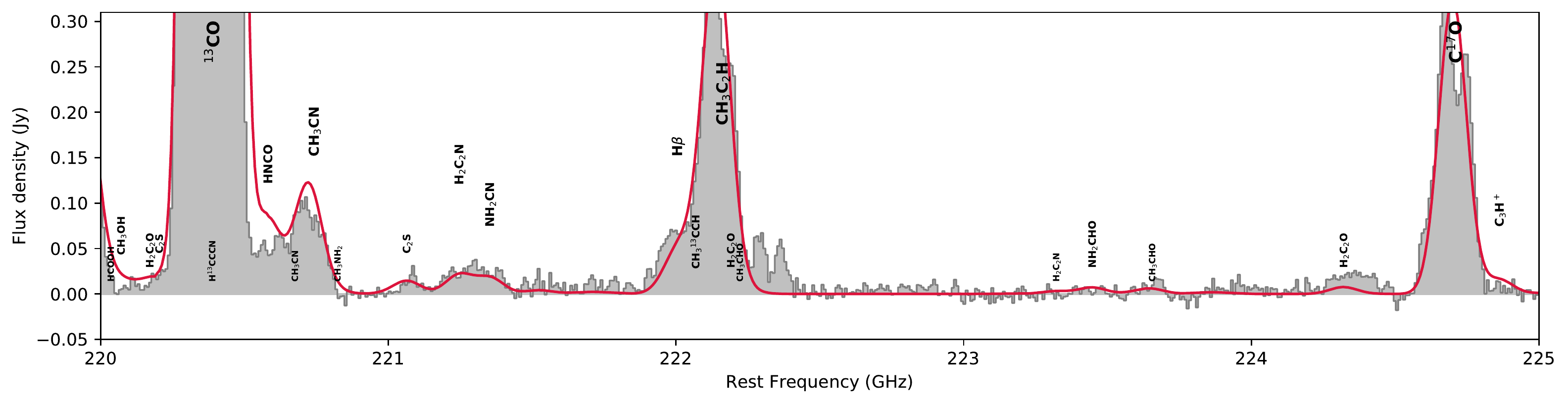}
\caption{Same as Fig.~\ref{fig.fullspectrum1}. \label{fig.fullspectrum4}}
\end{figure}

\begin{figure}
\includegraphics[width=\textwidth]{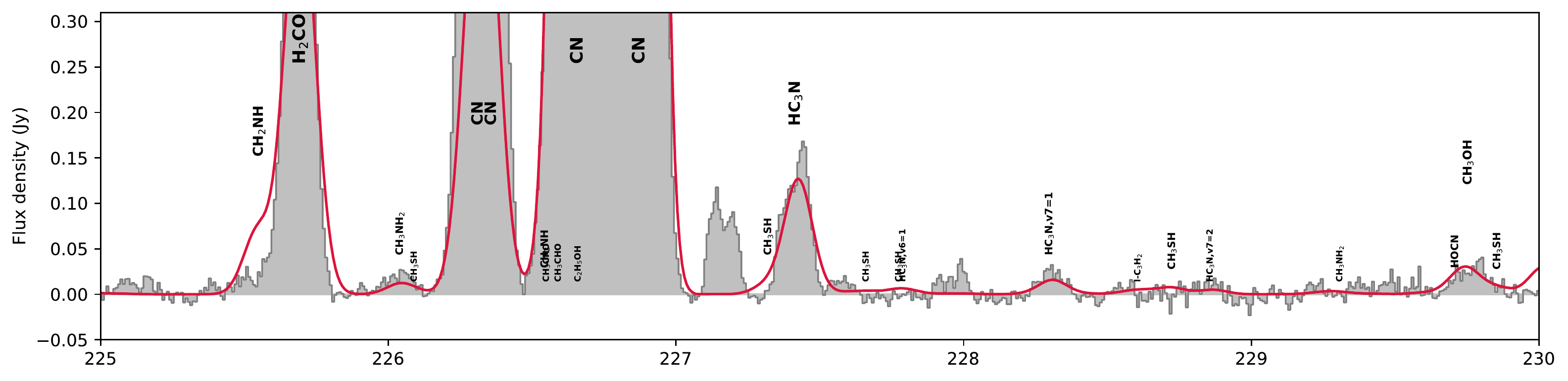}
\includegraphics[width=\textwidth]{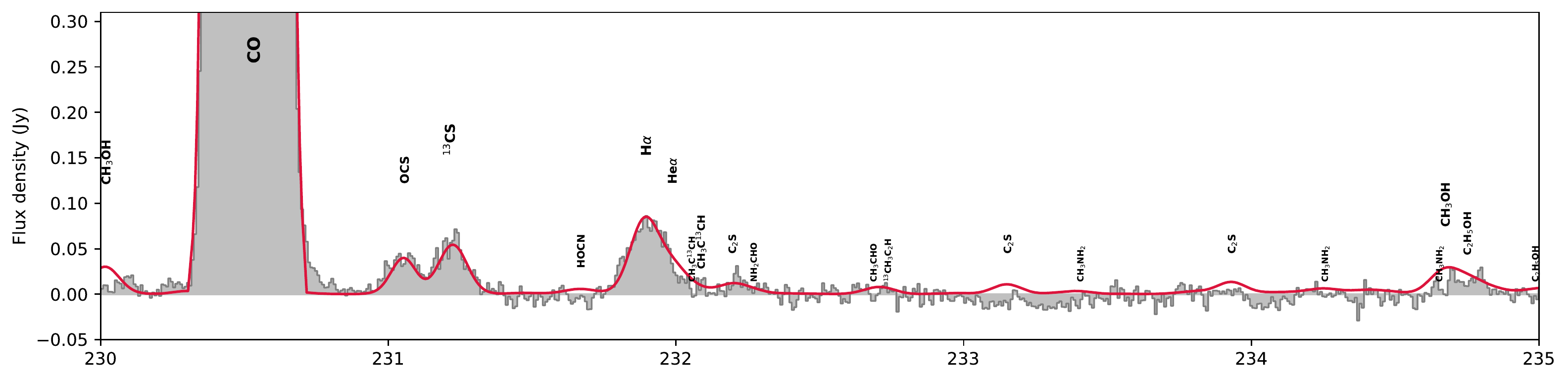}
\includegraphics[width=\textwidth]{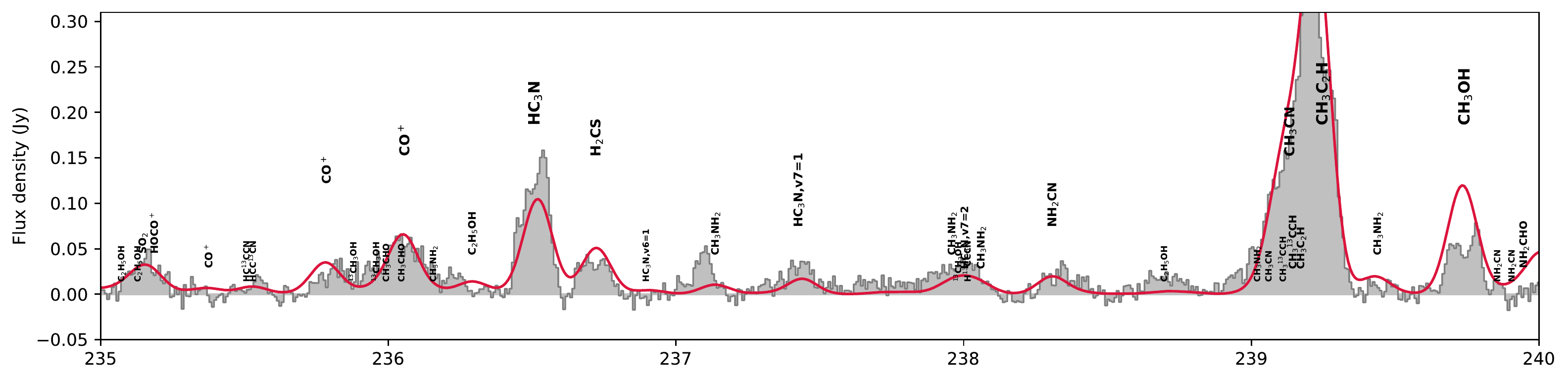}
\includegraphics[width=\textwidth]{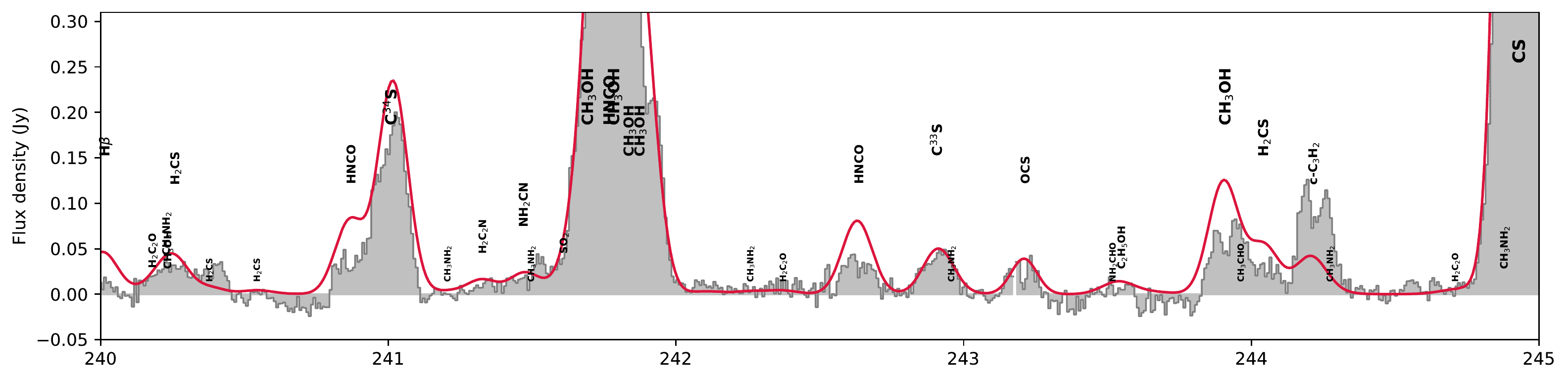}
\includegraphics[width=\textwidth]{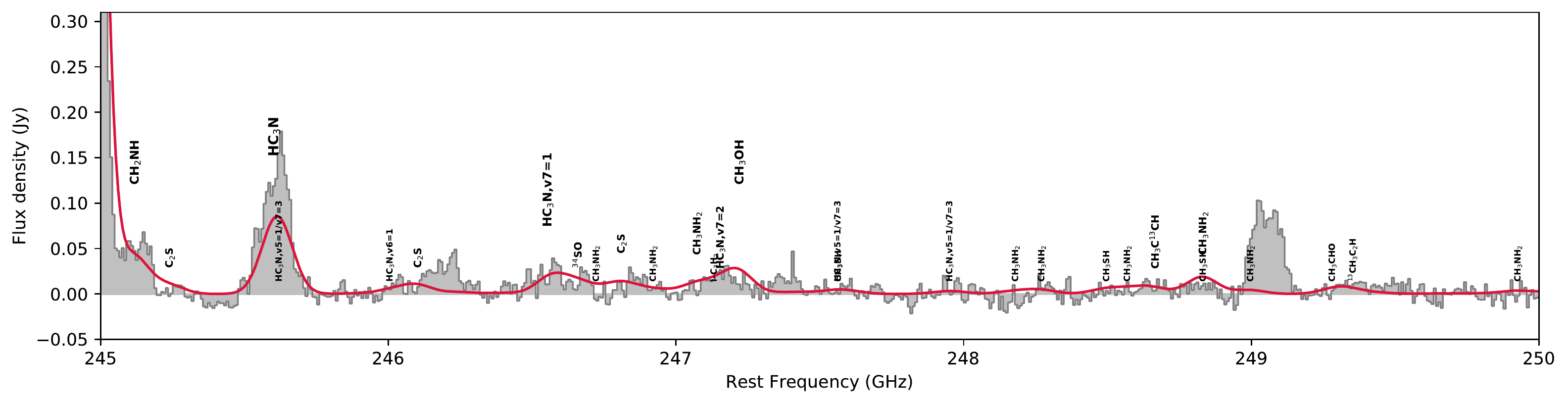}
\caption{Same as Fig.~\ref{fig.fullspectrum1}. \label{fig.fullspectrum5}}
\end{figure}

\begin{figure}
\includegraphics[width=\textwidth]{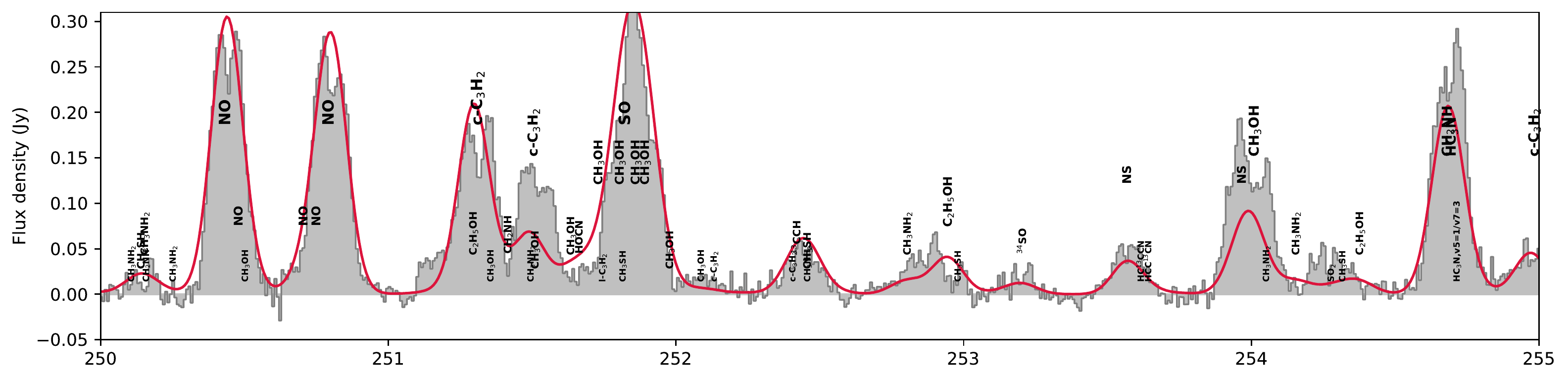}
\includegraphics[width=\textwidth]{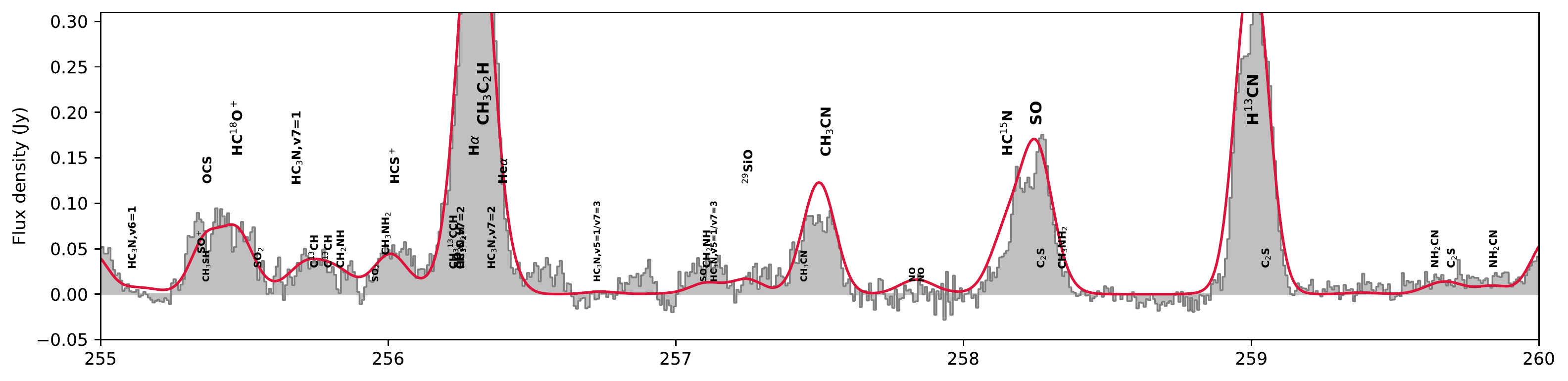}
\includegraphics[width=\textwidth]{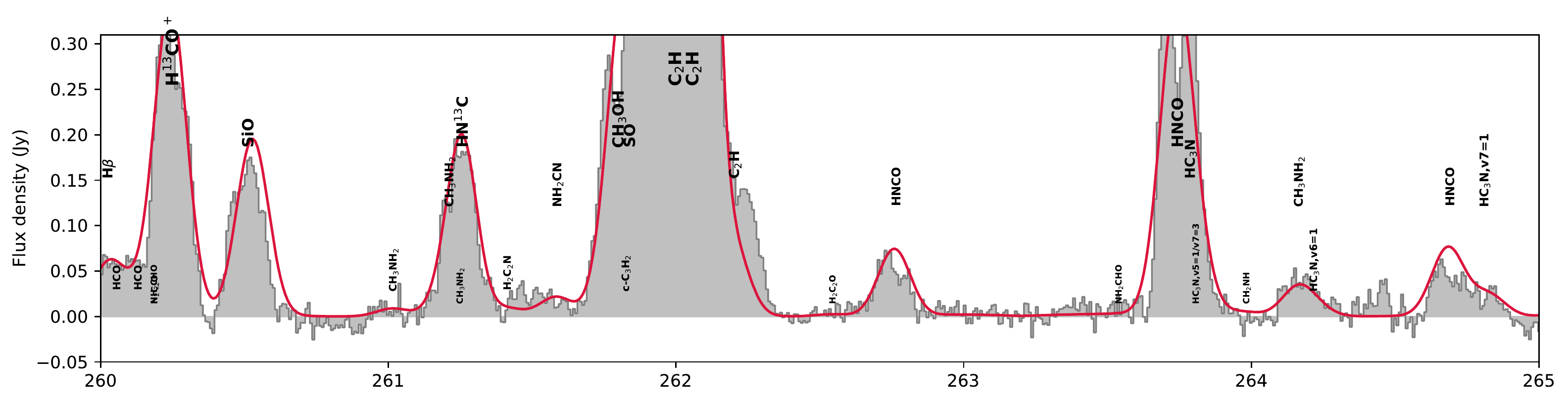}
\includegraphics[width=\textwidth]{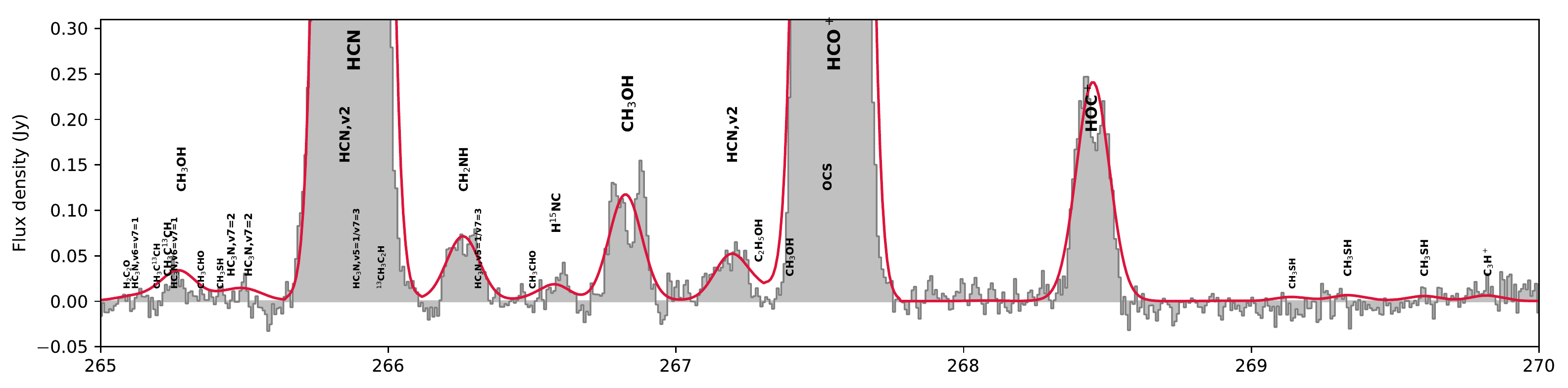}
\includegraphics[width=\textwidth]{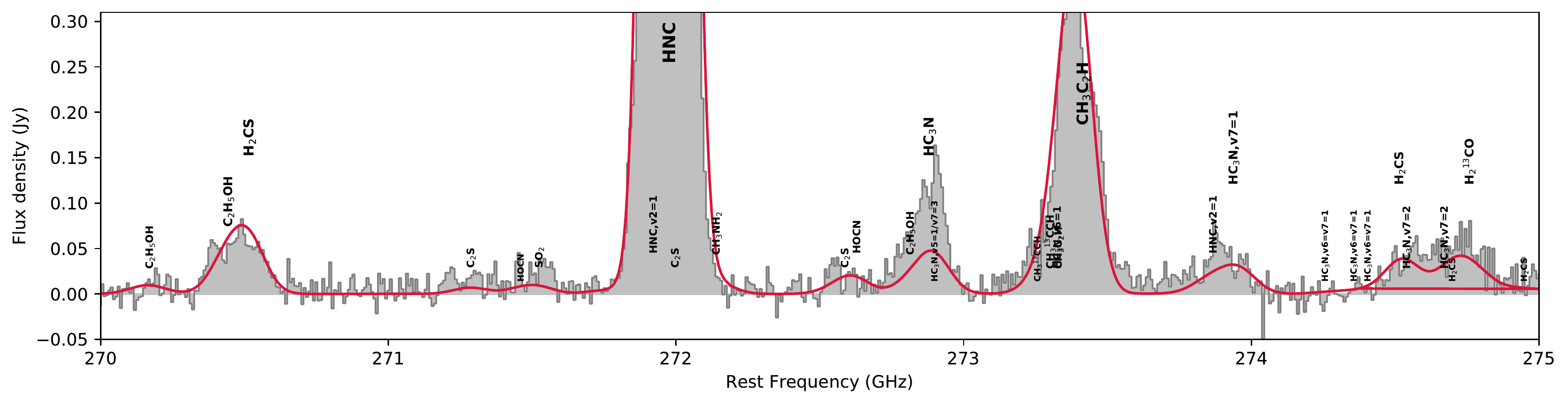}
\caption{Same as Fig.~\ref{fig.fullspectrum1}. \label{fig.fullspectrum6}}
\end{figure}

\begin{figure}
\includegraphics[width=\textwidth]{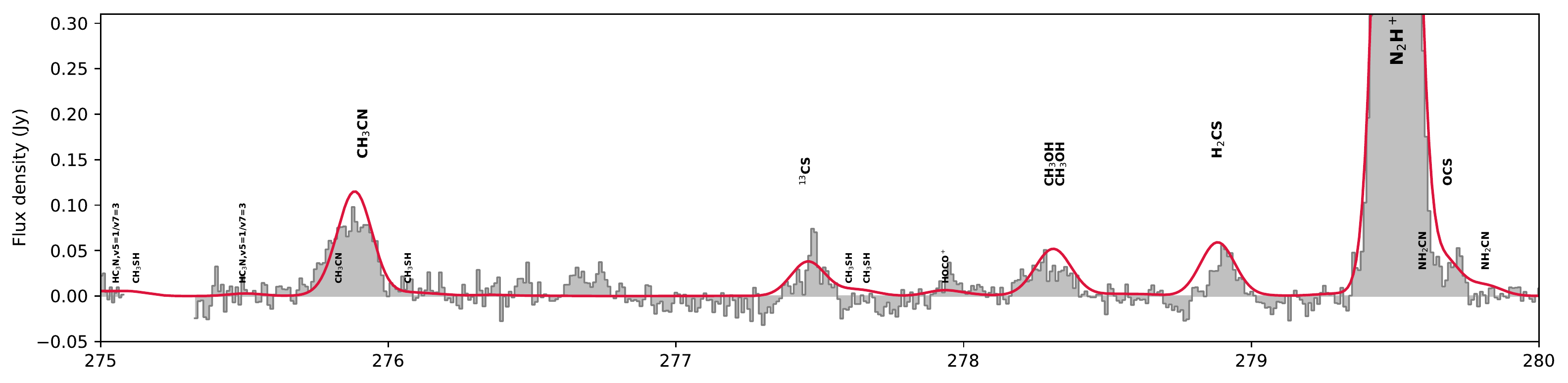}
\includegraphics[width=\textwidth]{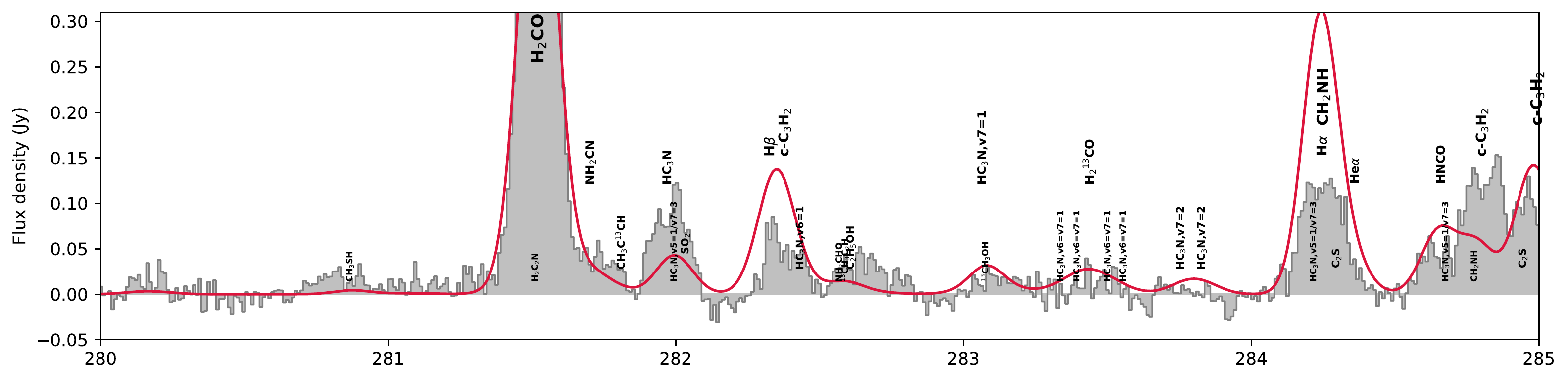}
\includegraphics[width=\textwidth]{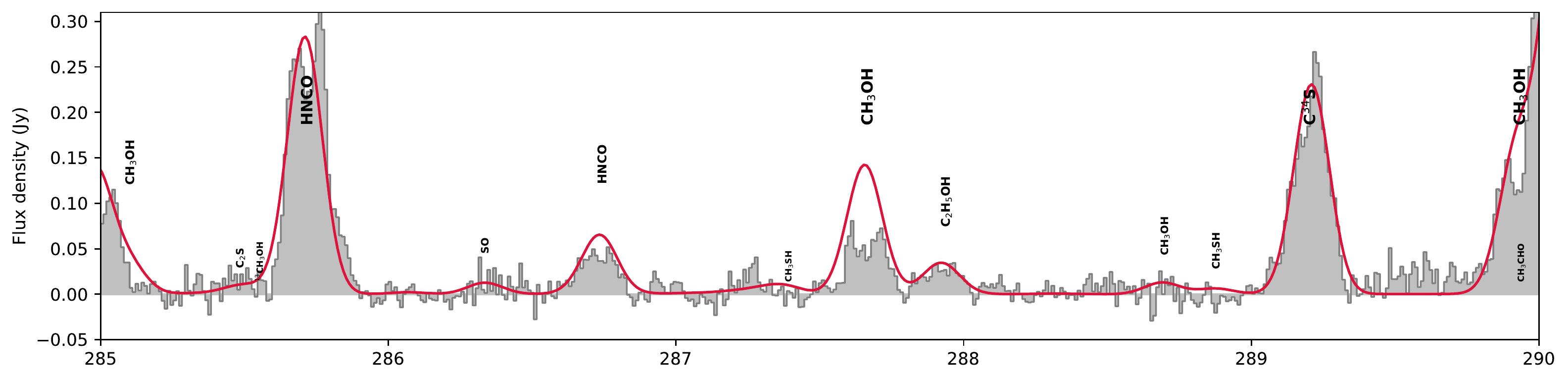}
\includegraphics[width=\textwidth]{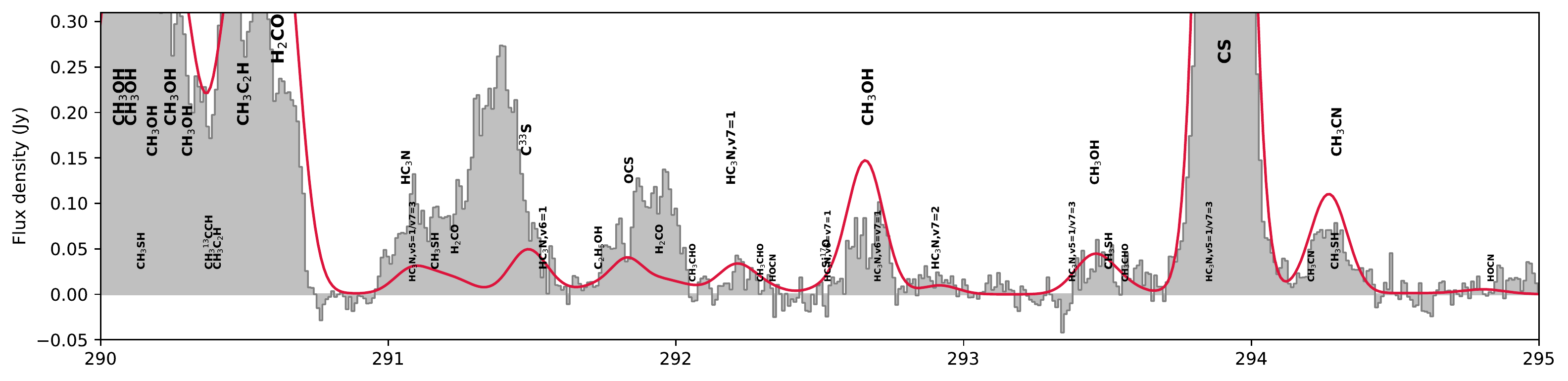}
\includegraphics[width=\textwidth]{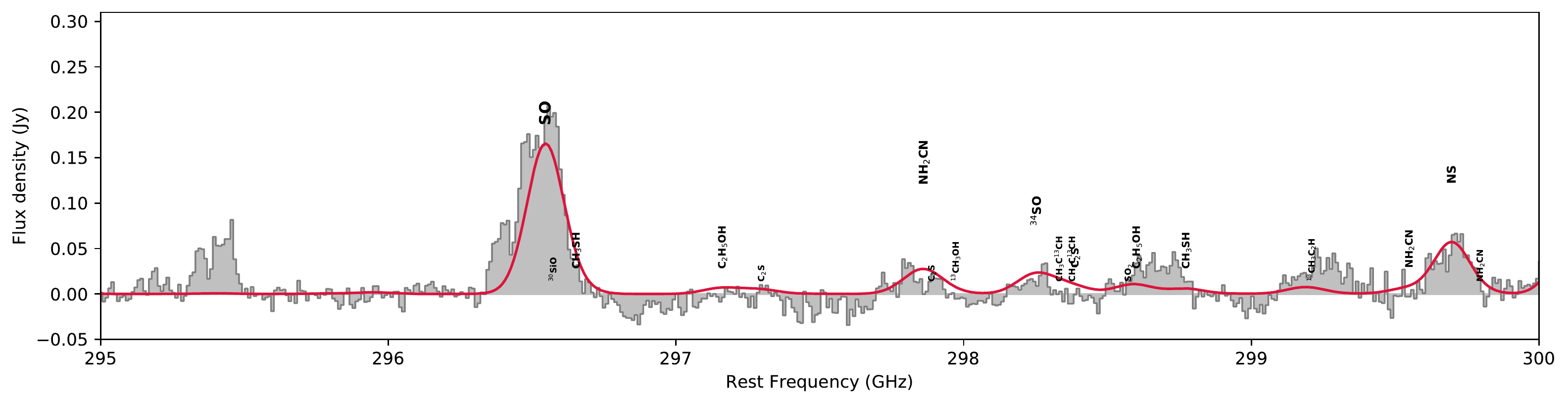}
\caption{Same as Fig.~\ref{fig.fullspectrum1}. \label{fig.fullspectrum7}}
\end{figure}

\begin{figure}
\includegraphics[width=\textwidth]{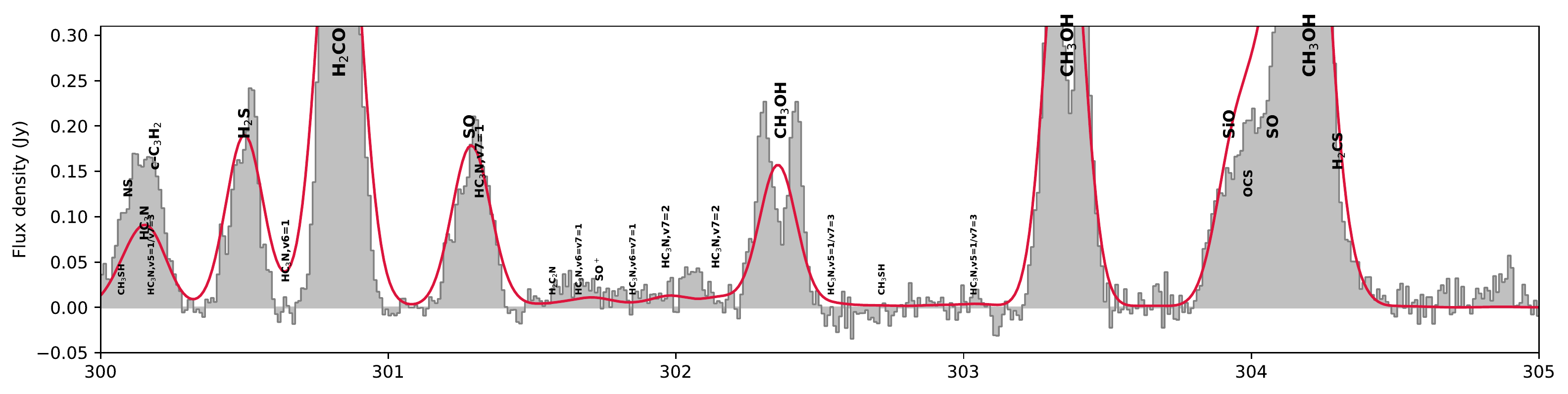}
\includegraphics[width=\textwidth]{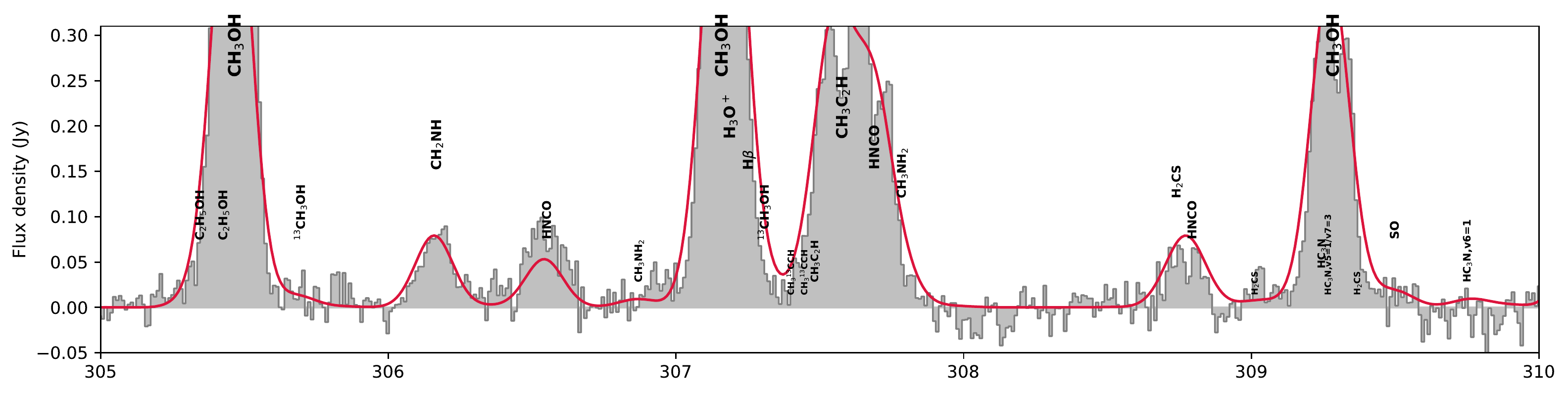}
\includegraphics[width=\textwidth]{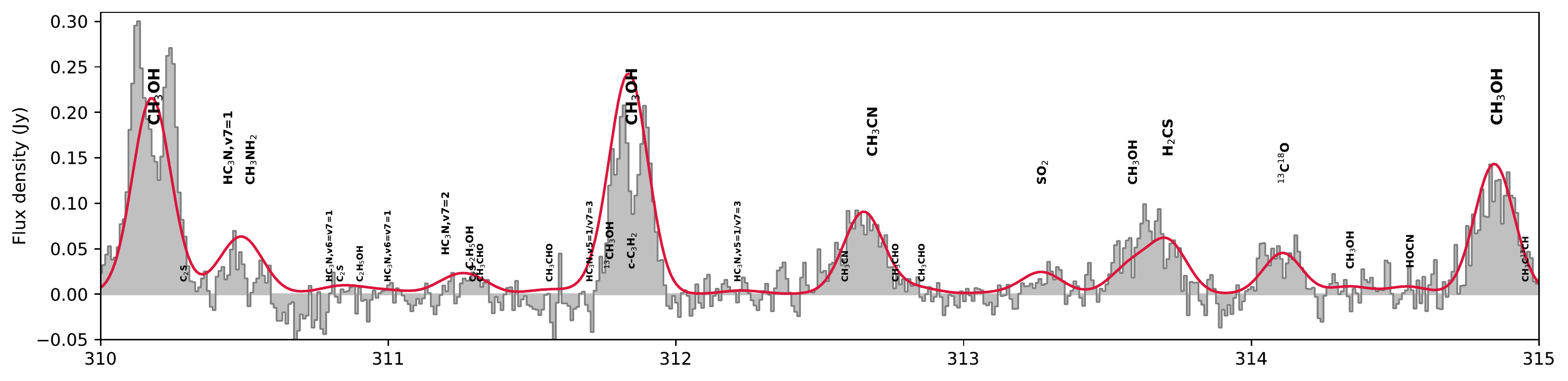}
\includegraphics[width=\textwidth]{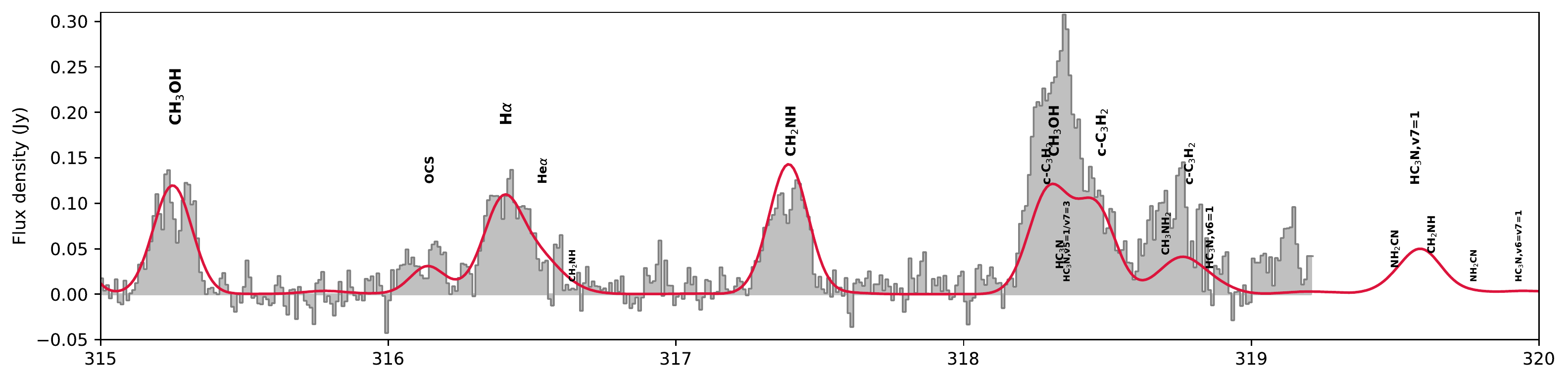}
\includegraphics[width=\textwidth]{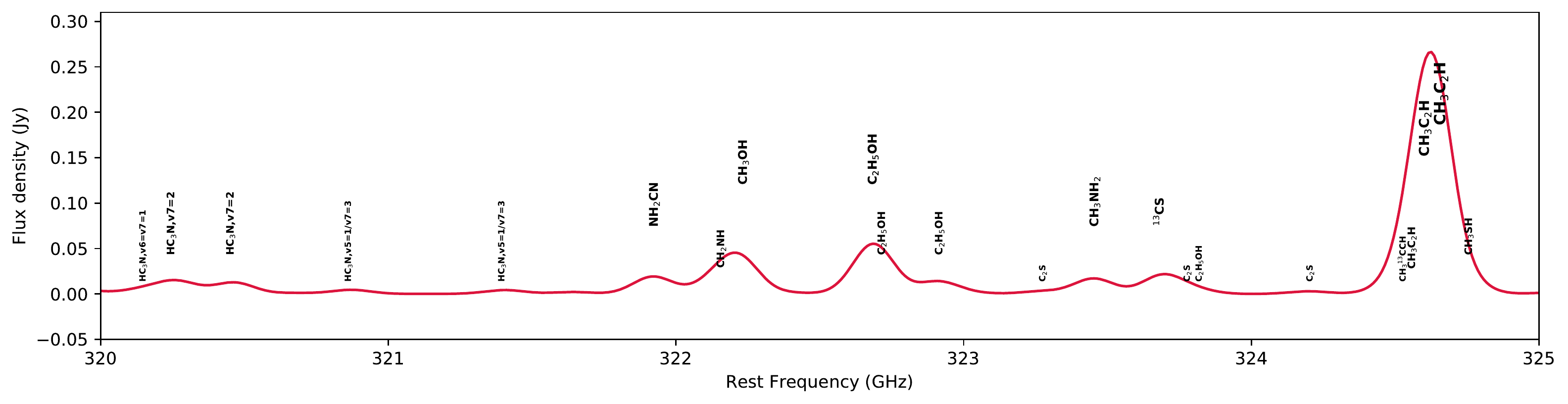}
\caption{Same as Fig.~\ref{fig.fullspectrum1}. \label{fig.fullspectrum8}}
\end{figure}

\begin{figure}
\includegraphics[width=\textwidth]{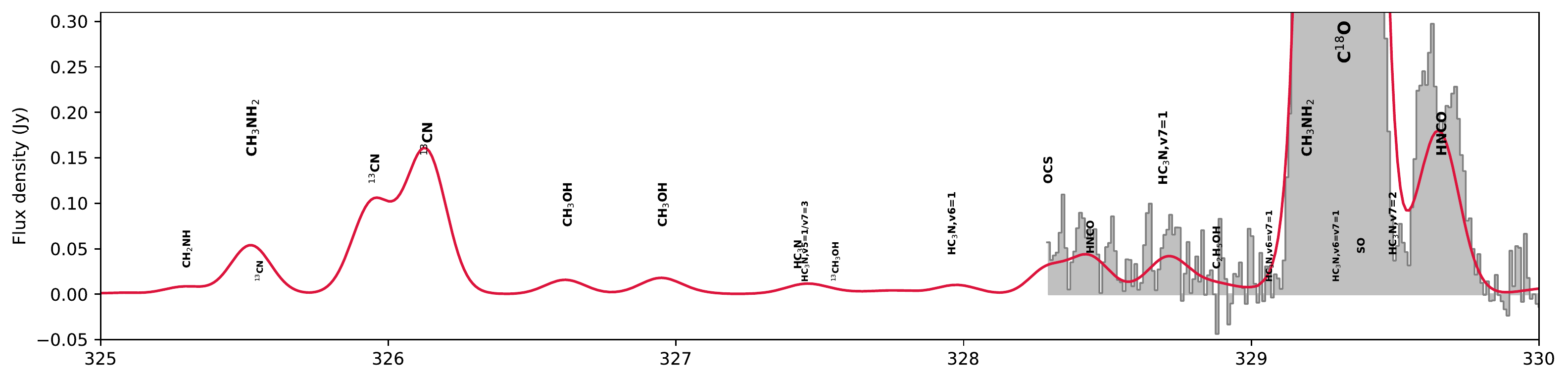}
\includegraphics[width=\textwidth]{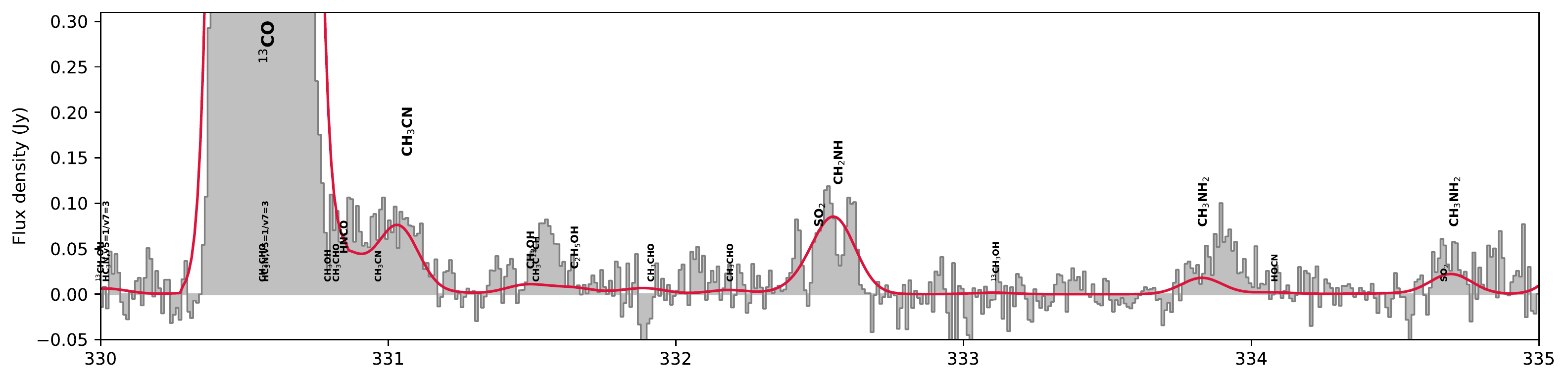}
\includegraphics[width=\textwidth]{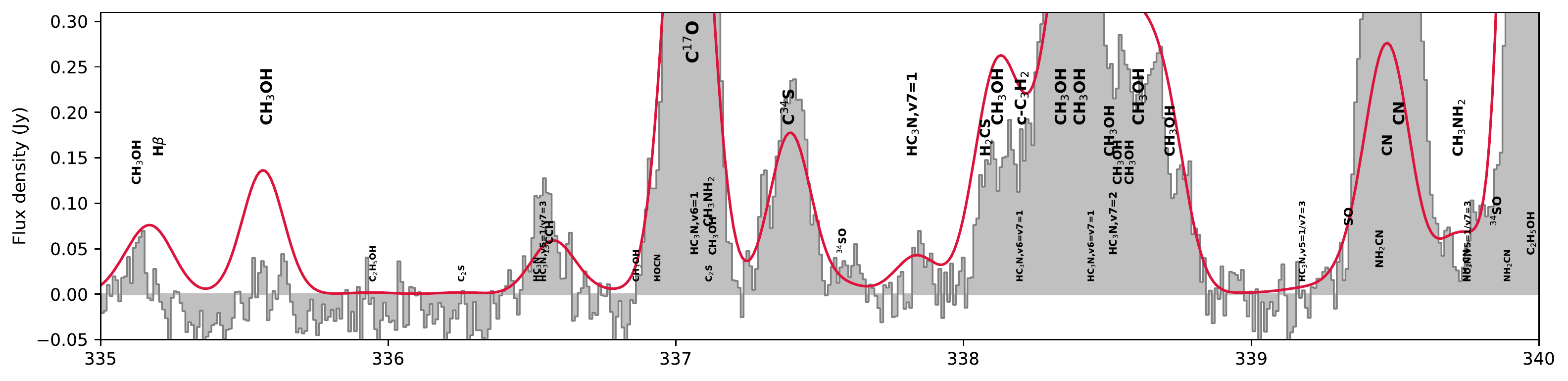}
\includegraphics[width=\textwidth]{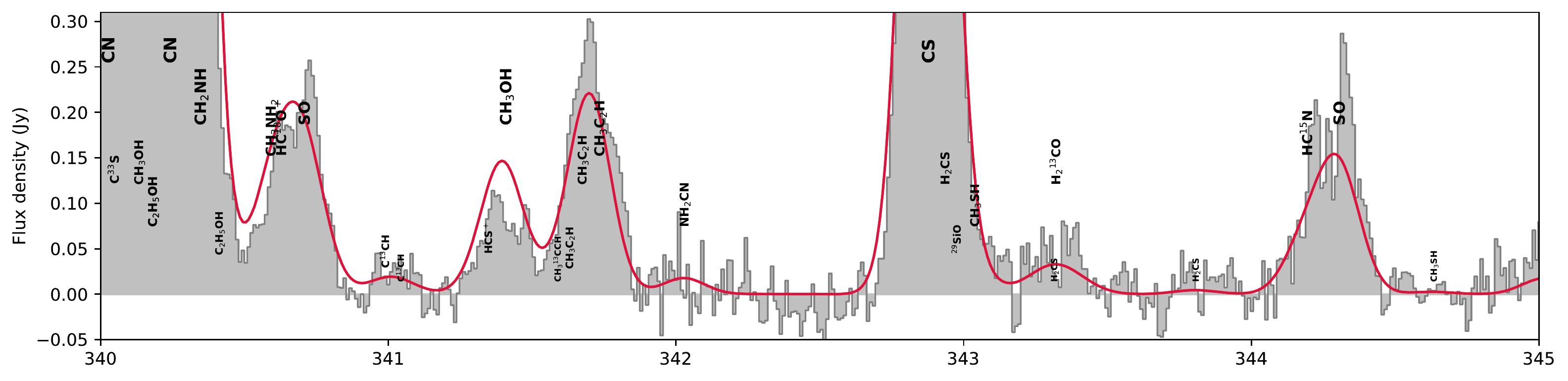}
\includegraphics[width=\textwidth]{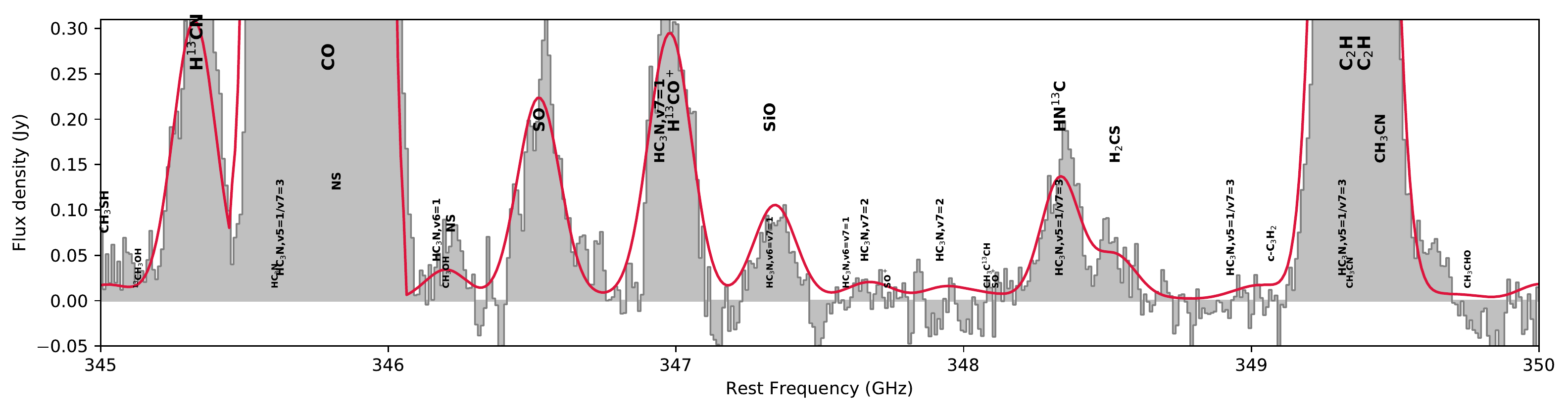}
\caption{Same as Fig.~\ref{fig.fullspectrum1}. \label{fig.fullspectrum9}}
\end{figure}

\begin{figure}
\includegraphics[width=\textwidth]{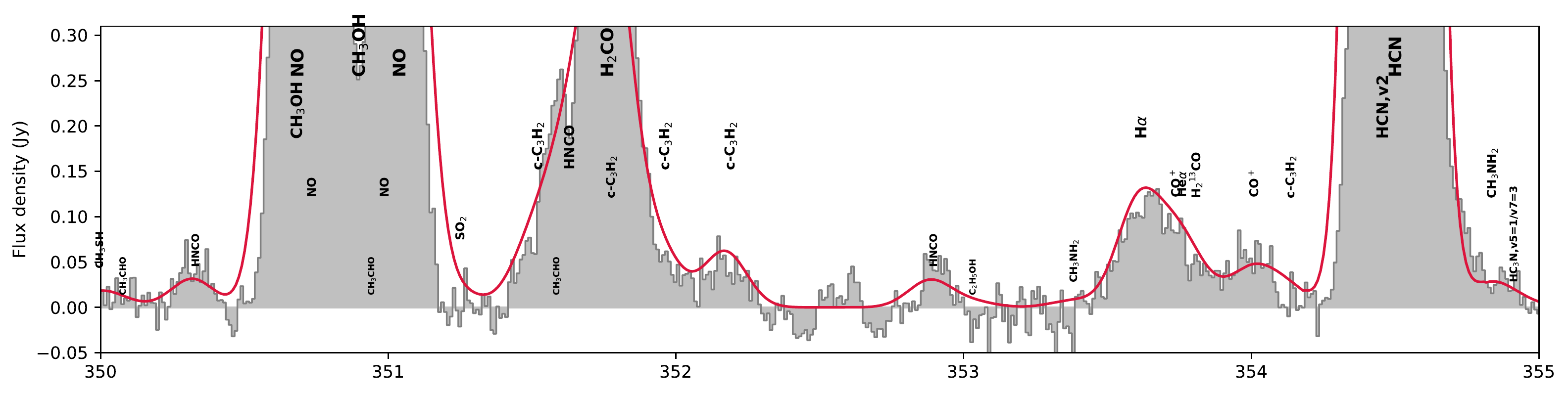}
\includegraphics[width=\textwidth]{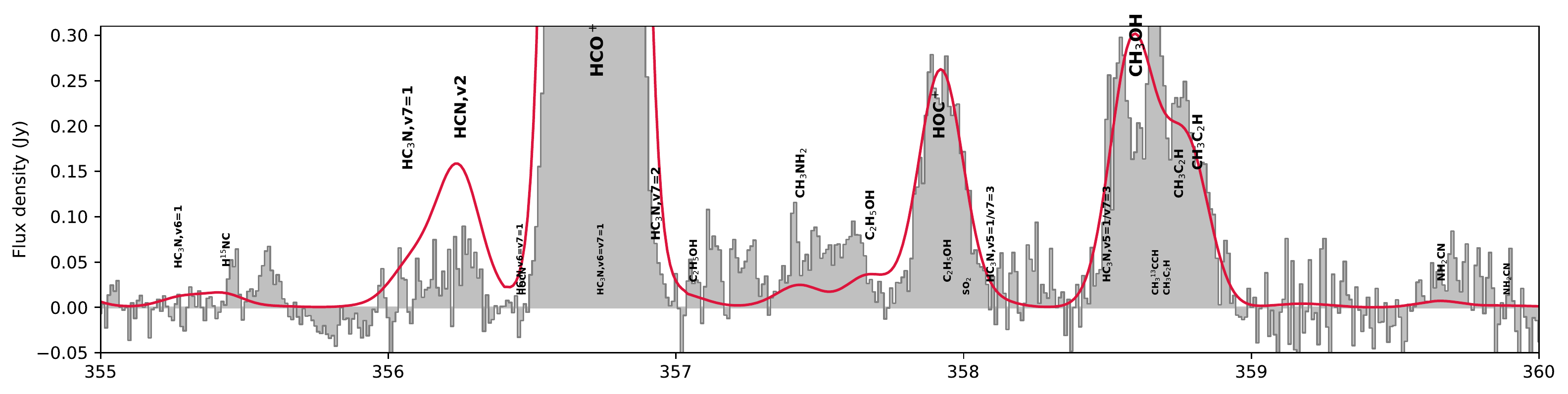}
\includegraphics[width=\textwidth]{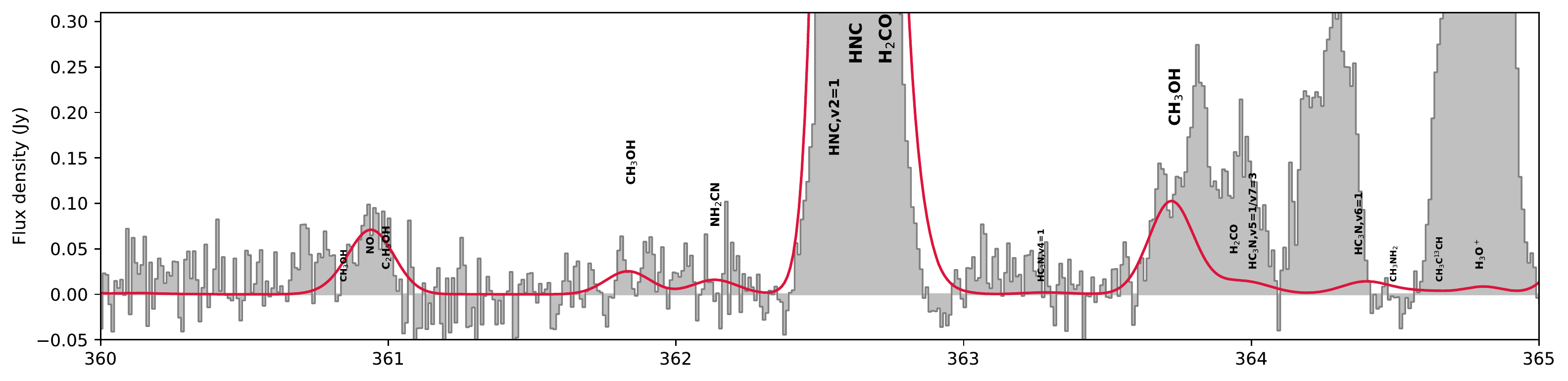}
\includegraphics[width=\textwidth]{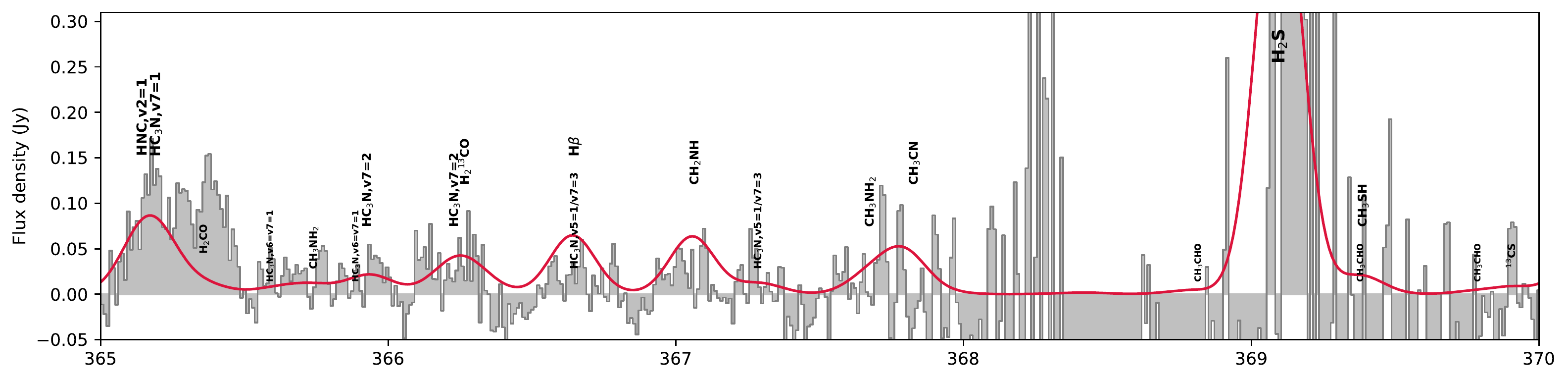}
\includegraphics[width=\textwidth]{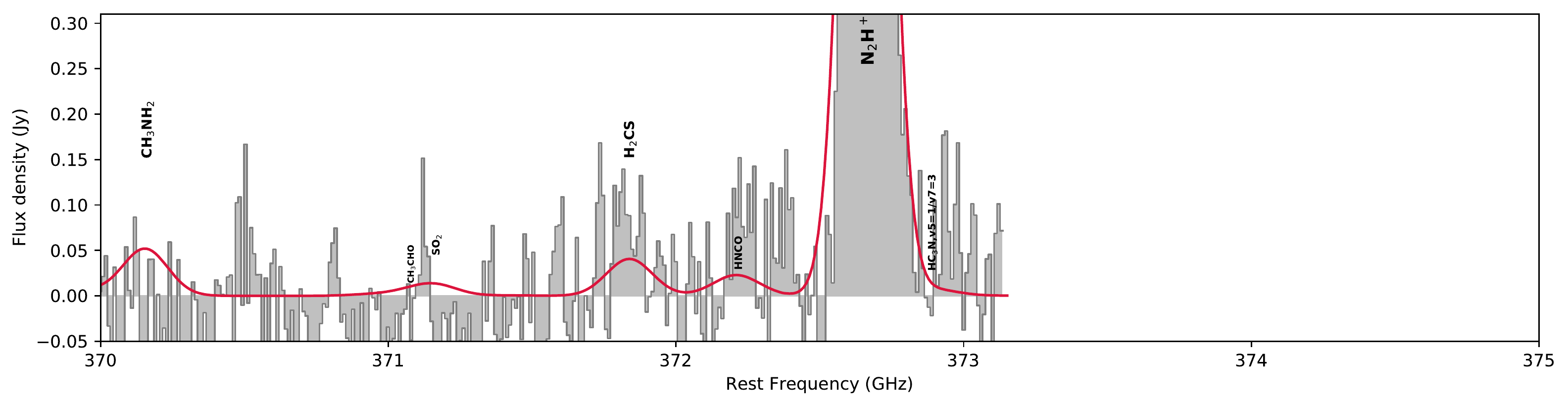}
\caption{Same as Fig.~\ref{fig.fullspectrum1}. \label{fig.fullspectrum10}}
\end{figure}

Fig.~\ref{fig.fullspectrum50GHz} presents the full spectrum analyzed in this article (gray histogram) with the best LTE model fit (red line) as well as the labels for each individual transition with flux density above 100~mJy according to the LTE model. Figs.~\ref{fig.fullspectrum1} to ~\ref{fig.fullspectrum10} present a zoomed version of Fig.~\ref{fig.fullspectrum50GHz} in 5~GHz windows and labeling transitions down to 2~mJy.
We note that despite what was indicated in Table~\ref{tab.observingsetup}, the spectral window centered at 368.7~GHz could not be imaged with the 12~m data due to the poor atmospheric transmission. Despite the poorer quality of the data, this spectral window is included in Fig.~\ref{fig.fullspectrum10}, which actually shows a bright spectral feature due to H$_2$S. The quality of the data can only be used to confirm the presence of the line but was not included in the fit.

%\textcolor{red}{\\\bf --------REVISION v3.0 DOWN TO HERE (Sergio)--------}

\end{document}